\documentclass[3p,12pt]{elsarticle}

\usepackage{lineno}
\usepackage{hyperref}
\usepackage{graphicx}
\usepackage{amssymb}
\usepackage[thinc]{esdiff}
\usepackage{rotating}
\usepackage[english]{babel}
\usepackage{csquotes}
 
\hypersetup{
    colorlinks=true,
    linkcolor=blue,
    filecolor=magenta,      
    urlcolor=cyan,
    pdftitle={Sharelatex Example},
    pdfpagemode=FullScreen,
    }
\journal{Physics Reports}

\begin{document}
\begin{frontmatter}
\title{The PVLAS experiment: a 25 year effort to measure vacuum magnetic birefringence}
\author[a1]{A.~Ejlli}
\author[a2,a3]{F.~Della~Valle}
\author[a4]{U.~Gastaldi}
\author[a5]{G.~Messineo}
\author[a6]{R.~Pengo}
\author[a6]{G.~Ruoso}
\author[a4,a7]{G.~Zavattini\corref{cor1}}\cortext[cor1]{Corresponding author} 
\ead{zavattini@fe.infn.it}

\address[a1]{School of Physics and Astronomy, Cardiff University, Queen's Building, The Parade, Cardiff CF24 3AA (United Kingdom)}
\address[a2]{Dip. di Scienze Fisiche della Terra e dell'Ambiente, via Roma 56, I-53100 Siena (Italy)}
\address[a3]{INFN - Sez. di Pisa, Largo B. Pontecorvo 3, I-56127 Pisa (Italy)}
\address[a4]{INFN - Sez. di Ferrara, Via G. Saragat 1, I-44100 Ferrara (Italy)}
\address[a5]{Dept. of Physics, University of Florida, Gainesville, Florida 32611, (USA)}
\address[a6]{INFN - Laboratori Nazionali di Legnaro, Viale dell'Universit\`a 2, I-35020 Legnaro (Italy)}
\address[a7]{Dip. di Fisica e Scienze della Terra, Via G. Saragat 1, I-44100 Ferrara (Italy)}



\begin{abstract}
This paper describes the 25 year effort to measure vacuum magnetic birefringence and dichroism with the PVLAS experiment. The experiment went through two main phases: the first using a rotating superconducting magnet and the second using two rotating permanent magnets. The experiment was not able to reach the predicted value from QED. Nonetheless the experiment set the current best limits on vacuum magnetic birefringence and dichroism for a field of $B_{\rm ext} = 2.5$~T, namely, $\Delta n^{\rm (PVLAS)} = (12\pm17)\times10^{-23}$ and \mbox{$|\Delta\kappa|^{\rm (PVLAS)} = (10\pm28)\times10^{-23}$.} The uncertainty on $\Delta n^{\rm (PVLAS)}$ is about a factor 7 above the predicted value of $\Delta n^{\rm (QED)} = 2.5\times10^{-23}$ @ 2.5~T. 
\end{abstract}
\date{May 2020}
\end{frontmatter}
\tableofcontents

\newpage
\section{Introduction}
The velocity of light in vacuum is considered today to be a universal constant and is defined as $c = 299\,792\,458$~m/s in the International System of Units:
\begin{equation}
c = \frac{1}{\sqrt{\varepsilon_0\mu_0}}.
\end{equation}
It is related to the vacuum magnetic permeability $\mu_0$ and the vacuum permittivity $\varepsilon_0$ which
describe the properties of classical electromagnetic vacuum.\footnote{Today (after the redefinition of the SI system on May 20$^{\rm th}$ 2019) the values of $\varepsilon_0$ and $\mu_0$ are both derived from the measurement of the fine structure constant
\[\displaystyle
\alpha =\frac{e^2}{4\pi\varepsilon_0\hbar c} = \frac{e^2c\mu_0}{4\pi\hbar} 
\]
being the values of $\hbar$, $c$ and $e$ defined.} Classically this relation derives directly from Maxwell's equations in vacuum. Due to their linearity $c$ does not depend on the presence of other electromagnetic fields (photons, static fields). 

Today Quantum Electrodynamics (QED) describes electrodynamics to an incredibly accurate level having been tested in many different systems at a microscopic level: $(g-2)_{e,\mu}$ \cite{Hanneke:2008tm,Bennett:2008dy}, Lamb-shift \cite{PhysRev.72.241}, Delbr\"uck scattering \cite{Schumacher1999} etc. One fundamental process predicted since 1935 \cite{Heisenberg1936,Euler:1935,EulerKockel1935} (before the formulation of QED), namely  4-field interactions with only photons present in both the initial and final states, still needs attention. In the above mentioned measurements either the accuracy is such that the 4-field interaction must be taken into account as a correction to the first order effect being observed or, as is the case of Delbr\"uck scattering, this contribution must be distinguished from a series of other effects. The 4-field interaction considered to first order will lead to two effects: light-by-light (LbL) scattering and vacuum magnetic (or electric) linear birefringence (VMB) due to low energy coherent Delbr\"uck scattering. This first effect occurs at a microscopic level whereas VMB describes a macroscopic effect related to the index of refraction \cite{Toll1952,Erber1961,Klein1964_BR,Baier:1967zza,Baier:1967zzc} and is a direct manifestation of quantum vacuum. In recent years, with the ATLAS experiment at the LHC accelerator, LbL scattering at high energies has been observed \cite{atlas2017,atlas2019} via $\gamma\gamma$ pair emission during Pb-Pb peripheral collisions for $\hbar\omega\gg m_ec^2$. Furthermore, optical polarimetry of an isolated neutron star has led Mignani et al. to publish evidence of VMB \cite{Mignani2016}. 

It remains that this purely quantum mechanical effect still needs a direct laboratory verification in the low energy regime, $\hbar\omega\ll m_ec^2$, at the macroscopic level.

As will be discussed in the following sections, not only does the index of refraction depend on the presence of external fields but it depends also on the polarisation direction of the propagating light. In the presence of an external magnetic field perpendicular to the propagation direction of a beam of light, one finds
 \begin{eqnarray}
 \Bigg \{ \begin{array}{ll}
 n_{\parallel \vec{B}} &= 1 + 7 A_{e}B_{\rm ext}^2\\
 n_{\perp \vec{B}} &= 1 + 4 A_{e}B_{\rm ext}^2
\end{array}
\label{indexB}
\end{eqnarray}
where the subscripts $(\parallel\vec{B})$ and $(\perp\vec{B})$ indicate the polarisation direction with respect to the external field.
Similarly, in the presence of an electric field
\begin{eqnarray}
 \Bigg\{ \begin{array}{ll}
 n_{\parallel \vec{E}} &= 1 + 4 A_{e}E^2_{\rm ext}/c^2\\
 n_{\perp \vec{E}} &= 1 + 7 A_{e}E^2_{\rm ext}/c^2.
\end{array}
\label{index}
 \end{eqnarray}
As will be discussed in Section~\ref{Theory}, the parameter $A_e$ describes the non linearity of Maxwell's equations due to vacuum fluctuations:
\begin{equation}
A_e =\frac{2}{45\mu_{0}}\frac{\hbar^{3}}{m_e^4c^{5}}\,\alpha^{2} = 1.32\times 10^{-24}\;{\rm T}^{-2}
\end{equation}
where $\alpha$ is the fine structure constant, and $m_e$ the mass of the electron.

The first modern proposal to measure QED non-linearities due to vacuum polarisation at very low energies dates back to 1979 \cite{Iacopini:1979ci} and a first attempt was performed at CERN during the beginning of the '80s to study the feasibility of such a measurement. Since then the idea to detect the induced birefringence due to an external magnetic field using optical techniques has been an experimental challenge. Optical elements and lasers have since improved tremendously but as of today, in spite of the unceasing efforts \cite{Nezrick1998,Chen2007,Cadene2014EPJD,Fan2017}, the direct measurement of VMB is still lacking. Another ongoing attempt to tackle the same physics is in Reference~\cite{Sarazin2016EPJD}: detecting the refraction of light-by-light in vacuum is their goal. Searches for direct photon-photon elastic scattering are reported in References~\cite{Moulin1996,Bernard2000}. Note that heuristic approaches to this very same matter started before and prescinded from any theoretical justification \cite{Watson1929,Hughes1930,Vavilov1930,Farr1932,Banwell1940,Jones1960,Jones1961}. More literature and a general treatment of non linear vacuum properties can be found in Reference~\cite{Battesti2013}.

Following two precursor experiments, the one at CERN and the other at the Brookhaven National Laboratories (BNL) briefly described in Sections~\ref{CERN} and \ref{BFRT}, the PVLAS (Polarizzazione del Vuoto con LAser) experiment, financed by INFN (Istituto Nazionale di Fisica Nucleare, Italy), performed a long lasting attempt starting from 1993. This experiment went through two major phases: the first with a rotating superconducting magnet and the second with two rotating permanent magnets.

In this paper we will describe at length the 25 year development of the PVLAS experiment and present the final results which represent today the best limit on VMB, closest to the expected value determined in equations~(\ref{indexB}). In the description of the various phases of the experiment many details are given which generally are excluded in scientific papers. We believe this is an opportunity to gather all this information together in a single publication. The last three years of activity of PVLAS coincided with the PhD thesis of one of the authors. More details on the experiment can be found in that work \cite{Ejlli2017}.

In Section~\ref{Theory} we will present the physics related to PVLAS including the possibility of searching for physics beyond the Standard Model. In Section~\ref{Method} we will describe the general experimental method of the experiment including systematic effects. In Section~\ref{Phases} each attempt with its peculiarities will be described: there we will discuss the limitations of each effort and the results obtained. Finally in Sections~\ref{sec:commissioning}, \ref{sec:FE_results} and \ref{FinalResults} we will present the calibration method, systematics-hunting, noise issues and results of the last phase of the PVLAS experiment.

\section{Theoretical considerations}
\label{Theory}

\subsection{\bf Classical electromagnetism}

Maxwell's equations in a medium are given by
\begin{eqnarray}
\begin{array}{ll}
\vec{\nabla}\cdot\vec{D} = \rho \quad\quad\vec{\nabla}\times\vec{E} = -\diffp{\vec{B}}{t}\\
\vec{\nabla}\cdot\vec{B} = 0\quad\quad\vec{\nabla}\times\vec{H} = \vec{J}+\diffp{\vec{D}}{t}
\end{array}
\label{Maxwellmatter}
\end{eqnarray}
where $\vec{E}$ and $\vec{B}$ are respectively the electric field and the magnetic induction, $\vec{D}$ and $\vec{H}$ are respectively the electric displacement field and the magnetic intensity and $\rho$ and $\vec{J}$ are the free charge density and free current density. 
The relations between $\vec{E}$ and $\vec{D}$ and between $\vec{B}$ and $\vec{H}$ are given by
\begin{equation}
\vec{D} = \varepsilon_0\vec{E} + \vec{P}\qquad\qquad\vec{H} = \frac{\vec{B}}{\mu_0} - \vec{M}
\end{equation}
 where $\vec{P}$ and $\vec{M}$ are the polarisation and magnetisation vectors, respectively, with which one can describe the polarisation and magnetisation properties of the medium.

These equations can be derived from the Lagrangian density ${\cal L}_{\rm Matt}$ in matter
\begin{equation}
    {\cal L}_{\rm Matt} = \frac{1}{2\mu_0}\left(\frac{E^2}{c^2}-B^2\right) + \vec{E}\cdot\vec{P} + \vec{B}\cdot\vec{M}-\rho\varphi+
    \vec{J}\cdot\vec{A}
\end{equation}
by applying the Euler-Lagrange equations
\begin{equation}
    \diffp{}{t}\diffp{{\cal L}_{\rm Matt}}{{\left(\diffp{q_i}{{t}}\right)}}+\sum_{k = 1}^{3}{}\diffp{}{{x_k}}\diffp{{\cal L}_{\rm Matt}}{{\left(\diffp{q_i}{{x_k}}\right)}}-\diffp{{\cal L}_{\rm Matt}}{{q_i}} = 0.
\label{LagrangeMatt}
\end{equation}
The generalised coordinates $q_0 = \varphi$ and $q_{1,2,3} = \vec{A}$ are the scalar and vector potentials, and the fields $\vec{E}$ and $\vec{B}$ are defined as 
\begin{eqnarray}
&&\vec{E} = -\vec{\nabla}\varphi - \frac{\partial\vec{A}}{\partial t}\\
&&\vec{B} = \vec{\nabla}\times\vec{A}.
\end{eqnarray}
Considering first of all $q_0 = \varphi$, equations~(\ref{LagrangeMatt}) lead to
\begin{equation}
    \vec{\nabla}\cdot\left(\varepsilon_0\vec{E} + \vec{P}\right) = \rho
\end{equation}
from which we define $\vec{D} = \varepsilon_0\vec{E} + \vec{P}$. 
Similarly by applying equations~(\ref{LagrangeMatt}) with respect to $q_1 = A_x$ one finds
\begin{equation}
    \left[\diffp{{\left(\varepsilon_0\vec{E} + \vec{P}\right)}}{t} - \vec{\nabla}\times\left(\frac{1}{\mu_0}\vec{B} - \vec{M}\right)\right]_x = \vec{J}_x
\end{equation}
and {\em idem} for $q_2$ and $q_3$.
We can then also define $\vec{H}=\frac{1}{\mu_0}\vec{B}-\vec{M}$.

 The definitions of $\vec{D}$ and $\vec{H}$, besides depending on $\vec{P}$ and $\vec{M}$, also contain the two fundamental parameters $\varepsilon_0$ and $\mu_0$.

In the absence of matter, free charges and currents, resulting in $\vec{P} = 0$, $\vec{M} = 0$, $\vec{J} = 0$ and $\rho = 0$, the Lagrangian density simplifies to 
\begin{equation}
    {\cal L}_{\rm Cl} =\frac{1}{2\mu_0}\left(\frac{E^2}{c^2}-B^2\right) 
\end{equation}
and one finds that $\vec{D} = \varepsilon_0\vec{E}$ and $\vec{B} = \mu_0\vec{H}$.  Maxwell's equations in vacuum then become
\begin{eqnarray}
\begin{array}{ll}
\vec{\nabla}\cdot\vec{E} = 0 \quad\quad\vec{\nabla}\times\vec{E} = -\diffp{\vec{B}}{t}\\
\vec{\nabla}\cdot\vec{B} = 0\quad\quad\vec{\nabla}\times\vec{B} = \varepsilon_0\mu_0 \diffp{\vec{E}}{t}.
\end{array}
\label{Maxwell}
\end{eqnarray}
Equations~(\ref{Maxwell}) admit as solutions electromagnetic waves freely propagating in vacuum at a velocity given by
\begin{equation}
    c = \frac{1}{\sqrt{\varepsilon_0\mu_0}}.
\end{equation}
Due to the linear behaviour of Maxwell's equations in vacuum, $c$ does not depend on the presence of external fields.

In general, given a Lagrangian density ${\cal L}$, the vectors $\vec{D}$ and $\vec{H}$ can also be determined through the constitutive relations
\begin{equation}
\vec{D} = \diffp{{\cal L}}{{\vec{E}}}\qquad\qquad
\vec{H} = -\diffp{{\cal L}}{{\vec{B}}}
\end{equation}
and the polarisation vector $\vec{P}$ and magnetisation $\vec{M}$ can be written as
\begin{equation}
\vec{P} = \diffp{{\cal L}}{{\vec{E}}} - \epsilon_0\vec{E}\qquad\qquad
\vec{M} = \diffp{{\cal L}}{{\vec{B}}}+\frac{\vec{B}}{\mu_0}.
\label{PolMag}
\end{equation}

\subsection{\bf Light-by-light interaction at low energies}
\label{EHsection}

This classical scenario changed drastically with the introduction of three new facts at the beginning of the 20$^{\rm th}$ century:
\begin{itemize}
    \item Einstein's energy-mass relation ${\cal E} = mc^2$;
    \item Heisenberg's uncertainty principle $\Delta{\cal E}\Delta t \ge \hbar/2 $;
    \item Dirac's relativistic equation of the electron admitting negative energy states today identified as anti-matter.
\end{itemize}
These three facts together allow vacuum to fluctuate changing completely the idea of vacuum and allowing for non linear electrodynamic effects in vacuum. Today vacuum is considered as a minimum energy state. 
Citing from O. Halpern's letter (1933) \cite{Halpern:1933dya}
\begin{quote}
    .... Here purely radiation phenomena are of particular interest inasmuch as they might serve in an attempt to formulate observed effects as consequences of hitherto unknown properties of corrected electromagnetic equations. We are seeking, then, scattering properties of the ``vacuum".
\end{quote}
In 1935, soon after Halpern's intuition, two of Heisenberg's students H. Euler and B. Kockel \cite{EulerKockel1935} determined a relativistically, parity-conserving effective Lagrangian density which, to second order in the invariants of the electromagnetic field tensor $F^{\mu\nu}$ (see for example Reference~\cite{Puri})
\begin{equation}
    F = \left(B^2 - \frac{E^2}{c^2}\right)\qquad{\rm and}\qquad G = \frac{\vec{E}}{c}\cdot\vec{B},
\end{equation}
takes into account electron-positron vacuum fluctuations:
\begin{equation}
    {\cal L}_{\rm EK} = {\cal L}_{\rm Cl} + \frac{A_e}{\mu_0}\left[\left(\frac{E^2}{c^2} - B^2\right)^2+7\left(\frac{\vec{E}}{c}\cdot\vec{B}\right)^2\right]
\label{EK}
\end{equation}
where 
\begin{equation}
    A_e = \frac{2}{45\mu_0}\frac{\hbar^3}{m_e^4c^5}\alpha^2 = 1.32\times10^{-24}\;{\rm T}^{-2}.
    \label{Ae}
\end{equation}
This Lagrangian was derived in the approximation of low energy photons $\hbar\omega\ll m_ec^2$.

The effective Lagrangian density ${\cal L}_{\rm EK}$ leads to non linear effects even in the absence of matter thereby violating the superposition principle, one of the building blocks of Maxwell's theory in vacuum. Indeed by applying the Euler-Lagrange equations~(\ref{LagrangeMatt}) with respect to $q_0 = \varphi$ one obtains
\begin{equation}
    \vec{\nabla}\cdot\left[\varepsilon_0\vec{E}+4
    A_e\left(\frac{E^2}{c^2}-B^2\right)\varepsilon_0
    \vec{E}+14\varepsilon_0A_e\left(\vec{E}\cdot\vec{B}\right)\vec{B}\right]=0
\end{equation}
where one can identify 
\begin{equation}
    \vec{D} = \varepsilon_0\vec{E}+4
    A_e\left(\frac{E^2}{c^2}-B^2\right)\varepsilon_0
    \vec{E}+14\varepsilon_0A_e\left(\vec{E}\cdot\vec{B}\right)\vec{B} 
    \label{D}
\end{equation}
consistent with $\vec{D} = \diffp{{\cal L}_{\rm EK}}{{\vec{E}}}$ and
$\vec{\nabla}\cdot\vec{D} = 0$.
The relation between $\vec{D}$ and $\vec{E}$ is no longer linear in the field $\vec{E}$. In a similar way by applying the Euler-Lagrange equations~(\ref{LagrangeMatt}) with respect to $q_1 = A_x$ one again finds that
\begin{equation}
    \diffp{{D_x}}{t}-\left(\vec{\nabla}\times\vec{H}\right)_x = 0
\end{equation}
and {\em idem} for $q_2$ and $q_3$. These equations represent Amp\`ere-Maxwell's law in a medium where
\begin{equation}
    \mu_0\vec{H} = \vec{B}+ 4A_e\left(\frac{E^2}{c^2}-B^2\right)\vec{B}-14A_e\left(\frac{\vec{E}}{c}\cdot\vec{B}\right)\frac{\vec{E}}{c}.
    \label{H}
\end{equation}

Vacuum therefore behaves as a non linear polarisable and magnetisable medium, where $\vec{P}$ and $\vec{M}$ are given by
\begin{eqnarray}
\vec{P} &=& 4A_e\left(\frac{E^2}{c^2}-B^2\right)\varepsilon_0\vec{E}+14\varepsilon_0A_e\left(\vec{E}\cdot\vec{B}\right)\vec{B} \\
\vec{M} &=& -4A_e\left(\frac{E^2}{c^2}-B^2\right)\frac{\vec{B}}{\mu_0}+14A_e\left(\frac{\vec{E}}{c}\cdot\frac{\vec{B}}{\mu_0}\right)\frac{\vec{E}}{c}.
\end{eqnarray}

Using the Lagrangian density (\ref{EK}) one can still describe electromagnetism in the absence of matter using Maxwell's equations but in the form (\ref{Maxwellmatter}): i.e. in a medium which is both magnetised and polarised by an external field due to the presence of virtual electron-positron pairs.  

A direct consequence of the non linear behaviour of equations~(\ref{D}) and (\ref{H}) is that the velocity of light now depends on the presence of external fields in contradiction with Maxwell's equations in classical vacuum.
Given a certain configuration of external fields, for example in which $\vec{B} = \mu_0\,\mu(\vec{E},\vec{B})\,\vec{H}$ and $\vec{D} = \varepsilon_0\,\varepsilon(\vec{E},\vec{B})\,\vec{E}$, the index of refraction $n$ is
\begin{equation}
    n = \sqrt{\varepsilon\mu}\ne 1.
    \label{ref_ind}
\end{equation}

To summarise, vacuum fluctuations determine the following important facts:
\begin{itemize}
\item  in vacuum $\vec{D} \ne \varepsilon_0\vec{E}$ and $\vec{B} \ne \mu_0\vec{H}$;
\item  Maxwell's equations are no longer linear and the superposition principle is violated;
\item  in vacuum Light-by-Light scattering can occur and the velocity of light is $v_{\rm light} < c$ in the presence of other electromagnetic fields;
\item electromagnetism in vacuum is described by Maxwell's equations in a medium.
\end{itemize}

Detecting this manifestation of quantum vacuum fluctuations at a macroscopic level leading to a dependence of the velocity of light on an external field has been the primary goal of the PVLAS experiment.

The effective Lagrangian density (\ref{EK}) was generalised in 1936 by W.~Heisenberg and H.~Euler \cite{Heisenberg1936}. They determined an effective Lagrangian taking into account electron-positron pairs in a non perturbative expression to all orders in the field invariants $F$ and $G$ in a uniform \emph{external} background field. Furthermore they introduced the idea of a critical electric field
\begin{equation}
    E_{\rm cr} = \frac{m_e^2c^3}{\hbar e} = 1.32\times10^{18}\;{\rm V/m}.
\end{equation}
This field corresponds to the field intensity whose work over a distance equal to the reduced Compton wavelength of the electron amounts to the rest energy of the electron: for fields above $E_{\rm cr}$ real production of electron-positron pairs arise in vacuum \cite{Sauter}. Today $E_{\rm cr}$ is known as the Schwinger critical field. One can also define a critical magnetic field $B_{\rm cr}$ as
\begin{equation}
    B_{\rm cr} = \frac{m_e^2c^2}{\hbar e} = 4.4\times10^{9}\;{\rm T}.
\end{equation}
Furthermore Heisenberg and Euler set the following conditions on the field derivatives
\begin{equation}
\frac{\hbar}{m_e c} \vert\nabla E\vert \ll E, \qquad\qquad \frac{\hbar}{m_e c^2} \left| \frac{\partial E}{\partial t} \right| \ll E 
\label{eq:cond1}
\end{equation} 
\begin{equation}
\frac{\hbar}{m_e c} \vert \nabla B \vert \ll B, \qquad\qquad \frac{\hbar}{m_e c^2} \left| \frac{\partial B}{\partial t} \right| \ll B
\label{eq:cond2}
\end{equation}
and asked that the field intensities were much smaller than their critical values: $B \ll B_{\rm cr}$ and  $E \ll E_{\rm cr}$.

The resulting Heisenberg-Euler effective Lagrangian density for electromagnetic fields in the absence of matter is
\begin{small}
\begin{eqnarray}
    {\cal L}_{\rm HE}& =& \frac{1}{2\mu_0}\left(\frac{E^2}{c^2}-B^2\right) + \alpha\int_0^\infty{e^{-\xi}\frac{d\xi}{\xi^3}}\times\nonumber\\
    &&\times\;{\left\{i\xi^2\sqrt{\frac{\varepsilon_0}{\mu_0}}\left(\vec{E}\cdot\vec{B}\right)\frac{\cos{\left[\frac{\xi\sqrt{C}}{\sqrt{\varepsilon_0}E_{\rm cr}}\right]}+{\rm conj.}}{\cos{\left[\frac{\xi\sqrt{C}}{\sqrt{\varepsilon_0}E_{\rm cr}}\right]}-{\rm conj.}}+\varepsilon_0E_{\rm cr}^2+\frac{\xi^2}{3\mu_0}\left(B^2-\frac{E^2}{c^2}\right)\right\}}
\label{EH}
\end{eqnarray}
\end{small}
with
\begin{equation}
    C = \frac{1}{\mu_0}\left(\frac{E^2}{c^2}-B^2\right)+2i\sqrt{\frac{\varepsilon_0}{\mu_0}}\left(\vec{E}\cdot\vec{B}\right).
\end{equation}

The Euler-Kockel Lagrangian density (\ref{EK}) can be obtained from (\ref{EH}) through a second order expansion in the field invariants $F$ and $G$ (see also Reference~\cite{Weisskopf1936}).

A few years later, a number of researchers obtained the same effective Lagrangian density from QED \cite{Karplus:1950zza,Schwinger:1951nm}.

\subsubsection{Leading order vacuum birefringence and dichroism in Electrodynamics}

In general, the index of refraction of a medium is a complex quantity: $\tilde n = n + i\kappa$. The real part $n$ (known as the index of refraction \emph{tout court}) determines the velocity of propagation of light in the medium, whereas the imaginary part, known as the index of absorption $\kappa$, describes the absorption of the medium.

A medium is said to be birefringent if $n$ depends on the polarisation state of the propagating light. Both linear and circular birefringences exist: the first is a birefringence for linearly polarised light whereas the second is a birefringence for circularly polarised light (also know as optical activity).
Similarly a medium is said to be dichroic if the index of absorption $\kappa$ depends on the polarisation (both linear and circular).

Consider a linearly polarised beam of light propagating along a direction $\hat k$ through an external field perpendicular to $\hat k$. The relative dielectric constant and relative magnetic permeability will be obtained from equations~(\ref{D}) and (\ref{H}) where the electric and magnetic fields $\vec{E}$ and $\vec{B}$ are the sum of the external fields, $\vec{E}_{\rm ext}$ and $\vec{B}_{\rm ext}$, and the light fields $\vec{E}_{\gamma}$ and $\vec{B}_{\gamma}$. In the case of an external magnetic field $\vec{B}_{\rm ext}$ one has $\vec{E} = \vec{E}_\gamma$ and $\vec{B} = \vec{B}_{\rm ext} + \vec{B}_\gamma$. Furthermore, considering the case in which $| \vec{B}_{\rm ext}| \gg |\vec{B}_{\gamma}|$ one finds
\begin{eqnarray}
    \vec{D}_\gamma = \varepsilon_0\left[\vec{E}_\gamma - 4A_eB_{\rm ext}^2\vec{E}_\gamma+14A_e\left(\vec{E}_\gamma\cdot\vec{B}_{\rm ext}\right)\vec{B}_{\rm ext}\right]
    \label{D_light}\\
	\vec{H}_\gamma = \frac{1}{\mu_0}\left[\vec{B}_\gamma- 4A_eB_{\rm ext}^2\vec{B}_\gamma - 8A_e\left(\vec{B}_\gamma\cdot\vec{B}_{\rm ext}\right)\vec{B}_{\rm ext}\right].
	\label{H_light}
\end{eqnarray}
The last terms on the right of these equations determines a polarisation dependence of the relative dielectric constant $\varepsilon$ and magnetic permeability $\mu$. Indicating with the subscript $\parallel$ and $\perp$ the polarisation direction (electric field direction of the light) parallel and perpendicular to the external magnetic field respectively one finds
\begin{eqnarray}
 \left \{ \begin{array}{ll}
 \varepsilon_{\parallel} = 1 + 10A_eB_{\rm ext}^2\\
\mu_{\parallel} = 1 + 4A_e B_{\rm ext}^2\\
n_\parallel = 1 + 7A_eB_{\rm ext}^2
\end{array}\right.
\qquad\qquad
\left \{ \begin{array}{ll}
 \varepsilon_{\perp} = 1 -4A_eB_{\rm ext}^2\\
\mu_{\perp} = 1 + 12A_e B_{\rm ext}^2\\
n_\perp = 1+4A_eB_{\rm ext}^2
\end{array}\right.
\label{n_B}
\end{eqnarray}
where $n$ is determined from equation~(\ref{ref_ind}). Both $n_\parallel$ and $n_\perp$ are greater than unity and a birefringence is apparent:
\begin{equation}
n_\parallel - n_\perp = \Delta n^{\rm (EK)} = 3A_eB_{\rm ext}^2.
\label{birif_B}
\end{equation}
A measurement of the induced birefringence of vacuum due to an external magnetic field would therefore allow a direct verification of the ${\cal L}_{\rm EK}$ Lagrangian. Better still would be the independent measurement of $n_\parallel$ and $n_\perp$ which would completely fix the factors multiplying the relativistic field invariants in the non linear Lagrangian correction (see equation~(\ref{eq:pM}) below).

This birefringence is extremely small, reason for which it has never been directly observed yet. Indeed for a field $B_{\rm ext} = 1\;$T the induced birefringence is $\Delta n^{\rm (EK)} = 3A_eB_{\rm ext}^2 = 3.96\times10^{-24}$.

Similarly, by considering linearly polarised light propagating in an external electric field $\vec{E}_{\rm ext}$, the corresponding relations to (\ref{n_B}) are
\begin{eqnarray}
 \left \{ \begin{array}{ll}
 \varepsilon_{\parallel} = 1 + 12A_e E_{\rm ext}^2/c^2\\
\mu_{\parallel} = 1 - 4A_e E_{\rm ext}^2/c^2\\
n_\parallel = 1 + 4A_e E_{\rm ext}^2/c^2
\end{array}\right.
\qquad\qquad
\left \{ \begin{array}{ll}
 \varepsilon_{\perp} = 1 +4A_e E_{\rm ext}^2/c^2\\
\mu_{\perp} = 1 + 10A_e E_{\rm ext}^2/c^2\\
n_\perp = 1+7A_e E_{\rm ext}^2/c^2.
\end{array}\right.
\label{n_E}
\end{eqnarray}
Again both $n_\parallel$ and $n_\perp$ are greater than unity and the birefringence is
\begin{equation}
n_\parallel - n_\perp = \Delta n^{\rm (EK)} = -3A_e\frac{E_{\rm ext}^2}{c^2}.
\label{birif_E}
\end{equation}

Maximum electric fields of about 100 MV/m can be obtain in radio-frequency accelerator cavities leading to a value of $E^2/c^2 \approx 0.1$~T$^2$ whereas constant magnetic fields up to $\approx 10$~T are relatively common leading to a $B^2 \approx 100$~T$^2$. Furthermore, as will be discussed in Section~\ref{sec:polarimetry}, for measuring vacuum birefringence the length of the field is also an important factor. For this reason VMB experiments have been attempted only with external magnetic fields. More details on the Kerr effect in vacuum can be found in Reference \cite{Robertson2019}.

From ${\cal L}_{EK}$, and today from QED, it is also possible to determine the Light-by-Light differential and total elastic cross section \cite{Breit1934,Euler:1935,Karplus:1950zz,DeTollis:1965vna}.
In the center of mass and in the low energy photon limit $\hbar\omega\ll m_e c^2$ the differential cross section for unpolarised light is
\begin{equation}
\frac{{\rm d}\sigma}{{\rm d}\Omega} = \left|f(\vartheta,E_\gamma)\right|^2 = \frac{139}{4\pi^290^2}\alpha^4\left(\frac{\hbar\omega}{m_ec^2}\right)^6\left(\frac{\hbar}{m_ec}\right)^2{\left(3+\cos^2\vartheta\right)^2}.
\end{equation}
Integrating over one hemisphere, since the two final-state photons are identical, results in the total cross section for unpolarised light
\begin{equation}
\sigma_{\rm LbL} = \frac{973}{10125\pi}\alpha^4\left(\frac{\hbar\omega}{m_ec^2}\right)^6\left(\frac{\hbar}{m_ec}\right)^2 = \frac{973\mu_0^2}{20\pi}\left(\frac{\hbar^2\omega^6}{c^4}\right)A_e^2
\end{equation}
proportional to $A_e^2$.
For light with wavelength $\lambda = 1064$~nm the total elastic cross section in the center of mass is
\begin{equation}
\sigma_{\rm LbL} = 1.8\times10^{-69}\;{\rm m}^2.
\end{equation} 
Measurements performed by Bernard \emph{et al.} \cite{Bernard2000} have reached $\sigma_{\rm LbL}^{\rm (exp.)} = 1.48\times10^{-52}$~m$^{2}$ for $\lambda = 805$~nm. At very high energies the ATLAS collaboration has observed Light-by-Light elastic scattering confirming ${\cal L}_{\rm EK}$ \cite{atlas2017,atlas2019}.

The connection between the total photon-photon cross section and vacuum birefringence through the parameter $A_e$ describing non linear QED effects, makes non linear QED searches via ellipsometric techniques very attractive. 

\begin{figure}[bht]
\begin{center}
\includegraphics[width=10cm]{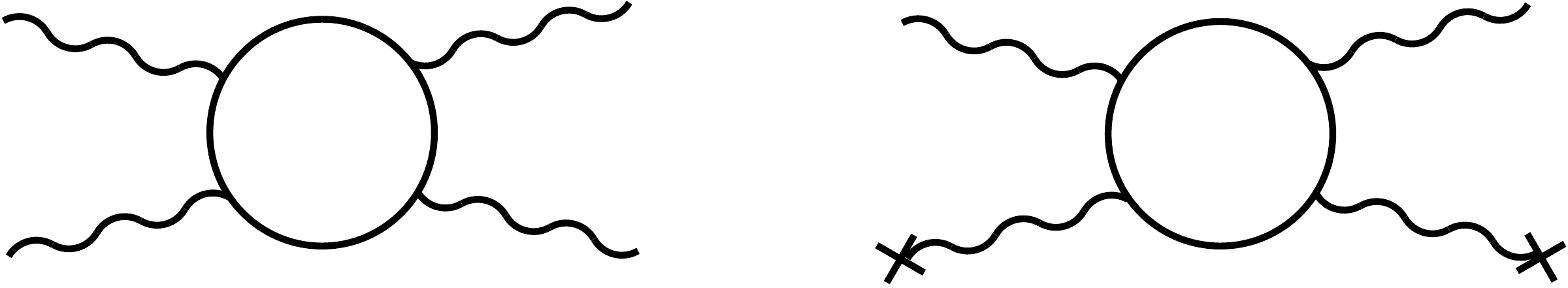}
\end{center}
\caption{Feynman diagrams representing Light-by-Light elastic scattering (left) and vacuum magnetic birefringence (right).}
\label{box}
\end{figure}

Today Light-by-Light scattering and vacuum magnetic birefringence are represented using the Feynman diagrams in Figure~\ref{box} left and right respectively.

Let us now consider the imaginary part $\kappa$ of the complex index of refraction $\tilde n$. A value of $\kappa$ different from zero corresponds to a disappearance of photons from the propagating beam. In a dichroism, this interaction depends on the polarisation of the light resulting in $\Delta\kappa\ne0$. In QED vacuum is dichroic for external fields of the order of the critical fields $B_{\rm cr}$ and $E_{\rm cr}$ \cite{Klein:1964zza}. Another possible process resulting in $\Delta\kappa\neq0$ is photon splitting \cite{Adler1970,BialynickaBirula1970,Adler1971,Stoneham1979,Baier1996,Adler1996PRL} whereby an incident photon of energy $\hbar\omega$ is transformed into two photons of energy $\hbar\omega '$ and $\hbar\omega ''$ such that $\hbar\omega ' + \hbar\omega '' = \hbar\omega$.

\begin{figure}[bht]
\begin{center}
\includegraphics[height=1.8cm]{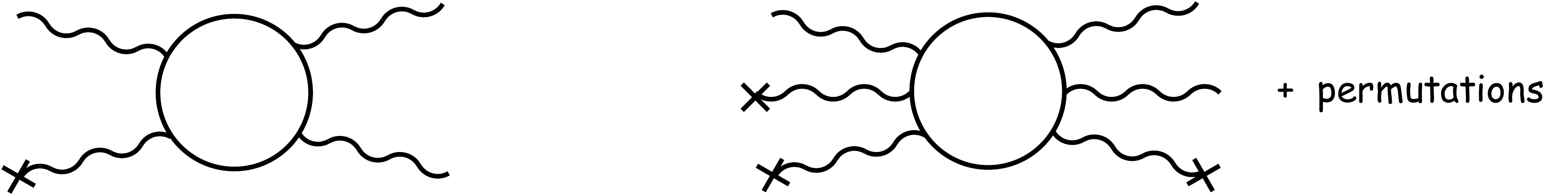}
\end{center}
\caption{Forbidden low energy photon-splitting process with only one interaction with the external field (left) and the lowest order photon splitting diagram (right).}
\label{splitting}
\end{figure}

For photon splitting, several authors \cite{Adler1970,BialynickaBirula1970,Adler1971} have shown that this is forbidden in the non-dispersive case with only one interaction with the external field. The Feynman diagram representing this forbidden process is shown in Figure~\ref{splitting}, left. It has also been shown that the first non zero term in photon splitting is with three interactions with the external field for a total of six couplings to the fermion loop (Figure~\ref{splitting}, right). Furthermore, the 6-vertex photon splitting process is polarisation dependent generating a dichroism (polarisation dependent absorption) $\Delta\kappa^{\rm (EH)} = \kappa_\parallel-\kappa_\perp$. Given linearly polarised light with polarisation parallel $\parallel$ or perpendicular $\perp$ to the plane formed by $\vec{B}_{\rm ext}$ and $\vec{k}$, the absorption indeces are
\begin{equation}
    \kappa_{\perp\choose\parallel} = {0.030\choose0.014} \left(\frac{\alpha^3}{120\pi^2}\right)\left(\frac{\hbar\omega}{m_ec^2}\right)^4\left(\frac{B_{\rm ext}}{B_{\rm cr}}\right)^6\approx{30\choose14}\;7\times10^{-94}\left(\frac{\hbar\omega}{1~\rm eV}\right)^4\left(\frac{B_{\rm ext}}{1~\rm T}\right)^6.
\end{equation}
Clearly with such small values of $\kappa$ one can assume that the Heisenberg-Euler effective Lagrangian ${\cal L}_{\rm HE}$ does not generate a magnetic dichroism in vacuum in the optical range. Photon splitting has been observed for high-energy photons in the electric field of atoms \cite{Akhmadaliev1998}.

\subsubsection{Higher order corrections}

\begin{figure}[bht]
\begin{center}
\includegraphics[width=10cm]{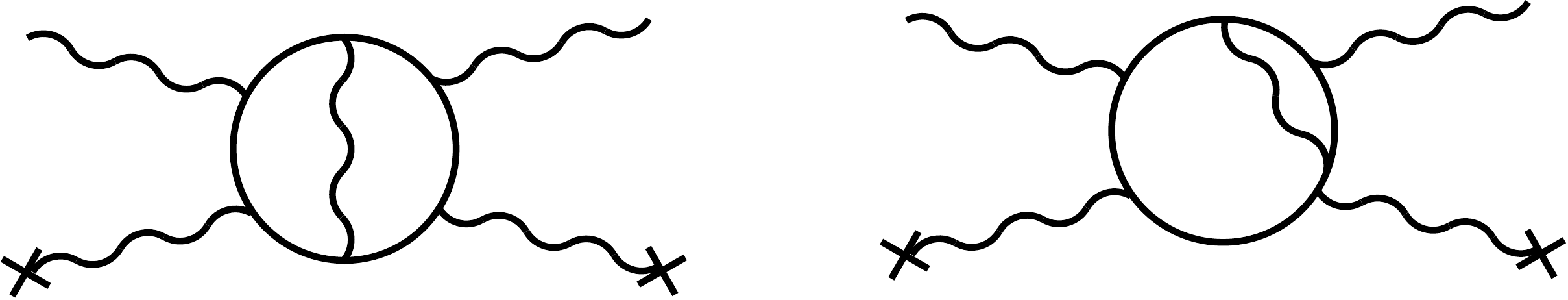}
\end{center}
\caption{Two Feynman diagrams representing the radiative corrections to vacuum magnetic birefringence.}
\label{fig:rad}
\end{figure}

V. I. Ritus (1975) \cite{Ritus:1975cf} determined the correction to ${\cal L}_{\rm HE}$ in an external field taking into account the radiative interaction between the vacuum electron-positron pairs. These radiative corrections, represented by the Feynman diagrams shown in Figure~\ref{fig:rad}, also contribute to VMB.

In general, given a Lagrangian density to second order in the invariants $F$ and $G$ with coefficients $\frac{A_e}{\mu_0}\eta_1$ and $\frac{A_e}{\mu_0}\eta_2$ respectively
\begin{equation}
 {\cal L} = {\cal L}_{\rm Cl} + \frac{A_e}{\mu_0}\left[\eta_1\left(\frac{E^2}{c^2} - B^2\right)^2+\eta_2\left(\frac{\vec{E}}{c}\cdot\vec{B}\right)^2\right]
 \label{eq:pM}
\end{equation}
 the vacuum magnetic birefringence resulting from equations~(\ref{D_light}) and (\ref{H_light}) is
\begin{equation}
n_\parallel - n_\perp = \Delta n = \left(\eta_2 - 4\eta_1\right)A_eB_{\rm ext}^2.
\label{pm}
\end{equation}

Ritus determined the complete $\alpha^3$ two-loop correction to the Lagrangian density which in the approximation for $E \ll E_{\rm cr}$ and $B \ll B_{\rm cr}$ results in
\begin{equation}
    {\cal L}^{\rm (\le 2~loop)} = {\cal L}_{\rm EK} + \frac{\alpha}{36\pi}\frac{A_e}{\mu_0}\left[160\left(\frac{E^2}{c^2} - B^2\right)^2+1315\left(\frac{\vec{E}}{c}\cdot\vec{B}\right)^2\right].
\label{EH2}
\end{equation}
According to equation~(\ref{pm}) the resulting radiative corrected vacuum birefringence is
\begin{equation}
\Delta n^{\rm (EK,rad)} = 3A_eB_{\rm ext}^2\left(1 + \alpha\frac{25}{4\pi}\right)
\end{equation}
where the correction term with respect to $\Delta n^{\rm (EK)}$ is $\alpha\frac{25}{4\pi} = 1.45\%$.

\subsubsection{Born-Infeld}

Other non linear electrodynamic theories have been proposed, one of which is the Born-Infeld theory (1934) \cite{Born1933Nature,BornInfeld1933,BornInfeld1934}. The basis of this theory is to limit the  electric field to a maximum value defined by a parameter $b$. The corresponding Lagrangian density is
\label{BIsecion}
\begin{equation}
    {\cal L}_{\rm BI} = \frac{b^2}{c^2\mu_0}\left[1-\sqrt{1-\frac{c^2}{b^2}\left(\frac{E^2}{c^2} - B^2\right)-\frac{c^4}{b^4}\left(\frac{\vec{E}}{c}\cdot\vec{B}\right)^2}\right]
\label{BI}
\end{equation}
which expanded to second order in the invariants $F$ and $G$ results in
\begin{equation}
    {\cal L}_{\rm BI} = {\cal L}_{\rm Cl} + \frac{c^2}{8b^2\mu_0}\left[\left(\frac{E^2}{c^2} - B^2\right)^2+4\left(\frac{\vec{E}}{c}\cdot\vec{B}\right)^2\right]+...
\label{BIexpand}
\end{equation}
One feature of this theory is that the self energy of a point charge is finite. In the case of the electron, by setting this self energy equal to the rest mass energy of the electron results in a maximum electric field \cite{Iacopini:1982cm}
\begin{equation}
b_0 = E_{\rm BI} = 1.19\times10^{20}\;{\rm V/m}.
\end{equation}
Other interesting consequences of this model can be found in Reference \cite{Fouche2016}. Interestingly from equation~(\ref{pm}) the Born-Infeld theory does not predict a birefringence hence the measurement of a vacuum magnetic birefringence would completely rule out the model. This theory, though, does predict variations from unity of $n_\parallel$ and $n_\perp$ and also predicts Light-by-Light scattering, independently from the value of the parameter $b$ \cite{Davila2014}. As briefly discussed in Section~\ref{sec:ligo} in principle the separate determination of $n_\parallel$ and $n_\perp$ could be possible using, for example, a gravitational wave antenna such as LIGO or VIRGO equipped with magnetic fields along the arms.

\subsection{\bf Axion like particles}
\label{sec:axions}

\begin{figure}[bht]
\begin{center}
\includegraphics[width=10cm]{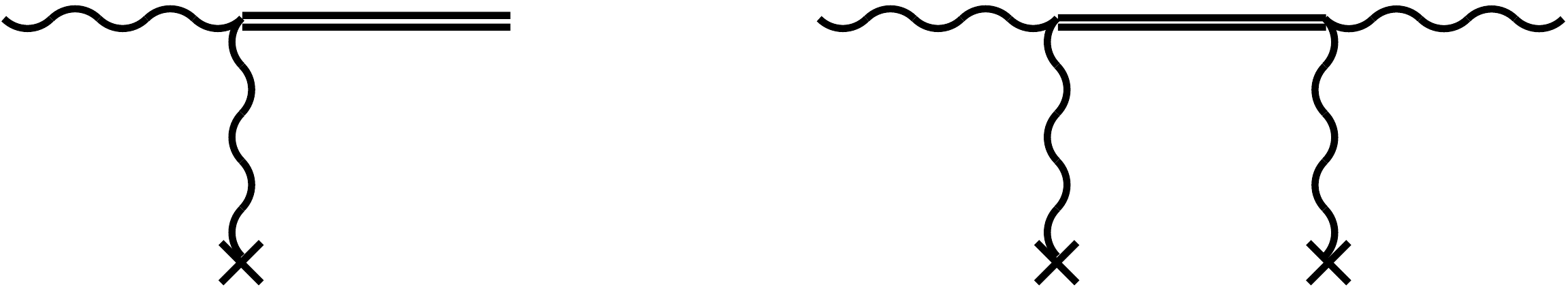}
\end{center}
\caption{Production (left) and recombination (right) of a spin-zero particle coupled to two photons through the Primakoff effect \cite{Primakoff}.}
\label{fig:Primakoff}
\end{figure}

The propagation of light in an external electromagnetic field could also depend on the existence of hypothetical light neutral particles coupling to two photons. The involved processes are shown in Figure~\ref{fig:Primakoff}: the production diagram implies an absorption of light quanta, whereas a phase delay is produced by the recombination process. The search for such particles having masses below $\sim 1\;$eV has recently gained strong impulse after it was clear that such particles could be a viable candidate for particle dark matter.

In general, there are arguments to believe that there is new physics, mainly meaning new particles, beyond the standard model. The  indications for the existence of dark matter and dark energy, and the absence of an electric dipole moment of the neutron are among the experimental facts requesting an extension of the standard model. 

Light, weakly interacting, neutral pseudoscalar or scalar particles
 arise naturally in extensions of the standard model that introduce new fields and symmetries. In fact, in the presence of a spontaneous breaking of a global symmetry, such particles appear as massless Nambu-Goldstone bosons. If there is a small explicit symmetry breaking, either in the Lagrangian or due to quantum effects, the boson acquires a mass and is called a pseudo-Nambu-Goldstone boson.
 Typical examples are familons \cite{Wilczek:1982rv}, and Majorons \cite{Chikashige:1980ui} associated, respectively, with the spontaneously broken family and lepton-number symmetries.

Another popular example of a pseudo Nambu-Goldstone boson is the axion. Its origin stems from the introduction by Peccei and Quinn (PQ) \cite{Peccei:1977ur,Peccei:1977hh} of a new global symmetry to solve the strong CP problem of QCD, i.e. the absence of CP violation within the strong interactions. The high energy breaking of the PQ symmetry gives rise to a light pseudoscalar called the axion \cite{Weinberg1978,Wilczek1978,PDG2019}: the value of its mass is not predicted while the couplings to the standard model particles are well defined by the exact model implementing the PQ symmetry. Couplings are generally very weak and proportional to the mass of the axion. 

A more general class of Axion Like Particles (ALPs) has also been introduced: for the ALPs the mass and coupling constants are independent. Axions and ALPs have been searched for in dedicated experiments since their proposal \cite{axion_searches}, however to date no detection has been reported and only a fraction of the available parameter space has been probed. Indeed, nowadays there are experiments or proposals that study masses starting from the lightest possible value of 10$^{-22}$~eV up to several gigaelectronvolt. A most favorable window has been also identified in the mass range between 10~$\mu$eV and 1~meV.

The greater part of the experimental searches relies on the axion-photon coupling mediated by a two photon vertex of Figure~\ref{fig:Primakoff}. Other searches  are based  on the axion-electron interaction, present through an axion spin interaction only in some models like the Dine-Fishler-Sredincki-Zhitnitsky (DFSZ) \cite{Dine:1981rt,Zhitnitsky:1980tq}
one. A comprehensive review of the experimental efforts to search for ALPs and axion can be found in \cite{Kuster:1105861,Irastorza:2018dyq}.

Due to its very small coupling and mass,  the axion could be a valid candidate as a dark matter component, since large quantities could have been produced at an early stage of the Universe. Axion haloscopes search for the conversion of cosmological axions with the assumption that axions are the dominant component of the local dark matter density. The  current leading experiment following this line of research is ADMX (Axion Dark Matter eXperiment) searching for the resonant conversion of axions in a microwave cavity immersed in a strong static magnetic field \cite{PhysRevLett.120.151301}.

Axions and ALPs can be produced in hot astrophysical plasmas and could transport energy out from stars, thus contributing to stellar lifetimes. Limits on axion mass and coupling can be set by studying stellar evolution. In the case of the Sun, solar axions could also be detected on earth based apparatuses. While these experiments rely on solar/stellar models and on dark matter models, there are pure laboratory experiments where the axion is produced and directly detected in a totally model independent manner. However, due to the smallness of the coupling, only ALPs parameter space is studied with presently available techniques.

Search for axions or ALPs using laboratory optical techniques was experimentally pioneered by the BFRT collaboration \cite{Semertzidis:1990qc} and subsequently continued by the PVLAS collaboration with an apparatus installed at INFN National Laboratories in Legnaro (LNL) \cite{Bakalov1998HI,Zavattini2007PRD,Bregant2007PRD}. As will be discussed below, the measurement in a PVLAS-type apparatus of the real and imaginary part of the index of refraction of a vacuum magnetised by an external field could give direct information on the mass and coupling constant of the searched for particle. Other laboratory optical experiments are the so called ``light shining through a wall" (LSW) apparatuses, where a regeneration scheme is employed \cite{VanBibber:1987rq,Hoogeveen:1990vq,Sikivie:1012940,Ruoso:1992nx,Ehret:2010mh,OSQAR2015}. 

Regarding polarisation effects the Lagrangian densities describing the interaction of axion-like particles with two photons, for convenience expressed in natural Heaviside-Lorentz units\footnote{In natural Heaviside-Lorentz units 1~T $=\sqrt{\frac{\hbar^{3}c^{3}}{e^{4}\mu_{0}}}= 195$~eV$^2$ and 1~m $=\frac{e}{\hbar c}=5.06\times10^{6}$~eV$^{-1}$.}, can be written as
\begin{equation}
{\cal L}_{a} = g_a\phi_{a} \vec{E}\cdot \vec{B} \qquad\qquad{\rm and}\qquad\qquad
{\cal L}_{s} = g_s\phi_{s} \left(E^{2}-B^{2} \right)
\label{lagalp}
\end{equation}
where $g_a$ and $g_s$ are the coupling constants to two photons of the pseudoscalar field $\phi_{\rm a}$ or scalar field $\phi_{\rm s}$, respectively.
Therefore, in the presence of an external uniform magnetic field $\vec{B}_{\rm ext}$ the component of the electric field of light $\vec{E}_{\rm \gamma}$ parallel to $\vec{B}_{\rm ext}$ will interact with the pseudoscalar field. For the scalar case the opposite is true: an interaction is only possible for the component of $\vec{E}_{\rm \gamma}$ normal to $\vec{B}_{\rm ext}$.

For photon energies above the mass $m_{a,s}$ of such particle candidates, a real production can follow: the oscillation of photons into such particles decreases  the amplitude of only one of the polarisation components of the propagating light resulting in a dichroism $\Delta\kappa$. On the other hand, even if the photon energy is smaller than the particle mass, virtual production can take place, causing a reduction of the speed of light of one component with respect to the other resulting in a birefringence $\Delta n$. The phase difference $\Delta\varphi=\varphi_{\parallel}-\varphi_{\perp}$ and the difference in relative amplitude reduction $\Delta\zeta = \zeta_{\parallel} - \zeta_{\perp}$ accumulated in an optical path $L_B$ inside the magnetic field region resulting respectively from $\Delta\kappa$ and $\Delta n$ are
\begin{equation}
\Delta\varphi = \Delta n\frac{2\pi L_B}{\lambda}\qquad\qquad\Delta\zeta = \Delta\kappa\frac{2\pi L_B}{\lambda}.
\end{equation}

In the pseudoscalar case $n^{a}_{\parallel}>1$, $\kappa^a_\parallel>0$, $n^{a}_{\perp}=1$ and $\kappa^a_\perp=0$ whereas in the scalar case $n^{s}_{\perp}>1$, $\kappa^s_\perp>0$, $n^{s}_{\parallel}=1$ and $\kappa^s_\parallel=0$. It can be shown that the dichroism $\Delta \kappa$ and the birefringence $\Delta n$ due to the existence of such bosons can be expressed in both the scalar and pseudoscalar cases as \cite{Maiani:1986md,Cameron1993PRD,Sikivie:1983ip,Raffelt:1987im}:
\begin{eqnarray}
&&|\Delta\kappa|=\kappa_{\parallel}^{a} = \kappa_{\perp}^{s} = \frac{2}{\omega L_B}\left(\frac{g_{a,s}B_{\rm ext}L_B}{4}\right)^{2}\left(\frac{\sin x}{x}\right)^{2}
\label{dichroism}\\
&&|\Delta n|= n_{\parallel}^{a}-1 = n_{\perp}^{s}-1 = \frac{1}{2}\left(\frac{g_{a,s}B_{\rm ext}}{2m_{a,s}}\right)^2\left(1-\frac{\sin2x}{2x}\right)
\label{pseudo}
\end{eqnarray}
where, in vacuum, $x=\frac{L_Bm_{a,s}^{2}}{4\omega}$, $\omega$ is the photon energy and $L_B$ is the magnetic field length. 
In the approximation $x\ll1$ (small masses) expressions (\ref{dichroism}) and (\ref{pseudo}) become

\begin{eqnarray}
&&|\Delta\kappa| = \kappa_{\parallel}^{a} = \kappa_{\perp}^{s} = \frac{2}{\omega L_B}\left(\frac{g_{a,s}B_{\rm ext}L_B}{4}\right)^2
\label{eq:assionedic}\\
&&|\Delta n| = n^{a}_{\parallel}-1 = n^{s}_{\perp}-1 = \frac{1}{3}\left(\frac{g_{a,s}B_{\rm ext}m_{a,s}L_B}{4\omega}\right)^2
\label{eq:assionebirif}
\end{eqnarray}
where it is interesting to note that $\Delta\kappa$ in this case is independent of $m_{a,s}$.
{\em Vice versa} for $x \gg 1$
\begin{eqnarray}
&&|\Delta\kappa| = \kappa_{\parallel}^{a} = \kappa_{\perp}^{s} < 2\omega\left(\frac{g_{a,s}B_{\rm ext}}{m_{a,s}^{2}}\right)^2\\
&&|\Delta n| = n^{a}_{\parallel}-1 = n^{s}_{\perp}-1 = \frac{1}{2}\left(\frac{g_{a,s}B_{\rm ext}}{m_{a,s}}\right)^2.
\end{eqnarray}
Note that the birefringence induced by pseudoscalars and scalars are opposite in sign: $n_{\parallel}^a > n_{\perp}^a = 1$ whereas $n_{\parallel}^s = 1 < n_{\perp}^s$. 

The detection of an ALPs-induced birefringence and dichroism would allow the determination of the mass and coupling constant of the ALPs to two photons.

\subsection{\bf Millicharged particles}

Consider now vacuum fluctuations of hypothetical particles with charge $\pm\epsilon e$ and mass $m_{\epsilon}$ as discussed in References \cite{Ahlers:2006iz,Gies:2006ca}. Photons traversing a magnetic field may interact with such fluctuations resulting in a phase delay and, for photon energies $\hbar\omega>2m_\epsilon c^2$, in a millicharged pair production. Therefore a birefringence and a dichroism will result if such hypothetical particles existed. The cases of Dirac fermions (Df) and of scalar (sc) bosons here are considered separately. The indices of refraction for light polarised respectively parallel and perpendicular to the external magnetic field have two different mass regimes defined by the dimensionless parameter $\chi$:
\begin{equation}
\chi\equiv\frac{3}{2}\frac{\hbar\omega}{m_{\epsilon}c^{2}}\frac{\epsilon e B_{\rm ext}\hbar}{m_{\epsilon}^{2}c^{2}}.
\label{chi}
\end{equation}
In the case of fermions, it can be shown that \cite{Tsai:1975iz,Ahlers:2006iz}
\begin{equation}
\Delta n^{\rm (Df)}=A_{\epsilon} B_{\rm ext}^{2}
\left\{\begin{array}{ll}
3 & \rm { for  } \chi \ll 1 \\
\displaystyle-\frac{9}{7}\frac{45}{2}\frac{\pi^{1/2}2^{1/3}\left[\Gamma\left(\frac{2}{3}\right)\right]^{2}}{\Gamma\left(\frac{1}{6}\right)}\chi^{-4/3} & \rm{ for   }\chi\gg 1
\end{array}\right.
\label{DeltanDF}
\end{equation}
where
\begin{equation}
A_{\epsilon}=\frac{2}{45\mu_0}\frac{\hbar^3}{m_\epsilon^4c^5}\epsilon^4\alpha^2
\end{equation}
in analogy to equation~(\ref{Ae}). Note that in the limit of large masses ($\chi\ll1$) expression (\ref{DeltanDF}) reduces to equation~(\ref{n_B}) with the substitution of $\epsilon e$ with $e$ and $m_{\epsilon}$ with $m_{e}$. For small masses ($\chi\gg1$) the birefringence depends on the parameter $\chi^{-4/3}$ therefore resulting in a net dependence of $\Delta n^{\rm (Df)}$ with $B_{\rm ext}^{2/3}$ rather than $B_{\rm ext}^{2}$ as in equation~(\ref{n_B}).
For dichroism one finds \cite{Tsai:1974fa,Ahlers:2006iz}
\begin{equation}
\Delta \kappa^{\rm (Df)}=\frac{1}{8\pi}\frac{\epsilon^3e\alpha\lambda B_{\rm ext}}{m_\epsilon c}
\left\{\begin{array}{ll}
\sqrt{\frac{3}{32}}\,e^{-4/\chi} & \rm { for  } \chi \ll 1 \\
\displaystyle\frac{2\pi}{3\,\Gamma(\frac{1}{6})\Gamma(\frac{13}{6})}\,\chi^{-1/3} & \rm{ for   }\chi\gg 1.
\end{array}\right.
\end{equation}

Very similar results are found for the case of scalar millicharged particles \cite{Ahlers:2006iz}. Again there are two mass regimes defined by the same parameter $\chi$ of expression (\ref{chi}). In this case the magnetic birefringence is
\begin{equation}
\Delta n^{\rm (sc)}=A_{\epsilon} B_{\rm ext}^{2}\left\{\begin{array}{ll}
\displaystyle-\frac{6}{4} & \rm{ for  } \chi \ll 1 \\
\displaystyle\frac{9}{14}\frac{45}{2}\frac{\pi^{1/2}2^{1/3}\left[\Gamma\left(\frac{2}{3}\right)\right]^{2}}{\Gamma\left(\frac{1}{6}\right)}\chi^{-4/3} & \rm{ for   }\chi\gg 1
\end{array}\right.\nonumber
\end{equation}
and the dichroism is
\begin{equation}
\Delta \kappa^{\rm (sc)}=\frac{1}{8\pi}\frac{\epsilon^3e\alpha\lambda B_{\rm ext}}{m_\epsilon c}
\left\{\begin{array}{ll}
-\sqrt{\frac{3}{8}}\,e^{-4/\chi} & \rm { for  } \chi \ll 1 \\
\displaystyle-\frac{\pi}{3\,\Gamma(\frac{1}{6})\Gamma(\frac{13}{6})}\,\chi^{-1/3} & \rm{ for   }\chi\gg 1.
\end{array}\right.
\end{equation}
As can be seen, there is a sign difference with respect to the case of Dirac fermions, both for the induced birefringence and the induced dichroism.

Vacuum magnetic birefringence and vacuum magnetic dichroism limits can therefore constrain the existence of such millicharged particles.

\subsection{\bf Chameleons and Dark energy}
An open issue of modern cosmology is the understanding of the cosmic acceleration \cite{Riess_1998,Perlmutter_1999}. The presence of a scalar field sourcing the dark energy responsible for this acceleration is envisaged in several theories \cite{Caldwell2009}. To comply with experimental bounds, a screening mechanism preventing the scalar field to act as a fifth force is however necessary \cite{JAIN20101479}. The chameleon mechanism provides a way for this suppression via nonlinear field self-interactions and interactions with the ambient matter \cite{PhysRevLett.93.171104,PhysRevD.69.044026}. It can be seen that the chameleon fields behave as Axion Like Particles with respect to photons in an experiment of PVLAS type, with coupling constant to photons $M_\gamma$. Brax and co-workers \cite{PhysRevD.76.085010} have calculated the effect on the rotation and ellipticity measurements in the presence of a chameleon field. One feature of the chameleon model is that the ellipticity is predicted to be much larger than the rotation. This can be viewed as a generic prediction of chameleon theories and it is due to the fact that chameleons could be reflected off the cavity mirrors. The difficulty in calculating the expected effects is that these are related to the geometrical size of the cavity, the magnetic field and the density of matter in the laboratory vacuum. For these reasons in this paper we will not try to extract chameleon information from the PVLAS data. More details can be found in References~\cite{PhysRevD.76.085010,Burrage2018}.

\section{The experimental method}
\label{Method}

\subsection{\bf Polarimetric scheme}
\label{sec:polarimetry}

Birefringence and dichroism are local properties of a medium and can be determined by detecting their effect on the propagation of light. Here we will discuss the polarimetric scheme adopted by PVLAS in the attempt to measure magnetically induced vacuum birefringence and dichroism.

\begin{figure}[bht]
\begin{center}
\includegraphics[width=10 cm]{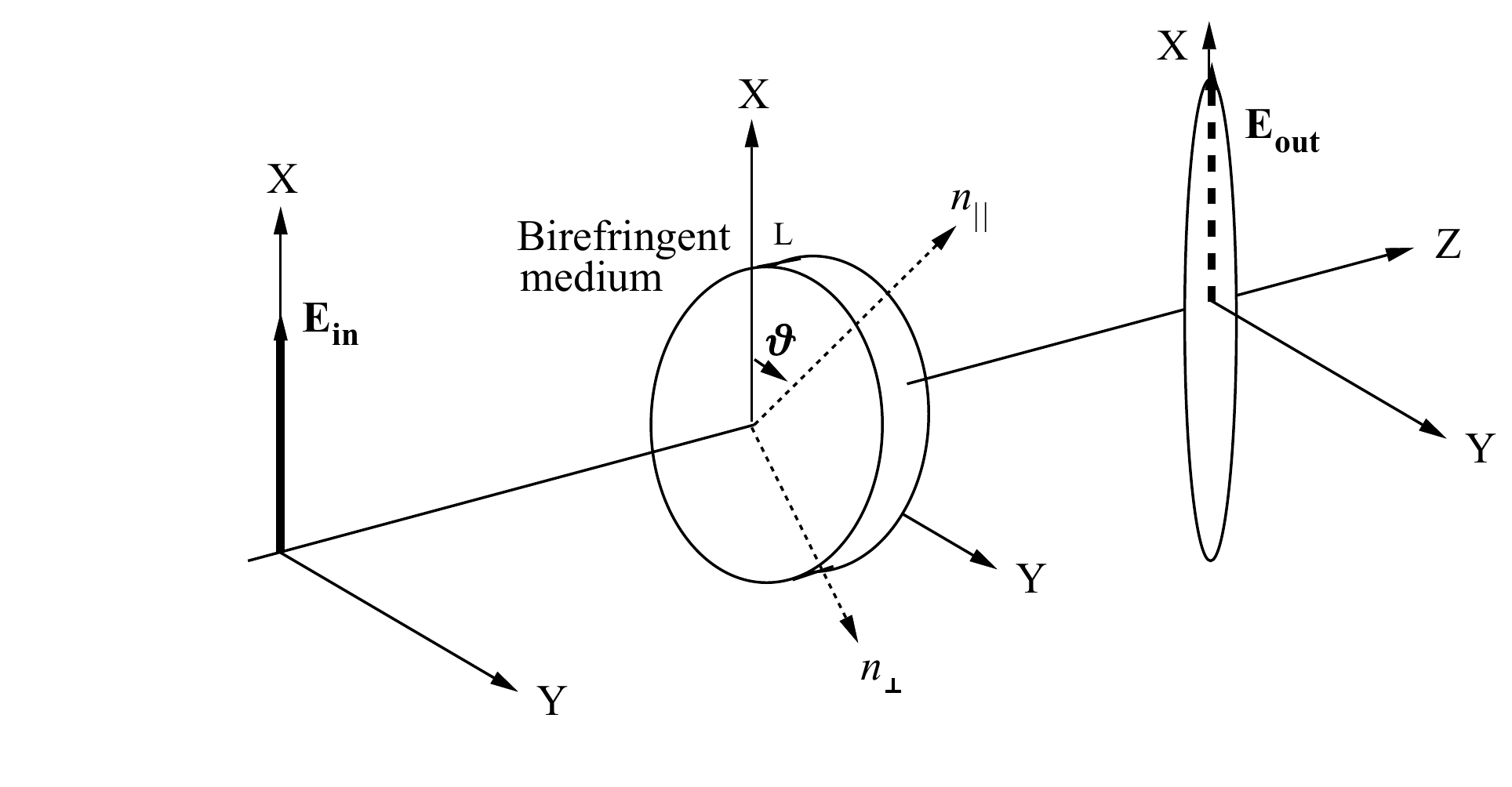}
\end{center}
\caption{Reference frame for the calculations below. The parameters $n_\parallel$ and $n_\perp$ are the indices of refraction for light polarised parallel and perpendicularly to the axis of the medium.}
\label{fig:frame}
\end{figure}

Consider a monochromatic linearly polarised beam of light propagating along the $\hat{\rm Z}$ axis. Let us also assume that the polarisation (electric field) is directed vertically along the $\hat{\rm X}$ axis and let this beam propagate through a uniformly birefringent medium of thickness $L$ whose slow $\parallel$ and fast $\perp$ axes are perpendicular to $\hat{\rm Z}$. Finally let the slow axis of the medium form an angle $\vartheta$ with the $\hat{\rm X}$ axis. This reference frame is shown in Figure~\ref{fig:frame}. The components of the electric field along the $\parallel$ and $\perp$ axes of the propagating beam will acquire a phase difference $\Delta\varphi$ at the output of the medium given by
\begin{equation}
\Delta\varphi = \varphi_\parallel - \varphi_\perp =  \frac{2\pi}{\lambda}\left(n_\parallel - n_\perp\right)L.
\end{equation}
More in general, the total optical path difference $\Delta{\cal D}$ between the $\parallel$ and $\perp$ components of the electric field is
\begin{equation}
\Delta{\cal D} = \int{\Delta n(z) dz}.
\end{equation}

Given the reference frame in Figure~\ref{fig:frame} the input electric field can be written as $\vec{E}_{\rm in} = E_{\rm in} e^{i\varphi(t)}{1\choose 0}$ where $\varphi(t)$ contains the time dependent phase of the wave which, from now on, we will neglect. To determine the output electric field one can project $\vec{E}_{\rm in}$ along the $\parallel$ and $\perp$ axes, propagate the beam through the medium and finally project back to the $\hat{\rm X}$,$\hat{\rm Y}$ reference frame. Assuming $\Delta\varphi\ll1$, the output field will acquire a component along the $\hat{\rm Y}$ axis:
\begin{equation}
\vec{E}_{\rm out} \approx E_{\rm in} {1+i\frac{\Delta\varphi}{2}\cos2\vartheta\choose i\frac{\Delta\varphi}{2}\sin2\vartheta} = E_{\rm in} {1+i\frac{\pi}{\lambda}{\Delta\cal D}\cos2\vartheta\choose i\frac{\pi}{\lambda}{\Delta\cal D}\sin2\vartheta}
\end{equation}
describing an ellipse.
The ratio of the amplitudes of the output electric field along the $\hat{\rm Y}$ and $\hat{\rm X}$ axes, $E_{y,\rm out}/E_{x,\rm out}$, is defined as the ellipticity $\psi_\vartheta$ of the polarisation:
\begin{equation}
\psi_\vartheta = \psi\sin2\vartheta \approx \frac{\Delta\varphi}{2}\sin2\vartheta = \frac{\pi}{\lambda}\int{\Delta n(z) dz}\sin2\vartheta = \frac{\pi}{\lambda}{\Delta\cal D}\sin2\vartheta.
\label{eq:ellipticity}
\end{equation}
Setting $\vartheta = \pi/4$ the measurement of the electric field component along $\hat{\rm Y}$ gives a direct determination of $\Delta{\cal D}$.

Note here that the two components of the electric field along the $\hat{\rm X}$ and $\hat{\rm Y}$ axes oscillate with a phase difference of $\pi/2$. This fact is important inasmuch as it will allow the distinction between an ellipticity $\psi$ and a rotation $\phi$. Indeed an electric field whose polarisation is rotated by an angle $\phi\ll 1$ with respect to the $\hat{\rm X}$ axis can be written as
\begin{equation}
\vec{E}_{\rm out} = E_{\rm in} {\cos\phi\choose \sin\phi} \approx E_{\rm in} {1\choose \phi}
\end{equation}
where the $\hat{\rm X}$ and $\hat{\rm Y}$ components of the electric field oscillate in phase.

A similar treatment may be made in the presence of a dichroism. Assuming absorption indices $\kappa_\parallel$ and $\kappa_\perp$ along the $\parallel$ and $\perp$ axes, the electric field after the medium will be
\begin{equation}
\vec{E}_{\rm out} \approx E_{\rm in} {1-\frac{\Delta\zeta}{2}\cos2\vartheta\choose -\frac{\Delta\zeta}{2}\sin2\vartheta}
\end{equation}
where $\Delta\zeta = \frac{2\pi\Delta\kappa L}{\lambda}$. A rotation is therefore apparent given by
\begin{equation}
\phi_\vartheta = \phi\sin2\vartheta = -\frac{\Delta\zeta}{2}\sin2\vartheta = -\frac{\pi}{\lambda}\int{\Delta\kappa(z)\;dz}\sin2\vartheta= -\frac{\pi}{\lambda}{\Delta\cal A}\sin2\vartheta
\label{eq:rotation}
\end{equation}
where in analogy to the optical path difference $\Delta{\cal D}$ we have introduced $\Delta{\cal A} = \int\Delta\kappa(z)\,dz$.

\begin{figure}[bht]
\begin{center}
\includegraphics[width=14cm]{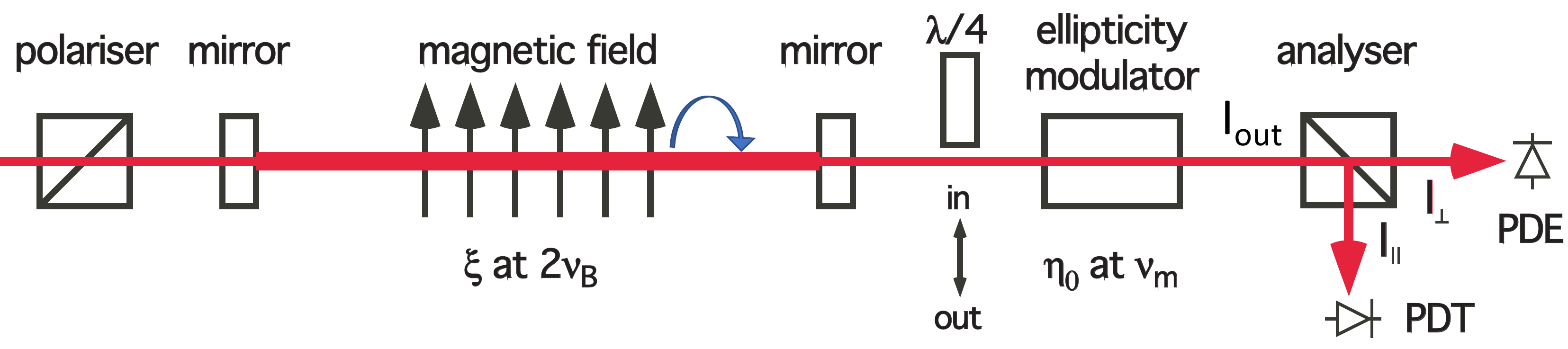}
\end{center}
\caption{Scheme of the PVLAS polarimeter. A rotating magnetic field between the cavity mirrors generates a time dependent ellipticity.} 
\label{fig:scheme}
\end{figure}

The general scheme of a sensitive polarimeter is shown in Figure~\ref{fig:scheme}. Linearly polarised light is sent to a Fabry-Perot optical cavity. The beam then passes through a dipolar magnetic field forming an angle $\vartheta$ with the polarisation direction. In general either the intensity of the magnetic field $B_{\rm ext}$ or its direction may vary in time so as to modulate the induced ellipticity and/or rotation. A variable known ellipticity $\eta(t)= \eta_0\cos(2\pi\nu_mt + \vartheta_{m})$ generated by a modulator is then added to the polarisation of the beam transmitted by the Fabry-Perot. For rotation measurements (as will be discussed below) a quarter-wave plate (QWP) may be inserted between the output mirror of the cavity and the modulator. Finally the beam passes through a second polariser set to extinction. Both the powers $I_\perp$ and $I_\parallel$, of the ordinary and extraordinary beams are collected by photodiodes. The ellipticity and/or rotations induced by the magnetic field can be determined from a Fourier analysis of the detected currents.

When considering monochromatic light, the Jones' matrices \cite{Jones1941} may be used to describe how an ellipticity and/or a rotation evolves when light passes consecutively through several media. Here we will assume the presence of both a linear birefringence and a linear dichroism both having the same axes. These will generate an ellipticity $\psi$ and a rotation $\phi$. Defining $\xi/2 = i\psi + \phi$, the Jones' matrix of these effects is diagonal  in the $(\parallel,\perp)$ reference frame:
\begin{equation}
\mathbf{X}_{\parallel,\perp}=
\left(
\begin{array}{cc}
e^{\xi/2}  & 0 \\
 0 & e^{-\xi/2}
  \end{array}
 \right).
 \label{Chi}
\end{equation}
With respect to the $\hat{\rm X}$ and $\hat{\rm Y}$ axes $\mathbf{X}_{\parallel,\perp}$ must be rotated by an angle $\vartheta$ resulting in
\begin{equation}
\mathbf{X}(\vartheta)=
\left(
\begin{array}{cc}
e^{\xi/2}\cos^2\vartheta +e^{-\xi/2}\sin^2\vartheta  & \frac{1}{2}\sin{2\vartheta}\left(e^{\xi/2} - e^{-\xi/2}\right) \\
 \frac{1}{2}\sin{2\vartheta}\left(e^{\xi/2} - e^{-\xi/2}\right) & e^{-\xi/2}\cos^2\vartheta +e^{\xi/2}\sin^2\vartheta  
 \end{array}
 \right).
 \label{ChiRotated}
 \end{equation}

Three matrices of the type of equation~(\ref{ChiRotated}) will be of particular interest for us to describe this scheme: the first has $|\psi|\ll 1$ and $|\phi| = 0$ describing the effect of a pure birefringence, the second $|\psi|=0 $ and $|\phi|\ll 1$ describing a pure rotation and the third, describing the ellipticity modulator, has $|\psi| = |\eta| \ll 1$ and $|\phi| = 0$ with $\vartheta = \pi/4$. These three matrices are respectively
\begin{equation}
\mathbf{BF}(\vartheta)=
\left(
\begin{array}{cc}
1 + i\psi\cos2\vartheta  & i\psi\sin{2\vartheta} \\
 i\psi\sin{2\vartheta} & 1 - i\psi\cos2\vartheta  
 \end{array}
 \right)
 \label{BF}
 \end{equation}
\begin{equation}
\mathbf{DC}(\vartheta)=
\left(
\begin{array}{cc}
1 + \phi\cos2\vartheta  & \phi\sin{2\vartheta} \\
\phi\sin{2\vartheta} & 1 -\phi\cos2\vartheta  
 \end{array}
 \right)
 \label{DC}
 \end{equation}
\begin{equation}
\mathbf{MOD}=
\left(
\begin{array}{cc}
1   & i\eta \\
 i\eta & 1
 \end{array}
 \right)
 \label{MOD}
 \end{equation}
where $\mathbf{BF}(\vartheta)\cdot\mathbf{DC}(\vartheta)=\mathbf{X}(\vartheta)$.

Neglecting for the moment the Fabry-Perot cavity, the polarimeter configured for ellipticity measurements can be described by the composition of the above matrices (\ref{BF}), (\ref{DC}) and (\ref{MOD}). The output electric field after the analyser will be ($\psi\ll1, \phi\ll1, \eta\ll1$)
\begin{equation}
\vec{E}_{\rm out}^{\rm (ell)} = E_{\rm in}\,\mathbf{A}\cdot\mathbf{MOD}\cdot\mathbf{X}(\vartheta){1\choose 0} \approx E_{\rm in}\,\mathbf{A}\cdot\mathbf{MOD}\cdot\mathbf{BF}\cdot\mathbf{DC}{1\choose 0}
\label{eq:e_out_ell}
 \end{equation}
where 
\begin{equation}
\mathbf{A}=
\left(
\begin{array}{cc}
0   & 0 \\
0 & 1
 \end{array}
 \right)
 \end{equation}
represents the analyser. The extinguished power after the analyser is therefore
\begin{equation}
I_\perp^{\rm (ell)} =  I_{\rm out}\left|i\eta(t) + (i\psi+\phi)\sin2\vartheta(t)\right|^2 \approx I_\parallel\left[\eta(t)^2 + 2\eta(t)\psi\sin2\vartheta(t) + ...\right]\nonumber
\label{eq:ell_sig}
 \end{equation}
where we have approximated $I_{\rm out} \approx I_\parallel$. In the absence of losses
\begin{equation}
    I_\parallel\approx I_{\rm out} = I_{\rm in} = \frac{\varepsilon_0 c}{2}\int E_{\rm in}^2\,d\Sigma.
\end{equation}
In equation~(\ref{eq:ell_sig}) the dots indicate higher order terms in $\phi$ and $\psi$ and we assume that the magnetic field direction is rotating such that $\vartheta(t) = 2\pi\nu_B t + \vartheta_{B}$. In general the field intensity may be varied with a fixed $\vartheta$ and the induced ellipticity would then be $\psi(t)\sin2\vartheta$ but since the PVLAS experiment has always rotated the magnetic field, expression (\ref{eq:ell_sig}) makes it clear that the sought for effect will appear at \emph{twice} the rotation frequency of the magnetic field.

 Being both the ellipticity terms $i\psi\sin2\vartheta(t)$ and $i\eta(t)$ imaginary quantities, these will beat linearising the signal which would otherwise be quadratic in $\psi$. The rotation $\phi$ generated between the polariser and the analyser, though, will \emph{not} beat with the modulator since it is real.

In the scheme of Figure~\ref{fig:scheme}, to perform rotation measurements one must insert before the modulator the quarter-wave plate aligned with one of its axes parallel to the input polarisation. The matrix describing this optical element with the \emph{slow} axis aligned with the polarisation is
\begin{equation}
 \mathbf{Q}={1\over\sqrt{2}}\left(\begin{array}{cc}1+i&0\\0 &1-i\end{array}\right).
 \label{eq:QWP}
\end{equation}

The effect of this wave plate is to add a phase $\pi/2$ to $E_\parallel$ with respect to $E_\perp$ such that $i\psi\rightarrow+\psi$ and $\phi\rightarrow-i\phi$. On the other hand with the fast axis of the QWP aligned with the polarisation the signs of the transformations will change. Using the matrix in equation~(\ref{eq:QWP}) the electric field after the analyser will be
\begin{equation}
\vec{E}_{\rm out}^{\rm (rot)} \approx E_{\rm in}\mathbf{A}\cdot\mathbf{MOD}\cdot\mathbf{Q}\cdot\mathbf{BF}\cdot\mathbf{DC}{1\choose 0}.
\label{eq:e_out_rot}
 \end{equation}
The rotation $\phi$ transformed to ellipticity $-i\phi$ will now beat with the modulator, whereas the ellipticity $i\psi$ transformed to a rotation $\psi$ will not. The extinguished power at the output of the polarimeter will be
\begin{equation}
I_\perp^{\rm (rot)} =  I_{\rm out}\left|i\eta(t) + (\psi-i\phi)\sin2\vartheta(t)\right|^2 \approx I_\parallel\left[\eta(t)^2 - 2\eta(t)\phi\sin2\vartheta(t) + ...\right].
\label{eq:rot_sig}
 \end{equation}
By inserting and extracting the QWP one can then switch between rotation and ellipticity measurements.

The heterodyne method is employed to measure $\psi$ or $\phi$. By setting $\eta(t) = \eta_0\cos(2\pi\nu_mt + \vartheta_{m})$, with $\nu_B\ll\nu_m$, the sought for values of the quantities $\psi$ or $\phi$ can be extracted from equations~(\ref{eq:ell_sig}) or (\ref{eq:rot_sig}) from the amplitude and phase of three components in a Fourier transform of the extinguished power $I_\perp$: the component $I_{2\nu_m}=I_\parallel\eta_0^2/2$ at $2\nu_m$ and the sideband components $I_+$ and $I_-$ at $\nu_m\pm 2\nu_B$. When a lock-in amplifier is used to demodulate $I_\perp$ at the frequency $\nu_m$, instead of $I_+$ and $I_-$ there is a single component $I_{2\nu_B} = I_+ + I_- = 2I_\parallel\eta_0\psi$ at $2\nu_B$ (or $I_{2\nu_B} = 2I_\parallel\eta_0\phi$ in the case of rotations). The resulting ellipticity and rotation can be written as functions of measured quantities:
\begin{equation}
\psi,\phi = \frac{I_{2\nu_B}}{2\sqrt{2I_\parallel I_{2\nu_m}}} = \frac{I_{2\nu_B}}{2\eta_0I_\parallel} = \frac{I_{2\nu_B}}{I_{2\nu_m}}\frac{\eta_0}{4}.
\label{signal}
\end{equation}

The ellipticity and the rotation come with a well defined phase $2\vartheta_{B}$ such that $\psi(t)$ or $\phi(t)$ are maximum for $\vartheta = \pi/4\;$ (mod $\pi$). 

\subsection{\bf The Fabry-Perot interferometer as an optical path multiplier}
Consider now the presence of the Fabry-Perot cavity whose mirrors have reflectivity, transmissivity and losses respectively $R$, $T$ and $P$ such that $R+T+P = 1$. We are assuming the two mirrors to be identical. If ${\cal D}$ is the optical path length between the two mirrors let $\delta = \frac{4\pi {\cal D}}{\lambda}$ be the round trip phase acquired by the trapped light. To understand the principle of the Fabry-Perot let us neglect for the moment polarisation effects due to the magnetic field and to the cavity itself. The electric field at the output of the Fabry-Perot will be
\begin{equation}
E_{\rm out} = E_{\rm in} Te^{i\delta/2}\sum_{j = 0}^\infty R^{j}e^{ji\delta} = E_{\rm in}T\frac{e^{i\delta/2}}{1-Re^{i\delta}}
\label{eq:FPn}
\end{equation}
where $E_{\rm in}$ is the incident electric field.
It is clear that for $\delta = 2m\pi$
\begin{equation}
E_{\rm out} = \pm\left(\frac{T}{T+P}\right)E_{\rm in}
\label{Etrasm}
\end{equation}
and in the ideal case in which $P\ll T$ then $E_{\rm out} = \pm E_{\rm in}$ and $I_{\rm out} = I_{\rm in}$. If $\delta = 2m\pi + \delta'$ with $\delta'\ll(1-R)$, the output field will become
\begin{equation}
E_{\rm out} = E_{\rm in}T\frac{(1-R)\cos\frac{\delta'}{2}+i(1+R)\sin\frac{\delta'}{2}}{1+R^2 - 2R\cos\delta'} \approx E_{\rm in}\frac{T}{T+P}e^{i\frac{1+R}{1-R}\frac{\delta'}{2}}.
\end{equation}
The Fabry-Perot will amplify the small phase shift $\frac{\delta'}{2}$ by a factor $\frac{1+R}{1-R} \approx \frac{2}{1-R}$ with $R\lesssim1$. Now $\frac{\delta'}{2}$ represents the single pass phase deviation from $\pi$ acquired by the light between the two mirrors. Therefore the phase shift $\Delta\varphi$ of the output field will be [compare with equation~(\ref{eq:ellipticity})]
\begin{equation}
\Delta\varphi = \frac{1+R}{1-R}\frac{\delta'}{2} = \frac{2\pi}{\lambda}\left(\frac{1+R}{1-R}\right){\cal D}'.
\label{eq:sfasamento}
\end{equation}
The Fabry-Perot amplifies the optical path length variation of the light between the mirrors by a factor $N$ where
\begin{equation}
N = \frac{2}{1-R} = \frac{2{\cal F}}{\pi}
\end{equation}
with ${\cal F} = \pi/(1-R)$ being the \emph{finesse} of the cavity. Today values of $N \sim 10^5-10^6$ can be reached \cite{DellaValle2014OE}.

Remembering that the index of refraction is actually a complex quantity, then also $\delta$ will be. Its imaginary part will depend on the presence of an index of absorption $\kappa$: $i\bar{\delta} = i\frac{4\pi\kappa L}{\lambda}$. Assuming now $\delta = 2m\pi + i\bar{\delta}$, expression (\ref{eq:FPn}) will become
\begin{equation}
E_{\rm out} = \pm E_{\rm in}T\frac{e^{-\bar{\delta}/2}}{1-Re^{-\bar{\delta}}} \approx \pm E_{\rm in}\frac{T}{T+P}\left(1-\frac{1+R}{1-R}\frac{\bar{\delta}}{2}\right).
\label{eq:FPk}
\end{equation}
The relative amplitude reduction $\zeta = \bar\delta/2$ is also multiplied by $N$:
\begin{equation}
\zeta = \frac{1+R}{1-R}\frac{\bar\delta}{2} = \frac{2\pi}{\lambda}\left(\frac{1+R}{1-R}\right){\cal A}.
\end{equation}

Hence, given the relations (\ref{eq:ellipticity}) 
and (\ref{eq:rotation}), 
a Fabry-Perot can be used to amplify both an ellipticity and a rotation \cite{Rosenberg1964,Pace1994,Jacob1995APL,Zavattini2006APB}. Indeed, if the region between the mirrors is birefringent the Fabry-Perot can be described using the Jones formalism as
\begin{equation}
\mathbf{FP} = Te^{i\delta/2}\sum_{j = 0}^\infty \left[Re^{i\delta}\mathbf{BF}^2\right]^j\cdot\mathbf{BF} = Te^{i\delta/2}\left[\mathbf{I} - Re^{i\delta}\mathbf{BF}^2\right]^{-1}\cdot\mathbf{BF}.
\label{eq:jones_fp}
\end{equation}
taking into account the ellipticity accumulated within the Fabry-Perot. Applying $\mathbf{FP}$ to $\vec{E}_{\rm in} = E_{\rm in}{1\choose0}$ and considering $\delta = 2m\pi$ with $N\psi\ll1$ one finds
\begin{eqnarray}
\vec{E}_{\rm out} \approx E_{\rm in}\,\frac{T}{T+P}
\left(
\begin{array}{c}
1+iN\psi\cos2\vartheta\\
iN\psi\sin2\vartheta
 \end{array}
\right).
\end{eqnarray}
The optical path \emph{difference} between the two polarisation components is multiplied by the amplification factor $N$ of the Fabry-Perot and consequently the total accumulated ellipticity. Equation~(\ref{eq:ell_sig}) will therefore be modified to
\begin{equation}
I_\perp^{\rm (ell)} = I_\parallel\left[\eta(t)^2 + 2\eta(t)N\psi\sin2\vartheta(t) + ...\right].
\label{eq:ell_sig_FP}
 \end{equation}
The same is true in the case of rotation measurements. The differential absorption of the light along the $\parallel$ and $\perp$ directions is amplified by $N$ and therefore also the induced rotation $\phi$. The total rotation $N\phi$ is then transformed by the QWP: $N\phi \rightarrow -iN\phi$ (with the quarter-wave plate in Figure~\ref{fig:scheme} inserted oriented as in equation~(\ref{eq:QWP})). 
\begin{equation}
I_\perp^{\rm (rot)} = I_\parallel\left[\eta(t)^2 - 2\eta(t)N\phi\sin2\vartheta(t) + ...\right].
\label{eq:rot_sig_FP}
 \end{equation}
In both cases we have that
\begin{equation}
I_\parallel = I_{\rm in}\left(\frac{T}{T+P}\right)^2
\label{i_parallel}
\end{equation}

Experimentally two other considerations must be made during ellipticity measurements: the polariser and analyser have an intrinsic non zero extinction ratio $I_\perp/I_\parallel = \sigma^2$ and an ellipticity noise $\gamma(t)$ (which may or may not have contributions depending on $N$) is present between the two polarisers. As will be discussed in Section~\ref{sec:intrinsic noise}, this latter issue is dominated by the Fabry-Perot mirrors when the finesse is very high. Considering the case of no magnetically induced dichroism, therefore $\phi = 0$, equation~(\ref{eq:ell_sig_FP}) is modified to
\begin{equation}
I_\perp^{\rm (ell)} = I_\parallel\left[\sigma^2 + \eta(t)^2 + 2\eta(t)N\psi\sin2\vartheta(t) +  2\eta(t)\gamma(t) +...\right].
\label{eq:ell_sig_FP_tot}
 \end{equation}
The same is true in the case of rotation measurements. In the presence of a dichroism but in the absence of birefringence equation~(\ref{eq:rot_sig_FP}) will become 
\begin{equation}
I_\perp^{\rm (rot)} =I_\parallel\left[\sigma^2 + \eta(t)^2 - 2\eta(t)N\phi\sin2\vartheta(t) - 2\eta(t)\Gamma(t) + ...\right].
\label{eq:rot_sig_FP_tot}
 \end{equation}
where again one must also include the rotation noise $\Gamma(t)$.

By applying equation~(\ref{signal}) one then determines the total acquired ellipticity $\Psi = \frac{2{\cal F}}{\pi}\psi$ and rotation $\Phi = \frac{2{\cal F}}{\pi}\phi$ generated inside the cavity.

\subsection{\bf Fabry-Perot systematics}
\label{sec:Fabry-Perot_Systematics}

\subsubsection{Phase offset and cavity birefringence}
\label{sec:MirrorBirefringence}

An issue arises if the light inside the Fabry-Perot cavity does not satisfy exactly the condition $\delta = 2m\pi$ \cite{DellaValle2016EPJC}. Consider a Fabry-Perot with a birefringent medium between the mirrors as in equation~(\ref{eq:jones_fp}). Applying $\mathbf{FP}$ to $\vec{E}_{\rm in} = E_{\rm in}{1\choose0}$ with $\psi\ll\delta\ll(1-R)$ one finds
\begin{eqnarray}
\vec{E}_{\rm out} \approx E_{\rm in}T\frac{e^{i\delta/2}}{1-Re^{i\delta}}
\left(
\begin{array}{c}
1\\
 \frac{1+Re^{i\delta}}{1-Re^{i\delta}}i\psi\sin2\vartheta
 \end{array}
\right)
\approx E_{\rm out}e^{i\delta/2}
\left(
\begin{array}{c}
1 \\
\frac{(1+iN\delta)Ni\psi\sin2\vartheta}{1+N^2\sin^2\delta/2}
\end{array}
\right)
\label{eq:E_perp}
\end{eqnarray}
where
\begin{equation}
|E_{\rm out}|^2 = E_{\rm in}^2\left(\frac{T}{T+P}\right)^2\left(\frac{1}{{1+N^2\sin^2\delta/2}}\right)
\label{eq:i_out}
\end{equation}
It is apparent now that the component of $\vec{E}_{\rm out}$ along the $\hat{\rm Y}$ direction (see Figure~\ref{fig:frame}) is no longer only imaginary. It also has a real part indicating a rotation:
\begin{equation}
i\Psi = i\frac{N\psi}{1+N^2\sin^2\delta/2}\qquad\qquad\Phi = -\frac{N^2\psi\delta}{1+N^2\sin^2\delta/2}
\end{equation}
In general in the presence of both a rotation $\phi$ and an ellipticity $\psi$, $\vec{E}_{\rm out}$ along the $\hat{\rm Y}$ will have an imaginary part and a real part leading to
\begin{eqnarray}
i\Psi &=& i\frac{N}{1+N^2\sin^2\delta/2}\left[\psi+N\phi\delta\right]
\label{eq:EllipticityAiry}\\
\Phi &=& \frac{N}{1+N^2\sin^2\delta/2}\left[\phi-N\psi\delta\right].
\label{eq:RotationAiry}
\end{eqnarray}
There is therefore a cross talk between ellipticities and rotations in FP cavities not perfectly tuned to the maximum of a resonance. With a purely birefringent medium this cross talk mimics a dichroism and {\em vice versa}.

Until now we have assumed the mirrors of the cavity to be isotropic. In practice this is not the case. All mirrors present a small birefringence `map' over the reflecting surface \cite{Bouchiat:1982,Carusotto:1989,MicossiAPB1993}. A roundtrip of light inside the Fabry-Perot requires one reflection on each mirror where each mirror acts like a wave plate with a retardation $\alpha_{1,2}$ and an orientation. We recall from Reference~\cite{Brandi1997APB} that the effect of two birefringent wave plates is equivalent to that of a single wave plate with a phase difference $\alpha_{\rm EQ}$ given by 
\begin{equation}
\alpha_{\rm EQ}=\sqrt{(\alpha_1-\alpha_2)^2 +4\alpha_1\alpha_2\cos^2\phi_{\rm WP}}
\label{alfaEQ}
\end{equation}
where $\phi_{\rm WP}$ is the azimuthal angle of the second mirror's slow axis with respect to the first mirror's slow axis. Furthermore the slow axis of the equivalent wave plate has an angle $\phi_{\rm EQ}$ with respect to the slow axis of the first mirror given by
\begin{equation}
\cos2\phi_{\rm EQ}=\frac{\alpha_1/\alpha_2+\cos2\phi_{\rm WP}}{\sqrt{(\alpha_1/\alpha_2-1)^2+4(\alpha_1/\alpha_2)\cos^2\phi_{\rm WP}}}.
\label{phiEQ}
\end{equation}

During the operation of the polarimeter the input polarisation is aligned to either the slow or the fast axis of the equivalent wave plate of the two mirrors so as to be able to work at maximum extinction. In the following we will assume that the polarisation is aligned with the slow axis of the cavity.

With these considerations one can introduce, in the Jones' matrix of the birefringent Fabry-Perot cavity of equation~(\ref{eq:jones_fp}), the diagonal matrix describing the equivalent wave plate of the cavity mirrors:
\begin{equation}
\mathbf{M}=\left(\begin{array}{cc}e^{i\,\frac{\alpha_{\rm EQ}}{2}}&0\\0&e^{-i\,\frac{\alpha_{\rm EQ}}{2}}\end{array}\right).
\end{equation}
Assuming initially that $\psi = 0$, the polarisation auto-states of the Faby-Perot cavity are given by
\begin{equation}
\left(\begin{array}{c}\left[1-R\,e^{i(\delta+\frac{\alpha_{\rm EQ}}{2})}\right]^
{-1}\\
0\end{array}\right)\qquad{\rm and}\qquad
\left(\begin{array}{c}0\\
\left[1-R\,e^{i(\delta-\frac{\alpha_{\rm EQ}}{2})}\right]^{-1}\end{array}\right).
\label{pol_modes}
\end{equation}
The above equations show that the resonance curves of the two polarisation modes are no longer centred at $\delta=0$ and are separated by the quantity $\alpha_{\rm EQ}$: there is a frequency difference between the two resonances
\begin{equation}
\Delta\nu_{\alpha_{\rm EQ}} = \alpha_{\rm EQ}\frac{c}{4\pi d} = \frac{\alpha_{\rm EQ}}{2\pi}\nu_{\rm fsr}
\label{AiryDifference}
\end{equation}
where we have introduced the free spectral range of the cavity $\nu_{\rm fsr}$ corresponding to the frequency difference between two adjacent longitudinal Fabry-Perot modes: $\Delta\delta = 2\pi$.

In the PVLAS experiment the laser is phase-locked to the resonance frequency of the cavity by means of a feedback electronic circuit based on the Pound-Drever-Hall locking scheme \cite{Pound:1946,Drever:1983qsr} in which the error signal is carried by the light reflected from the cavity \emph{through the input polariser}. As a consequence, the laser is locked to the $\parallel$ polarisation auto-state of the cavity in equation~(\ref{pol_modes}) with $\delta=-\alpha_{\rm EQ}/2$ (corresponding to the $\hat{\rm X}$ axis) while the orthogonal component which will contain the ellipticity and/or rotation information is off resonance by $\delta = \alpha_{\rm EQ}$.

Two issues arise. Firstly, when analysing the extinguished beam one must take into account the fact that its power is reduced by the factor in equation~(\ref{eq:i_out}), with the substitution $\delta \rightarrow \alpha_{\rm EQ}$:
\begin{equation}
k(\alpha_{\rm EQ})=\frac{1}{1+N^2\sin^2(\alpha_{\rm EQ}/2)}\leq1
\label{kappa}
\end{equation}
with respect to the parallel polarisation power $I_\parallel$. By varying the input polarisation direction and the relative angle $\phi_{\rm WP}$ between the two mirrors' axes, it is possible to minimise the effect of the wave plates of the mirrors by aligning the slow axis of one mirror against the fast axis of the other corresponding to $\cos\phi_{\rm WP} = 0$ in equation~(\ref{alfaEQ}). This ensures that the two curves are as near as possible with $\alpha_{\rm EQ}$ at its minimum and $k(\alpha_{\rm EQ})$ at its maximum.

\begin{figure}[bht]
\begin{center}
\includegraphics[width=12cm]{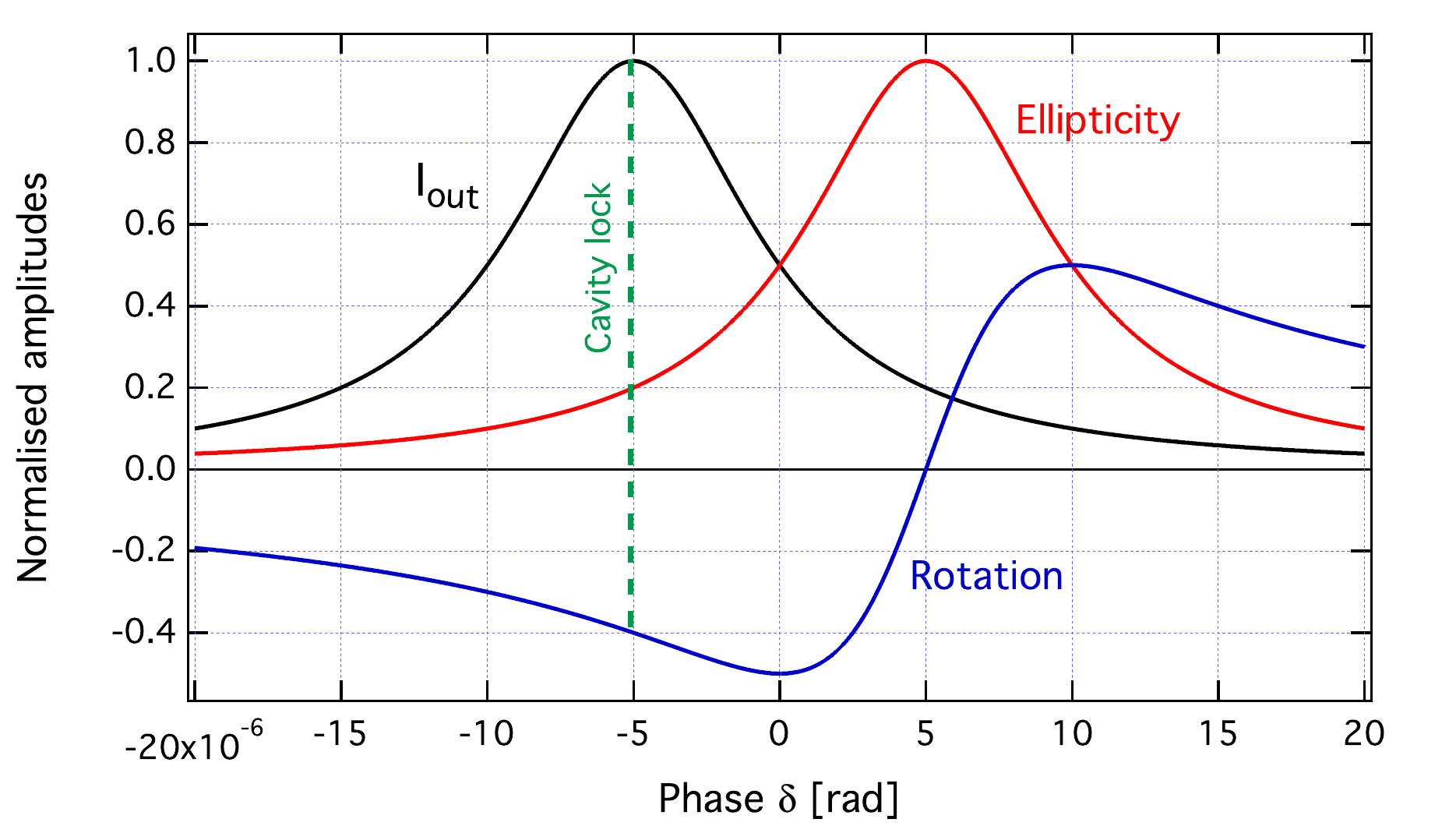}
\end{center}
\caption{Example of the transmitted power $I_{\rm out}$, and of the amplitudes of the ellipticity and of the rotation in the case of a pure birefringence, as functions of the Fabry-Perot cavity round-trip phase $\delta$, for $\alpha_{\rm EQ}=10^{-5}$~rad and $N=4\times10^5$. The Airy curves are normalised to unity and the rotation bears the same normalisation coefficient as the ellipticity. The transmitted power is centred at $\delta=-\alpha_{\rm EQ}/2$ whereas the other two curves at $\delta=\alpha_{\rm EQ}/2$. The value of the ellipticity at $\delta=-\alpha_{\rm EQ}/2$ is a factor $k(\alpha_{\rm EQ})=0.2$ smaller than the maximum [see equation~(\ref{kappa})].}
\label{BirefringentAiry}
\end{figure}

Secondly, in the presence of a birefringent and/or dichroic medium between the mirrors, analogously to equations~(\ref{eq:EllipticityAiry}) and (\ref{eq:RotationAiry}), a symmetrical mixing appears between rotations and ellipticities. In fact, the electric field at the exit of the cavity is
\begin{equation}
\vec{E}_{\rm out}(\vartheta,\delta)=E_{\rm in}\,\left[\mathbf{I}-Re^{i\delta}\,\mathbf{X}(\vartheta)\cdot\mathbf{M}\cdot\mathbf{X}(\vartheta)\right]^{-1}\cdot Te^{i\frac{\delta}{2}}\mathbf{X}(\vartheta)\cdot\left(\begin{array}{c}1\\0\end{array}\right)
\end{equation}
resulting in
\begin{equation}
{E}_{{\rm out},\perp}\approx E_{\rm out}\,\frac{1+iN(\frac{\delta}{2}-\frac{\alpha_{\rm EQ}}{4})}{1+N^2\sin^2(\frac{\delta}{2} - \frac{\alpha_{\rm EQ}}{4})}N(i\psi + \phi)\sin2\vartheta(t)
\label{eq:E_perp_alpha}.
\end{equation}
$E_{\rm out}$ is given by equation~(\ref{eq:i_out}) with the substitution $\delta \rightarrow \delta - \alpha_{\rm EQ}/2$ [compare with equation~(\ref{eq:E_perp})]. The behaviour of the transmitted power $I_\parallel$ and of the total ellipticity and  rotation $\Psi$ and $\Phi$ can be studied by changing $\delta$, which experimentally can be done with the feedback circuit by changing the error reference voltage. In the following this will be referred to as an `offset'. A calculated example of these three curves is shown in Figure~\ref{BirefringentAiry} for a pure birefringence, $N = 4\times10^5$ and $\alpha_{\rm EQ} = 10^{-5}$~rad. In Section~\ref{sec:meas_birif_cavity} we will present some measurements.

From equation~(\ref{eq:rot_sig}) the power at the detector for small $\alpha_{\rm EQ}$'s, and $R\approx1$, and with the laser locked to the top of the resonance for a polarisation along the $\parallel$ direction ($\delta = -\alpha_{\rm EQ}/2$) one finds
\begin{equation}
I_\perp^{\rm (ell)}(t)=I_\parallel\left\{\sigma^2+\eta(t)^2+2\eta(t) k(\alpha_{\rm EQ})\left[\Psi_\vartheta^{\rm Tot}(t)+\Phi_\vartheta^{\rm Tot}(t)N\frac{\alpha_{\rm EQ}}{2}\right]\right\},
\label{i_perp_ell}
\end{equation}
for the measurements of ellipticity, and
\begin{equation}
I_\perp^{\rm (rot)}(t)=I_\parallel\left\{\sigma^2+ \eta(t)^2+2\eta(t) k(\alpha_{\rm EQ})\left[-\Phi_\vartheta^{\rm Tot}(t)+\Psi_\vartheta^{\rm Tot}(t)N\frac{\alpha_{\rm EQ}}{2}\right]\right\},
\label{i_perp_rot}
\end{equation}
for rotation measurements. In the equations, $\sigma^2$ is the extinction ratio of the polarisers. To simplify the expressions we have included the spurious ellipticity and rotation noises $\gamma(t)$ and $\Gamma(t)$, respectively, generated in the polarimeter, in the quantities
\begin{equation}
\Psi_\vartheta^{\rm Tot}(t) = N\psi\sin2\vartheta(t) + \gamma(t) = \Psi\sin2\vartheta(t) + \gamma(t)
\label{eq:gammacavity}
\end{equation}
and 
\begin{equation}
\Phi_\vartheta^{\rm Tot}(t) = N\phi\sin2\vartheta(t) + \Gamma(t) = \Phi\sin2\vartheta(t) + \Gamma(t),
\label{eq:Gammacavity}
\end{equation}
and as usual
\begin{equation}
I_\parallel=I_{\rm in}\left(\frac{T}{T+P}\right)^2k(\alpha_{\rm EQ}).
\end{equation}

Both $\gamma(t)$ and $\Gamma(t)$ may have contributions from inside or outside of the cavity: $\gamma(t) = \gamma_{\rm cavity}(t)$+ $\gamma_{\rm other}(t)$ and $\Gamma(t) = \Gamma_{\rm cavity}(t)$+ $\Gamma_{\rm other}(t)$. Note that the terms generated inside the cavity include the factor $N$ just like a birefringence signal. Since the cavity mirrors are birefringent with $N\alpha_{\rm EQ}/2\lesssim 1$, the total spurious DC ellipticity can be zeroed just by carefully orienting the input polariser so as to compensate $\gamma({\rm DC})$ (or $\Gamma({\rm DC})$) with the cavity. This will be further discussed in Section~\ref{sec:wide_band_noise} regarding noise issues.


From equations (\ref{i_perp_ell}) and (\ref{i_perp_rot}) it is apparent that in the case $\phi = 0$, the ratio of the `spurious' rotation $\Phi^{\rm (spurious)} = \Psi N\frac{\alpha_{\rm EQ}}{2}$ to the `true' ellipticity $\Psi$ is
\begin{equation}
R_{\Phi',\Psi}=\frac{\Phi^{\rm (spurious)}}{\Psi} = N\frac{\alpha_{\rm EQ}}{2},
\label{R_Phi_Psi}
\end{equation}
hence allowing a direct determination of $\alpha_{\rm EQ}$. With the PVLAS apparatus this is done by taking measurements with and without the quarter-wave plate inserted.
In the same way one can define the `spurious' ellipticity to `true' rotation ratio in the case $\psi = 0$ and $\phi\ne0$:
\begin{equation}
R_{\Psi',\Phi}=\frac{\Psi^{\rm (spurious)}}{\Phi} = -N\frac{\alpha_{\rm EQ}}{2}.
\label{R_Psi_Phi}
\end{equation}

\subsubsection{Frequency response}
\label{sec:PolarisationDynamics}

When doing sensitive polarimetry with a modulated signal and a Fabry–Perot cavity, since the mirrors in practice always feature some birefringence, one has to pay attention to the different frequency dependences of the two terms between square parentheses appearing in each of the two equations~(\ref{i_perp_ell}) and (\ref{i_perp_rot}) \cite{Ejlli2018}. Let us first discuss the case of a pure birefringence and let us neglect the noises $\gamma(t)$ and $\Gamma(t)$. The rotating transverse magnetic field can be schematised as a rotating birefringent medium of length $L_B$. In this case the result of a polarimetric measurement will give both a `true' ellipticity and a `spurious' rotation:
\begin{eqnarray}
    \Psi_\vartheta(t)^{\rm (true)} &=& N\psi\sin2\vartheta(t)
    \label{truepsi}\\
    \Phi_\vartheta(t)^{\rm (spurious)} &=& N^2\psi\frac{\alpha_{\rm EQ}}{2}\sin2\vartheta(t)
    \label{spuriousrot}
\end{eqnarray}
In the absence of a cavity birefringence, $\alpha_{\rm EQ} = 0$, the time dependent signal given by equation~(\ref{truepsi}) is filtered as a first-order filter \cite{Uehara1995} and the `spurious' rotation is zero. With a non-zero value of $\alpha_{\rm EQ}$ and considering low rotation frequencies with respect to the cavity line width $\Delta\nu_{\rm c}$ the `true' ellipticity still behaves as a first order filter whereas the now present `spurious' rotation behaves as a second order filter. This can be understood in the following way \cite{Ejlli2018}: the `true' ellipticity is accumulated during the multiple reflections in the cavity of the $\perp$ polarisation. It will therefore have the same frequency dependence as the Fabry-Perot cavity itself. Differently, the `spurious' rotation is generated only by the presence of a $\perp$ radiation inside the cavity due to the birefringence which already behaves as a first order filter. Hence the `spurious' rotation will behave as a second order filter.

For rotation frequencies comparable or greater than $\Delta\nu_{\rm c}$, a detailed analysis shows that these signals deviate from the pure first and second order filters \cite{Ejlli2018}. The resulting expressions for the electric field at the output of the Fabry-Perot cavity are
\begin{equation}
E_{\rm out,\parallel}=E_{\rm in}
\end{equation}
\begin{equation}
E_{\rm out,\perp}=E_{\rm in}\,\frac{\psi T}{1-R}\left[\frac{e^{2i\vartheta(t)}}{1-Re^{i(\alpha_{\rm EQ}-2\omega_B\tau)}}+\frac{e^{-2i\vartheta(t)}}{1-Re^{i(\alpha_{\rm EQ}+2\omega_B\tau)}}\right]
\end{equation}
where $\tau = N\frac{d}{2c}$ is the lifetime of the cavity. 

We treat now separately the cases of the ellipticity and rotation measurements (in the case of a pure birefringence).
\begin{flushleft}
\emph{Ellipticity measurements}
\end{flushleft}
The orthogonal electric field after the analyser $\mathbf{A}$ is:
\begin{equation}
E_\perp^{\rm (ell)} = \left(\mathbf{A}\cdot \mathbf{MOD} \cdot \vec{E}_{\rm out}\right)_{\perp}=i \eta E_{\rm out,\parallel} + E_{\rm out,\perp}
\end{equation}
The power $I_\perp^{\rm (ell)}$ associated to $E_\perp^{\rm (ell)}$ is demodulated at the frequency of the ellipticity modulator $\nu_m$. We consider only the term linear in the product $\eta_0\psi$. Skipping cumbersome calculations, which can be found in Reference~\cite{Ejlli2018}, the phase of the ellipticity is 
\begin{eqnarray}
\varphi^{\rm (ell)}(\nu)=\tan ^{-1}\left[\frac{-R\left[\left(1+R^2\right)\cos\alpha_{\rm EQ}-2R\cos\delta_\nu\right]\sin\delta_\nu}{1+R^2\left(1+\cos2\alpha_{\rm EQ}+\cos2\delta_\nu\right)-R\left(R^2+3\right)\cos\alpha_{\rm EQ}\cos\delta_\nu}\right]
\label{eq:fase_first}
\end{eqnarray}
where $\delta_\nu=2\pi\nu\tau$ with $\nu$ the signal frequency. The amplitude of the ellipticity is:
\begin{eqnarray}
|\Psi^{\rm (true)}(\nu)|= \sqrt{\frac{2\psi^{2}\left[2-4R\cos\alpha_{\rm EQ}\cos\delta_\nu+R^2(1+\cos2\alpha)\right]}{\left[1+R^2-2R\cos(\alpha_{\rm EQ}-\delta_\nu)\right]\left[1+R^2-2 R\cos(\alpha_{\rm EQ}+\delta_\nu)\right]}}
\label{eq:amp_first}
\end{eqnarray}
As can be easily verified, for $\nu\ll\Delta\nu_{\rm c}$ and $\alpha_{\rm EQ} = 0$, one finds 
\begin{eqnarray}
    &&|\Psi^{\rm (true)}(\nu)|\rightarrow \frac{N\psi}{\sqrt{1+N^2\sin^2\delta_\nu/2}}
    \label{eq:FirstOrderAmpl}\\
    &&\varphi^{\rm (ell)}(\nu)\rightarrow -\frac{N}{2}\sin\delta_\nu.
    \label{eq:FirstOrderPhase}
\end{eqnarray}
which are the amplitude and phase of a first order filter as expected.

\begin{flushleft}
\emph{Rotation measurements}
\end{flushleft}
To detect rotations, the quarter-wave plate is inserted before the ellipticity modulator, thus transforming rotations into ellipticities. The orthogonal electric field after the analyser is:
\begin{equation}
E_\perp^{\rm (rot)} = \left(\mathbf{A}\cdot \mathbf{MOD} \cdot \mathbf{Q}\cdot\vec{E}_{\rm out}\right)_\perp={1+i\over\sqrt{2}} \eta E_{\rm out,\parallel} +{1-i \over \sqrt{2}}E_{\rm out,\perp}.
\end{equation}
Here the phase of the spurious rotation is
\begin{equation}
\varphi^{\rm (rot)}(\nu)=-\tan ^{-1}\left[\frac{\left(1-R^2\right)\sin\delta_\nu}{\left(1+R^2\right)\cos\delta_\nu-2R\cos\alpha_{\rm EQ}}\right]
\label{eq:fase_sec}
\end{equation}
and its amplitude is:
\begin{equation}
|\Phi^{\rm (spurious)}(\nu)|= \sqrt{\frac{4\psi^2R^2\sin^2\alpha_{\rm EQ}}{\left[1+R^2-2R\cos(\alpha_{\rm EQ}-\delta_\nu)\right]\left[1+R^2-2 R\cos(\alpha_{\rm EQ}+\delta_\nu)\right]}}
\label{eq:amp_sec}
\end{equation}

In this case the rotation amplitude disappears for $\alpha_{\rm EQ}=0$ (no `spurious' rotation is generated in a non birefringent cavity). Nevertheless, in the limit $N\frac{\alpha_{\rm EQ}}{2}\ll1$ the phase of the rotation $\varphi^{\rm (spurious)}(\nu)$ and the amplitude of the rotation $\Phi^{\rm (spurious)}(\nu)$ reduce this time to the phase and amplitude of a second order filter:
\begin{eqnarray}
    &&|\Phi^{\rm (spurious)}(\nu)|\rightarrow N\frac{\alpha_{\rm EQ}}{2}\frac{N\psi}{{1+N^2\sin^2\delta_\nu/2}}
    \label{eq:SecondOrderAmpl}\\
    &&\varphi^{\rm (spurious)}(\nu)\rightarrow -{N}\sin\delta_\nu.
    \label{eq:SecondOrderPhase}
\end{eqnarray}

\begin{figure}[bht]
\centering
\includegraphics[width=8cm]{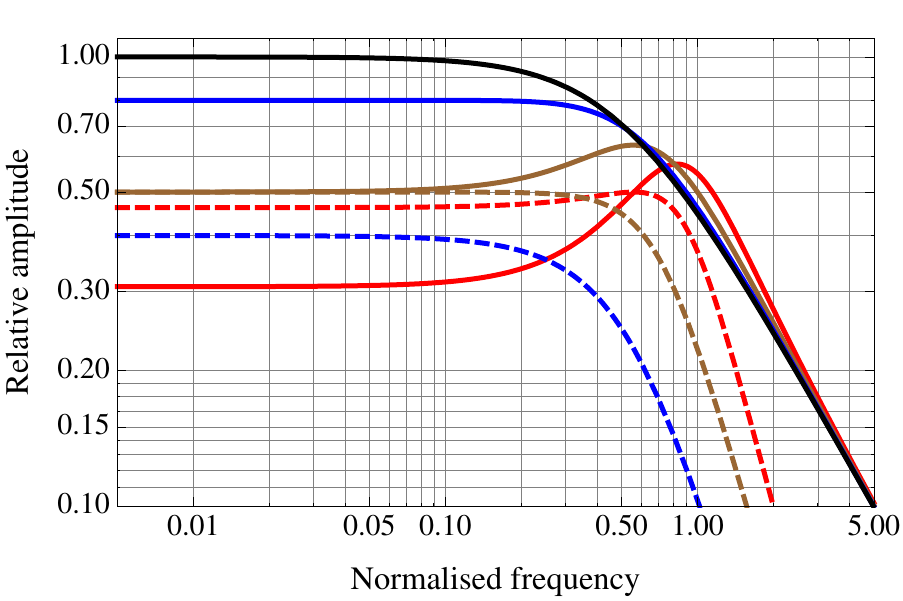}
\includegraphics[width=8cm]{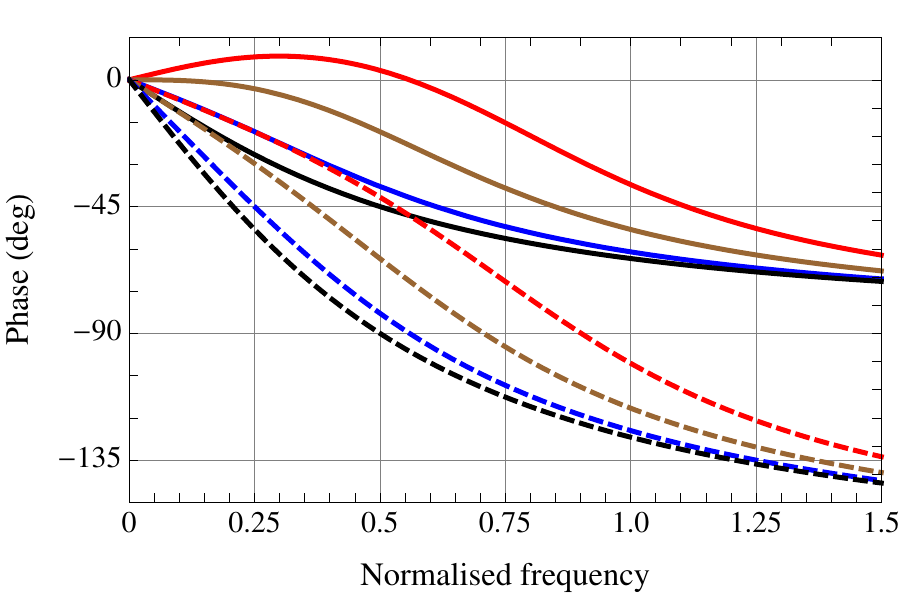}
\caption{Left: Calculated frequency response of the `true' ellipticity amplitude (continuous curves) using equation~(\ref{eq:amp_first}) and of the `spurious' rotation (dashed curves) using equation~(\ref{eq:amp_sec}) generated by magnetic birefringence in gas for ${\cal F}=662\times10^3$. The frequency scale is expressed in units of the cavity line-width $\Delta\nu_{\rm c}$; the vertical scale is normalised to the low-frequency `true' ellipticity amplitude given by in equation~(\ref{eq:FirstOrderAmpl}) (with $\alpha_{\rm EQ}=0$). The ellipticity curves are drawn for the values of the low-frequency ratio $R_{\Phi',\Psi}=0$ (black), 0.5 (blue), 1.0 (brown) and 1.5 (red); the `spurious' rotation curves have $R_{\Phi',\Psi}=0.5$, 1.0 and 1.5. Right: Calculated frequency response of the phase of the `true' ellipticity (continuous curves) and of the `spurious' rotation (dashed curves) considering the same values of $R_{\Phi',\Psi}$ as for the left panel. These curves have been arbitrarily chosen to start at zero phase and have negative slope. From reference \cite{Ejlli2018}, Figure 3.}
\label{fig:freq_resp_vs_alfa}
\end{figure}

In Figure~\ref{fig:freq_resp_vs_alfa} the amplitude and phase given by the equations~(\ref{eq:fase_first}), (\ref{eq:amp_first}), (\ref{eq:fase_sec}) and (\ref{eq:amp_sec}) for a few values of the parameter $R_{\Phi',\Psi}$ of equation~(\ref{R_Phi_Psi}) are shown.

\begin{figure}[bht]
\centering
\includegraphics[width=8cm]{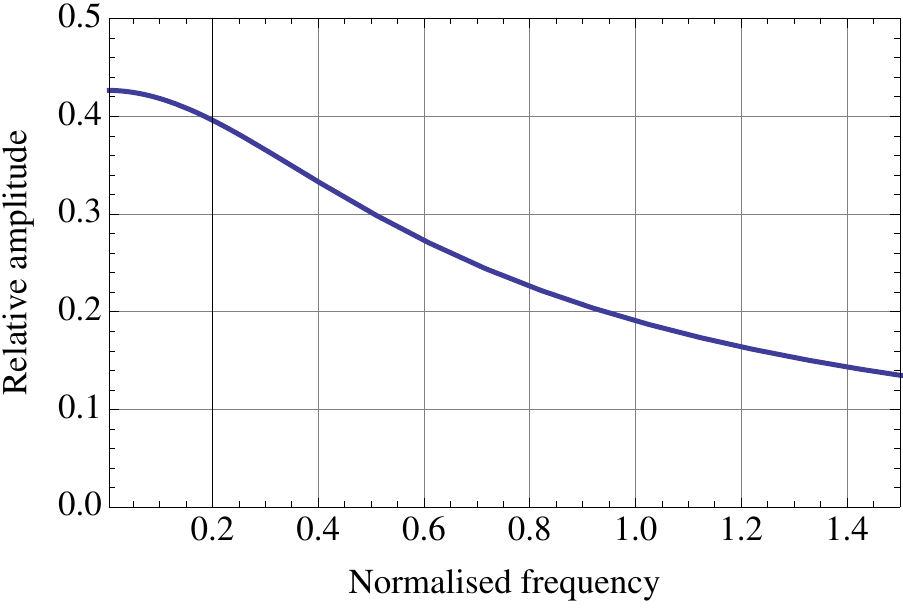}
\includegraphics[width=8cm]{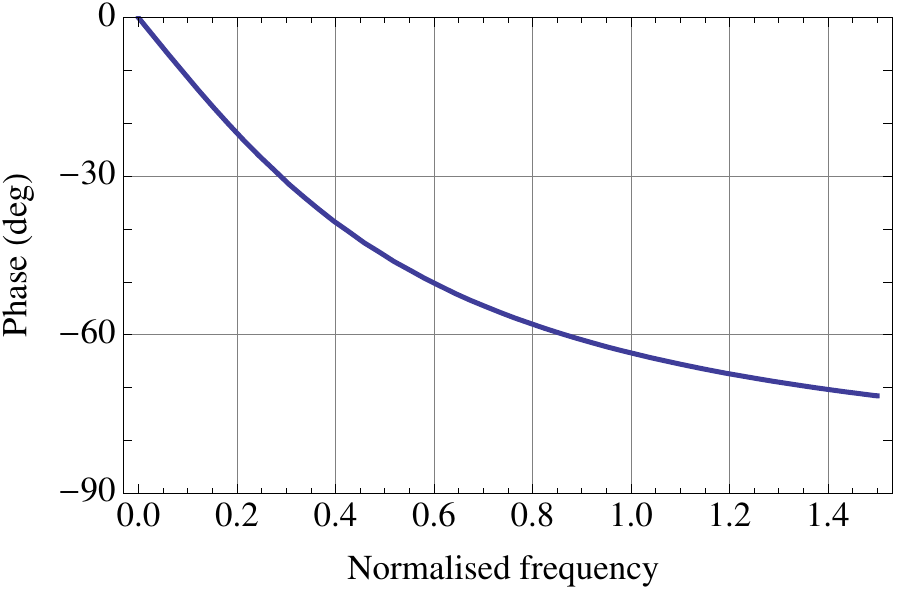}
\caption{Left: ratio of rotation and ellipticity in a dynamic FP cavity with birefringent mirrors in the presence of a pure birefringence. The frequency is expressed in units of the cavity line-width $\nu_{c}=\nu_{\rm fsr}/{\cal F}$ for ${\cal F}=660\times10^3$ and a value of $\alpha_{\rm EQ}=2~\mu$rad. Right: phase difference between rotation and ellipticity.}
\label{fig:ratio_vs_freq}
\end{figure}

The ratio $R_{\Phi',\Psi}$ of the `spurious' rotation to the `true' ellipticity [see equation~(\ref{R_Phi_Psi})] is now a function of frequency: 
\begin{equation}
{|\Phi^{\rm (spurious)}|\over |\Psi^{\rm (true)}|}=\sqrt{\frac{2 R^2\sin^2\alpha_{\rm EQ}}{2+R^2\cos2\alpha_{\rm EQ}+R^2-4 R\cos\alpha_{\rm EQ}\cos\delta_\nu}}
\end{equation}
A plot of the frequency dependence of $R_{\Phi',\Psi}$ is shown in Figure~\ref{fig:ratio_vs_freq} for ${\cal F}=660\,00$ and $\alpha_{\rm EQ}=2~\mu$rad. The static approximation holds up to $\sim5$~Hz ($\nu_B\approx2.5$~Hz). At higher frequencies the ratio is filtered as a first order filter.

This mathematics was developed for the case of a pure birefringence, but equally well applies to the case of a pure dichroism. The formulas are easily obtained from the ones already shown with the substitution of $\psi$ with $\phi$ and of $\Psi^{\rm (true)}$ with $\Phi^{\rm (true)}$ and of $\Phi^{\rm (spurious)}$ with $\Psi^{\rm (spurious)}$.

\subsection{\bf Calibration}
\label{sec:calibration-CM}

With all of the above attentions, the response of the apparatus can now be properly calibrated at the working frequency. Two quantities need to be extracted during calibration measurements: the correction factor $k(\alpha_{\rm EQ})$ and the absolute amplitude and phase calibration. The Cotton-Mouton effect \cite{Rizzo:1997cm}, namely the magnetic birefringence of gases, can provide this information: the ratio of the values of the rotation to ellipticity of equation~(\ref{R_Phi_Psi}) gives $\alpha_{\rm EQ}$; the observed ellipticity, corrected for $k(\alpha_{\rm EQ})$, define an absolute scale for the magneto-optical effects. To continuously monitor the mirror birefringence during the vacuum measurements, however, one can extract a value of $\alpha_{\rm EQ}$ by inducing a Faraday effect on the mirror and comparing the 'spurious' ellipticity to the 'true' rotation with the reciprocal equation~(\ref{R_Psi_Phi}).

The Cotton-Mouton -- or Voigt -- effect is perfectly analogous to the vacuum magnetic birefringence described by equation~(\ref{birif_B}), but is far more intense already at low gas pressures. The birefringence generated in a gas at pressure $P$ by a magnetic field $B_{\rm ext}$ is given by the expression
\begin{equation}
\Delta n = n_\parallel - n_\perp = \Delta n_u  B_{\rm ext}^2 P
\end{equation}
where $\Delta n_u$ is the unit birefringence generated by $P = 1$~atm and $B_{\rm ext} = 1$~T and hence the pressure is expressed in atmospheres and the magnetic field in tesla. Typical values of $\Delta n_u$ range from about $2.2\times10^{-16}$~T$^{-2}$atm$^{-1}$ for He to about  $-2.3\times10^{-12}$~T$^{-2}$atm$^{-1}$ for O$_2$ and to $\approx 10^{-11}$~T$^{-2}$atm$^{-1}$ for a few other simple molecules. In Table~\ref{tab:ListCM} we report some values of $\Delta n_u$ including the equivalent partial pressure $P_{\rm EQ}$ which would induce a birefringence equal to VMB. Other molecules and older values for the species listed here can be found in Reference~\cite{Rizzo:1997cm}.

\begin{table}[bht]
\centering
\begin{tabular}{c|r@{$\times$}l|r|c|c}
\hline\hline
Species & \multicolumn{2}{|c|}{$\Delta n_u$ (T$^{-2}$atm$^{-1}$)} & $\lambda$ (nm) & $P_{\rm EQ}$ (mbar) & Ref.   \\
\hline\hline
H$_2$   & $(8.28\pm0.57)$&$10^{-15}$  &  514 & $4.8\times10^{-7}$ & \cite{Rizzo:1997cm}         \\
\hline
He      & $(2.08\pm0.14)$&$10^{-16}$  & 1064 &$1.9\times10^{-5}$ & \cite{Bregant2009CM_He}       \\
\hline
        & $(2.22\pm0.16)$&$10^{-16}$  & 1064 &$1.8\times10^{-5}$ & \cite{PhysRevA.88.043815}        \\
\hline
        & $(2.7\pm0.3)$&$10^{-16}$    & 1064 &$1.5\times10^{-5}$ & \cite{DellaValle2014PRD} \\
\hline
H$_2$O  & $(6.67\pm0.21)$&$10^{-15}$  & 1064 &$6.0\times10^{-7}$ & \cite{DellaValle2014CPL}  \\
\hline
Ne      & $(6.9\pm0.2)$&$10^{-16}$    & 1064 &$5.8\times10^{-6}$ & \cite{Bregant2005CM_Ne,Bregant2009Erratum}       \\
\hline
CO      & $(-1.80\pm0.06)$&$10^{-13}$ &  546 &$2.2\times10^{-8}$ & \cite{Rizzo:1997cm}         \\
\hline
N$_2$   & $-(2.66\pm0.42)$&$10^{-13}$ & 1064 &$1.5\times10^{-8}$ & \cite{Chen2007,MEI2009216}  \\
\hline
O$_2$   & $(-2.29\pm0.08)$&$10^{-12}$ & 1064 &$1.8\times10^{-9}$ & \cite{Brandi1998JOSAB}        \\
\hline
Ar      & $(7.5\pm0.5)$&$10^{-15}$    & 1064 &$5.3\times10^{-7}$ & \cite{DellaValle2016EPJC}    \\
\hline
        & $(4.31\pm0.38)$&$10^{-15}$  & 1064 &$9.3\times10^{-7}$ & \cite{MEI2009216}           \\
\hline
CO$_2$  & $(-4.22\pm0.31)$&$10^{-13}$ & 1064 &$9.5\times10^{-9}$ & \cite{MEI2009216}           \\
\hline
Kr      & $(9.98\pm0.40)$&$10^{-15}$  & 1064 &$4.0\times10^{-7}$ & \cite{Bregant2004CM_KrXe,Bregant2009Erratum}       \\
\hline
Xe      & $(2.85\pm0.25)$&$10^{-14}$  & 1064 &$1.4\times10^{-7}$ & \cite{Bregant2004CM_KrXe,Bregant2009Erratum}       \\
\hline
        & $(2.59\pm0.40)$&$10^{-14}$  & 1064 &$1.5\times10^{-7}$ & \cite{Cadene2015JCP}        \\
\hline\hline
\end{tabular}
\caption{Unitary magnetic birefringence of common inorganic gaseous species. The equivalent partial pressure which would mimic a VMB signal are also reported.}
\label{tab:ListCM}
\end{table}

The  verification of the correct functioning of the apparatus is done as follows:
\begin{enumerate}
    \item the vacuum system is filled with a pressure $P$ of a pure gas;
    \item the finesse of the cavity is measured;
    \item using a low rotation frequency of the magnets, to avoid frequency response effects, both the ellipticity (QWP extracted) and the rotation (QWP inserted) are measured and the ratio $R_{\Phi',\Psi}$ is calculated. Using equation~(\ref{R_Phi_Psi}) and the value of the finesse the value of $\alpha_{\rm EQ}$ is determined;
    \item the phase of the ellipticity is determined;
    \item the amplitude correction factor $k(\alpha_{\rm EQ})$ using both the finesse value and $\alpha_{\rm EQ}$ is determined;
    \item the frequency response correction in amplitude is determined using equation~(\ref{eq:amp_sec});
    \item the Cotton-Mouton constant is extracted and compared to values found in literature.
\end{enumerate}

These measurements therefore give two calibration parameters: the amplitude and the phase of the ellipticity. The amplitude can be compared to theoretical calculations as well as to other experimental results, and calibrates the linear response of the polarimeter; the phase of the ellipticity is determined by the geometry and the electronic response of the apparatus. The phase of the ellipticity is directly related to the condition in which the polarisation of the light forms an angle $\vartheta = \pi/4$ with the magnetic field direction. It also depends on the settings of the lock-in filters and the frequency of the signal. Due to very slow drifts in the equivalent wave plate angle $\phi_{\rm EQ}$ this phase may change during a day of measurements (see Section~\ref{sec:cal_meas}).

The phase of the Cotton-Mouton effect (including its sign) defines what we call the physical phase for a field induced ellipticity: the vacuum magnetic birefringence must come with the same phase as the Cotton-Mouton measurement of a noble gas with $\Delta n_u >0$. In general, all the measured signals are projected onto both the physical and the non-physical axes. We explicitly note that the gas measurements are interpreted in terms of a pure birefringence. In fact, for gases, no linear dichroism is associated to a transverse magnetic field. A Faraday rotation which could result from a time variation of an eventual small longitudinal component of the magnetic field along the light path would appear at the frequency $\nu_B$ and not at $2\nu_B$.

\subsection{\bf Noise budget}
\label{sec:noise}

Let us determine the limiting sensitivity of such a polarimeter. Starting from equation~(\ref{signal}), if the rms intensity noise $S_{I_{-}}$ at the frequency $\nu_m-2\nu_B$ is uncorrelated to the rms intensity noise $S_{I_{+}}$ at $\nu_m+2\nu_B$ and $S_{I_{+}} = S_{I_{-}} \equiv S_{I_{\pm}}$, the demodulated rms intensity noise will be $S_{I_{{2\nu}_B}} = \sqrt{S_{I_{+}}^2 + S_{I_{-}}^2} = \sqrt{2}S_{I_{\pm}}$ due to the folding of the spectrum around $\nu_m$. Using equation~(\ref{signal}) the expected \emph{peak} ellipticity sensitivity $S_{\Psi_{{2\nu}_B}}$ of the polarimeter is 
\begin{equation}
S_{\Psi_{{2\nu}_B}}=\frac{S_{I_{{2\nu}_B}}}{I_\parallel\eta_0}.
\end{equation}
Several intrinsic effects contribute to $S_{I_{{2\nu}_B}}$, all of which can be expressed as a noise in the light power $I_\perp$. First consider the intrinsic rms shot noise spectral density due to the direct current $i_{\rm dc}$ in the detector
\begin{equation}
i^{\rm (shot)}=\sqrt{2e\,i_{\rm DC}}
\end{equation}
measured in ampere/$\sqrt{\rm hertz}$. Note that $i^{\rm (shot)}$ is independent on frequency.

According to equations~(\ref{eq:ell_sig_FP_tot}) or (\ref{eq:rot_sig_FP_tot}), the direct current inside the photodiode is $i_{\rm DC} = qI_\parallel\eta_0^2/2$, where $q$ is the efficiency of the detector PDE in units ampere/watt. Taking also into account the extinction ratio of the polarisers, which can be as low as $\sigma^2\lesssim10^{-7}$, this effect introduces an additional term in the detected DC power which is written as $I_\parallel\sigma^2$. This leads to an expression for the shot-noise spectral densities in the light power $I^{\rm (shot)}$ and in the ellipticity $S_\Psi^{\rm (shot)}$
\begin{equation}
I^{\rm (shot)}=\frac{i^{\rm (shot)}}{q} = \sqrt{\frac{2e\,I_\parallel}{q}\left(\sigma^2+\frac{\eta_0^2}{2}\right)}\qquad{\rm and}\qquad S^{\rm (shot)}_{\Psi}=\sqrt{\frac{2e}{qI_\parallel}\left(\frac{\sigma^2+\eta_0^2/2}{\eta_0^2}\right)}.
\end{equation}
Other effects contributing to the power and ellipticity noise spectral densities are the Johnson noise of the transimpedance $G$ of the photodiode
\begin{equation}
I^{\rm (J)}=\sqrt{\frac{4k_BT}{q^2G}},\qquad{\rm giving}\qquad S_\Psi^{\rm (J)}=\sqrt{\frac{4k_BT}{G}}\frac{1}{qI_\parallel\eta_0},
\end{equation}
the photodiode dark current
\begin{equation}
I^{\rm (dark)}=\frac{i_{\rm dark}}{q},\qquad{\rm with}\qquad S_\Psi^{\rm (dark)}=\frac{i_{\rm dark}}{qI_\parallel\eta_0},
\end{equation}
and the frequency dependent relative intensity noise $N^{\rm (RIN)}_{\nu}$ of the light emerging from the cavity
\begin{equation}
I^{\rm (RIN)}_{\nu}=I_\parallel\,N^{\rm (RIN)}_{\nu},
\end{equation}
giving
\begin{equation}
S^{\rm (RIN)}_{\Psi_{2\nu_B}}=N^{\rm (RIN)}_{\nu_m}\,\frac{\sqrt{(\sigma^2+\eta_0^2/2)^2+(\eta_0^2/2)^2}}{\eta_0}.
\label{eq:SRIN}
\end{equation}
In the last equation we consider that the contribution of $I_{\rm DC}$ and $I_{2\nu_B}$ in the Fourier spectrum add incoherently to the intensity noise at $\nu_m\pm2\nu_B$ with $\nu_B\ll\nu_m$.

\begin{figure}[bht]
\begin{center}
\includegraphics[width=12cm]{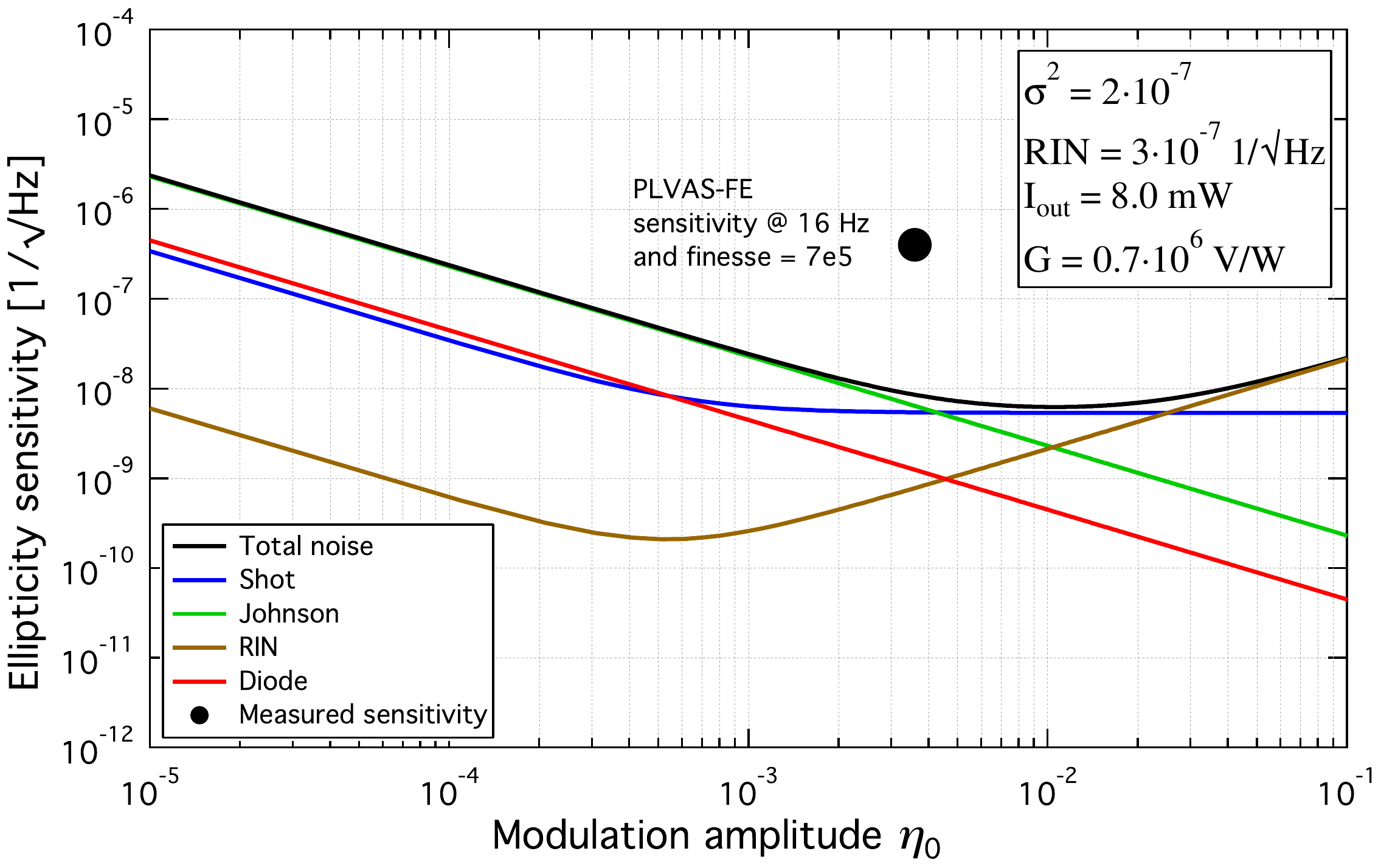}
\end{center}
\caption{Intrinsic peak noise components of the polarimeter as a function of the ellipticity modulation amplitude $\eta_0$. Superimposed is the PVLAS-FE peak ellipticity sensitivity at 16~Hz. }
\label{IntrinsicNoise}
\end{figure}

Figure~\ref{IntrinsicNoise} shows all the intrinsic contributions as functions of $\eta_0$ in typical PVLAS-FE operating conditions, with $q=0.7$~A/W, $I_\parallel=8$~mW, $\sigma^2=2\times10^{-7}$, $G=10^6~{\rm\Omega}$, $i_{\rm dark}=25~$fA$_{\rm rms}/\sqrt{\rm Hz}$, and $N^{\rm (RIN)}_{\nu_m}\approx3\times10^{-7}/\sqrt{\rm Hz}$ @ 50~kHz (resonance frequency of the PEM). The figure shows that the expected total ellipticity noise 
\begin{equation}
S_\Psi^{\rm (tot)} = \sqrt{{S_\Psi^{\rm (shot)}}^2+{S_\Psi^{\rm (J)}}^2+{S_\Psi^{\rm (dark)}}^2+{S_\Psi^{\rm (RIN)}}^2}
\label{eq:tot_noise}
\end{equation}
has a minimum for a modulation amplitude $\eta_0\approx10^{-2}$ close to shot-noise. 

In general a generic measured ellipticity noise $S_\Psi$ corresponds to a noise in optical path difference $S_{\Delta{\cal D}}$ accumulated between the two polarisers. The relation between ellipticity sensitivity $S_\Psi$ and optical path difference sensitivity $S_{\Delta{\cal D}}$ is
\begin{equation}
S_{\Delta{\cal D}} = \frac{\lambda}{N\pi}S_\Psi.
\label{eq:sensOPD}
\end{equation}

Considering the minimum estimated ellipticity noise budget $S_\Psi^{\rm (tot)}\approx8\times10^{-9}/\sqrt{\rm Hz}$ in Figure~\ref{IntrinsicNoise} this leads to $S_{\Delta{\cal D}} \approx 6\times10^{-21}\;{\rm m}/\sqrt{\rm Hz}$ having set $\lambda = 1064$~nm and $N \approx 450000$ (PVLAS-FE characteristics, see Section~\ref{Phases}).

The above noise considerations are also valid in the case of rotation measurements. One can therefore deduce $S_{\Delta{\cal A}}$ from a generic rotation noise $S_{\Phi}$ using the same equation as for $S_{\Delta{\cal D}}$:
\begin{equation}
S_{\Delta{\cal A}} = \frac{\lambda}{N\pi}S_\Phi.
\label{eq:sensAPD}
\end{equation}
In principle, if the noise sources are those shown in Figure~\ref{IntrinsicNoise}, the ellipticity and rotation sensitivities $S_{\Psi}$ and $S_{\Phi}$ should be equal. We note that the cross talk of ellipticity and rotation described by equations~(\ref{R_Phi_Psi}) and (\ref{R_Psi_Phi}) pertains also to noises: in the absence of a signal, an upper limit (namely an integrated noise floor) on one of the two quantities, translates into an upper limit also on the other one.

It is important to note here that the noises $S_{\Delta{\cal D}}$ and $S_{\Delta{\cal A}}$ in equations~(\ref{eq:sensOPD}) and (\ref{eq:sensAPD}) will improve increasing $N$ \emph{only} if $S_{\Psi}$ and $S_{\Phi}$ \emph{do not} depend on the equivalent number of passes $N$ of the cavity. 
The three noise sources presented above, namely shot-noise, Johnson noise and dark current noise, satisfy this condition depending in no way on the presence of the cavity. If, however, an ellipticity noise such as $\gamma_{\rm cavity}(t)$ [see equation~(\ref{eq:gammacavity}) and nearby text] originating from inside the cavity is present and if $\pi N\gamma_{\rm cavity}(t)/\lambda > S_\Psi^{\rm (tot)}$, no improvement in the sensitivity in $\Delta{\cal D}$ due to an increase in $N$ would be verified: both the magnetically induced signal and the noise would increase proportionally to $N$. The same argument is true in the presence of a rotation noise $\Gamma_{\rm cavity}(t)$ [see equation~(\ref{eq:Gammacavity})]. Furthermore, considering the case in which $S_\Psi = \pi N\gamma_{\rm cavity}(t)/\lambda$ dominates and considering $\alpha_{\rm EQ}\ne0$, the sensitivity $S_\Psi $ will also be affected by $k(\alpha_{\rm EQ})$. The integration time ${\cal T}$ to reach a given noise floor in ellipticity will be proportional to $1/k^{2}(\alpha_{\rm EQ})$.

Finally, the presence of a DC component of $\gamma(t)$ in equation~(\ref{eq:gammacavity}), indicated as $\gamma({\rm DC})$, may also contribute to an ellipticity noise $S_{\Psi}$ at the signal frequency $2\nu_B$ in the presence of a relative intensity noise $N^{\rm (RIN)}_\nu$ of $I_\parallel$ at $2\nu_B$. From equation~(\ref{i_perp_ell}) the condition for which this effect will not deteriorate $S_\Psi$ is that the product $N_\nu^{\rm (RIN)}\,\gamma({\rm DC})$ satisfies
\begin{equation}
   N_\nu^{\rm (RIN)}\,\gamma({\rm DC}) \ll S_\Psi.
\label{RIN_noise}
\end{equation}
In PVLAS-FE the relative intensity noise of $I_\parallel$ in the frequency range $10\div20\;$Hz is $N_\nu^{\rm (RIN)}$ $={S_{I_{\parallel}}}/{I_\parallel}\approx 10^{-4}/\sqrt{\rm Hz}$. By keeping $\gamma({\rm DC})\lesssim10^{-4}$ the contribution to $S_\Psi$ will be  $N_\nu^{\rm (RIN)}\,\gamma({\rm DC})\lesssim10^{-8}$. This was done with a very low-frequency feedback on the ellipticity at the output of the lock-in amplifier demodulating at $\nu_m$, by acting on the input polariser as discussed in more detail in Section~\ref{sec:meas_birif_cavity}.

In principle with the magnetic field parameter of the PVLAS-FE apparatus, $B_{\rm ext}^2 L_B \approx 10$~T$^2$m, the induced QED optical path difference to be measured is $\Delta{\cal D}^{\rm (QED)}  = 4\times10^{-23}$~m, a quantity measurable in ${\cal T} = \left(S_{\Delta{\cal D}}/\Delta{\cal D}^{\rm (QED)}\right)^2 \approx 6$~hours with a SNR = 1. Unfortunately shot-noise limited measurements with such high finesse values reaching such sensitivities in $S_{\Delta{\cal D}}$ have never been obtained due to the presence of $\gamma_{\rm cavity}(t)$ originating from the cavity mirrors. Indeed in the absence of the Fabry-Perot, shot-noise is achieved at output powers $I_\parallel\approx 10\;$mW. This subject will be treated in Section~\ref{sec:wide_band_noise}.

\subsection{\bf Gravitational antennae}
\label{sec:ligo}
The polarimetric scheme presented above is a differential measurement of the speed of light between two perpendicular polarisations. The measured sensitivity in optical path difference for a configuration with a very high finesse cavity obtained by several different experiments can be very approximately described by $S_{\Delta{\cal D}} \sim 3.5\times10^{-18}\nu^{-0.78}$ m/$\sqrt{\rm Hz}$ for frequencies $10^{-3}\;{\rm Hz} \le \nu \le 10^3\;{\rm Hz}$ as will be discussed in Section~\ref{sec:intrinsic noise} on page \pageref{sec:intrinsic noise}. A more detailed study of $S_{\Delta{\cal D}}$ for PVLAS-FE will be given in Section \ref{sec:intrinsic_noise}  on page \pageref{sec:intrinsic_noise}. There we will show that this noise originates from the Fabry-Perot mirrors.

\begin{figure}[b!]
\begin{center}
\includegraphics[width=14cm]{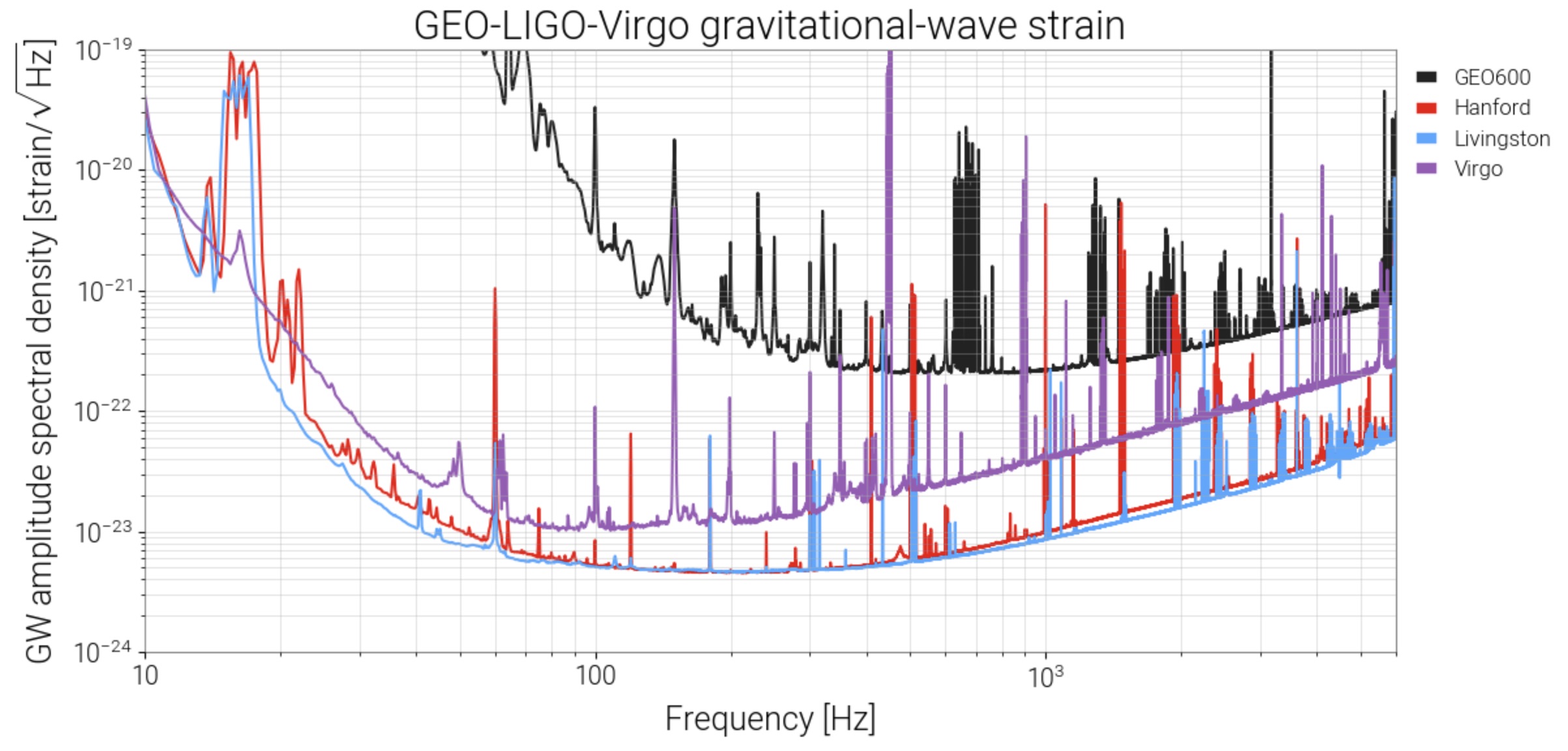}
\end{center}
\caption{Strain rms sensitivity curves of the GEO600, VIRGO, and LIGO gravitational antennae on Feb. 26, 2020 \cite{GWOpenscience}}
\label{strain}
\end{figure}

In the last few years impressive results have been obtained with Michelson-Morley type interferometers which measure the optical path difference between two spatially separated perpendicular beams. Indeed both LIGO detectors (Hanford, WA and Linvingston, LA) and the VIRGO detector (Cascina, PI, Italy) have detected gravitational waves and the Japanese interferometer KAGRA is on its way \cite{KAGRA}. It is interesting to compare the sensitivity in optical path length variation of the polarimetric technique and with the Michelson-Morley schemes.
At present the LIGO detectors have the better sensitivity in the strain $h$ of a gravitational wave shown in Figure~\ref{strain} \cite{LIGO}   . The strain of a gravitational wave is defined as $h = 2\Delta L/L$ where $\Delta L$ is the optical path \emph{amplitude} oscillation of \emph{each} of the two arms. Given a sinusoidal strain $h(t)$, the two arm lengths will change as $\Delta L_\parallel = \Delta L\cos\omega t$ whereas $\Delta L_\perp = -\Delta L\cos\omega t$. The interferometer measures the difference in optical path between the two arms: $\Delta L_\parallel - \Delta L_\perp = 2\Delta L = Lh$.

If the length of the two arms fluctuate with a \emph{peak} spectral density  
$S_{\Delta L_{\parallel,\perp}} = S_{\Delta L}$ the detected arm difference noise will be $\sqrt{S_{\Delta L_{\parallel}}^2+S_{\Delta L_\perp}^2}= \sqrt{2}S_{\Delta L}$ and the \emph{rms} sensitivity in $h$ will be $S^{\rm (rms)}_h = S_{\Delta L}/L$. In Figure~\ref{strain} are reported the rms measured sensitivities of the two Advanced LIGO detectors, the Advanced VIRGO detector and the GEO600 detector \cite{GWOpenscience}. The best rms sensitivity in strain $h$ measured by both the Advanced LIGO detectors is $S_{h}^{\rm (LIGO)} \approx 4\times10^{-24}/\sqrt{\rm Hz}$ at a frequency of about 200~Hz. The arm lengths of these detectors is $L_{\rm LIGO} = 4$~km and the rms sensitivity in differential optical path between the two arms is therefore $S_{\Delta\cal D}^{(\rm LIGO)} \approx S^{\rm (rms)}_{\Delta L} = 1.6\times10^{-20}$~m/$\sqrt{\rm Hz}$. 

Some thought has been put into the idea of using a gravitational wave interferometer to measure VMB \cite{GrassiStrini:1979kp,Calloni, Grote}. It is interesting to compare the capability in detecting VMB using polarimetry with respect to a gravitational wave interferometer. 
 
 Assume one arm of one of the LIGO detectors to be equipped with a magnet characterised by a $B^2L_B$ generating a VMB signal at 20~Hz. This can be done either by modulating the current in a magnet or by rotating a fixed field magnet. At this frequency the rms sensitivity is $S^{\rm (rms)}_h(20~{\rm Hz}) \approx 10^{-22}/\sqrt{\rm Hz}$. The deviation from unity of the index of refraction inside a magnetic field with a fixed direction will be $7A_eB^2_{\rm ext}$ or $4A_eB^2_{\rm ext}$ depending on the polarisation direction of the propagating light or will vary by $3A_eB^2_{\rm ext}$ in the case of a rotating fixed field magnet. The modulated current fixed direction configuration could allow the independent measurement of $n_\parallel$ and $n_\perp$ thereby determining independently the parameters $\eta_1$ and $\eta_2$ in the Lagrangian (\ref{eq:pM}). This would be of particular interest in the light of the Born-Infeld theory and in the separation of hypothetical axion signals from non linear electrodynamic effects in vacuum \cite{Calloni}.
 
 The difference in optical path between the two arms in the three cases will lead to an equivalent strain signal
 \begin{equation}
 h_{\rm equiv} = \frac{xA_eB^2L_B}{2L_{\rm LIGO}}
 \end{equation}
 where $x = \{7, 4, 3\}$.     
 Considering the peak sensitivity at 20~Hz, $S_h = \sqrt{2}S_h^{\rm (rms)}$, and assuming an integration time of ${\cal T} = 10^6$ s, the characteristics of the magnet to reach SNR = 1 would need to be
  \begin{equation}
B^2L_B \ge 2S_h^{\rm (20\,Hz)} \frac{L_{\rm LIGO}}{xA_e\sqrt{{\cal T}}} = \{121,\;212,\;283\}\;{\rm T}^2{\rm m},
 \end{equation} 
an extremely difficult configuration to construct also considering the diameter of the bore which would need to be used to avoid sensitivity issues. 

It is also interesting to compare the peak sensitivities $S_{\Delta{\cal D}}$, at 20~Hz, of PVLAS-FE and LIGO:
\begin{eqnarray}
&&S_{\Delta{\cal D}}^{\rm (PVLAS)} \approx 3.5\times 10^{-19}\;\frac{\rm m}{\sqrt{\rm Hz}}\quad @ \quad 20\;{\rm Hz}\\
&&S_{\Delta{\cal D}}^{\rm (LIGO)} = L_{\rm LIGO}S_h \approx 5.6\times 10^{-19}\;\frac{\rm m}{\sqrt{\rm Hz}}\quad @ \quad 20\;{\rm Hz}.
\end{eqnarray}
The PVLAS-FE peak sensitivity curve for $S_{\Delta{\cal D}}$ can be found in Figure~\ref{fig:intrinsic_noise} on page~\pageref{fig:intrinsic_noise}.

\section{PVLAS forerunners}
\label{Phases}

After the original paper by E. Iacopini and E. Zavattini (1979) \cite{Iacopini:1979ci} proposing an optical polarimetric method to measure vacuum magnetic birefringence, several attempts have been performed by several groups world wide to observe this minute effect \cite{Iacopini:2301707,Semertzidis:1990qc,Ni:1991iu,BAKALOV1994,Lee:1995wp,Ni:1996dg,Pengo1998QED,Battesti2008,Fan2017}. Here we will trace the attempts and difficulties encountered which have led to the final PVLAS-FE setup and hence to the results presented in detail in this paper on this intriguing quantum mechanical effect which still needs a direct laboratory confirmation. The CERN setup (1980 - 1983) \cite{Iacopini:2301707,Iacopini:1980ws,Carusotto:2301709} was a precursor attempt in understanding the method and where the principle difficulties could be. It was followed by the BFRT (Brookhaven, Fermilab, Rochester, Trieste) collaboration (1985 - 1993) \cite{Semertzidis:1990qc,Cameron1993PRD}, experiment run at the Brookhaven National Laboratories (BNL), and had the principle goal of putting limits on axion coupling to two photons through the Primakoff effect. As discussed in Section~\ref{sec:axions}, axions coupling to two photons also generate a birefringence but, contrarily to QED, also a dichroism. The detection of VMB due to QED was out of reach but the experiment was the first complete attempt on putting limits on magnetically induced vacuum magnetic birefringence and dichroism. The modulation of the effect was obtained by ramping the current of two superconducting magnets available at the time at BNL. In both these precursor attempts a non resonant multipass cavity was used to increase the effective optical path length within the magnetic field.

Following these two precursor setups which indicated the difficulties to be overcome, the PVLAS (Polarizzazione del Vuoto con Laser) collaboration formed, financed by the Istituto Nazionale di Fisica Nucleare (INFN), Italy, with the ambitious goal of measuring VMB \cite{BAKALOV1994,Bakalov1998QSO}. The real novelties in the PVLAS-LNL setup were two: the use of a rotating superconducting magnet in persistent mode \cite{Bakalov1998HI} to increase by an order of magnitude the frequency modulation of the effect and the use of a resonant Fabry-Perot cavity \cite{BAKALOV1994,Cantatore1995RSI,DeRiva1996RSI} to increase the number of equivalent passes through the magnetic field. A lack of detailed debugging of the experiment, limited mainly by the liquid helium availability to run the superconducting magnet, limited the final result of this attempt.

This led the PVLAS collaboration to move towards rotating permanent magnets \cite{Zavattini2012IJMPA,Zavattini2013JPCS,DellaValle:2013dwa,DellaValle2013NJP}: the PVLAS-FE setup. Another factor 10 in modulation frequency was gained thanks to the higher rotation frequency of the magnets and very detailed debugging of systematics was performed leading to the present best limit on VMB, less than a factor ten from its first detection.
 
 In the following sections we will describe the different experimental setups which led to PVLAS-FE indicating their differences and limits. The PVLAS-FE setup will then be described in particular detail in Section~\ref{sec:PVLASFE}; systematics-hunting and wide-band noise issues will be presented in Section~\ref{sec:commissioning}.

\subsection{\bf CERN proposal: 1980-1983}
\label{CERN}

\begin{figure}[bht]
\begin{center}
\includegraphics[width=12cm]{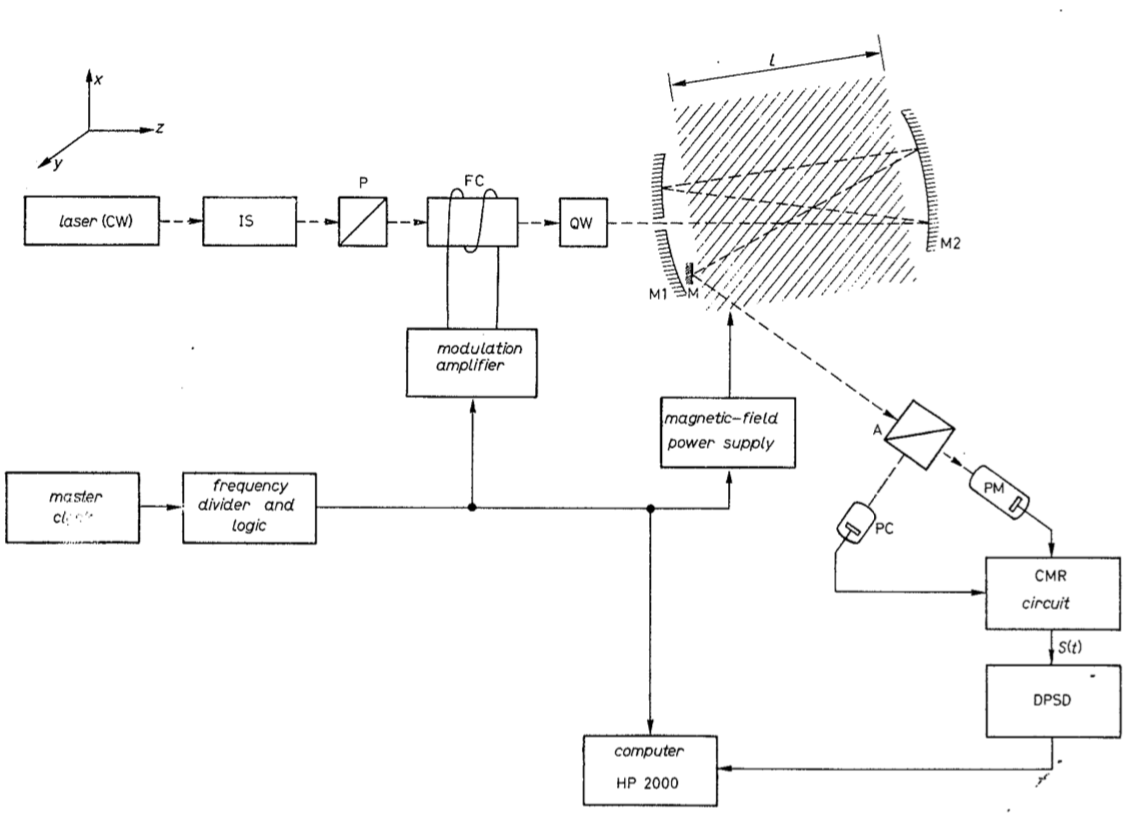}
\end{center}
\caption{Figure of the original polarimetric scheme taken from Reference \cite{Iacopini:1980ws}, Figure 1. A laser beam linearly polarised by P passes through a rotation modulator (Faraday modulator) FC and a quarter-wave plate QW transforming rotations into ellipticities. M1 and M2 compose a multi-pass cavity where the magnetic field should be present. A small mirror M extracts the beam from the cavity sending it to the analyser A.}
\label{fig:D2_scheme}
\end{figure}

The first proposal to measure VMB was presented at CERN in 1980 with the title ``Experimental Determination of Vacuum Polarization Effects on a Laser Light-beam Propagating in a Strong Magnetic Field" \cite{Iacopini:2301707,Iacopini:1980ws}. The original optical scheme of the polarimeter has remained substantially unchanged since. The scheme of the 1980-1983 setup is shown in Figure~\ref{fig:D2_scheme} and was used to measure the Cotton-Mouton effect of nitrogen and oxygen \cite{Carusotto:136650,Carusotto:1983tt}. In these first measurements the field intensity was varied in time. In the proposal the idea of wobbling the magnetic field direction by $\pm 25^\circ\;$is also mentioned. A laser beam linearly polarised passes through a rotation modulator (Faraday modulator) and a quarter-wave plate. When correctly aligned the quarter-wave plate transforms rotations into ellipticities. Two mirrors compose a multi-pass cavity where the magnetic field was to be present. The beam was sent into the cavity through a hole in the front mirror, generating with its reflections a Lissajous pattern over the mirror surfaces. Each reflection was spatially separated from the previous one in such a way that the beam could be extracted by a small mirror sending it to the analyser. Both the extinguished and transmitted powers were then recorded. The induced ellipticity was extracted using the heterodyne detection discussed in Section~\ref{sec:polarimetry}.

During the first Cotton-Mouton measurements, without a cavity, an ellipticity sensitivity $S_\psi = 1.5\times10^{-7}/\sqrt{\rm Hz}$ was obtained at the signal frequency $\nu_B = 0.397$~Hz while the expected noise should have been $S_\psi = 1.6\times10^{-8}/\sqrt{\rm Hz}$.

\begin{figure}[bht]
\begin{center}
\includegraphics[width=9cm]{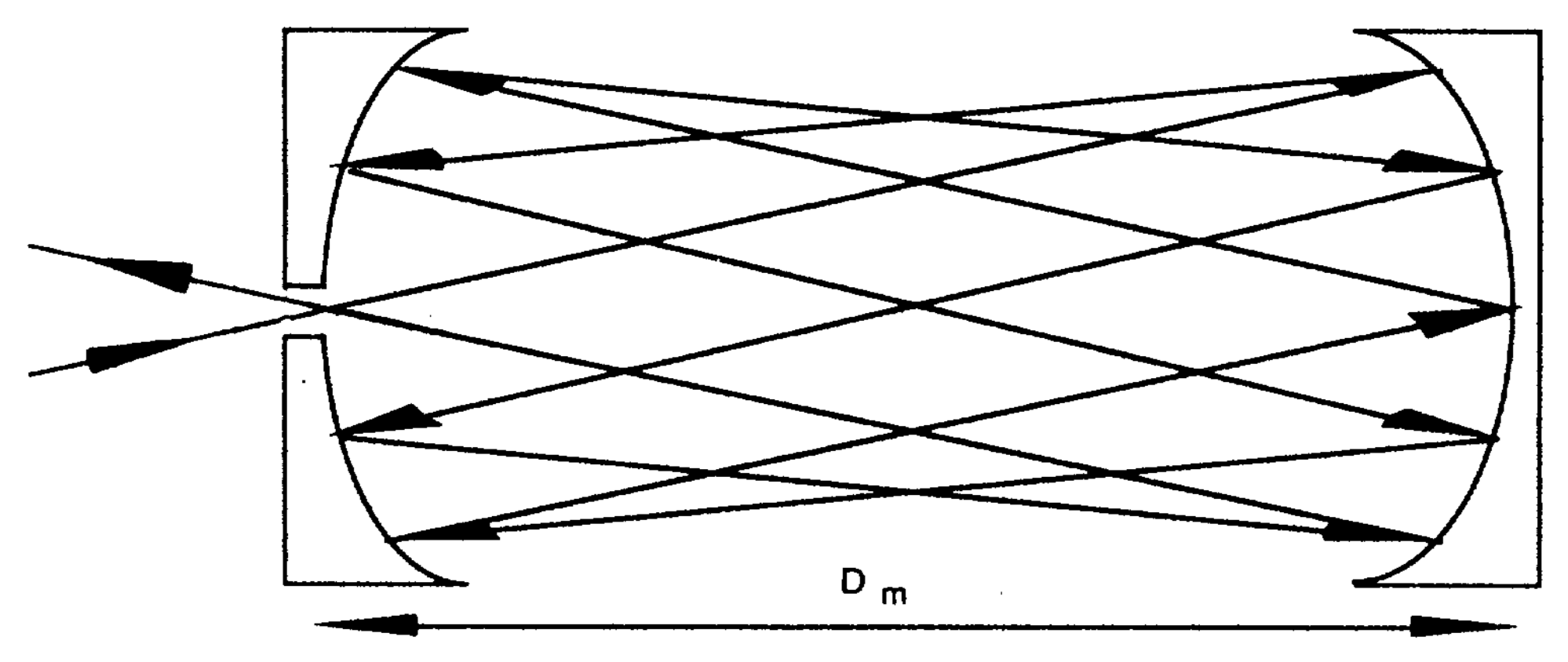}
\includegraphics[width=7cm]{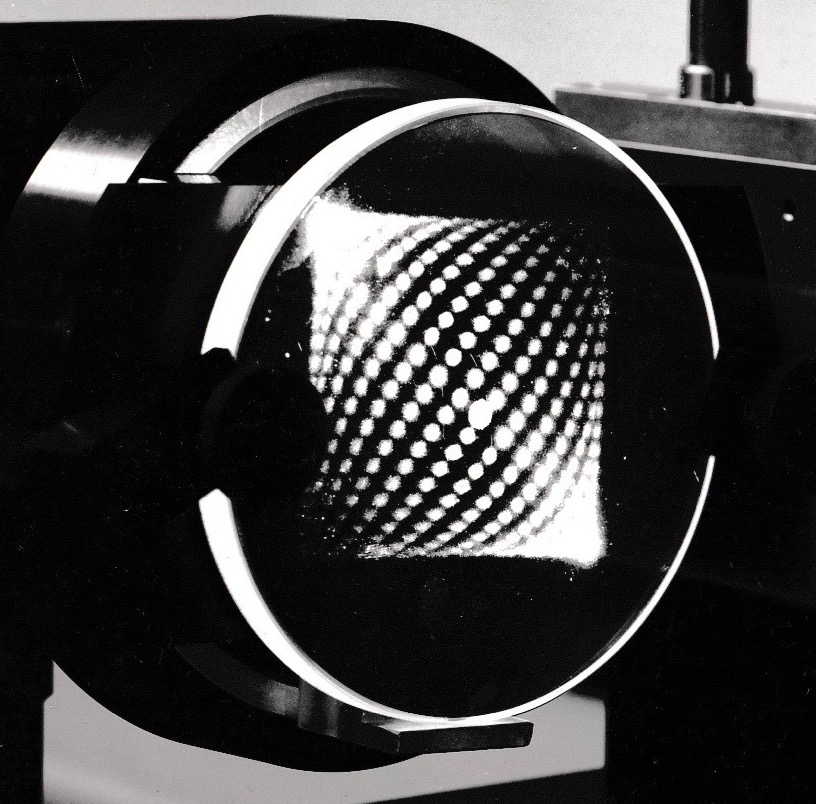}
\end{center}
\caption{Left: Scheme of the multi-pass cavity. Right: Lissajous pattern on the surface of one of the multi-pass mirrors. Photograph courtesy of E. Iacopini.}
\label{fig:Lissajous}
\end{figure}

After the first tests, in 1983 an Addendum to the proposal was presented with some improvements \cite{Carusotto:2301709}.
The main differences between the original CERN scheme presented in the Proposal D2 and its Addendum
can be summarised as follows:
\begin{itemize}
\item As in the Proposal, the optical cavity was a non resonant multi-pass cavity. In the Addendum the extraction of the beam changed: after a given number of reflections the beam could exit the cavity through the entrance hole. A scheme of such a cavity and a typical Lissajous pattern are shown in Figure~\ref{fig:Lissajous}. In the right panel the entrance and exit hole can be seen near the center of the mirror. The number of reflections was limited to a few hundreds and was determined substantially by the dimensions of the mirrors.
\item In Proposal D2 the modulator for heterodyne detection was placed before the multi-pass cavity as is shown in Figure~\ref{fig:D2_scheme}. It consisted of a Faraday cell powered by an alternating current followed by a quarter-wave plate. The Faraday cell generated a periodic rotation of the polarisation direction which was then transformed into a modulated ellipticity by means of the quarter-wave plate whose axis was aligned to the input polarisation.

During the second phase reported in the D2 Addendum the modulator was placed on the output beam coming from the multi-pass cavity. In this way the polarisation of the beam reflecting on the mirrors was fixed (to within the magnetically induced modulation). The quarter-wave plate was also substituted to render the ellipticity-rotation transformation less dependent on the wavelength of the light: the output beam was first reflected off of a gold mirror with an incident angle $\approx 73^\circ$ such that an ellipticity is completely transformed into a rotation and {\em vice versa}.  The magnetically induced ellipticity was therefore transformed into a rotation. The Faraday cell generating the known rotation would then beat with the `transformed ellipticity' signal generated by the magnetic field.
\item In the D2 Addendum, four possible solutions were presented for the magnetic field configuration including two different field intensity modulations at about 20 mHz, rotating the magnet over a $\pm 45^\circ$ angle with the magnet in a vertical position and finally four coils mounted in a dual dipole configuration such that powering the two dipoles with a phase difference of $90^\circ$ would generate a rotating constant intensity field.
\end{itemize}

Successively to the Addendum, new Cotton-Mouton measurements were performed on noble gases using a rotating 0.4~T normal conducting magnet \cite{Carusotto:1983tt}.

From their experience the proposers estimated that the sensitivity in ellipticity $S_\Psi$ (and hence in optical path difference $S_{\Delta{\cal D}}$) which could be reached, considering a multi-pass cavity with $N = 600$ and a wavelength $\lambda = 488$~nm, was
\begin{equation}
S_\Psi = 4\times 10^{-9}/{\sqrt{\rm Hz}} \quad \Rightarrow \quad S_{\Delta{\cal D}} = \frac{\lambda}{\pi N}S_\Psi = 10^{-18}\;{\rm m}/{\sqrt{\rm Hz}}
\end{equation}
a factor about 4 above the shot-noise limit. The source of this noise according to the proposers was due to beam stability issues. This sensitivity was never actually reached at CERN.

The original project to measure VMB required two 6 m long, 8 T magnets which needed to be designed. A first 1~m long prototype dipole magnet was built by Mario Morpurgo \cite{Morpurgo:1720288} and tested reaching, at the time, a record 7.6 T field intensity. This small prototype magnet was then used as the PVLAS-LNL rotating magnet as will be discussed in Section~\ref{sec:mag_morpurgo}.

The CERN activity did not go farther.

\subsection{\bf BFRT: 1986 - 1993}
\label{BFRT}

At the beginning of the '80s the existence of very light axions coupling to two photons was suggested by M. Dine et al. \cite{Dine:1981rt} and A. R. Zhitnitsky \cite{Zhitnitsky:1980tq}. As was discussed in Section  \ref{sec:axions} the existence of such an `invisible' axion, as it was often referred to, was shown to generate both a birefringence and a dichroism in the presence of an external magnetic field \cite{Maiani:1986md,Raffelt:1987im} and could therefore be searched for using polarimetric techniques just in the same way as searches for VMB due to vacuum fluctuations. At the time the masses of such axions were estimated to be in the range $10^{-6}\;{\rm eV} < m_a < 10^{-3}\;{\rm eV}$, an interesting window since for a magnetic field length $L_B$ and optical photons such that $\omega\gg {L_Bm_a^2}/4$ the induced dichroism is independent of mass [compare equation~(\ref{eq:assionedic})]. Other than searching for this fundamental boson, this experiment was a perfect test-bench for understanding the limits of the polarimetric scheme for future VMB measurements.

The BFRT collaboration attempted a search for axions in this mass range using the same optical scheme presented a few years before at CERN \cite{Iacopini:2301707,Carusotto:2301709}. This experiment was the first real attempt to study and determine the sensitivity of a polarimeter based on a multi-pass cavity together with heterodyne detection. The experiment was performed at the Brookhaven National Laboratories (BNL) where two superconducting magnets were available. The fields were modulated from a central value of $B_0 = 3.25$~T with an amplitude of $B_\Delta = 0.62$~T at a frequency of 20 mHz. Each magnet was 4.4 m long for a total magnetic field equivalent parameter of $(B_{\rm ext}^2 L_B)_{\rm equiv} = 2\times2B_0B_\Delta L_B = 35.5$~T$^2$m.

\begin{figure}[bht]
\begin{center}
\includegraphics[width=14cm]{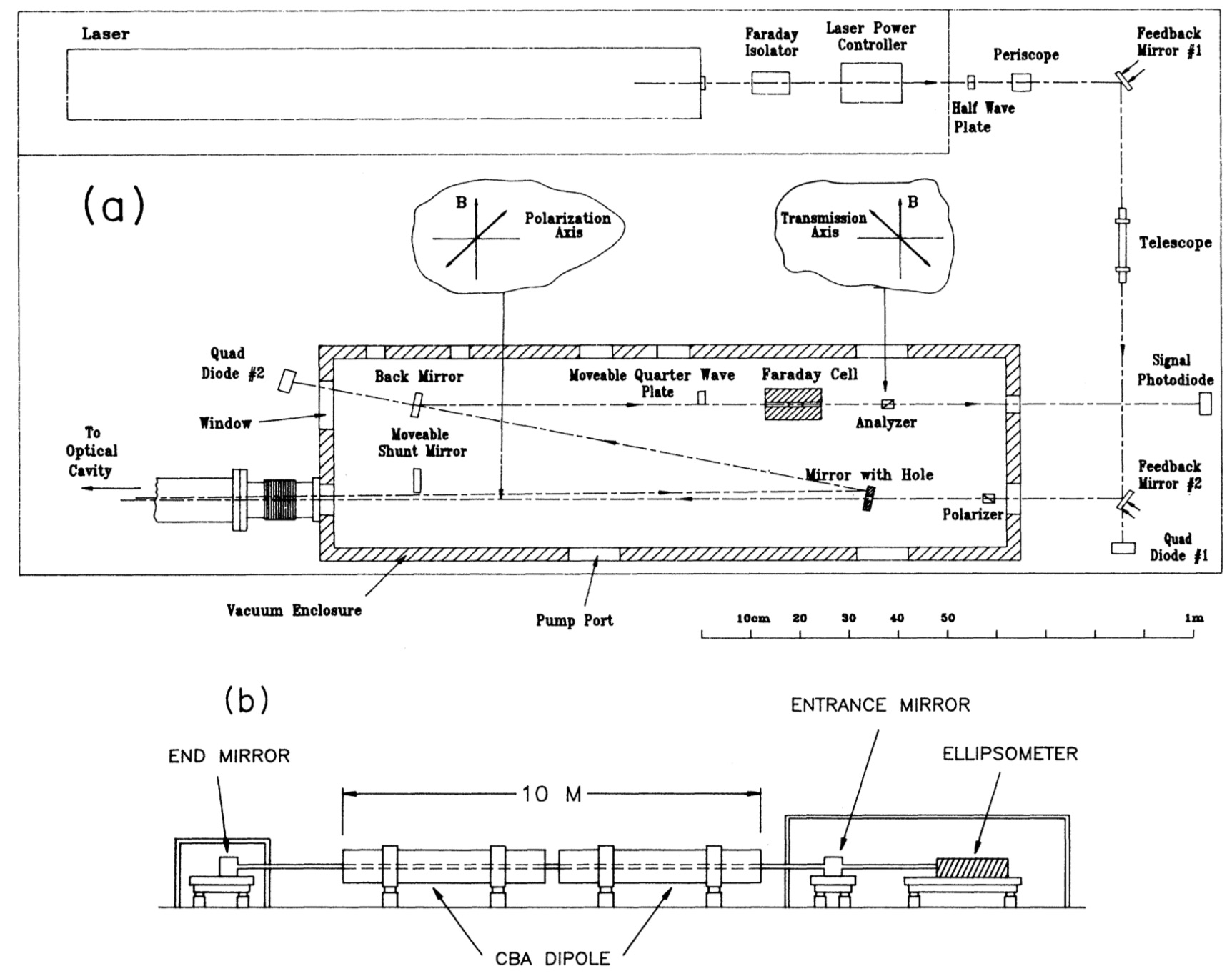}
\end{center}
\caption{Scheme of the BFRT experiment. Panel (a): scheme of the optics. Panel (b): general layout of the experiment. Figure taken from Reference  \cite{Cameron1993PRD}, Figure 4.}
\label{fig:BFRT_scheme}
\end{figure}

Several improvements were implemented among which a feedback loop on the steering mirrors to compensate the movement of the beam due to the varying magnetic field. Indeed the alignment of the mirrors of the cavity changed during the magnetic field cycles. In this setup the whole polarimeter, shown as a dashed box in Figure~\ref{fig:BFRT_scheme}(a) \cite{Cameron1993PRD}, was in an ultra-high vacuum chamber. The Faraday cell was water cooled to limit the thermal stress induced noise in the Faraday crystal and a QWP was used before the Faraday cell to transform rotations into ellipticities.

\begin{table}[bht]
\centering
\begin{tabular}{c|c|c|c}
\hline\hline
& number of passes $N$ & noise $S\;\left[\frac{1}{\sqrt{\rm Hz}}\right]$ & normalised noise $S'/N\;\left[\frac{1}{\sqrt{\rm Hz}}\right]$\\
\hline\hline
& 578   & \quad$1.5\times10^{-6}$ & \quad$2.6\times10^{-9}$ \\
\cline{2-4}
ellipticity & 34   &\quad$7.9\times10^{-8}$ &\quad$2.2\times10^{-9}$ \\
\cline{2-4}
& shunt  & \quad$2.4\times10^{-8}$ & \\
\hline\hline
& 254      & \quad$6.7\times10^{-8}$ & \quad$2.4\times10^{-10}$ \\
\cline{2-4}
rotation & 34      & \quad $3.3\times10^{-8}$  & \quad$6.3\times10^{-10}$ \\
\cline{2-4}
& shunt     & \quad$2.5\times10^{-8}$ & \\
\hline\hline
\end{tabular}
\caption{Ellipticity and rotation sensitivities of the BFRT experiment in various experimental conditions. In the fourth column the noise with the shunt mirror has been subtracted in quadrature from the measured noise with the cavity: $S' = \sqrt{S^2-S^2_{\rm shunt}}$. The rotation and ellipticity noises with the shunt mirror are taken from tables IV and V of Reference \cite{Cameron1993PRD}, respectively; the other values from table II of the same paper.}
\label{tab:BFRT_sens}
\end{table}

In Table~\ref{tab:BFRT_sens} are reported the ellipticity and rotation sensitivities of the BFRT polarimeter for different values of $N$ with the magnetic field ON \cite{Cameron1993PRD}. It is interesting to note that there is a contribution to the noise in ellipticity proportional to the number of passes in the cavity (fourth column) whereas for the rotation noise this is not the case. This fact indicates a dominant ellipticity $\gamma_{\rm cavity}(t)$ noise at $\nu \approx 20$~mHz due to the presence of the cavity. During rotation measurements, therefore, an increase in $N$ led to an improvement in rotation sensitivity \emph{per pass} whereas during ellipticity measurements this was not the case. Furthermore the rotation sensitivity \emph{per pass} was significantly better than during ellipticity measurements. Finally, with a shunt mirror inserted, used to bypass the cavity, both ellipticity and rotation noises were the same (`shunt' lines in Table~\ref{tab:BFRT_sens}) but significantly above shot-noise given the powers used $I\sim0.1\div1$~W. Given that all optical elements including the mirrors have a small non uniform birefringence, the BFRT collaboration explained this ellipticity noise as originating from the laser beam pointing instability on the cavity mirrors and on the QWP.

Comparing the optical path difference sensitivities obtained from the ellipticity sensitivities, using equation~(\ref{eq:sensOPD}) with $N = 34$ and $N = 578$, one finds $(\lambda = 514.5\;{\rm nm})$
\begin{equation}
S^{\rm (34)}_{\Delta{\cal D}} = 3.8\times10^{-16}\;\frac{{\rm m}}{\sqrt{\rm Hz}}\qquad {\rm and}\qquad
S^{\rm (578)}_{\Delta{\cal D}} = 4.3\times10^{-16}\;\frac{{\rm m}}{\sqrt{\rm Hz}}.
\end{equation}
whereas during rotation measurements ($N = 34$ and $N = 254$)
\begin{equation}
S^{\rm (34)}_{\Delta{\cal A}} = 1.6\times10^{-16}\;\frac{{\rm m}}{\sqrt{\rm Hz}}\qquad {\rm and}\qquad
S^{\rm (254)}_{\Delta{\cal A}} = 4.3\times10^{-17}\;\frac{{\rm m}}{\sqrt{\rm Hz}}.
\end{equation}

Increasing the number of passes $N$ did not improve $S_{\Delta{\cal D}}$ but did improve $S_{\Delta{\cal A}}$ (but not proportionally to $N$).

During data taking with the ramping field, ellipticity signals were present with and without the cavity. These signals though, were not interpreted as physical signals deriving from magnetically induced birefringence in that the feedback signal compensating the mirror movements also had a peak in the Fourier spectrum at the frequency of the modulated magnetic field and were insensitive to the polarisation direction. Furthermore the amplitude and phase of the ellipticity signals were not stable. Also during rotation measurements systematic signals were reported during some of the runs.

Both the ellipticity and rotation data were finally interpreted as upper limits at 95\%~c.l.
\begin{eqnarray}
\Delta n^{\rm (BFRT)} &<& 1.1\times10^{-18}\qquad{\rm with}\qquad N = 34\\
\Delta \kappa^{\rm (BFRT)} &<& 4.4\times10^{-20}\qquad{\rm with }\qquad N = 254.
\end{eqnarray}

To determine limits on VMB and on the axion-photon coupling $g_{a}$ the two relevant parameters for the magnetic field are respectively
$(B_{\rm ext}^2 L_B)_{\rm equiv} = 35.5$~T$^2$m and $(B_{\rm ext} L_B)_{\rm equiv} = 17.6$~Tm 
and the wavelength of the light $\lambda = 514.5$~nm.
The following limits on VMB, reported as $\Delta n/B^2$, and on the axion-photon coupling $g_{a}$ were set by the BFRT collaboration:
\begin{eqnarray}
\frac{\Delta n^{\rm (BFRT)}}{\left(B_{\rm ext}^2\right)_{\rm equiv}} &<& 2.75\times10^{-19}\;{\rm T}^{-2}\\\nonumber\\
g_{a}^{\rm (BFRT)} &<& 3.6\times10^{-7} \;{\rm GeV}^{-1}\qquad{\rm for}\qquad m_a < 1\;{\rm meV}.
\end{eqnarray}

It can be noted that contributions from systematics limited the integration time.
Given the sensitivity in $\Delta{\cal D}$ the necessary integration time to reach QED vacuum magnetic birefringence was unthinkable:
\begin{equation}
{\cal T} =\left(\frac{S^{\rm (578)}_{\cal D}}{3A_e(B_{\rm ext}^2L_B)_{\rm equiv}}\right)^2 = 10^{14}\;{\rm s.}
\end{equation}

\subsection{\bf PVLAS-LNL: 1992 - 2007}
Learning from the BFRT experience, two major changes were introduced in the PVLAS-LNL (Polarizzazione del Vuoto con LASer, c/o INFN - Laboratori Nazionali di Legnaro, Italy) experiment: the use of a vertically rotating superconducting magnet (in persistent current mode) and of a Fabry-Perot resonant cavity. The cavity length was 6.4~m and the rotating field allowed signal frequencies a factor 30 higher than those of the BFRT setup with $2\nu_B \approx 0.6$~Hz. Not having superfluid helium, the field was limited to 5.5 T. Runs with various field intensities were used resulting in $B_{\rm ext}^2 L_B = (5 \div 30)$~T$^2$m.

\begin{figure}[bhtp]
\begin{center}
\includegraphics[width=10cm]{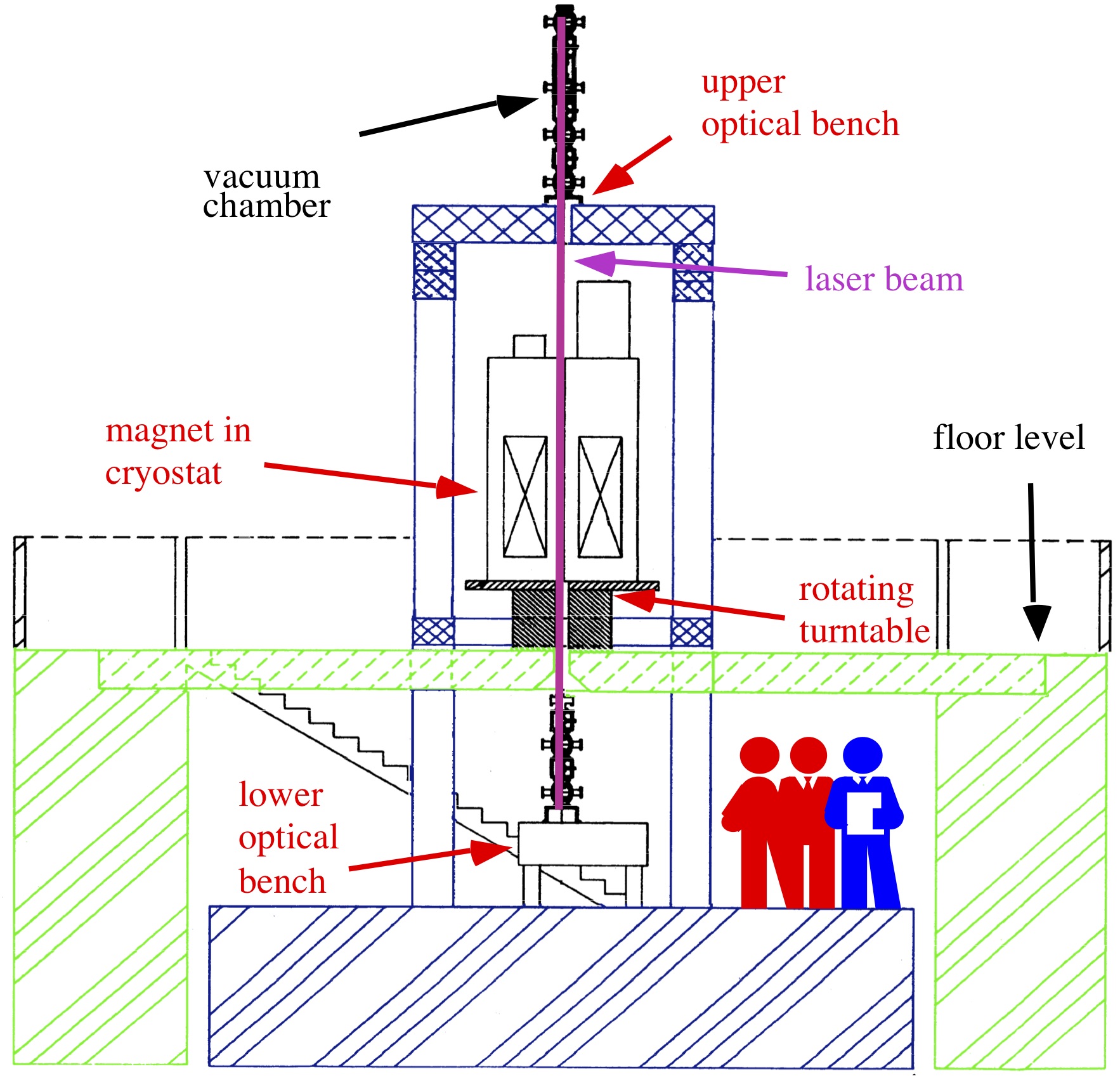}
\includegraphics[width=12cm]{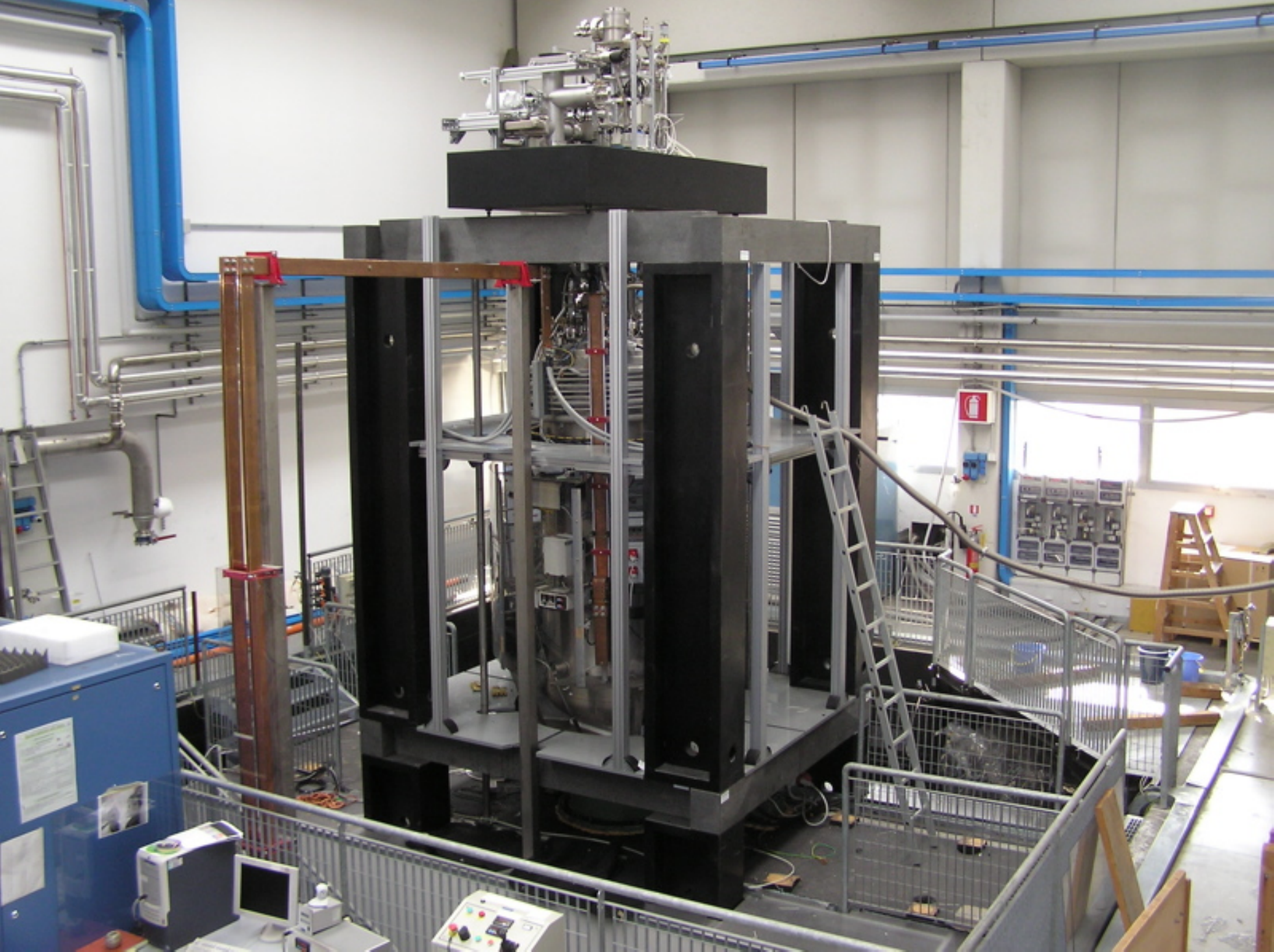}
\end{center}
\caption{Top: schematic drawing of the PVLAS-LNL setup. Bottom: a photograph of the apparatus above the floor level. The lower optical bench is below the beam supporting the cryostat and sits about 3~m below ground level on the concrete `raft'.}
\label{fig:PVLAS-LNL}
\end{figure}

\subsubsection{Infrastructures}

A drawing and a photograph of the setup are shown in Figure~\ref{fig:PVLAS-LNL}. The vertical construction of the setup, other than allowing the rotation of the magnet, was such that both the top and bottom optical benches were subject to the same ground movement. Following this idea, a dedicated infrastructure was designed and constructed in order to install the superconducting dipole magnet and its rotating cryostat. The experimental hall was equipped with a square pit, $8\times8$~m$^2$, 3~m deep from the floor level of the main building. Precautions were taken to avoid any mechanical vibrations coming from the rotating magnet to reach the optical benches. For this purpose the lower optical bench, the 7~m high vertical structure and the upper optical bench, all made of black granite, sat on a concrete `raft', forming the floor of the pit, supported by pillars embedded 14 m deep into sand. A reinforced concrete beam supported a rotating table driven by a hydraulic motor on which the cryostat containing the magnet sat. In the photograph of Figure~\ref{fig:PVLAS-LNL} the ladder leaning against the structure is standing on the concrete beam supporting the rotating turntable. During the rotation of the magnet no ellipticity or rotation signal was observed correlated with the uncharged magnet rotation.

\subsubsection{Superconducting magnet}
\label{sec:mag_morpurgo}

The magnet was the original 1~m long superconducting dipole magnet developed by Mario Morpurgo \cite{Morpurgo:1720288} as a prototype for the CERN D2 Proposal. It was manufactured at CERN and commissioned on July 1982. Its main characteristics are listed in the Table~\ref{tab:magnete1}.

\begin{table}[bht]
\begin{center}
\begin{tabular}{c|c}
\hline\hline
Cold bore useful diameter           & 0.1 m     \\\hline
Magnetic field length               & 1 m       \\\hline
Overall magnet length               & 1.3 m     \\\hline
Current (corresponding to a central field of 8 T) & 3810 A\\\hline
Max field on the conductor          & 8.8 T     \\\hline
Field uniformity in the useful bore & $\pm$ 2.5\% \\\hline
Average current density in the winding & 105 A/mm$^2$ \\\hline
Stored energy                       & 1.5 MJ     \\\hline\hline
\end{tabular}
\end{center}
\caption{Main characteristics of the PVLAS-LNL magnet built at CERN by Mario Morpurgo as a prototype for the original D2 CERN proposal.}
\label{tab:magnete1}
\end{table}

The 2.3 ton dipole magnet was wound with a hollow Cu-Nb-Ti composite conductor. The conductor had a square cross-section $5.5\times5.5$~mm$^2$ with a central bore approximately 2.5~mm diameter for cooling.
The dipole was composed of two identical coils each with 12 pancakes of 22 turns each. The 12 pancakes were subdivided into three groups of four, and each group was separately impregnated under vacuum with epoxy resin. The dipole yoke was made out of soft iron. Aluminium alloy plates and bolts were used to clamp together the various parts of the yoke. Since the aluminium thermal contraction is larger compared to the other metals, the coils were strongly compressed when cooled.

\subsubsection{Rotating cryostat}

The design of the cryostat was realised at LNL in collaboration with CERN \cite{Pengo:1999ud}. It was characterised by a room temperature 30.5~mm central bore to let the light beam traverse the magnet and by the possibility to rotate around its axis. 

\begin{figure}[bht]
\begin{center}
\includegraphics[width=.6\textwidth]{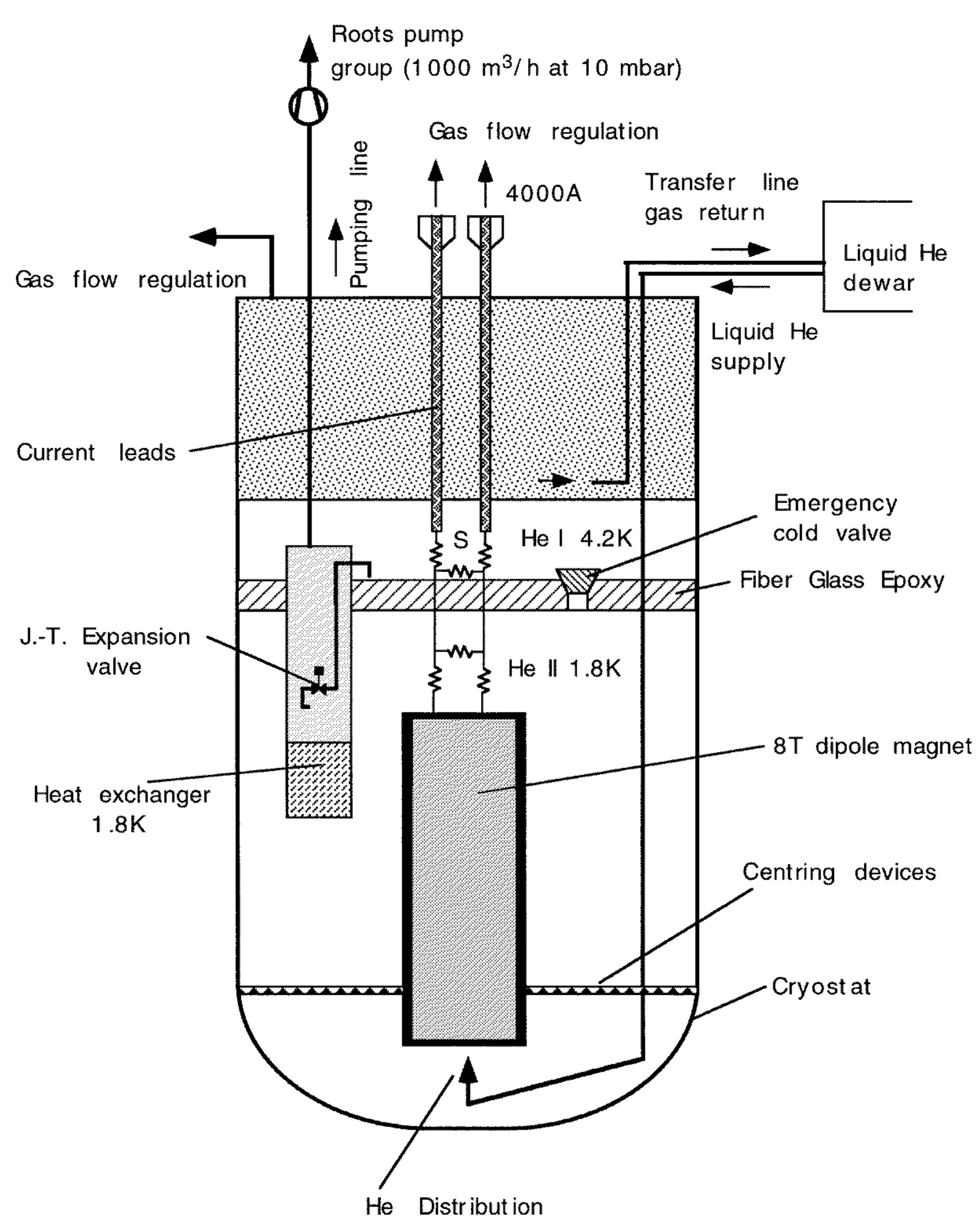}
\caption{Scheme of the rotating cryostat of the PVLAS-LNL experiment designed to cool the dipole magnet to superfluid helium. A 30.5~mm warm bore (not shown) vertically traversed the cryostat. From reference \cite{Bakalov1998QSO}, Figure 3.}
\label{fig:cryo}
\end{center}
\end{figure}

The scheme of the cryostat, shown in Figure~\ref{fig:cryo}, is of the Claudet type \cite{Claudet1974} with a fiber glass epoxy `lambda' plate to separate the superfluid helium at the bottom from the normal helium above. The outer jacket was superinsulated with mattresses of superinsulation. 
At the bottom of the cryostat and at its center, a few kilograms of activated charcoal were placed, in order to lower the pressure in the superinsulation region, thus improving the necessary insulation efficiency. Multi-layer superinsulation was also used for the central bore.
Inside the cryostat a stainless steel coil was wound around the magnet as a heat exchanger for the pre-cooling with liquid nitrogen (LN2) using helium gas as an exchange medium.

\begin{figure}[htb]
\begin{center}
\includegraphics[width=9.5cm]{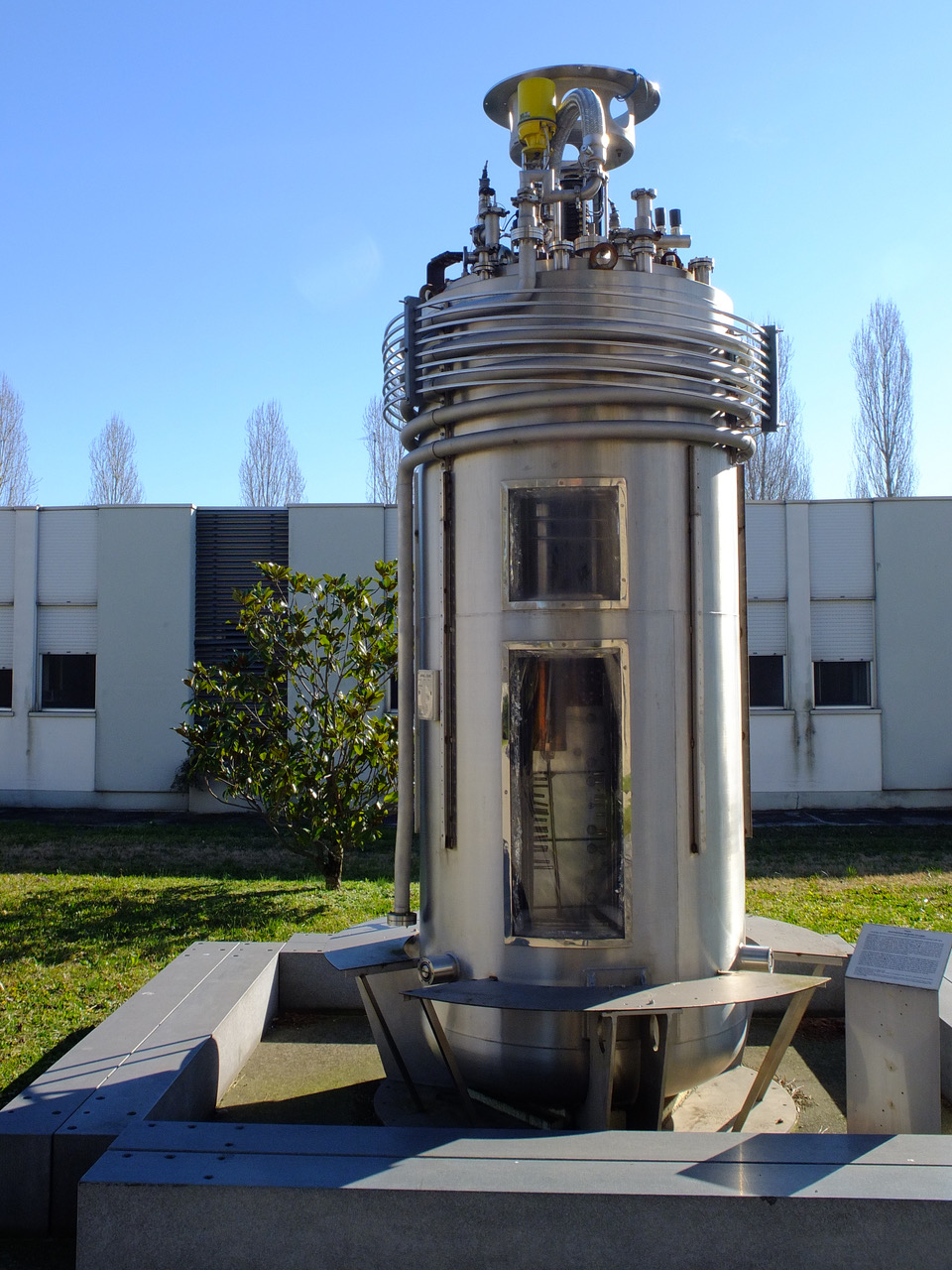}
\caption{Photograph of the cryostat of the PVLAS-LNL experiment. The cryostat is exhibited on the INFN National Laboratories of Legnaro (LNL) site near Padua, Italy. In the cutaway the prototype CERN magnet can be seen.} 
\label{fig:magset}
\end{center}
\end{figure}

The vacuum pumps for the superfluid helium production, one roots pump and two mechanical pumps, were installed on platforms outside of the cryostat and rotated with it. The pumps were never used during the measurements due to a limited availability of helium, hence the field never reached its maximum. Sliding brush contacts transmitted the electric power and the RS232 signals necessary for the instrumentation collecting data on board of the rotating cryostat. The helium boil-off vapors were collected at room temperature, both with the cryostat at rest and during the rotation, through a helium tight feedthrough and sent to the recovery system. 

In Figure~\ref{fig:PVLAS-LNL} the solid copper bars connecting the magnet to the 10 V, 5000 A power supply (blue cabinet on the left) can be seen. Inside the power supply a set of water cooled diodes and resistors were installed to dump the electromagnetic energy stored during the (fast) discharge. Once charged, the magnet was put in “quasi persistent” current mode through a very low resistance (0.24~$\mu\Omega$) silver-plated copper multi-blade socket \cite{Multi1999} and the power supply disconnected. 



\subsubsection{Fabry-Perot cavity}
The use of a resonant Fabry-Perot cavity allowed an increase in the number of equivalent passes, reaching $N \approx 45000$ for $\lambda = 1064$~nm and $N \approx 23000$ for $\lambda = 532$~nm. The light source was a Nd:YAG frequency doubled frequency tuneable laser capable of emitting 1.8~W at 1064~nm and 200~mW at 532~nm. The laser was frequency locked to the cavity via a modified Pound-Drever-Hall technique \cite{Cantatore1995RSI,DeRiva1996RSI,Bregant2002RSI}. Although the ellipticity noise was far from shot-noise at $2\nu_B\sim 0.6$~Hz, an improvement of about a factor 40 was obtained in the optical path difference sensitivity with respect to BFRT, reaching $S^{\rm (PVLAS-LNL)}_{\Delta{\cal D}} \approx 10^{-17}$~m/$\sqrt{\rm Hz}$. Given the increase in $N$, though, one would have expected a greater improvement in sensitivity. The independence of $S_{\Delta{\cal D}}$ on $N$ in the BFRT experiment and the limited improvement in sensitivity of PVLAS-LNL with respect to BFRT was a hint of the presence of a limiting noise originating from within the cavity also with the Fabry-Perot interferometer. It must be noted here that the beam stability inside the Fabry-Perot is determined only by the stability of the cavity mirrors and not by the pointing stability of the input beam, indicating that the origin of the noise of the Fabry-Perot was not due to input beam instabilities. The PVLAS-LNL sensitivities $S_{\Delta{\cal D}}$ and $S_{\Delta{\cal A}}$ were both more than a factor 100 above shot-noise at $2\nu_B = 0.6$~Hz and as was the case for the BFRT experiment the ellipticity sensitivity was slightly worse than the rotation sensitivity.

\subsubsection{Results}

By integrating in time, the ellipticity noise floor attainable is determined by $\Psi_{\rm floor} = S_\Psi/\sqrt{{\cal T}}$. Assuming ${\cal T} = 10^6$~s, in principle PVLAS-LNL could have reached a noise floor of $\Psi_{\rm floor} \approx 10^{-9}$ corresponding to ${\Delta{\cal D}} \approx 10^{-20}$~m a factor $\approx 80$ above the QED effect considering a field intensity of $B_{\rm ext} = 5.5$~T. Unfortunately systematic signals appeared within a few hours of integration in both the ellipticity and rotation configurations. Active magnetic shielding of the mirrors, vibration correlation studies along with measurements at different field intensities did not eliminate such systematics. Unfortunately a detailed study of these signals was made impossible given the limited availability of liquid helium for the magnet.

After several runs, the relatively stable phase and amplitude of the rotation and ellipticity and the observation of a trend of the rotation as a function of a buffer gas, attributable to an axion like particle \cite{IDM5,Nove2005,ZAVATTINI2007NPBPS,Zavattini:2006zz}, led the PVLAS collaboration to publish in 2006 a paper in which an unexplained polarisation rotation in the presence of the external magnetic field was reported, indicating a dichroism \cite{Zavattini2006PRL}. A year later, the collaboration retreated this observation after an upgrade of the apparatus \cite{Zavattini9,PRL_Erratum}. 

Only conjectures can be made today on the origin of the observed 2006 signal but, given the experience with PVLAS-FE, a plausible cause could have been the coupling between the diffused light inside the quartz tube and its movement induced by the rotating magnet (see Section~\ref{sec:tube-acceleration} for details). Furthermore, given that for magnetic fields intensities $B_{\rm ext}\gtrsim2.3$~T the iron yoke of the superconducting magnet saturated generating a stray field, other upgrades were implemented after the publication of the rotation. These included the complete recabling of the experiment and the elimination of iron support structures close to the experiment.

Although the result in the 2006 paper \cite{Zavattini2006PRL} was incorrect, it contributed to a revived interest in axion-like-particle searches by optical techniques.

Clearly a magnetic field source with unlimited time would have been necessary in any future effort for debugging and for systematics hunting, allowing to take maximum advantage of the apparatus. This led the PVLAS collaboration to end the PVLAS-LNL experiment and to upgrade to a new version using permanent magnets. Although limited to a lower field of $B_{\rm ext} = 2.5$~T they would allow extended debugging.

\subsection{\bf PVLAS-Test: 2009 - 2012}

The experience from PVLAS-LNL led the collaboration to attempt a prototype bench-top setup with two small 2.3 T 20 cm long permanent magnets with a total $\int{B_{\rm ext}^2\,dL} = 1.85$~T$^2$m. This setup was financed both by INFN and Italian Ministry of Research (MIUR). Other than having a permanent field allowing detailed debugging, permanent magnets can be rotated at still higher frequencies with respect to PVLAS-LNL. Furthermore, the advantage of using two magnets was for systematic hunting with the field ON but no induced ellipticity. Indeed with the two magnetic fields in a perpendicular configuration while rotating, the net effect of a magnetic birefringence is zero.

\begin{figure}[bht]
\begin{center}
\includegraphics[width=16cm]{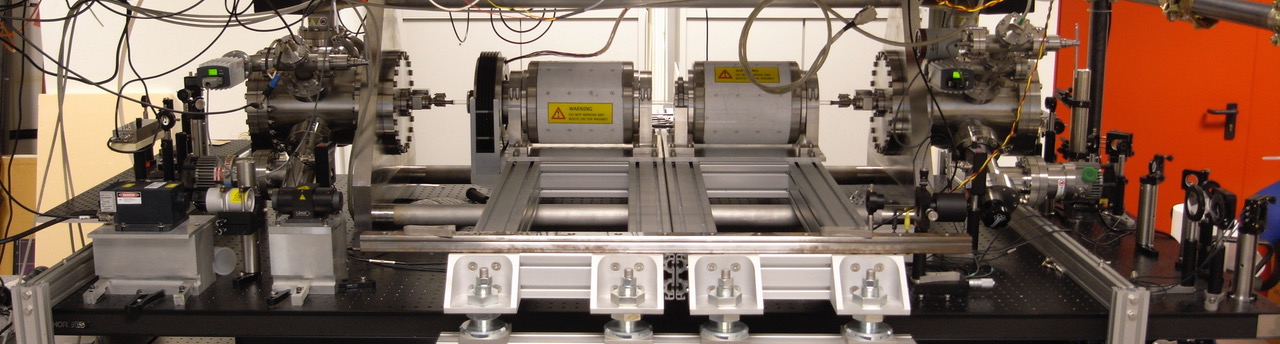}
\end{center}
\caption{Photograph of the PVLAS-Test apparatus in Ferrara. The laser, front left, sent a beam towards the right where the steering mirrors were. The input vacuum chamber containing the polariser and input mirror is on the right. A single 12~mm outer diameter glass tube traversed the two magnets connecting the two vacuum chambers. The output chamber contained the output mirror, the PEM and the analyser. The output detection bench can be seen on the far left.}
\label{fig:pvlas-test}
\end{figure}

The whole optical apparatus was placed on a single vibration isolated optical bench whereas the magnets were sustained by a support sitting on the laboratory floor mechanically disconnected as best we could from the optics. A photograph of the PVLAS-Test setup can be seen in Figure~\ref{fig:pvlas-test}. The distance between the two vacuum chambers was about 1.5~m. Mirror performance also improved allowing an equivalent number of passes $N= 280000$. The magnets were rotated at about 3~Hz generating a signal at 6~Hz.

In parallel to this compact setup the construction of a new larger setup, discussed in the next section, was ongoing. The goal of this test setup was to validate the choices made for the new apparatus.

Although this setup was a test apparatus, the capabilities of using permanent magnets successfully was demonstrated by a factor 2 improvement in VMB limits in a factor 8 shorter integration time with respect to PVLAS-LNL. At times, though, spurious signals were present indicating the presence of unidentified systematic sources. The sensitivity in optical path difference at 6~Hz was $S_{\Delta{\cal D}} = 6\times10^{-19}$~m/$\sqrt{\rm Hz}$ more than a factor 10 better than PVLAS-LNL. Again, as will be discussed in Section~\ref{sec:intrinsic noise}, the gain in sensitivity was not due to the increase in finesse but to the higher signal frequency.

\section{The PVLAS-FE experiment}
\label{sec:PVLASFE}

\subsection{\bf Summary}
The many years of experience led to the final PVLAS-FE setup in which all of the previous experience was put together. Four general features were implemented: 
\begin{enumerate}
    \item the polarimeter was to be mounted on a single vibration isolated optical bench to reduce seismic noise coming from the ground;
    \item the rotation frequency of the magnets was to be as high as possible;
    \item the finesse was to be as high as possible supposedly to increase the SNR;
    \item all components of the polarimeter were to be non magnetic to avoid magnetic coupling between the optics and the rotating stray magnetic field.
\end{enumerate}

Here each of the choices made for the setup will be discussed in detail and justified. Magnet rotations up to about 10 Hz were imagined during the design phase allowing a further reduction of the expected cavity noise contribution: at the time it was clear that the higher the signal frequency, the lower was the noise. What was still not clear was the proportionality of the ellipticity sensitivity $S_\Psi$ with $N$, or in other words, the \emph{independence} of the optical path difference noise $S_{\Delta{\cal D}}$ on $N$ for large $N$. For this reason the highest achievable finesse was still a goal. At the time the experiment was designed, fields up to 2.5~T were available with permanent magnets over a diameter of about 2~cm. Due to the available space, a total magnetic field length longer than about 2~m was difficult. The two magnet scheme was implemented to take advantage of the effective signal cancellation with perpendicular fields tested in the PVLAS-Test setup. Each of the PVLAS-FE magnets had a field length of $L_B = 0.82$~m.


The final sensitivity in optical path difference $\Delta{\cal D}$ of the PVLAS-FE apparatus was $S^{\rm (PVLAS)}_{\Delta{\cal D}} = 3.5\times10^{-19}$~m/$\sqrt{\rm Hz}$ at $2\nu_B = 16$~Hz, a factor of about 10 worse than the required one to reach VMB detection in ${\cal T} \approx 10^6$~s. Furthermore measurements showed the \emph{independence} on $N$, for large $N$, of the optical path difference noise $S^{\rm (PVLAS)}_{\Delta{\cal D}}$, meaning that increasing the finesse would have been useless to gain in signal to noise ratio. An increase in frequency to compensate this missing factor $\approx 10$ was beyond our possibility. 

The final sensitivity in absorption length $S^{\rm (PVLAS)}_{\Delta{\cal A}}$ was about a factor 2 better than $S^{\rm (PVLAS)}_{\Delta{\cal D}}$ and was limited by the cross-talk from the ellipticity noise to rotation noise as discussed in Section~\ref{sec:MirrorBirefringence} and as determined from equation~(\ref{R_Phi_Psi}):
\begin{equation}
S^{\rm (PVLAS)}_{\Delta{\cal A}} = N\frac{\alpha_{\rm EQ}}{2}\;S^{\rm (PVLAS)}_{\Delta{\cal D}}.
\end{equation}

An integrated noise floor in optical path difference $\Delta {\cal D} = (1.0\pm1.4)\times10^{-22}$~m, limited by statistics and not by systematics, was our final value after a run time of about $5\times10^6$~s. Translated to vacuum magnetic birefringence this leads to 
\begin{equation}
\frac{\Delta{\cal D}}{L_B} = \Delta n^{\rm (PVLAS)} = (12\pm17)\times10^{-23}\;\; @ \;\;2.5{\;\rm T}.
\end{equation}
where $L_B = 0.82$~m. The 1~$\sigma$ uncertainty is a factor of about 7 from the predicted value of $3A_eB_{\rm ext}^2 = 2.5\times10^{-23}$ with $B_{\rm ext} = 2.5$~T.
Similarly for the dichroism limit, the PVLAS-FE final value was
\begin{equation}
\Delta\kappa^{\rm (PVLAS)} = (10\pm28)\times10^{-23} \;\; @ \;\;2.5{\;\rm T}.
\end{equation}
A detailed discussion of these results will be given in Section~\ref{sec:FE_results}.

\subsection{\bf General description of the apparatus}

In the following sections we will describe each of the choices made for the PVLAS-FE apparatus. The experiment was located on the ground floor of an experimental hall at the Department of Physics and Earth Sciences of the University of Ferrara, Ferrara, Italy, inside a temperature controlled ($23^\circ\pm1^\circ$) and relative humidity controlled ($\approx 56\%$) clean room of ISO-4 class.

\begin{figure}[bht]
\begin{center}
\includegraphics[width=16cm]{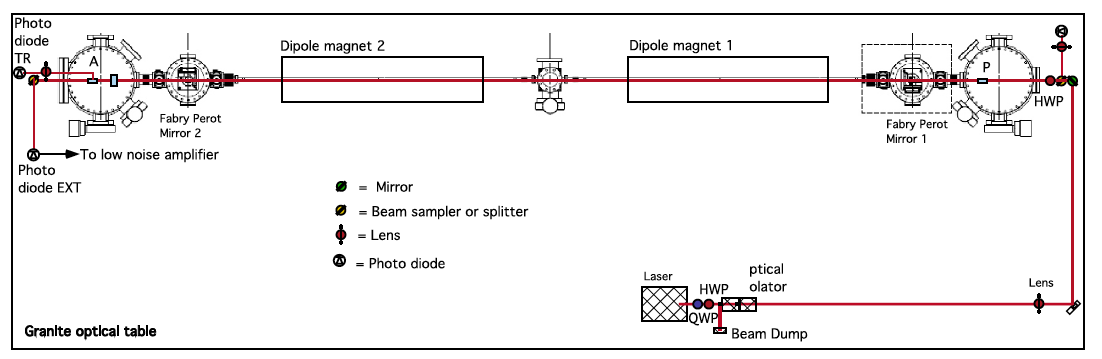}
\includegraphics[width=16cm]{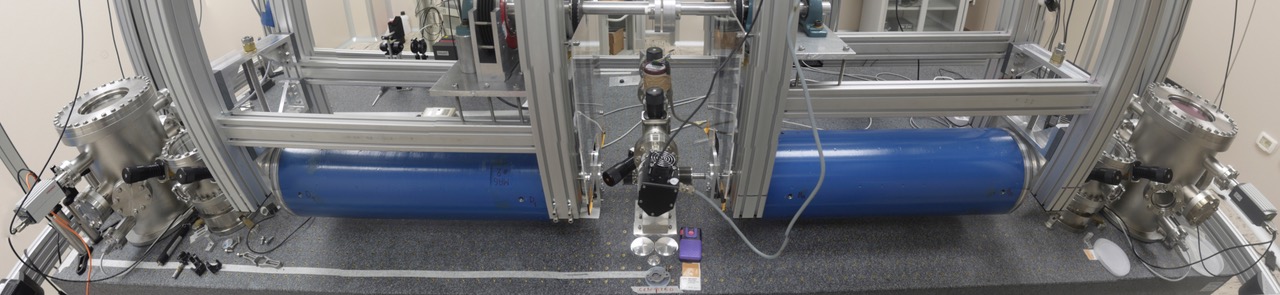}
\end{center}
\caption{Schematic view and photograph of the optical bench layout of the PVLAS-FE apparatus. 
HWP~=~Half-wave plate; P = Polariser; A = Analyser; QWP = quarter-wave plate; TR = transmission; EXT = extinction.}
\label{fig:pvlas-fe}
\end{figure}


A general scheme of the optical setup and a photograph of the apparatus are shown in Figure~\ref{fig:pvlas-fe}. A Nd:YAG laser  (Innolight Mephisto, 2~W power) emitted at $\lambda=1064$~nm. The beam first passed through a quarter-wave plate (QWP) reducing the initial ellipticity of the laser beam. A first half-wave plate (HWP) placed before a two stage Faraday isolator allowed the adjustment of the power being injected into the Fabry-Perot cavity. The beam then passed through a lens to match the laser waist with the cavity waist for optimal mode matching. Two steering mirrors followed by a second HWP brought the beam to the entrance of the vacuum system with the desired alignment and polarisation direction. Between the second steering mirror and this second HWP a glass window allowed the sampling of the reflected power from the cavity for phase locking the laser to the cavity via the Pound-Drever-Hall (PDH) technique. 
The same glass plate was also used to sample the beam power at the Fabry-Perot input. 
The sidebands for the PDH locking circuit were generated directly in the laser rather than with an external phase modulator \cite{Cantatore1995RSI,DeRiva1996RSI}. An automatic locking servo-circuit allowed operation of the apparatus with an almost unitary duty-cycle.

The second HWP  together with the rotatable polariser P allowed the alignment of the light polarisation with one of the axes of the equivalent wave plate of the cavity. The light path between the two mirrors passed through the bores of the two dipole magnets. 
At the cavity output an extractable QWP  was used to transform, when necessary, a polarisation rotation into an ellipticity (and {\em vice versa}). The light then passed through the resonant photo-elastic ellipticity modulator, PEM, (Hinds Instruments), and the analyser A, normally set to maximum extinction. The PEM was mounted on an axial rotation mount to set its axis at $45^\circ$ with respect to the polarisation direction, and on a translation stage to allow its extraction from the beam. The modulation amplitude was typically $\eta_0 \approx 3\times10^{-3}\div 10^{-2}$. Both the extraordinary and ordinary beams from the analyser A exited the vacuum enclosure: the former measured the power $I_\parallel\approx I_{\rm out}$ transmitted by the cavity, whereas the extinguished beam power, $I_\perp$, contained information on the ellipticity and rotation acquired by the light polarisation.
The extinction ratio was generally $\sigma^2 \lesssim 10^{-7}$. After a narrow-band optical filter, the extinguished beam was collected on an InGaAs low noise photodiode with gain $G = 10^6$~V/A and efficiency $q = 0.7$~A/W. The diode was placed about 2 m from the analyser to reduce contamination from diffused light. 


\subsection{\bf Optical bench and vibration isolation}

Special care was taken to limit any magnetic forces acting on the mechanical parts of the apparatus. 
Although the permanent magnets were designed following the Halbach configuration \cite{HALBACH19801}, which in principle cancels stray fields, a small stray field of about 10 mT was present near their surface, rapidly decaying with distance. 
The rotating magnets would then generate eddy currents and thus magnetic forces on nearby components: for this reason a granite optical bench was chosen as a support for the optics.
The bench, manufactured by Microplan, Quarona (VC), Italy, was 4.8~m long, 1.5~m wide and 0.5~m thick for a total weight of 4~tons. A granite `honeycomb' structure filled the inside of the bench to limit the total weight. The surface of the bench was equipped with a 5~cm~$\times$~5~cm matrix of threaded holes made of brass.

The BiAir$^{\circledR}$ membrane air spring legs sustaining the optical bench had a six degrees of freedom feedback system which maintained the position of the bench to within 10~$\mu$m. This was a fundamental characteristic of the support system since the glass tube passing through the magnets needed to be centred very carefully and in a repeatable way as will be discussed below. From the specifications of the manufacturer (Bilz Vibration Technology AG, Leonberg, DE) the filtering of the supporting system started at about 1~Hz.

\begin{figure}[bht]
\begin{center}
\includegraphics[width=16cm]{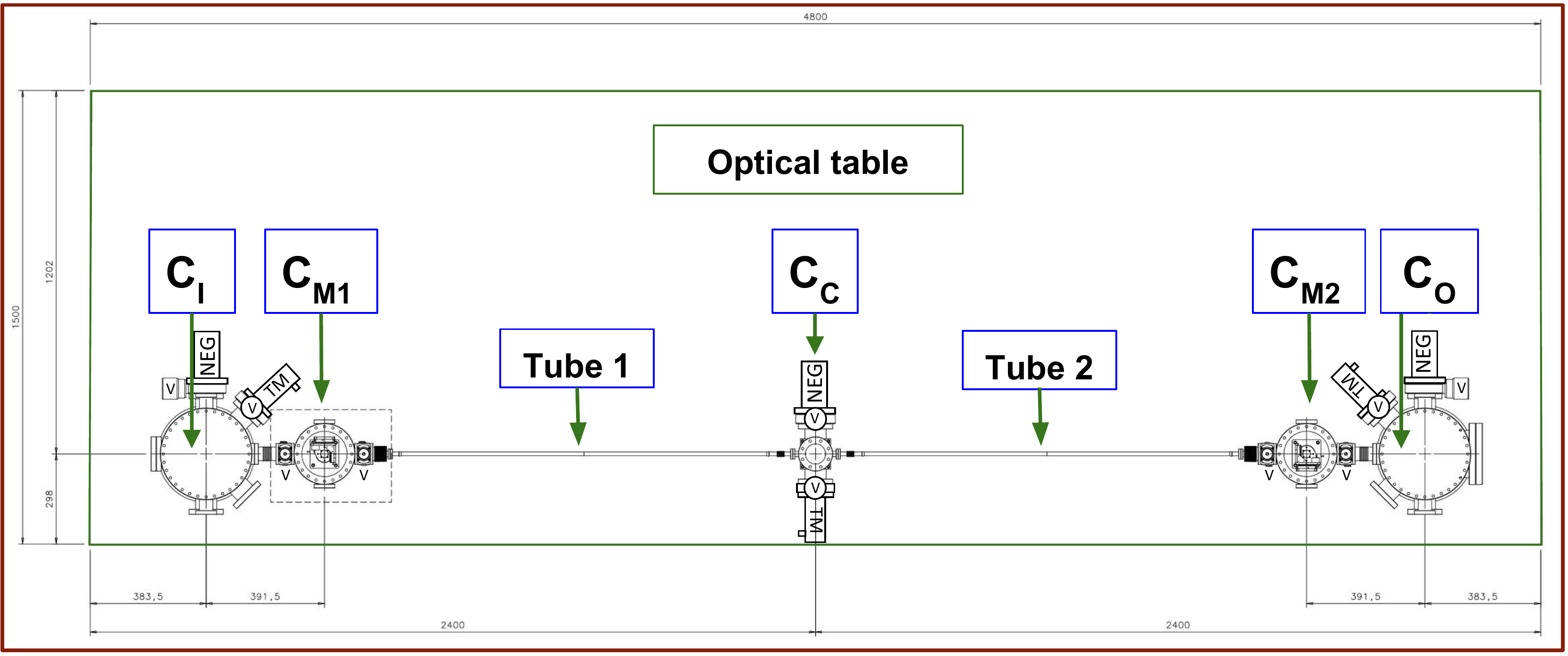}
\end{center}
\caption{Layout of the vacuum enclosures mounted on the optical bench: input vacuum chamber (C$_{\rm I}$) containing the polariser, first mirror vacuum chamber (C$_{\rm M1}$), central vacuum chamber C$_{\rm C}$ dedicated to pumping, second mirror vacuum chamber (C$_{\rm M2}$), output vacuum chamber (C$_{\rm O}$) containing the QWP, PEM and analyser. Mirror chambers are connected to the central chamber by diamagnetic tubes. Two residual gas analysers (not shown) were mounted on the chambers C$_{\rm C}$ and C$_{\rm O}$.}
\label{fig:vacuum}
\end{figure}

\subsection{\bf Vacuum system}

\begin{figure}[b!]
\begin{center}
\includegraphics[width=16cm]{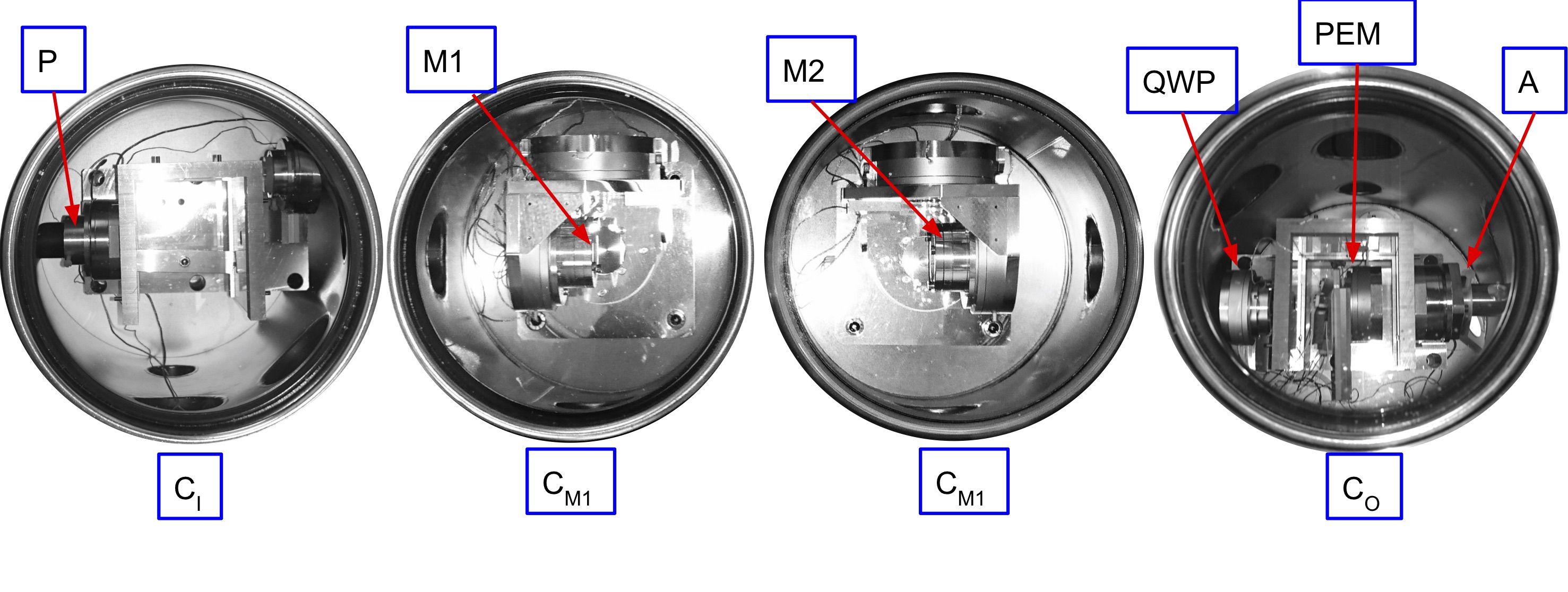}
\end{center}
\caption{Top view pictures of the inside of the vacuum chambers. Chambers C$_{\rm I}$ and C$_{\rm O}$ host a cage structure supporting 1064~nm absorption glasses to reduce diffused light.} 
\label{fig:top-view-chambers}
\end{figure}

The whole polarimeter, from polariser to analyser, was kept in a high vacuum system built employing ConFlat$^{\circledR}$ seals. A scheme of the vacuum system is shown in Figure~\ref{fig:vacuum}. Starting from the left one first finds the input chamber C$_{\rm I}$, the chamber for the first cavity mirror C$_{\rm M1}$, a vacuum tube passing through the bore of the first magnet, a central pumping station C$_{\rm C}$, a second vacuum tube passing through the bore of the second magnet, the chamber for the second cavity mirror C$_{\rm M2}$ and finally the output chamber C$_{\rm O}$. Short bellows where employed to allow the alignment of the chambers with the light path. The mirror chambers could be isolated by means of all-metal Viton-sealed manual gate valves: while adjusting in air the rest of the system, the mirrors were kept in vacuum. The chambers were made of 304 stainless steel with the exception of the bases of the mirrors chambers  
which were of solid titanium. All the equipment and fittings were in non magnetic materials. An all-metal gas line and a leak valve on C$_{\rm O}$ allowed to fill the chambers with sub-millibar pressures of ultra-pure gases  generating a Cotton-Mouton effect in the magnet bores.  
A top view of the inside of all the chambers is shown in Figure~\ref{fig:top-view-chambers}.


Three issues needed to be addressed in the design of the pumping system. The first was the use of dry pumps to prevent degradation of the reflecting surface of the mirrors and hence of the finesse of the cavity. Secondly, avoid vibrations from the pumps. 
The third and most important issue was the need to lower the partial pressures of all the gas species below the value that would mimic a vacuum magnetic birefringence through the Cotton-Mouton effect; such values are listed in Table~\ref{tab:ListCM} on page~\pageref{tab:ListCM}.  
The chambers C$_{\rm I}$, C$_{\rm O}$ and C$_{\rm C}$ were equipped with dry Turbo Molecular (TM) pumps and Non Evaporable Getter (NEG) pumps. 
No direct pumping was done in the mirror chambers C$_{\rm M1}$ and C$_{\rm M2}$. Each of the TM and NEG pumps could be isolated by means of a ultra-high vacuum manual gate or butterfly valve. The TM pumps had also a downstream valve to prevent contamination during maneuvers on the scroll primary pumps. 
Two residual gas analysers were installed in the central chamber C$_{\rm C}$ and in the output chamber C$_{\rm O}$.

\begin{figure}[bt]
\begin{center}
\includegraphics[width=10cm]{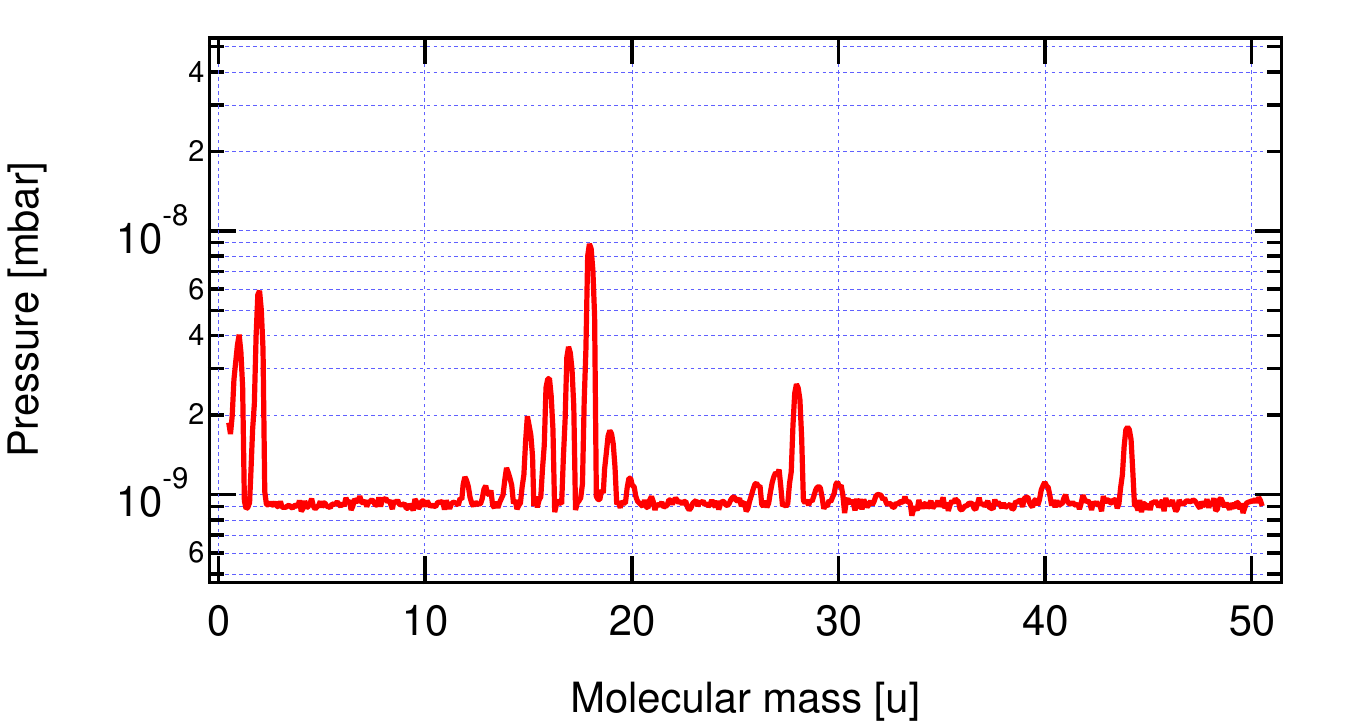}
\end{center}
\caption{Typical residual gas mass spectrum during data taking, measured in the output chamber C$_{\rm O}$. The main residual gases (from left to right) are hydrogen, methane, water vapour, carbon monoxide, and carbon dioxide. The absence of the oxygen peak indicates the absence of nitrogen. All gas pressures were well below the VMB equivalent pressures reported in Table~\ref{tab:ListCM} on page \pageref{tab:ListCM}.}
\label{fig:rga}
\end{figure}

Starting from atmospheric pressure, pumping was initially 
done slowly 
to avoid too much air flux over the surface of the mirrors. 
After a short transient period, the turbo molecular pumps were isolated and switched off to guarantee a quiet operation of the cavity. Due to the inability of NEG pumps to pump noble gases and due to a small production of methane from the NEG pumps themselves, the turbo pump of the central station was normally left running. Thanks to its distance from the mirrors, it caused no detectable vibrations on the optics. This system guaranteed a stable vacuum of $\approx 2\times 10^{-8}$~mbar in the three pumping chambers with a typical residual gas mass spectrum shown in Figure~\ref{fig:rga}. 

\subsection{\bf Vacuum tubes through the magnets}

Two different non magnetic materials were used for the vacuum tubes passing through the magnets: borosilicate glass and silicon nitride ceramics. To connect the tubes to the metal flanges, standard metal-glass junctions were soon dismissed as they did not allow a fast disassembly. After trying different solutions, we finally chose to employ home made ConFlat$^{\circledR}$ adapter flanges compressing a Viton o-ring against the external surface of the tubes. At the beginning, a non rotating carbon fiber sleeve, held by the support structure of the magnets, surrounded the glass tubes. These sleeves were intended to shield the tubes from air turbulence and periodic light coming from the rotating inner surface of the magnet's bore, which could give origin to synchronous noise. In a second phase, the carbon fiber tubes were removed, and the external walls of the glass tubes were painted in black. Given the 20~mm diameter of the bore of the magnets, borosilicate tubes with an outer diameter of 15~mm and 18~mm with a wall thickness of 1.5~mm were used, whereas the silicon nitride ceramic tubes had an outer diameter of 18~mm and a wall thickness of 3.75~mm.

Being diamagnetic bodies inserted in a non perfectly uniform magnetic field, these tubes were subject to rotating forces synchronous with the rotation of the magnets. As will be discussed in detail in Section~\ref{sec:diffused_light}, the coupling of the resulting tube's movements and of the diffused light reflected off of the inner surface of the tubes was the principal cause of spurious ellipticity signals at the various harmonics of the rotating magnets. To monitor the movement of the tubes, a 3-axis accelerometer was installed on the mirror end of each tube. A micrometric positioning system for the tubes was also installed to minimise the acceleration.
To block the diffused light in the Fabry-Perot reflected at a grazing angle off of the tube's inner surface, we inserted baffles inside the glass tubes. Silicon nitride tubes have an intrinsic roughness reducing the effect of diffused light. Moreover, the mass and the stiffness of the silicon nitride tube resulted in a reduced movement amplitude.

\subsection{\bf Optical vacuum mounts}

\begin{figure}[bht]
\begin{center}
\includegraphics[angle=-1,height=6cm]{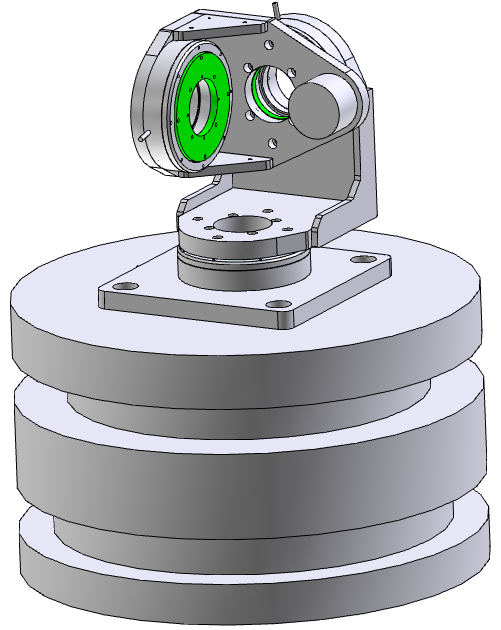}
\end{center}
\caption{Three axis mirror mount for the cavity mirrors fixed on the titanium bases of the chambers $C_{\rm M1}$ and $C_{\rm M2}$. The rotating stages were based on the piezo `slip-stick' principle and maintained their position in the absence of power.} 
\label{fig:mirror_tilter}
\end{figure}

For the mounting and alignment of all the optical elements in vacuum we opted for UHV non magnetic mounts driven by `slip-stick' piezo motors (SmarAct GmbH) to avoid any magnetic coupling with the rotating magnets.
The piezo motors kept their position even when switched off.
Both translation stages and rotation stages were implemented. The minimum angular step size we used was $\approx 10\;\mu$rad. The polariser and analyser motors were also equipped with encoders. 

Each cavity mirror was mounted on a three axis rotation mount shown in Figure~\ref{fig:mirror_tilter}. The center of rotation of the tip-tilt stages coincided with the reflecting surface of each mirror. The rotation around the cavity axis was used to minimise the equivalent wave plate of the cavity. No active alignment system was implemented for the optical cavity.

\subsection{\bf The rotating permanent magnets}
\label{sec:magnet_rotation}

\begin{table}[thb]
\begin{center}
\begin{tabular}{l|c}
\hline\hline
Magnetic system design &
cylindrical\\
\hline
Magnetic field direction &
normal to bore axis\\
\hline
Overall length &
934 mm\\
\hline
Outer diameter &
280 mm\\
\hline
Bore diameter &
20 mm\\
\hline
Net weight &
450 kg\\
\hline
Magnetic material &
high coercitivity Nd-Fe-B\\
\hline
Maximum field intensity $B_{\rm ext}$ &
2.5 T\\
\hline
Squared field integral &
5.12 T$^2$m\\
\hline
Magnetic field length $L_B$ &
0.82 m\\
\hline\hline
\end{tabular}
\end{center}
\caption{Main characteristics of the dipole magnets designed and built by Advanced Magnetic Technologies \& Consulting LLC, Troitsk, Russian Federation.}
\label{Tab:mag2}
\end{table}

The permanent magnets of the PVLAS-FE experiment aimed at taking advantage of the recently developed Nd-Fe-B sintered magnet technology. The set-up comprised the construction and installation of two identical dipole magnets of the Halbach type \cite{HALBACH19801}, with $B= 2.5$~T. The main technical characteristics of each magnet are listed in Table~\ref{Tab:mag2}.

\begin{figure}[bht]
\begin{center}
\includegraphics[width=.7\textwidth]{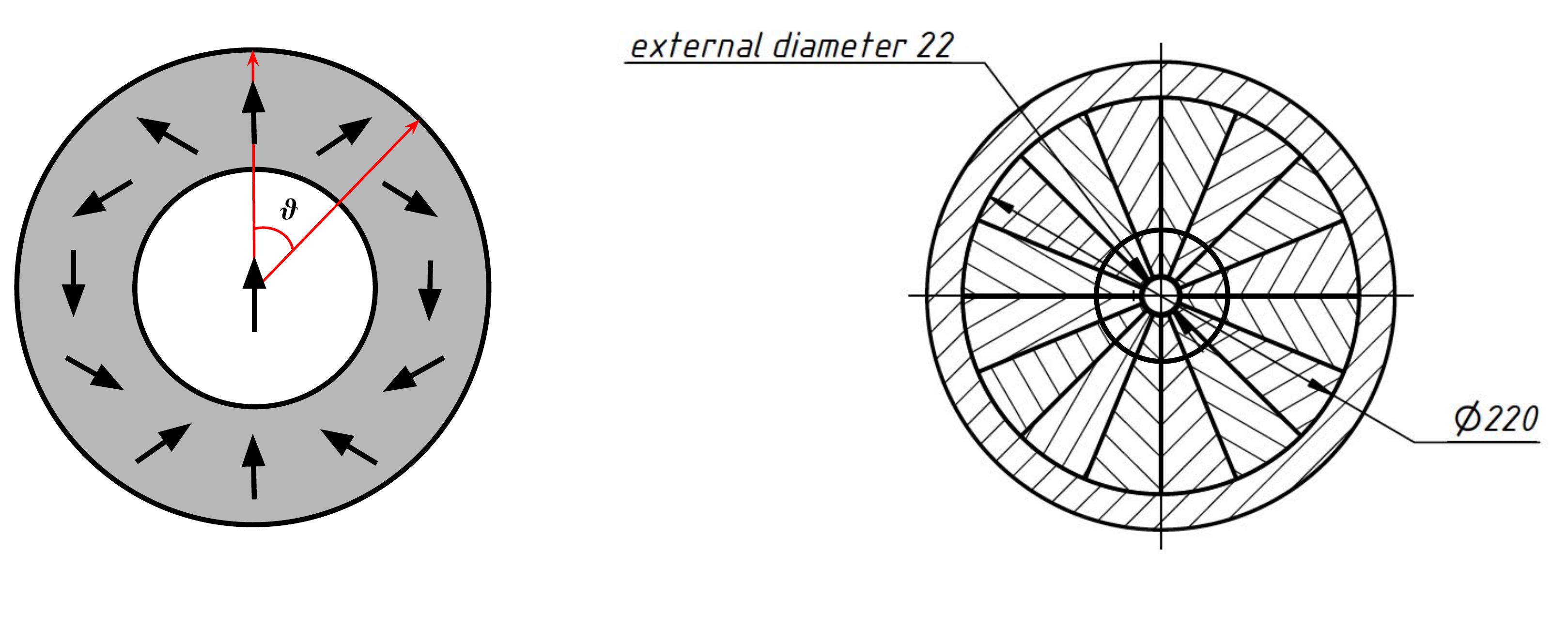}
\caption{Left: Magnetisation directions in the Halbach configuration of a dipole field. Right: Two rings, 16 segment Halbach configuration of the PVLAS-FE magnets with the external cylindrical enclosure.} 
\label{fig:Halbach-AMC}
\end{center}
\end{figure}

\begin{figure}[bht]
\begin{center}
\includegraphics[width=.5\textwidth]{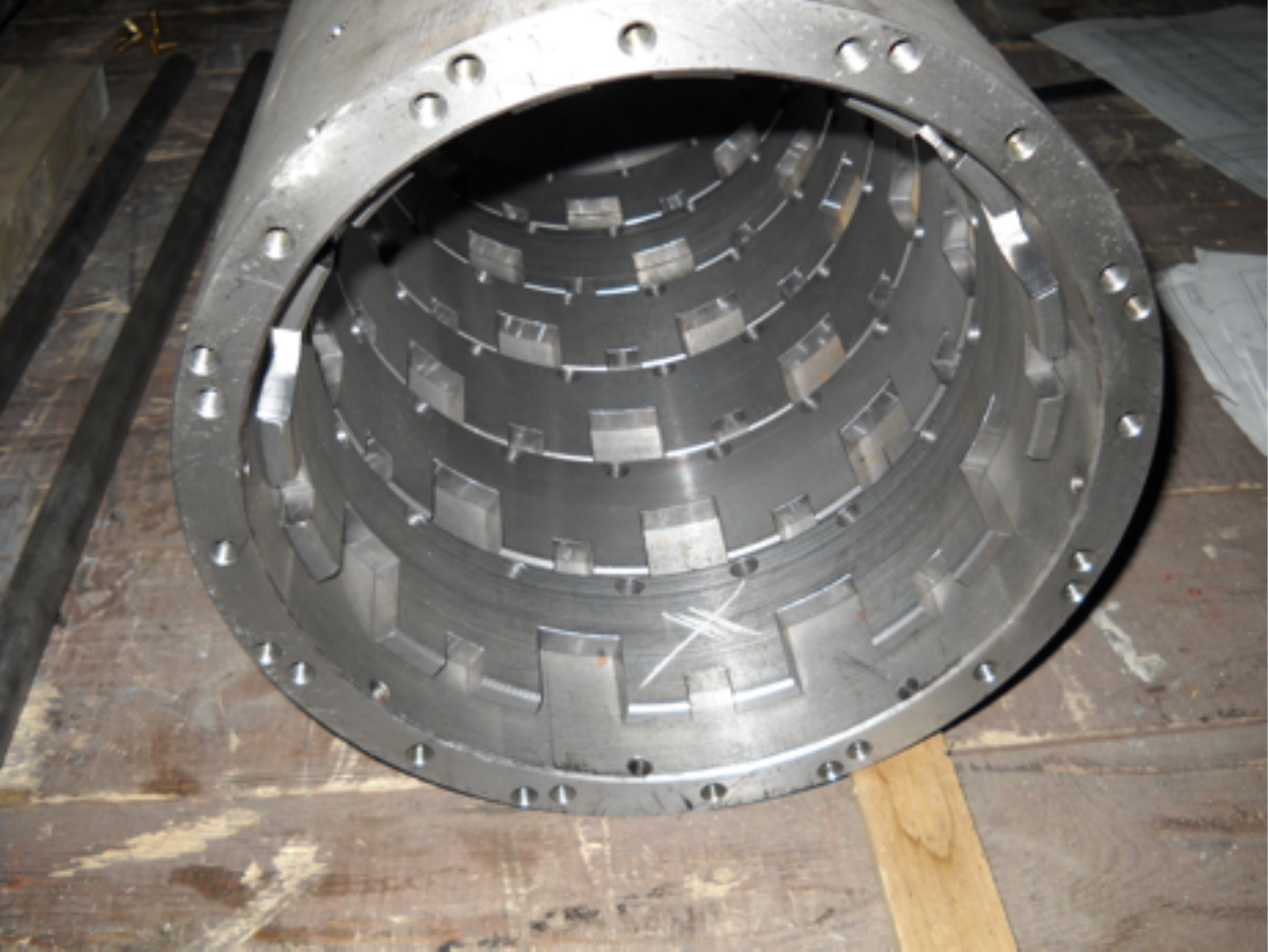}
\caption{Inside of the cylindrical enclosure hosting the Halbach elements. In the axial direction the magnet was composed of twelve layers each 7~cm thick.} 
\label{fig:magp1}
\end{center}
\end{figure}

\begin{figure}[bht]
\begin{center}
\includegraphics[width=8cm]{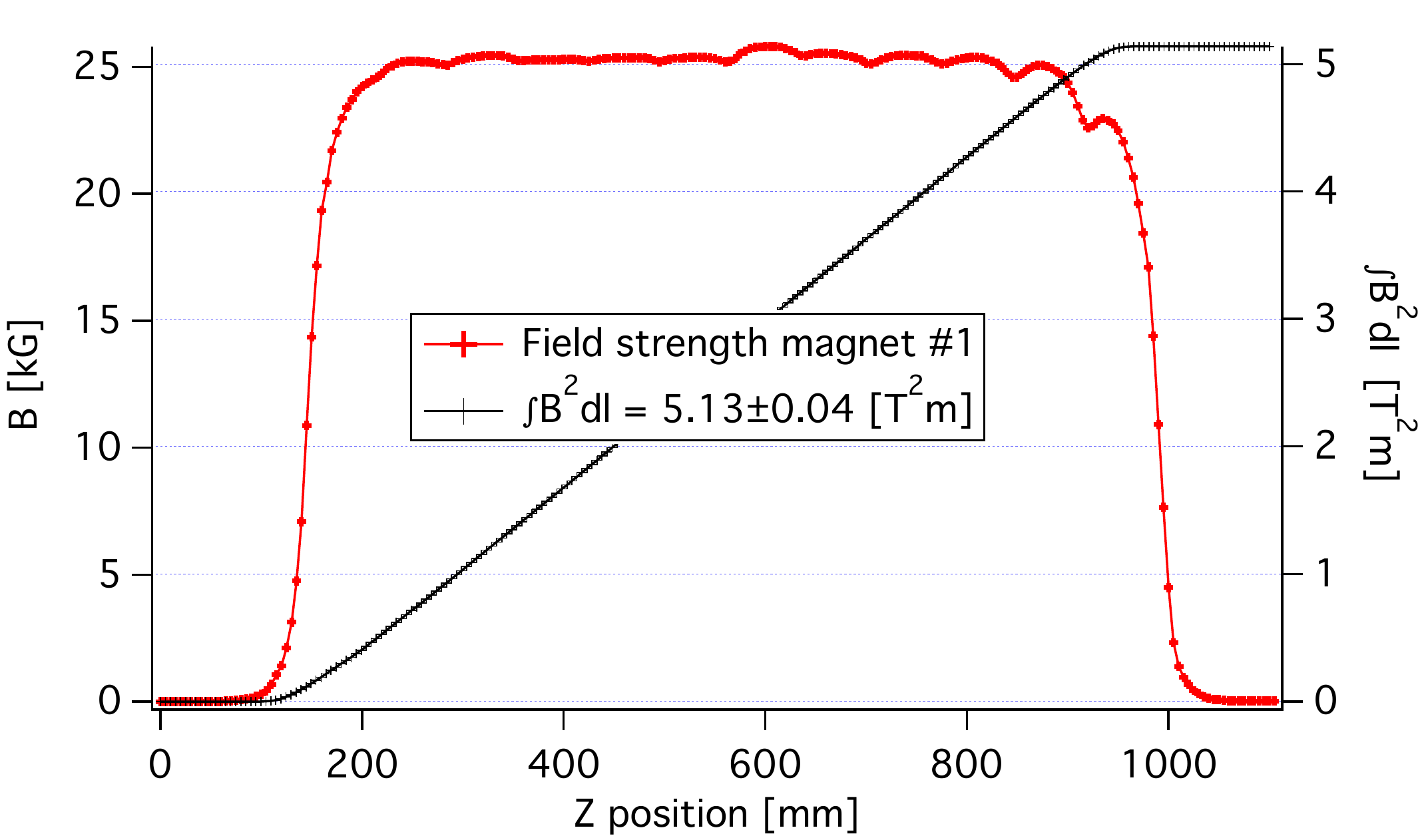}
\includegraphics[width=8cm]{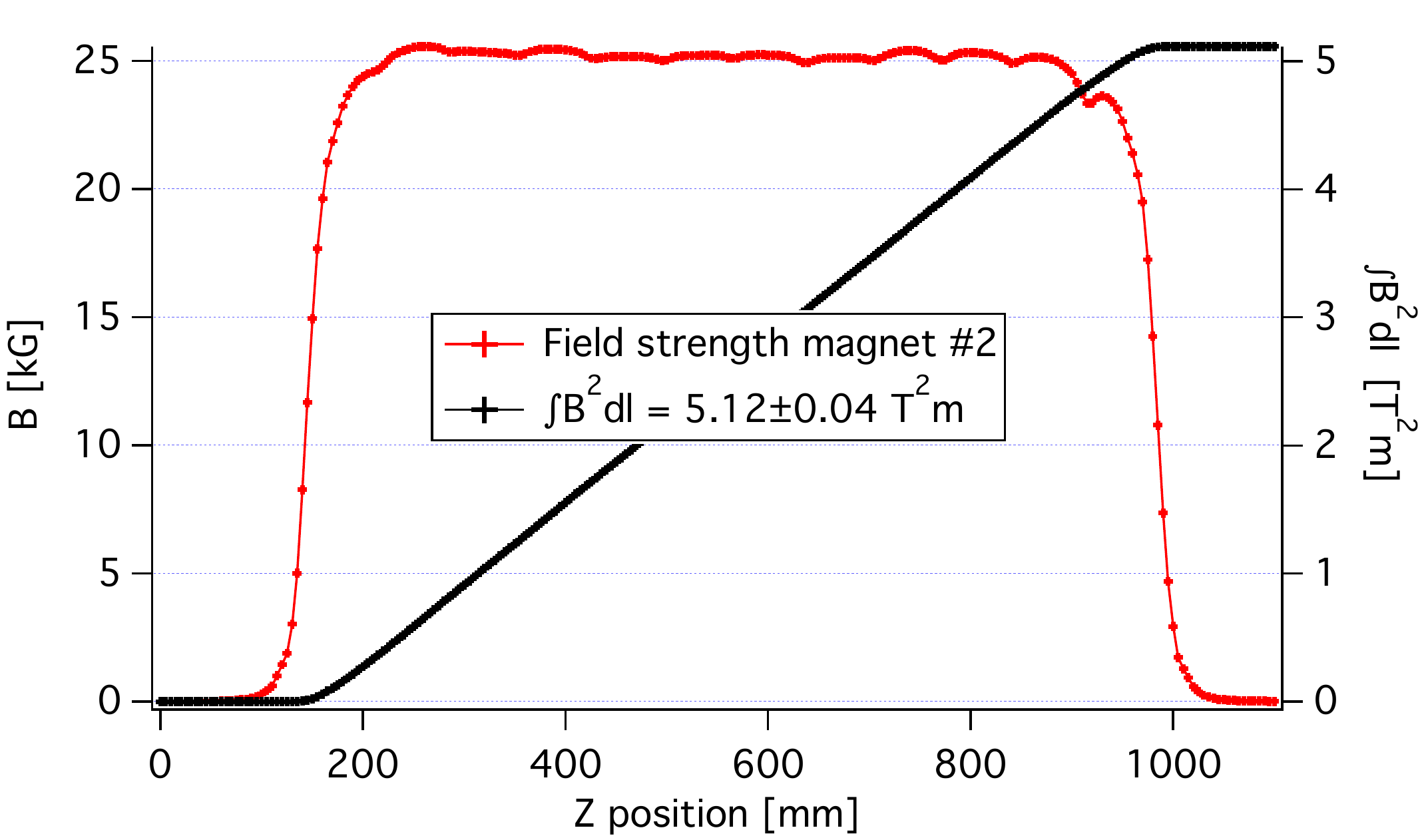}
\caption{Dipolar field profile of the two PVLAS-FE magnets. The values of $\int{B_{\rm ext}^2}\,dL$ are also reported.} 
\label{fig:mag_profile}
\end{center}
\end{figure}

The segments of the Halbach structure were divided into two concentric rings. Each ring was composed of 16 sectors of Nd-Fe-B pre-magnetised material. In Figure~\ref{fig:Halbach-AMC} one can see a drawing of the concentric structure of the PVLAS-FE magnets and a theoretical magnetisation direction of the various sectors. Twelve layers each 70~mm thick were assembled axially in a cylindrical soft magnetic steel case shown in Figure~\ref{fig:magp1}. To minimize the stray field each end flange was a four layers stack alternating aluminium and steel. 
Each magnet was balanced by the manufacturer according to G 2.5 accuracy class ISO 1940-1.
The measured transverse component of the magnetic field for the two magnets is shown in Figure~\ref{fig:mag_profile}.



Just as in the PVLAS-Test setup, the magnets were supported by an aluminium non magnetic structure set on the concrete floor 
of the experimental hall. There was no direct contact between the optical bench and the magnet support structure. The support structure allowed the horizontal movement of the magnets for their extraction and their orientation for optimal alignment with the vacuum tubes. The two magnets were kept in rotation by toothed belt transmissions driven by two independent brushless motors whose rotation frequencies were determined by phase-locked independent signal generators. In this way the rotation of the magnets was controlled in phase: the angular position of the magnetic field was exactly known at any time. In order to allow for systematics monitoring and debugging, the two magnets were generally rotated at two slightly different frequencies $\nu_\alpha$ and $\nu_\beta$. 


\subsection{\bf The Fabry-Perot cavity}

The length of the PVLAS-FE high-finesse Fabry-Perot cavity was $d=3.303\pm0.005$~m. This length defines the free spectral range of the interferometer
\begin{equation}
\nu_{\rm fsr}=\frac{c}{2d}=45.38\pm0.07{\rm~MHz}.
\end{equation}
Plano-concave dielectric mirrors with a radius of curvature of $R_{\rm M}=2$~m were manufactured by ATFilms (Boulder, CO, USA) using super-polished fused silica substrates 25.4~mm in diameter and 6~mm thick. The reflecting surface of the mirrors was designed for the highest possible finesse whereas the plane surface had a 1064~nm anti-reflective coating. 

Given the geometrical parameters, the minimum waist of the cavity $w_0$ was in the center of the cavity and had a value \cite{Kogelnik:66}
\begin{equation}
w_0=\sqrt{\frac{\lambda d}{2\pi}\sqrt{\frac{1+g}{1-g}}}=\sqrt{\frac{\lambda}{2\pi}\sqrt{d(2R_{\rm M}-d)}}=0.507{\rm~mm}
\end{equation}
where $g=1-\frac{d}{R_{\rm M}}$. On the mirrors, the beam radius $w_{\rm M}$ was 
\begin{equation}
w_{\rm M}=\sqrt{\frac{\lambda d}{\pi}\sqrt{\frac{1}{1-g^2}}}=1.21{\rm~mm}.
\end{equation}
The separation of the transverse modes for the symmetric cavity is given by
\begin{equation}
\Delta\nu=\,=\frac{\nu_{\rm fsr}}{\pi}\,\arccos\sqrt{g^2}=12.44{\rm~MHz}
\end{equation}
guaranteeing a good separation of the lowest index transverse modes. A single lens with focal length $f=75$~cm was employed to mode match the laser and the Fabry-Perot cavity whose waists were separated by 4.7~m. 


\begin{figure}[bht]
\centering
\includegraphics[width=10cm]{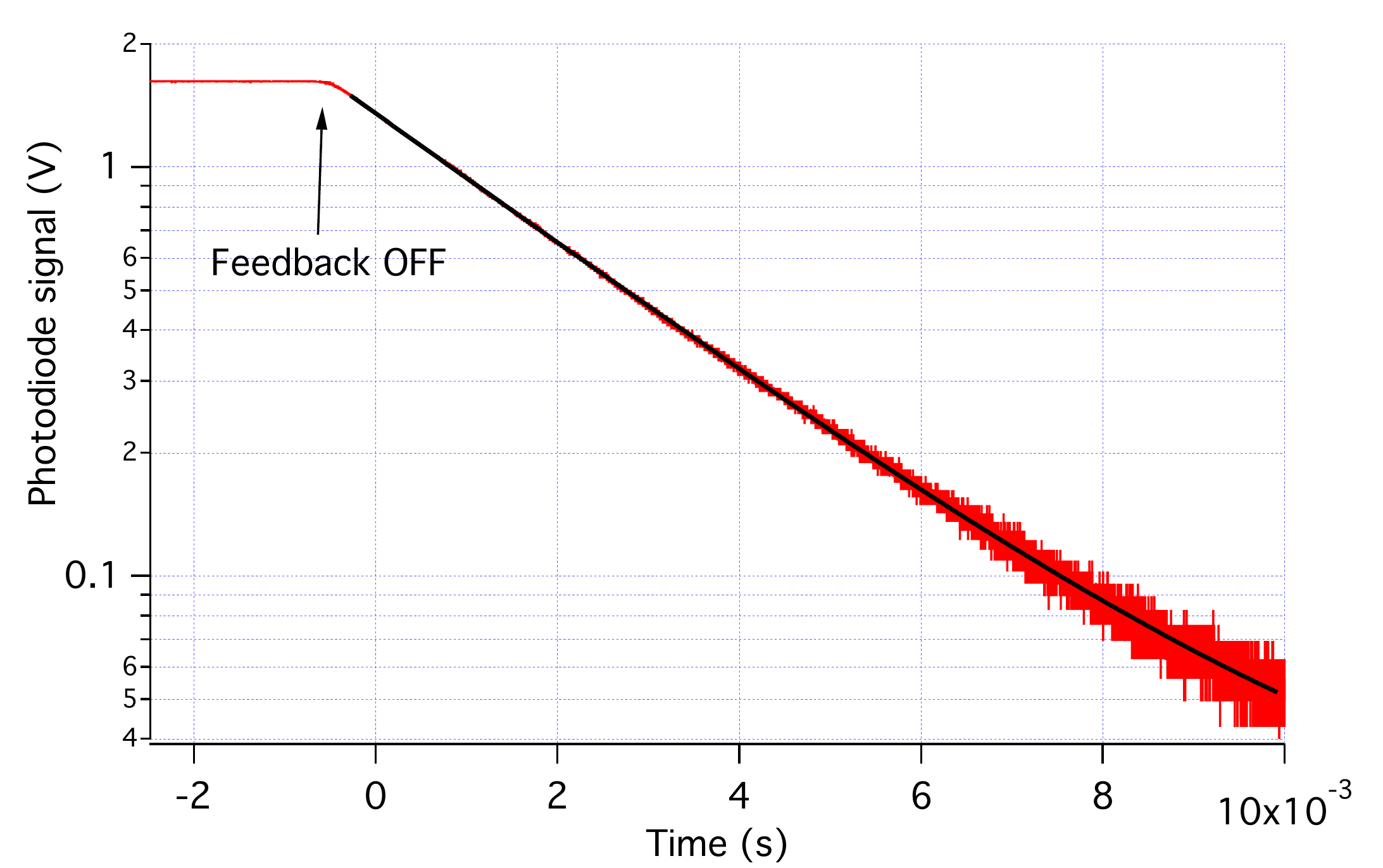}
\caption{Decay of the light power transmitted by the FP cavity following an abrupt unlocking of the laser from the cavity. The curve is fitted with an exponential function $a+be^{-t/\tau}$, with $\tau=2.70\pm0.02$~ms and $a$ compatible with the oscilloscope offset. Given the cavity length $d = 3.303$~m the corresponding finesse was ${\cal F} = 770000$.}
\label{fig:decay-time}
\end{figure}

The mirrors were installed in the vacuum chambers straight from the manufacturer's box without any selection or cleaning. The finesse ${\cal F}$ of the cavity was determined by measuring the decay time $\tau$ of the power exiting the cavity after switching off the locking circuit:
\begin{equation}
\tau = \frac{{\cal F} d}{\pi c}.
\label{eq:decay time}
\end{equation}
The longest decay time ever measured was $\tau=2.70\pm0.02$~ms, and is shown in Figure~\ref{fig:decay-time}. It corresponded to a finesse ${\cal F}=770\,000\pm6\,000$ \cite{DellaValle2014OE}. Such a finesse corresponds to a FWHM cavity line width
\begin{equation}
\nu_c= \frac{1}{2 \pi \tau}=58.9\pm0.4{\rm~Hz}.
\label{eq:fsr-dec-time}
\end{equation}
For this finesse one has
\begin{equation}
1-R=\frac\pi{\cal F}=P+T\approx4.1~{\rm ppm}
\label{eq:ExperimentalFinesse}
\end{equation}
where $R$, $T$ and $P$ are the reflectivity, the transmittivity and the losses of the mirrors, respectively. By measuring the light powers $I_{\rm out}$ and $I_{\rm R}$ transmitted and reflected by the Fabry-Perot cavity \cite{DellaValle2014OE}
\begin{equation}
\frac{I_{\rm out}}{I_{\rm in}}= 0.31 \pm 0.02\qquad{\rm and}\qquad\frac{I_{\rm R}}{I_{\rm in}}=0.25\pm 0.02
\end{equation}
the following values were obtained for the transmittivity $T$, the losses $P$ and the uncoupled power $I_{\rm nc}$:
\begin{equation}
T= (2.4\pm0.2)~{\rm ppm};\qquad P = (1.7\pm0.2)~{\rm ppm};\qquad I_{\rm nc}/I_{\rm in}=0.09\pm 0.04.
\label{Cavity_parameters}
\end{equation}
Most of the VMB measurements were performed with a set of mirrors from the same batch with a finesse about 10\% lower, namely ${\cal F}\approx700\,000$, corresponding to $\nu_c=65$~Hz. Given this value, magneto-optical effects in the cavity were enhanced by a factor
\begin{equation}
N=\frac{2{\cal F}}\pi=446\,000.
\end{equation}
Assuming that the parameter $T$ is an intrinsic property of the mirrors, one can estimate that the actual losses with ${\cal F}=690\,000$ are slightly higher than in equation~(\ref{Cavity_parameters}), namely $P\approx 2.1$~ppm. These two mirrors remained in vacuum (and low pressure gases for calibration) for about three years without losing in reflectivity thanks to the cleanliness of the pumping system and to the isolation valves of the mirrors' vacuum chambers.

During measurements the maximum input power was $I_{\rm in}\approx 50$~mW with a power density on the mirrors of the order of $\approx0.2$~MW/cm$^2$, a value well below the damage threshold of the mirrors as declared by the manufacturer.

\begin{figure}[hbt]
\centering
\includegraphics[width=16cm]{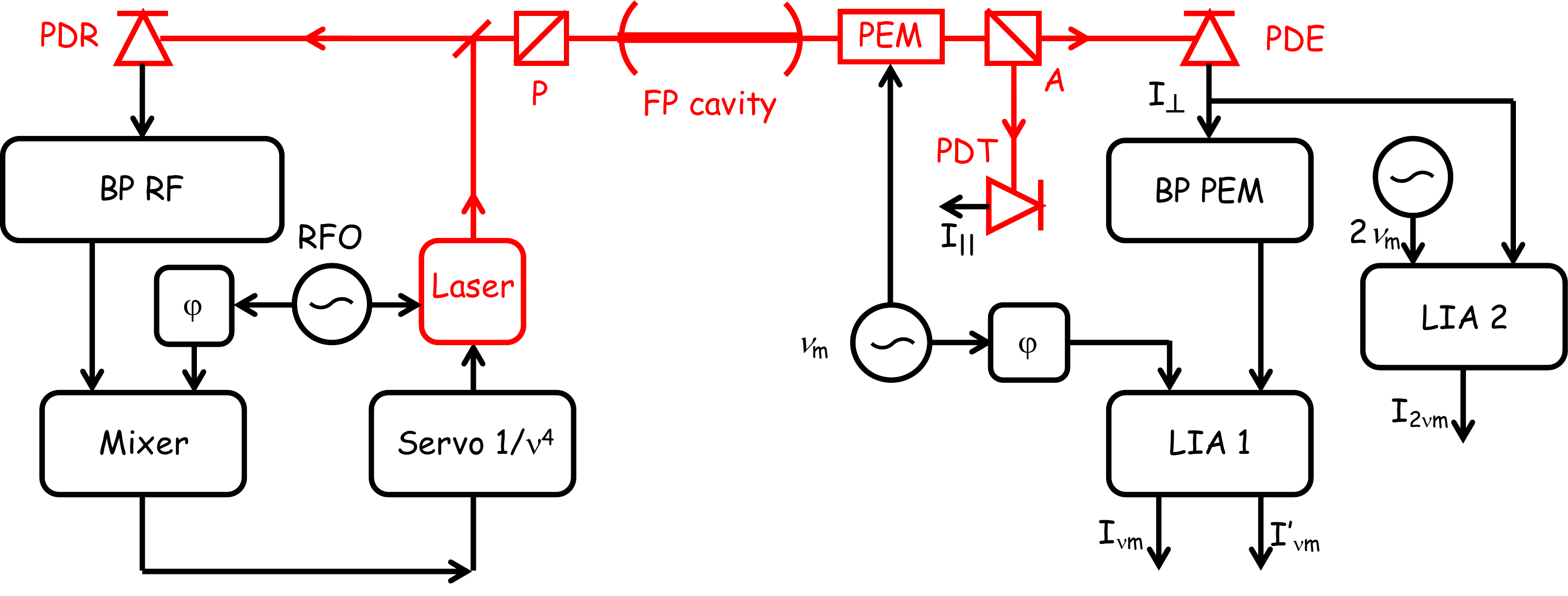}
\caption{General optical and electronic scheme of the PVLAS-FE experiment. BP~=~Band Pass filter; RFO~=~radio frequency oscillator for the PDH locking circuit; LIA~=~Lock-In Amplifier. The signals $I_{\nu_m}$ and $I'_{\nu_m}$ from LIA 1 are the in-phase and quadrature outputs of the demodulation at the RF.}
\label{Elettronica}
\end{figure}

\subsection{\bf Data acquisition}

A general optical and electronic scheme of the electronics of the experiment is shown in Figure~\ref{Elettronica}.

As seen in Section~\ref{sec:magnet_rotation}, the two magnets rotated independently at frequencies $\nu_{\alpha}$ and $\nu_{\beta}$. Normally $\nu_\alpha\neq\nu_\beta$, so that the measurements taken with one magnet were a counter check for the results of the other. The two frequencies were chosen so as to have no common low-order harmonics. The acquisition was started by a trigger of frequency $\nu_T$ equal to a common submultiple of $\nu_\alpha$ and $\nu_\beta$. In practice, $\nu_T=|\nu_\alpha-\nu_\beta|$. When the acquisition started, the magnetic fields of the two magnets had the same direction.

Two acquisition systems were used, a 4-channel spectrum analyser and a 16-channel acquisition board, both synchronised by means of a 10~MHz reference signal with the generators used for the magnets rotation.

The spectrum analyser only acquired the minimum data set necessary to implement the calculations in equation~(\ref{signal}). For this reason, this system was usually employed only for quick tests. To take full advantage of the measurement time, a uniform window was selected. The choice of the rotation frequencies of the magnets was hence limited to multiples of the frequency width of the bin of the spectrum analyser. In this condition the interesting signals appeared in a single bin. 

In the case of the acquisition board, data sampling was synchronous with the rotation of the magnets. The numbers of samples per magnet turn, $N_\alpha$ and $N_\beta$, were both integers; the smaller one was equal to 32 at the beginning, and became 16 later. The values of $N_\alpha$ and $N_\beta$ are related to the rate of sampling $\nu_S$ as
\begin{equation}
N_\alpha\nu_\alpha=N_\beta\nu_\beta=\nu_{\rm S}.
\end{equation}
During acquisition, a very low-frequency feedback kept low the frequency component $I_{\nu_m}$ at the modulation frequency $\nu_m$ of the PEM due to a $\gamma{\rm (DC)}$. This was done by continuously acting on the polariser angular position (and at same time on the analyser position to preserve the extinction condition). The imposed condition was $\gamma{\rm (DC)}<10^{-5}$.

\subsection{\bf Data analysis}
\label{DataAnalysis}

The acquired data were first of all scanned searching for anomalies. If needed, all the data between two trigger signals could be removed, corresponding to an integer number of revolutions of both magnets, thereby preserving phase continuity. Also, thanks to the complete control of the magnet phase, data blocks acquired in different times in the same experimental conditions could be sewn together to make longer data sets. Then the ellipticity (or the rotation) {\em vs} time was calculated using equation~(\ref{signal}). Finally the frequency spectrum was obtained by a Fast Fourier Transform (FFT), whereas the amplitude $\Psi_{2\nu_B}$ ($\Phi_{2\nu_B}$) and the phase $\varphi_{2\nu_B}$ of the bin at the frequency $2\nu_B$ were estimated by a discrete Fourier transform (DFT) at that frequency.

We assumed that in any small enough frequency interval around $2\nu_B$ the ellipticity histogram followed a Rayleigh distribution $P(\rho) = (\rho/\sigma^2) e^{-\frac{\rho^2}{2\sigma^2}}$, in which the parameter $\sigma$ represents the standard deviation of two identical independent Gaussian distributions of two variables $x$ and $y$ with expectation value equal to zero, hence $\rho = \sqrt{x^2+y^2}$. In our case, $x$ and $y$ represent the projections of the ellipticity (or of the rotation) along the physical and the non-physical axes, as defined by the CM calibration. The average value of $P(\rho)$ over the considered interval is related to $\sigma$ by $\langle P \rangle = \sigma\sqrt{\pi/2}$ allowing the determination, for each data run, of the standard deviation $\sigma_{2\nu_B}$ of the ellipticity (or of the rotation) around $2\nu_B$. 

The values of $\Delta n$ and $\sigma_{\Delta n}$ (or of $\Delta\kappa$ and $\sigma_{\Delta\kappa}$) were determined for each block as:
\begin{eqnarray}
\left(\Delta n\pm\sigma_{\Delta n}\right) &=& \frac{\lambda}{\pi N L_B}\left(\Psi_{2\nu_B}\pm\sigma_{2\nu_B}\right)\\
\left(\Delta\kappa\pm\sigma_{\Delta\kappa}\right) &=& \frac{\lambda}{\pi N L_B}\left(\Phi_{2\nu_B}\pm\sigma_{2\nu_B}\right).
\end{eqnarray}
where the values $\Psi_{2\nu_B}$,$\Phi_{2\nu_B}$ and $\sigma_{2\nu_B}$ were corrected for $k(\alpha_{\rm EQ})$ and for the frequency response of the system. The values of $\Delta n$ or $\Delta\kappa$ were then projected along the physical and non-physical axes. In the absence of a physical signal, the phase $\varphi_{2\nu_b}$ did not coincide with the phase of the CM signal measured during the calibration phase. It is clear, though, that in this case the meaningful numbers in the above expressions were $\sigma_{\Delta n}$ and $\sigma_{\Delta\kappa}$ and that $\sigma_{\Delta n},\sigma_{\Delta\kappa}\gtrsim\Delta n,\Delta\kappa$.

Finally, the weighted average of the values of $\Delta n$ and $\Delta \kappa$ were determined from the values for each run. In doing this a linear dependence of the ellipticity and the rotation with the length of the field region is assumed. As we will see, when putting limits on ALPs with different magnet lengths, this is not the case.

Notice that the mixing of ellipticities and rotations due to the cavity birefringence is described by the parameters $N$ and $\alpha_{\rm EQ}$ (see Section~\ref{sec:Fabry-Perot_Systematics}). Once they were known, from the birefringence noise measurements we could determine a limit also on the dichroism $\Delta\kappa$ and {\em vice versa}.

\section{PVLAS-FE commissioning}
\label{sec:commissioning}

The commissioning of the PVLAS-FE apparatus was divided in three phases: calibration measurements with large signals via the Cotton-Mouton effect, understanding and reduction of `in-phase' systematic noise signals appearing at harmonics of the rotating magnets and `wide-band' noise studies. The `in-phase' systematic signals are of particular interest in that they limit the ultimate noise floor which can be achieved. These issues will be treated separately in the following sections.

\subsection{\bf Calibration measurements}
\label{sec:cal_meas}

The calibration procedure of the PVLAS-FE apparatus was described in Section~\ref{sec:calibration-CM}. Here we will present some measurements which were usually performed before any vacuum measurement.

\subsubsection{Characterisation of the cavity birefringence}

\label{sec:meas_birif_cavity}
As mentioned in the calibration procedure the first parameter to be determined after the finesse ${\cal F}$ is $\alpha_{\rm EQ}$. It is also desirable to minimise this value so as to maximise the factor $k(\alpha_{\rm EQ})$. This is particularly important since the noise floor $\Delta{\cal D}$ achievable in a given integration time goes as $k(\alpha_{\rm EQ})^{-2}$. This minimum condition is obtained when the slow axis of one mirror is aligned with the fast axis of the other. This configuration ensures that the resonance peaks for the two polarisation states inside the cavity are as close as possible. 

If in equations~(\ref{alfaEQ}) and (\ref{phiEQ}) $\alpha_1$ were equal to $\alpha_2$ and $\phi_{\rm WP}=90^\circ$, the equivalent wave plate retardation would be zero. In this case, on a plot like that of Figure~\ref{BirefringentAiry} the resonance curves of the two polarisation auto-states would be superimposed (red and black curves). If $\alpha_1\neq\alpha_2$, the effect of the equivalent wave plate can only be minimised or maximised but never extinguished. To reach a good extinction and maximise $k(\alpha_{\rm EQ})$ the laser polarisation was always aligned to one of the axes of the equivalent wave plate of the cavity. If this were not the case, a large ellipticity would be observed in the polarisation of the extinguished beam with a large Fourier component at the frequency $\nu_m$ of the ellipticity modulator (PEM). To be more precise, there are also ellipticities generated by other optical elements between the input polariser and the output analyser which we called $\gamma_{\rm other}(t)$ in Section~\ref{sec:MirrorBirefringence}. In principle this signal would not affect the magnetic birefringence signal at $2\nu_B$ but in practice a large signal at $\nu_m$ would lead to a noise contribution to $S_\Psi$ due to the laser's relative intensity noise at $2\nu_B$ according to equation~(\ref{RIN_noise}). It is therefore necessary to keep the signal at $\nu_m$ as low as possible. The alignment procedure means that the input polarisation is aligned to the composition of all the static ellipticities generated by the birefringences existing in the polarimeter in the path from the polariser to the analyser. Of all these induced ellipticities the largest is indeed due to the cavity. The Fourier component at $\nu_m$ can therefore be maintained at zero by rotating the input polariser whereby the cavity ellipticity cancels all others.
 
In order to study the equivalent wave plate of the cavity, we performed the measurement of the ellipticity and of the rotation generated by the Cotton-Mouton effect in a gas as a function of the relative azimuthal position of the two mirrors. In this experimental condition there is no dichroism ($\phi=0$) and a rotation is therefore due solely to the presence of $\alpha_{\rm EQ}$. In these measurements, the magnets rotated at $\nu_B=4$~Hz; this corresponded to a negligible correction factor due to the cavity first order filtering \cite{Uehara1995} for the signal at $\nu=2\nu_B = 8$~Hz:
\begin{equation}
E_{\rm T}(2\nu_B)=E_{\rm out}  \frac{T}{\sqrt{1 +R^2 - 2R\cos\delta}} = 0.97
\end{equation}

\begin{figure}[bht]
\begin{center}
\includegraphics[width=12cm]{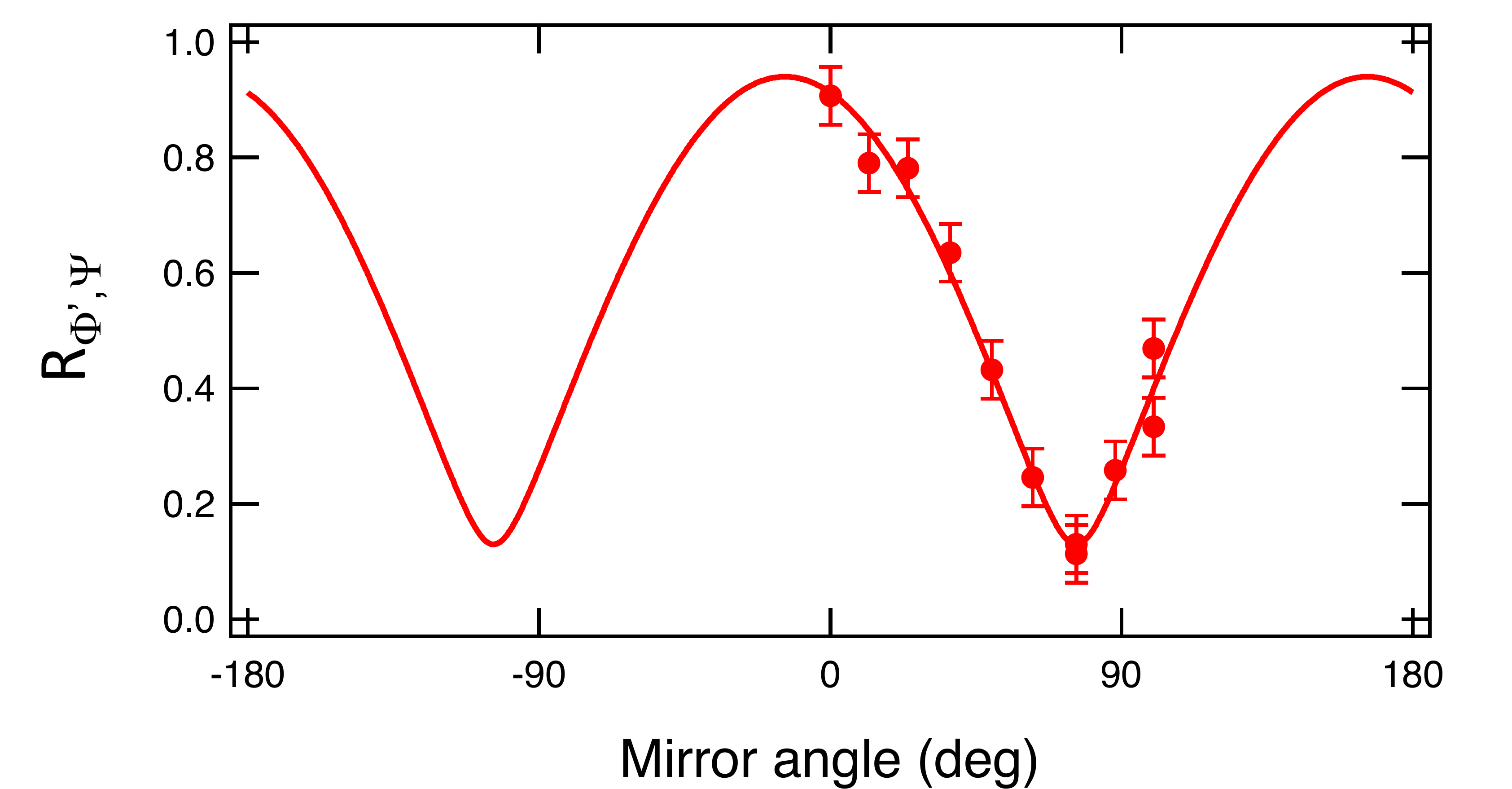}
\end{center}
\caption{Ratio of the `spurious' rotation to the `true' ellipticity, $R_{\Phi',\Psi} = N\alpha_{\rm EQ}/2$ of equation~(\ref{R_Phi_Psi}), plotted as a function of the azimuthal angle $\phi_{\rm WP}$ of the input mirror for a Cotton-Mouton effect of $230~\mu$bar of Ar gas. The continuous line is a best fit with parameters $\alpha_1$, $\alpha_2$ and $\phi_{\rm WP}$ of equations~(\ref{alfaEQ}) and (\ref{phiEQ}). From Reference \cite{DellaValle2016EPJC}, Figure 8.}
\label{MirrorWP}
\end{figure}

In Figure~\ref{MirrorWP}, we show the ratio $R_{\Phi',\Psi}$, given by equation~(\ref{R_Phi_Psi}), plotted as a function of the azimuthal angle $\phi_{\rm{WP}}$ of the first mirror (the second mirror was never moved). Each rotation step, of about $15^\circ$, was followed by cavity realignment through the adjustment of the two tilt stages of the mirror, by optimisation of the extinction ratio through the rotation of both the polariser and analyser and by a measurement of the finesse. The experimental points were fitted with equation~(\ref{R_Phi_Psi}), where $\alpha_{\rm EQ}$ is given by equation~(\ref{alfaEQ}). The best fit produced values for the quantities $N\alpha_1/2$, $N\alpha_2/2$, and for the angular position of the maxima with respect to the initial angular position of the input mirror ($\phi_{\rm{WP}}=0$). With $N/2\approx2.2\times10^5$, the phase differences of the two mirrors were calculated to be 
\begin{equation}
\alpha' = (2.4\pm0.1)~\mu {\rm rad}\qquad {\rm and} \qquad \alpha'' = (1.9\pm0.1)~\mu {\rm rad}.
\label{alphas}
\end{equation}
It was not possible to associate $\alpha'$ and $\alpha''$ uniquely to the two mirrors with this single measurement.
With the above values for the $\alpha$'s and by varying the relative angular position of the two mirrors $\phi_{\rm{WP}}$, the value of $\alpha_{\rm{EQ}}$ could vary between $0.6~\mu$rad and $4.3~\mu$rad, which is equivalent to saying that the maximum of the Airy curve of the ellipticity resonance would be set between 5~Hz to 31~Hz away from the resonance of the input polarisation. Correspondingly, the $k(\alpha_{\rm EQ})$ parameter could be varied between $\approx1$ and $\approx0.5$. 

\begin{figure}[bht]
\begin{center}
\includegraphics[width=12cm]{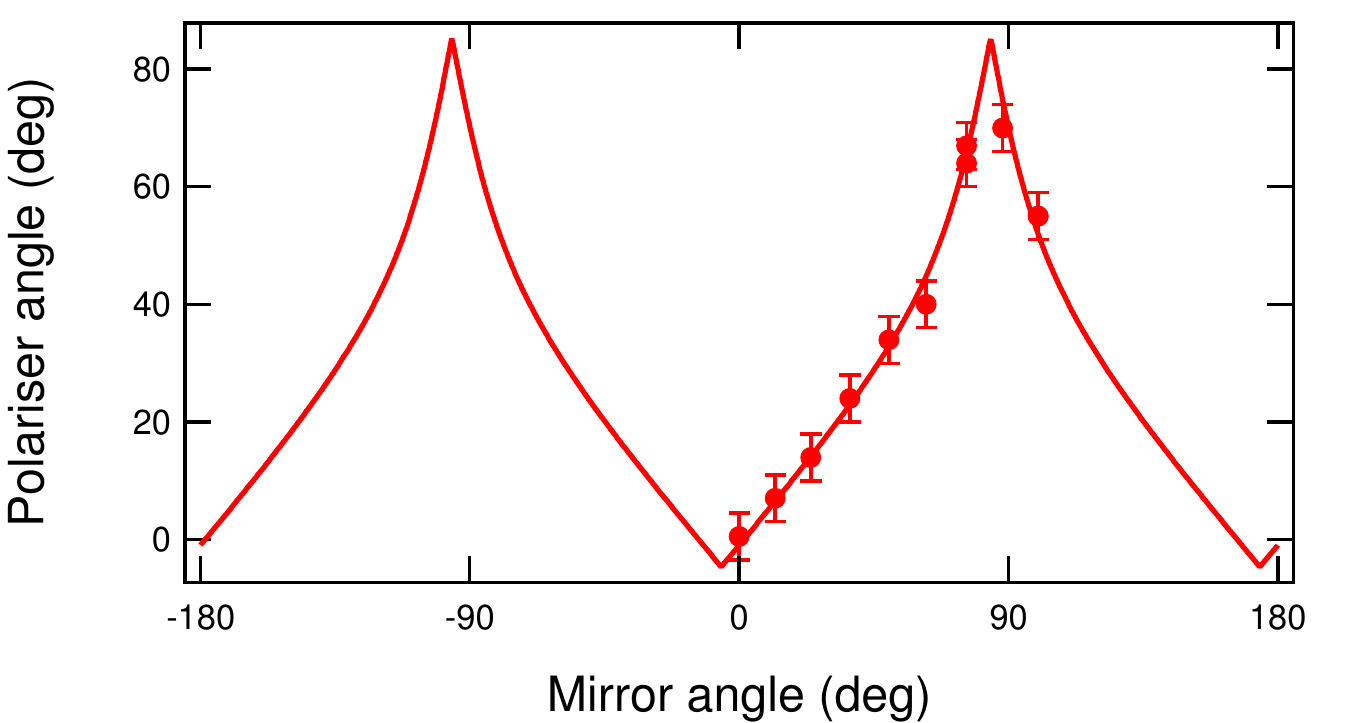}
\end{center}
\caption{Input polariser angle as a function of the azimuthal angle $\phi_{\rm WP}$ of the input mirror in a Cotton-Mouton measurement with $230~\mu$bar of Ar. Data are fitted with equation~(\ref{phiEQ}). From Reference \cite{DellaValle2016EPJC}, Figure 9.}
\label{fig:MirrorAxis}
\end{figure}

As described above, for each rotation step of the input mirror, the ellipticity and the rotation generated by the Cotton-Mouton effect were measured. These values depend only on the value of $\alpha_{\rm EQ}$ according to equations~(\ref{i_perp_ell}) and (\ref{i_perp_rot}). In these measurements the best extinction ratio, and therefore the lowest signal at $\nu_m$, was obtained by rotating the polariser and the analyser by a measured amount. The extinction condition ensured the alignment of the polarisation with the axis of the equivalent wave plate of the whole polarimeter (not just of the cavity). Figure~\ref{fig:MirrorAxis} shows the azimuthal angle $\phi_{\rm EQ}$ of the polariser for which the best extinction ratio was obtained, as a function of the  mirror angular position $\phi_{\rm WP}$. The data points were fitted with equation~(\ref{phiEQ}). The best fit gave a value of $\alpha_1/\alpha_2=0.62\pm0.08$, allowing the assignment of the phase delay of each mirror. This value is slightly different from $\alpha''/\alpha'$ of equation~(\ref{alphas}) obtained by the fit in Figure~\ref{MirrorWP}, but is compatible within the fit uncertainties. However, the zero references of $\phi_{\rm{WP}}$ in the two fits appear to be different by about $10^\circ$, well beyond the fit uncertainty. This is indeed evidence of a contribution of other birefringent elements (mirror substrates and PEM) between the polariser and the analyser. The apparent discrepancy of the two measurements is due to their different character: the positioning of the polariser in the measurement of the extinction ratio was made following the indications of the $\nu_m$ signal in the Fourier transform of the extinguished beam, which is the DC component of the demodulated power corresponding to $\gamma({\rm DC}) = \gamma_{\rm cavity}(\rm DC)+\gamma_{\rm other}(\rm DC)$, whereas the measurement of the CM effect is performed at the frequency $2\nu_B$, twice the rotation frequency of the magnet and depends on $\alpha_{\rm EQ}$.

\begin{figure}[bht]
\begin{center}
\includegraphics[width=14cm]{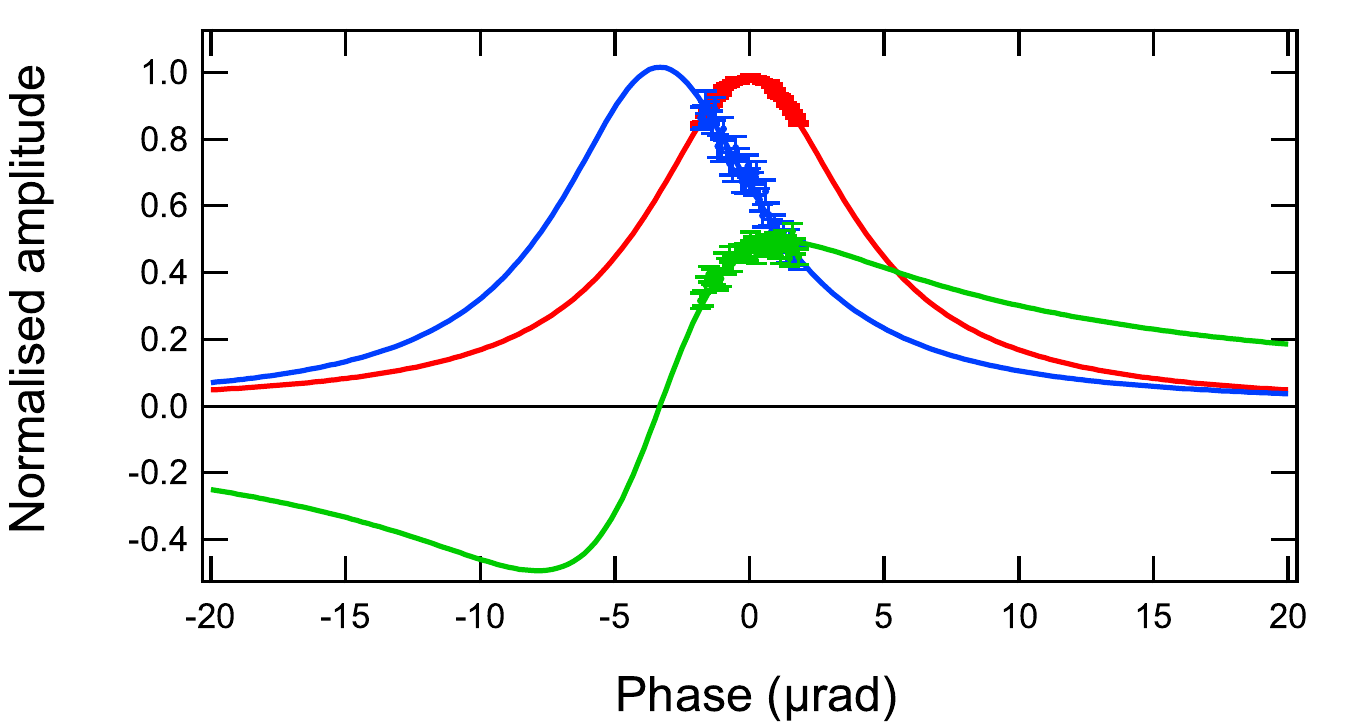}
\end{center}
\caption{Ellipticity (blue), transmitted power (red) and rotation (green) data measured in the Cotton-Mouton effect of $2.2~\mu$bar of O$_2$ gas, plotted as a function of the `offset' of the laser locking feedback circuit. The continuous lines are the fits obtained with formulas (\ref{eq:EllipticityAiry}), (\ref{eq:i_out}) and (\ref{eq:RotationAiry}), giving $|\alpha_{\rm EQ}|=(3.3\pm0.1)~\mu$rad. 
}
\label{BirefringentFP}
\end{figure}

A direct visual demonstration of the birefringence of the cavity was obtained thanks to the capability of the apparatus to modify the `offset' of the feedback which locked the frequency of the laser to the resonance frequency of the cavity. This allowed for polarimetric measurements off resonance with phase values of 
\begin{equation}
\Delta\delta \approx \pm2\pi\left(\frac{\Delta\nu_c}{4\nu_{\rm fsr}}\right)
\end{equation}
allowing experimental verification of the mathematics presented in Section~\ref{sec:MirrorBirefringence}. In these measurements, the azimuthal coordinate of the first mirror is kept fixed and $\Delta\delta$ is changed. Figure~\ref{BirefringentFP} shows the 2015 experimental data that correspond to the model of Figure~\ref{BirefringentAiry}. The solid lines are the fits obtained with the formulas (\ref{eq:i_out}), (\ref{eq:EllipticityAiry}), and (\ref{eq:RotationAiry}). In the three fits, a common value has been used for the resonance width. The ellipticity (in blue in the figure) and rotation data (in green) are forced to have the same resonance frequency: this corresponds to a maximum for the ellipticity and to a zero-crossing for the rotation. From the fits, one determines the calibration factor between the feedback `offset' and the phase $\Delta\delta$. The result for the phase delay between the Airy curves is $\alpha_{\rm EQ}=(3.3\pm0.1)~\mu$rad (with undetermined sign). This corresponds to a difference in the resonance frequencies of the two orthogonal polarisation of about $\Delta \nu = 24$~Hz and $k(\alpha_{\rm EQ})=0.65$.

As a final remark, we note that none of the experimental procedures described in this Section allowed us to define the sign of $\alpha_{\rm EQ}$: the plots of Figures~\ref{MirrorWP} and \ref{fig:MirrorAxis} fix only the relative sign of the wave plates of each mirror. The method of Figure~\ref{BirefringentFP} is in principle capable of defining the sign of $\alpha_{\rm EQ}$, but only as long as one knows whether the QWP used in the rotation measurements is aligned with the slow or the fast axis with respect to the polarisation direction. This was not the case. 


\subsubsection{Frequency response measurements}

As we will see, the wide-band noise decreases with a certain power law as a function of the frequency. As a consequence, and in principle, the higher the working frequency, namely twice the rotation frequency of the magnets, the better the signal to noise ratio (SNR) of the measurement (see Section~\ref{sec:wide_band_noise}). 
One has to note, however, that the Fabry-Perot cavity behaves like a low-pass filter. Well before $2\nu_B$ approaches the cutoff frequency of the cavity, the calibration discussed in Section~\ref{sec:calibration-CM} depends on the frequency response of the cavity. Furthermore, as was discussed in Section~\ref{sec:PolarisationDynamics}, there is a significant deviation of the frequency response from a simple first order filter, as expected from a Fabry-Perot cavity, due to the cavity birefringence. This difference will be necessary when explaining the intrinsic noise of the polarimeter in Section~\ref{sec:wide_band_noise}. 


We confirmed the frequency dependences presented above with two different experiments in each of which the ellipticity and the rotation were measured. The first one was the Cotton-Mouton effect in $880~\mu$bar of Ar gas, measured as a function of discrete frequencies between $\nu_B=0.5$~Hz and $\nu_B=23$~Hz with a measurement every 0.5~Hz. Each ellipticity and rotation measurement was integrated for a time of 256~s. 
For this measurement, a single magnet was employed. The phase of the magnet was defined by a trigger signal generated by a contrast sensor in correspondence of the passage of a mark drawn on the external surface of the rotating magnet. 

For the second experiment, as we are not aware of the existence of any magnetic dichroism in gases in the optical range, we used the Faraday effect in the coatings of the mirrors. We placed a solenoid coil to set a magnetic field with a component perpendicular to the reflecting surface of one of the cavity mirrors, thus generating a Faraday effect \cite{Iacopini:1983rp}. The effect is at the first harmonic of the oscillating magnetic field and is linear in the magnitude of the magnetic field. At the position of the mirror, at a distance of about 15~cm along the axis of the solenoid, the field  was $\approx1$~G. In this second experiment the ellipticity played the role of the 'spurious' effect described by equations~(\ref{eq:fase_sec}) and (\ref{eq:amp_sec}).

To perform the Faraday effect measurements, the Frequency Response function of an Agilent 35670A Dynamic Signal Analyzer was employed. The amplitude and phase of a voltage signal from a small resistance in series with the solenoid was used as a phase reference and to normalise the amplitude of the observed rotation. We explored the frequency range from 0 to 50~Hz with 400 frequency bins and a sweep time of 8~s. 
For the rotation, the total integration time was $\approx2$~hours, which corresponds to an integration time of 18~s per bin. The ellipticity, which was approximately three times smaller, was integrated for a total time of $\approx5.5$~hours, corresponding to an integration time of 50~s per bin.

A small frequency-dependent phase correction ($\sim1^\circ$) was subtracted from the measured phases of both the Cotton-Mouton effect and the Faraday effect, due to the frequency response of the lock-in amplifier used to demodulate the signal of the extinguished power.

\subsubsection{Cotton-Mouton measurements}

\begin{figure}[hbt]
\centering
\includegraphics[width=8cm]{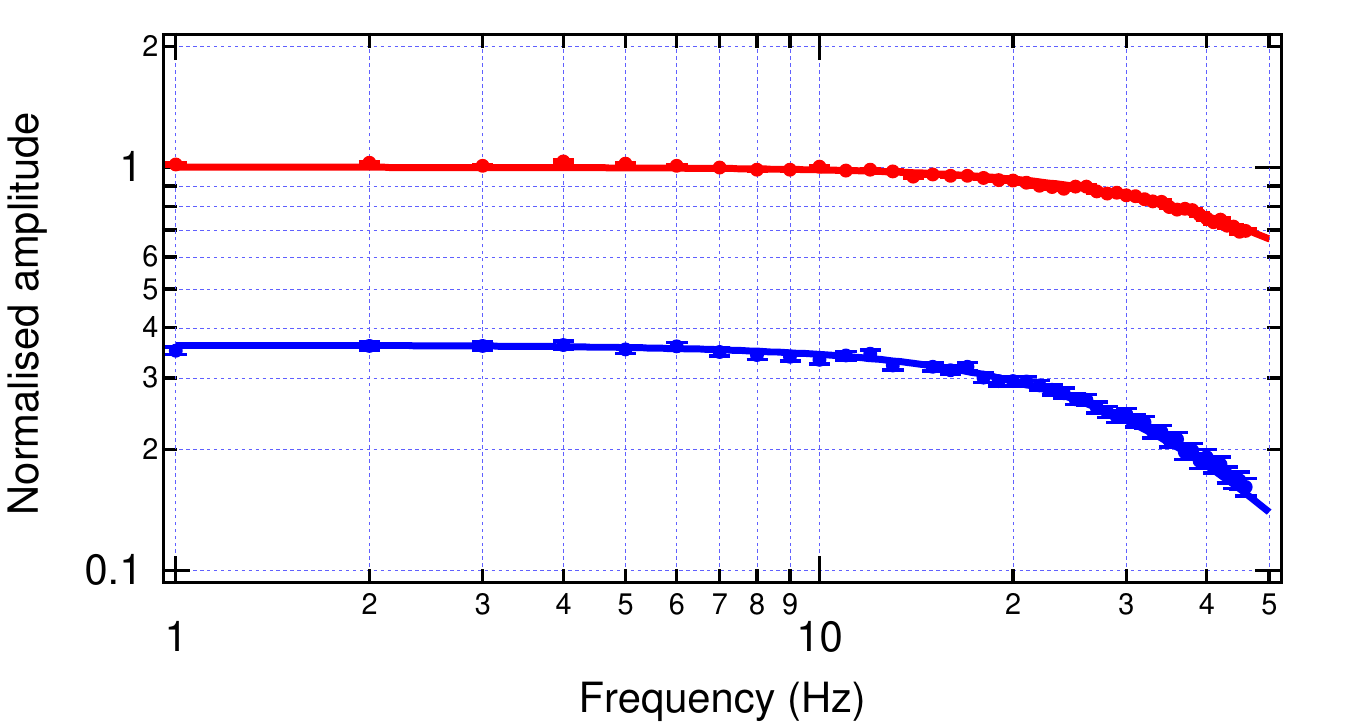}
\includegraphics[width=8cm]{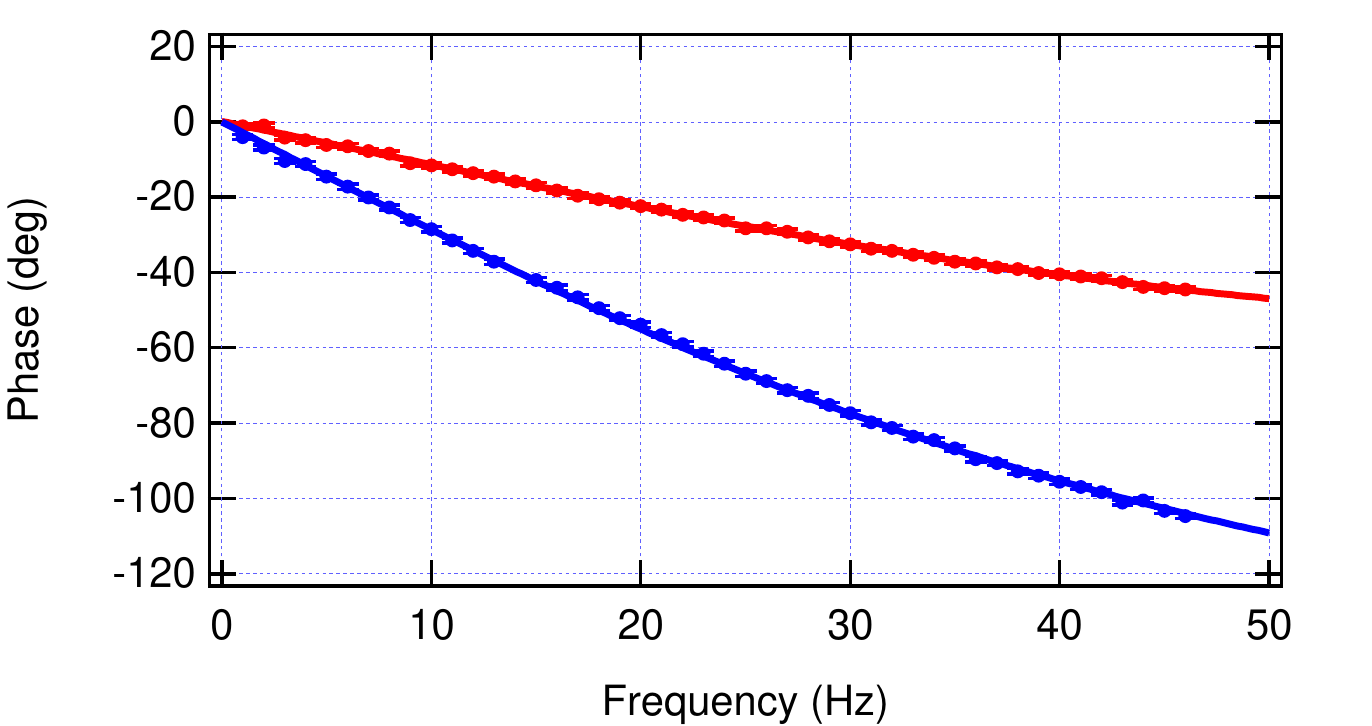}
\caption{Measured frequency response of the Cotton-Mouton effect with $880~\mu$bar of Ar gas as a function of the signal frequency. Left: amplitude of the `true' ellipticity (red) and of the `spurious' rotation (blue) signals. 
Right: phase of the ellipticity (red) and the rotation (blue) signals. The phase and amplitude data are simultaneously fitted with equations~(\ref{eq:fase_first}), (\ref{eq:amp_first}), (\ref{eq:fase_sec}) and (\ref{eq:amp_sec}). From reference \cite{Ejlli2018}, Figure 6.}
\label{fig:CME_freq_resp}
\end{figure}

The data of the frequency response of the ellipticity and the rotation generated during the Cotton-Mouton effect measurement are presented in Figure~\ref{fig:CME_freq_resp}. A constant phase, measuring the zero-frequency relative position of the signals and the trigger, has been subtracted from the phase data, so as to have both curves starting at zero phase. The data are fitted simultaneously with the four functions given in equations (\ref{eq:fase_first}), (\ref{eq:amp_first}), (\ref{eq:fase_sec}) and (\ref{eq:amp_sec}), and the values of the reflectivity $R$, and hence of the finesse ${\cal F}$ of the mirrors, and of the phase difference $\alpha_{\rm EQ}$ of their equivalent wave plate have been obtained:
\begin{equation}
\nonumber {\cal F} = (640 \pm 4)\times 10^3\qquad{\rm and}\qquad
 \alpha_{\rm EQ} = (1.78\pm 0.01)~\mu{\rm rad.}
\end{equation}
with a normalised $\chi^2_{\rm o.d.f.}=181/174$. The uncertainties used in the fit were the piecewise standard deviations of the residuals obtained by fitting the four curves separately.

In a first tentative of a global fit, the residuals of the phase data exhibited a marked linear behaviour of a few degrees over the whole frequency interval. This behaviour was attributed to the fact that, during the measurements, the polarisation direction of the light entering the Fabry-Perot cavity was varied by small quantities to compensate for the slow drift of the static birefringence of the cavity. We added then two linear functions to the two phase fit functions. The values of the slopes obtained through the fit were $(0.1^\circ\pm0.01^\circ)$~Hz$^{-1}$ for the phase of the ellipticity, and $(0.05^\circ\pm0.01^\circ)$~Hz$^{-1}$ for the phase of the rotation. Note that the duration of the ellipticity and rotation measurements were, respectively, eight hours and four hours, leading to an identical drift of $160~\mu$deg/s in the two measurements. This strongly supported our interpretation. This drift is associated with the thermalisation of the mirror shined upon by the laser beam in the presence of the gas. This process slows down only after many hours of continuous operation.  

It is worth noting that the value of $\alpha_{\rm EQ}$ was small enough that fitting simultaneously the four data sets with the expressions of the first and second order filters (\ref{eq:FirstOrderAmpl}) and (\ref{eq:SecondOrderAmpl}) still produced a reasonable fit, with a similar $\chi^2$ probability, but at the expense of an unreasonable 20\% reduction of the value of ${\cal F}$ and of completely incompatible drifts of the ellipticity and rotation phases.

\subsubsection{Faraday effect measurements}

\begin{figure}[bht]
\centering
\includegraphics[width=8cm]{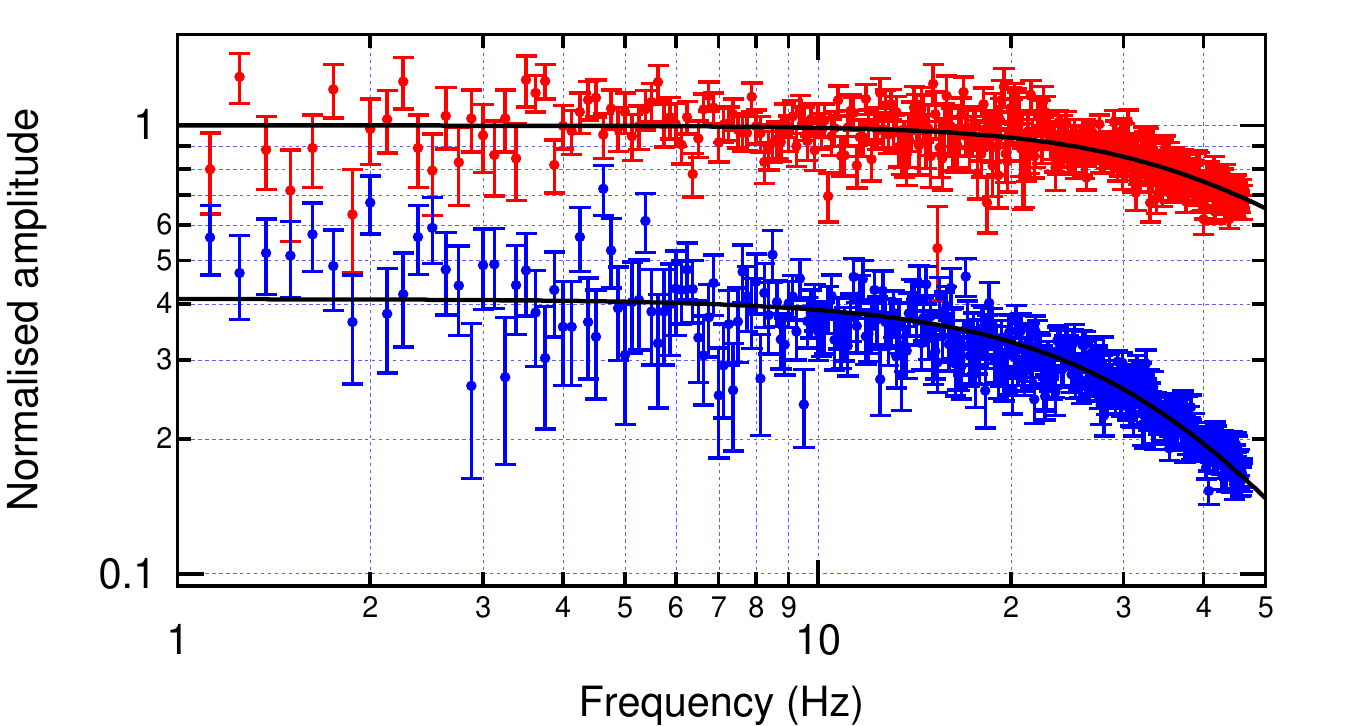}
\includegraphics[width=8cm]{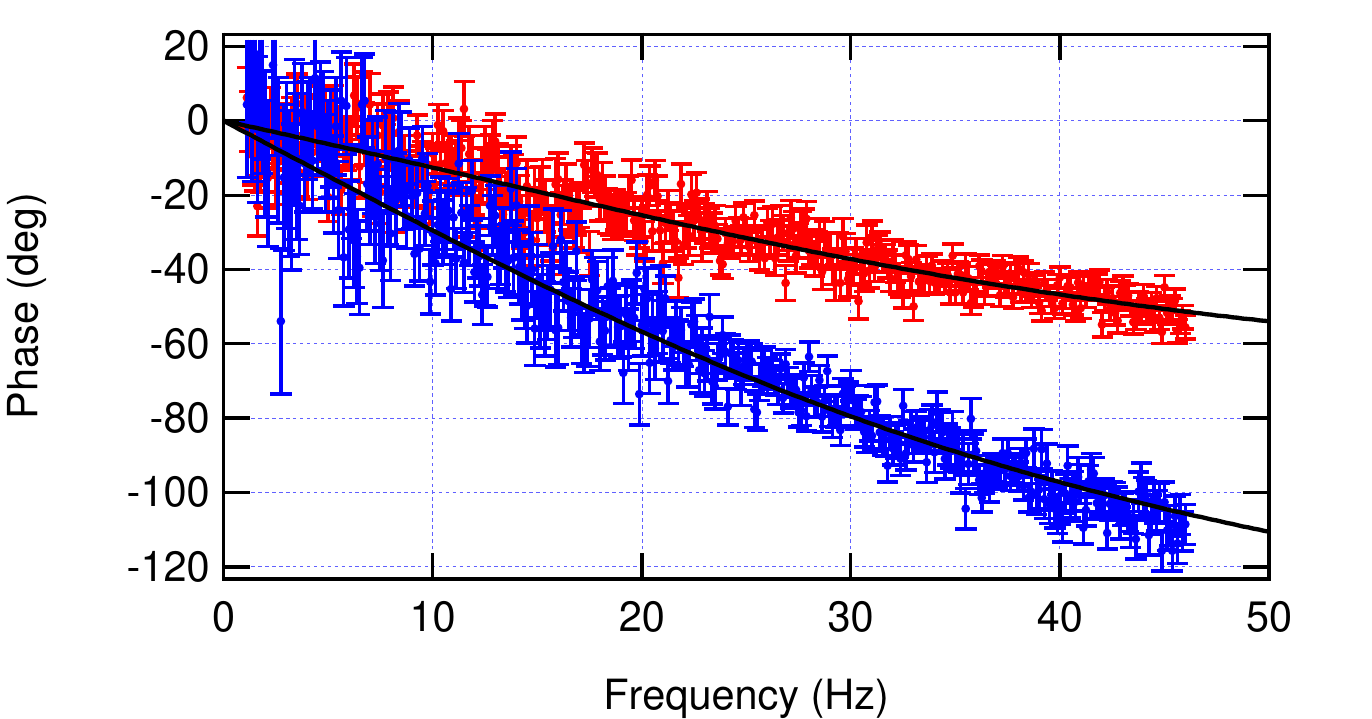}
\caption{Left: relative amplitude of the `true' rotation (red) and of the `spurious' ellipticity (blue) measured as a function of frequency for the Faraday effect on the reflecting surface of a mirror of the Fabry-Perot cavity. Right: phases of the rotation (red) and of the ellipticity (blue). The continuous lines are the global fits obtained with equations~(\ref{eq:fase_first}), (\ref{eq:amp_first}), (\ref{eq:fase_sec}) and (\ref{eq:amp_sec}). From reference \cite{Ejlli2018}, Figure 7.}
\label{fig:Faraday_freq_resp}
\end{figure} 

The data of the frequency response of the ellipticity and rotation generated in the Faraday effect are shown in Figure~\ref{fig:Faraday_freq_resp}. A constant phase, measuring the zero-frequency relative position of the signals and the trigger, has been subtracted from the phase data, so as to have both curves starting at zero phase. The data are fitted simultaneously with the four fit functions (\ref{eq:fase_first}), (\ref{eq:amp_first}), (\ref{eq:fase_sec}) and (\ref{eq:amp_sec}). The fit gives a unique value for the mirror reflectivity $R$ and the phase delay $\alpha_{\rm EQ}$:
\begin{equation}
\nonumber {\cal F}=(691 \pm 0.08)\times 10^3\qquad{\rm and}\qquad
 \alpha_{\rm EQ}=(1.87\pm 0.02)~\mu{\rm rad}
\end{equation} 
with a normalised $\chi^2_{\rm o.d.f.}=1472/1434$. The value of $\alpha_{\rm EQ}$ is 5\% larger than the one found in the Cotton-Mouton experiment. This small difference could be accounted for by the fact that the two data sets were taken in different days and that we know that $\alpha_{\rm EQ}$ is subject to small drifts. As in the case of the Cotton-Mouton measurement, the uncertainties used in the fit are the piecewise standard deviations of the residuals obtained by fitting the four curves separately.

Differently from the Cotton-Mouton case, no phase drift correction was necessary. This is consistent with the interpretation of the feature observed in the Cotton-Mouton effect: in fact, in the case of the Faraday measurements, the phase is electronically defined and therefore does not depend on the position of the polariser. 

By fitting the four curves with the expressions of the first and second order filters (\ref{eq:FirstOrderAmpl}) and (\ref{eq:SecondOrderAmpl}) we obtained ${\cal F}=594\times10^3$, with a $\chi^2$ probability of $5\times10^{-3}$, justifying the necessity of introducing the parameter $\alpha_{\rm EQ}$.

\subsection{\bf In-phase spurious signals}

In this section and  the next we will discuss noise sources afflicting the PVLAS-FE apparatus. We will discuss separately `in-phase' noise sources appearing at harmonics of the rotating magnets and `wide-band' noise present independently of the rotation of the magnets.

Spurious signals in ellipticity and rotation were observed in all the experimental setups of PVLAS, with similar characteristics of apparent non-repeatability. In the older setups, their sources were not identified due to insufficient debugging. Here we will present the different phenomena we explored as possible sources of spurious signals. 

The ellipticity coming from a true magnetic birefringence inside the cavity was calibrated in a CM measurement with a large signal-to-noise ratio, as described in Section~\ref{sec:calibration-CM}. In particular, the signals coming from the magnets rotating at two different frequencies had the same amplitude and phase once corrected for the dynamical response of the Fabry-Perot cavity. Moreover, the phase of the ellipticity was independent of the gas species (modulo $180^\circ$). 
As a last point, `good' signals occupied a single bin in the Fourier spectrum.

\begin{figure}[bht]
\centering
\includegraphics[width=10cm]{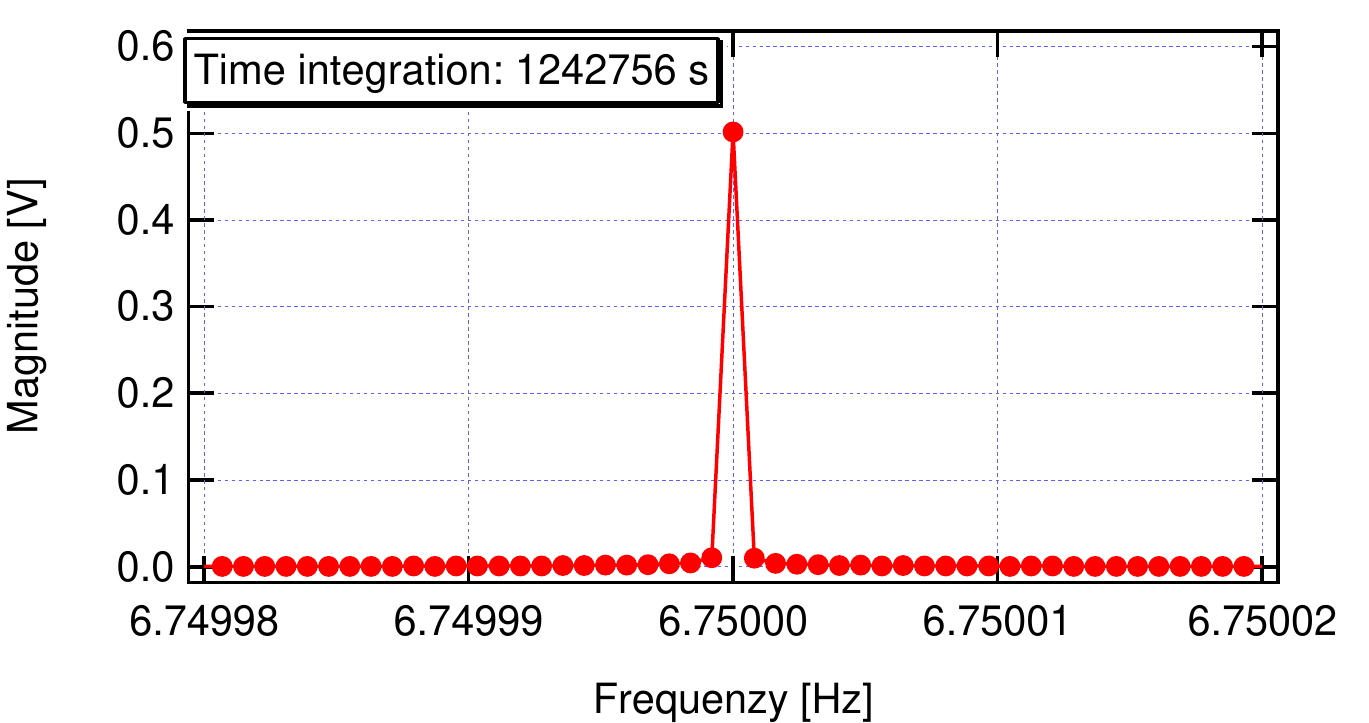}
\caption{Fourier transform of a $1.2\times10^6\;$s data set of the stray magnetic field of one of the rotating magnets.}
\label{fig:fft_mag}
\end{figure}

As mentioned before, one of the strengths of the PVLAS-FE experiment was the possibility of obtaining the Fourier transform of very long data sets; in a short data set disturbances not exactly at $2\nu_B$, for example due to a mechanical excitation, were likely to show up as a peak simulating a birefringence signal. In a long data set though such a peak would occupy several bins indicating a false signal. In Figure~\ref{fig:fft_mag} we report the Fourier spectrum of the magnetic stray field of one of the rotating magnet integrated for a time of $1.2\times10^6$~s. The harmonic occupies a single bin in the Fourier spectrum, as expected. A magnetically induced ellipticity must behave in the same way.

In the following we discuss some of the possible causes of spurious peaks, and describe the tests done, the successes and the questions still open. We will not discuss obvious issues like the CM signals related to the residual atmosphere probed by the laser beam in traversing the magnetic field: the residual pressure in the vacuum chambers was low enough to make this systematic undetectable and after the debugging no birefringence signal was observed. 

\subsubsection{Stray fields and pick-ups}
\label{sec:stray field}

A possible origin for spurious signals 
could have been the stray field of the rotating magnets. The magnetic stray field was $\lesssim0.1$~G along the magnet axis at a distance of $\approx40$~cm outside the magnet extremity, which was about the distance of the magnets from the cavity mirrors. The oscillating stray fields could also have acted on other elements of the apparatus or could be picked-up by one or more electronic circuits. The signal could therefore end up in the ellipticity directly, or via ground loops, or by means of some other mechanism. Notice again, however, that the spurious signals which might have been confused with an ellipticity were only those at the second harmonic of the rotation frequency of the magnetic field.

\begin{figure}[bt]
\centering 
\includegraphics[width=8cm]{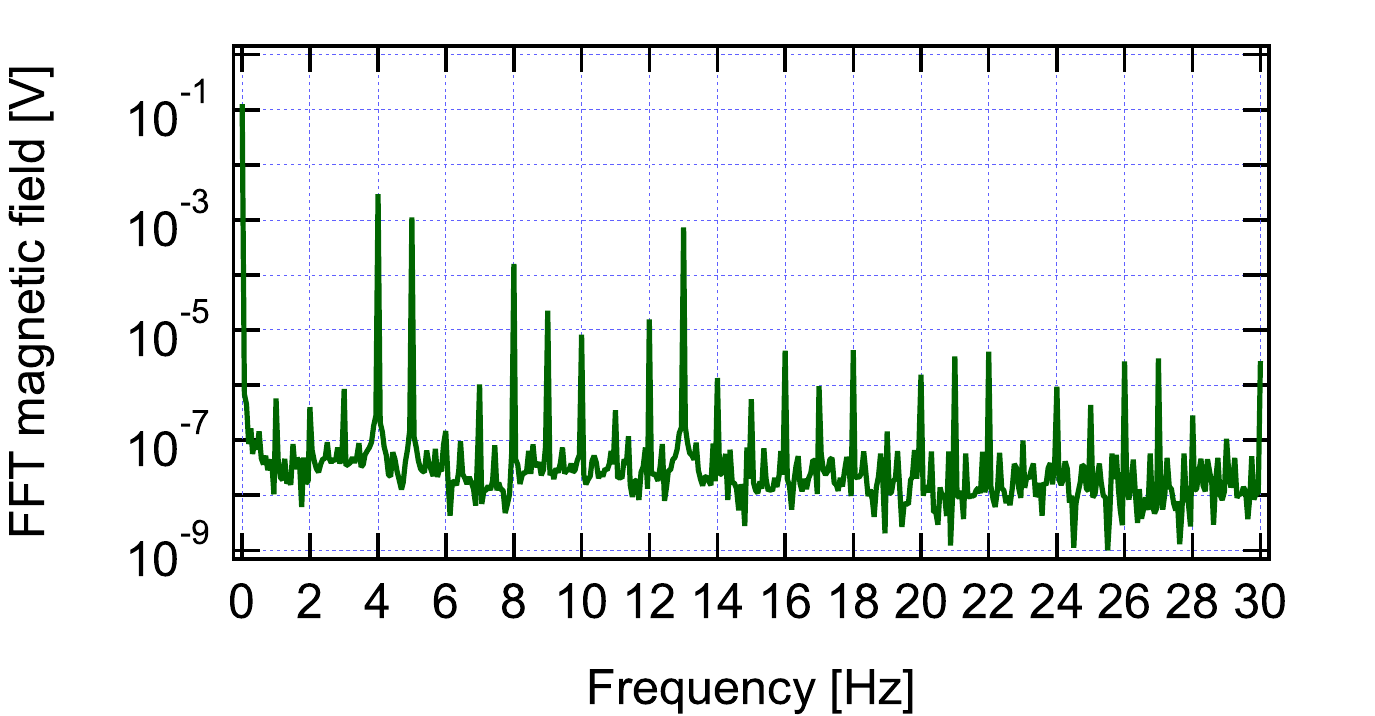}
\includegraphics[width=8cm]{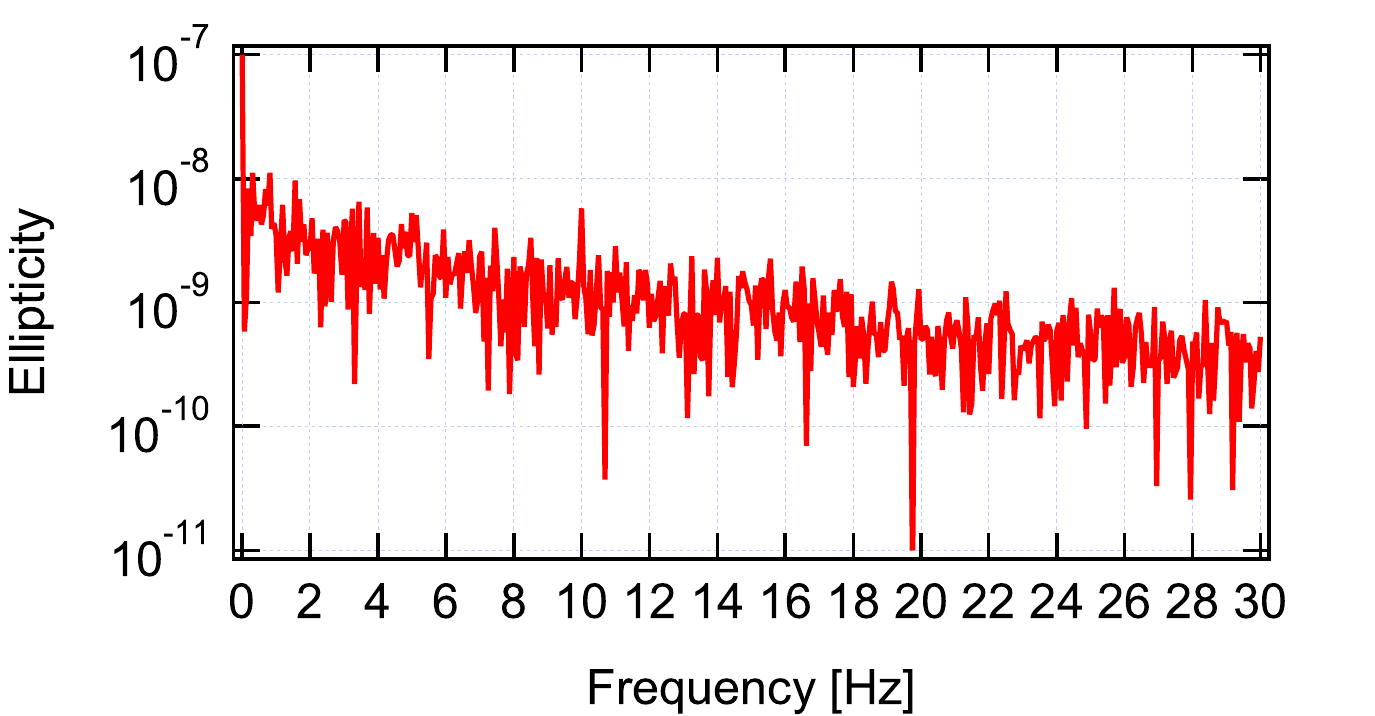}
\caption{Left: Fourier transform of the transverse horizontal magnetic field component at the position of the laser due to the rotating magnets and to a Faraday cell. The two magnets were rotating at 4~Hz and 5~Hz; the Faraday cell was fed with a sinusoidal current at 13~Hz. Right: ellipticity spectrum in vacuum showing a spurious signal only at 10~Hz. The integration time was ${\cal T} \approx 10^5$~s.}
\label{fig:Faraday_laser}
\end{figure}

In a first series of tests we tried to identify possible targets of the stray field. To this end a coil 
was employed to place a magnetic field on the various components of the experiment. This auxiliary magnetic field was at least a factor 10 more intense than the stray field from the rotating permanent magnets. The coil was positioned in proximity of an optical element or of an electronic instrument, and peaks in the Fourier spectrum at the frequencies $\nu_F$ and $2\nu_F$ of the coil were searched for in the ellipticity spectrum. 

In Figure~\ref{fig:Faraday_laser} we show the result of one of the tests performed. In this case, the laser was investigated. The left graph is the Fourier spectrum of the horizontal transverse component of the magnetic field at the position of the laser measured with a commercial magnetometer. The rotation frequencies of the two magnets, 4~Hz and 5~Hz with their harmonics, are visible together with the frequency of the alternating current in the Faraday coil at 13~Hz. A small non linearity of the magnetic field sensor was responsible for a slight frequency mixing. The sensor head of the magnetometer was aligned in the horizontal direction transverse with respect to the light path; quite similar spectra were recorded for the other two directions of the magnetic sensor. The corresponding ellipticity spectrum in the right panel of the Figure~\ref{fig:Faraday_laser} shows a single peak only at 10~Hz, excluding the stray magnetic field on the laser as a source of ellipticity. 

\begin{figure}[bht]
\begin{center}
\includegraphics[width=10cm]{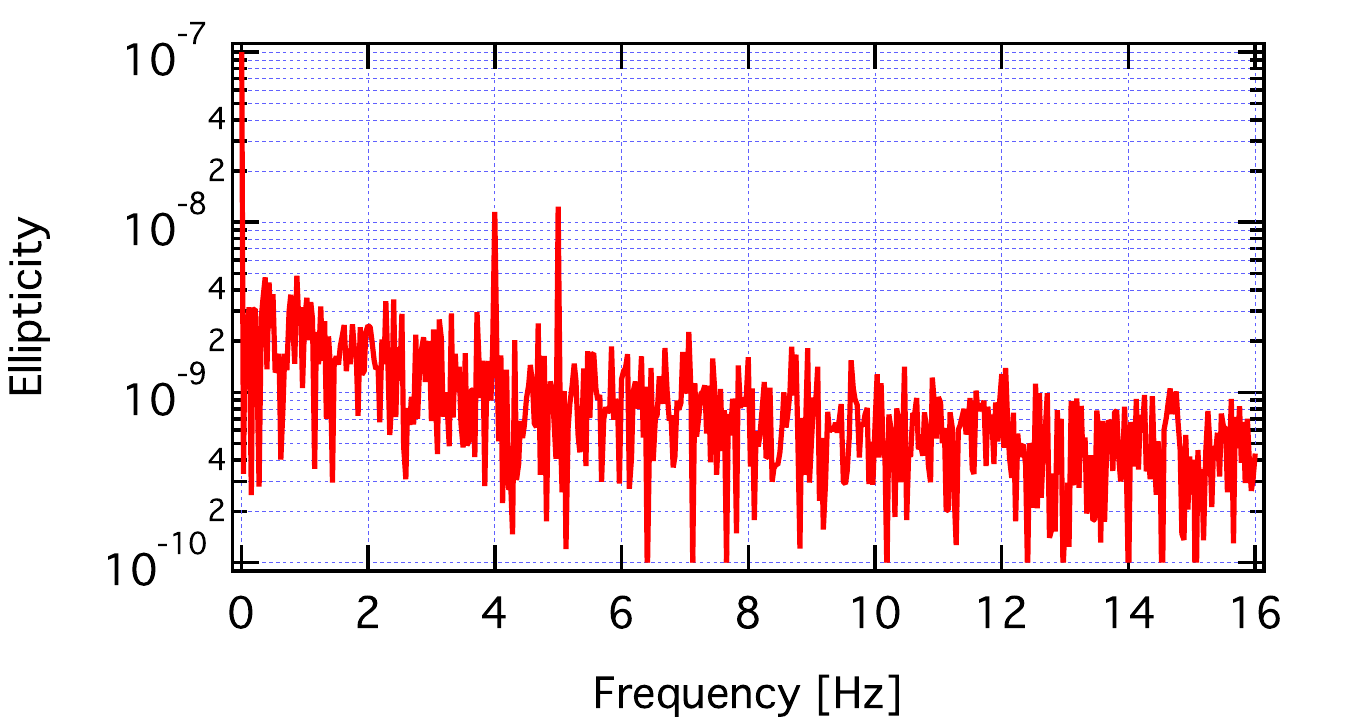}
\caption{Ellipticity spectra showing peaks at the rotation frequencies of the two magnets. These are due to Faraday rotations in the mirrors which are fed to the ellipticity channel by the factor $R_{\Phi',\Psi}$. Integration time ${\cal T}\approx9\times10^{5}$~s.}
\label{fig:FaradayEffect}
\end{center}
\end{figure}
       
This search gave a negative result for all the elements investigated except when the solenoid aimed at the mirrors of the cavity. This, however, was nothing new, since small ellipticity signals were always observed at the frequency $\nu_B$ (see Figure~\ref{fig:FaradayEffect}). These were due to the Faraday effect on the dielectric layers of the mirrors producing a rotation that the birefringent cavity transformed into ellipticity. Since the rotation axis of the magnets did not pass exactly through the center of the mirrors, there was always a small component of the stray magnetic field $B_{\rm long}$ perpendicular to the surface of the mirrors. Indeed a Faraday rotation should have an amplitude
\begin{equation}
\Phi_{\rm F}=N\,C_{\rm Ver}\,B_{\rm long}
\label{eq:verdet}
\end{equation}
where $N$ is the amplification factor of the Fabry-Perot and $C_{\rm Ver}$ is the effective Verdet constant describing the rotation per reflection per gauss. Remember rotations were transformed into ellipticities with a conversion factor $|R_{\Psi',\Phi}| \approx 0.5$. The Verdet constants of the materials composing the dielectric layers of the mirrors was measured by Iacopini et al. in 1983 \cite{Iacopini:1983rp} who found a value for the induced rotation per reflection per gauss of $C_{\rm Ver} = 0.37\times 10^{-9}$~rad/G. 
To reproduce the observed data, the longitudinal component of the stray magnetic field at the position of the mirrors should have been $B_{\rm long}\sim2\times 10^{-4}$~G, a perfectly plausible value.

\begin{figure}[hbt]
\centering
\includegraphics[width=10cm]{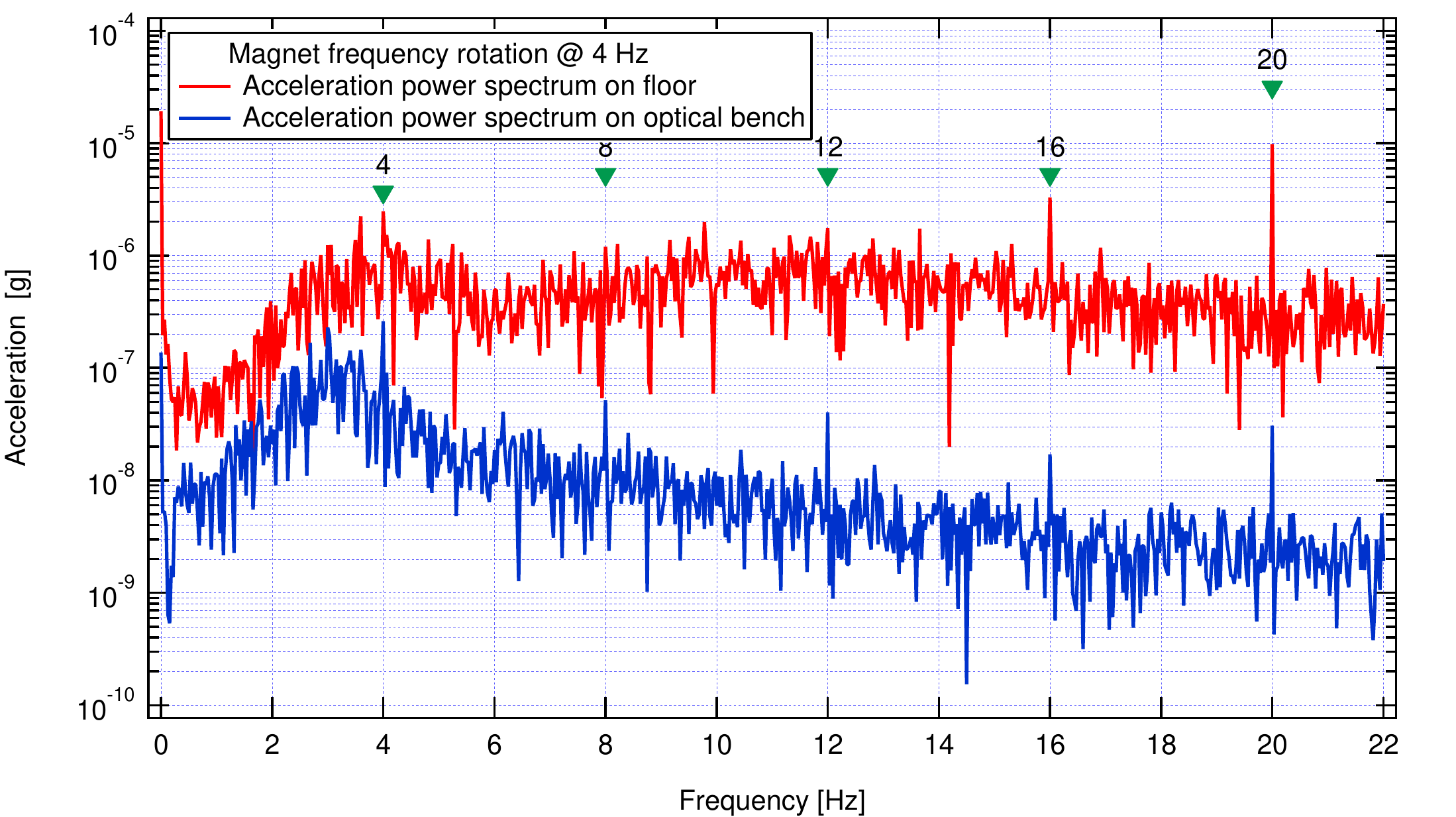}
\caption{Vertical component of the acceleration of the floor and of the optical bench before \emph{in situ} balancing of the rotating magnet. Harmonics of the rotating magnet (4~Hz) are also present on the optical bench.}
\label{fig:acc_ground_bench}
\end{figure}

\subsubsection{Mechanical noise from the rotating magnets}

As another possible source of noise and spurious signals, we investigated the mechanical vibrations transmitted by the rotating magnets to the optical components through the ground and the seismic isolation of the optical bench. The PVLAS-FE experiment was designed with the structures supporting the magnets separated from the optical bench, but both systems stood on the same concrete ground plate. The vibrations excited by a small unbalancing of the magnets were transmitted to the ground plate and filtered by the pneumatic air springs of the optical bench. This mechanism could have played a significant role in the generation of the spurious peaks, in particular when the rotation frequency of the magnets was increased. In Figure~\ref{fig:acc_ground_bench} we show the vertical component of the acceleration of the optical bench and of the ground, measured with the magnets rotating at 4~Hz. We balanced the magnets {\em in situ}, reducing the acceleration measured on the supporting structures down to $10^{-4}$~m/s$^2$ for a rotation frequency $\nu_B=4$~Hz. The results of this operation were quite unambiguous: the amplitude of the acceleration of the bench at the rotation frequency of the magnet decreased below the noise, whereas the second harmonic seemed not to be affected by the procedure, indicating a different origin of this acceleration (see Section~\ref{sec:tube-acceleration} below).

\begin{figure}[bht]
\centering
\includegraphics[width=10cm]{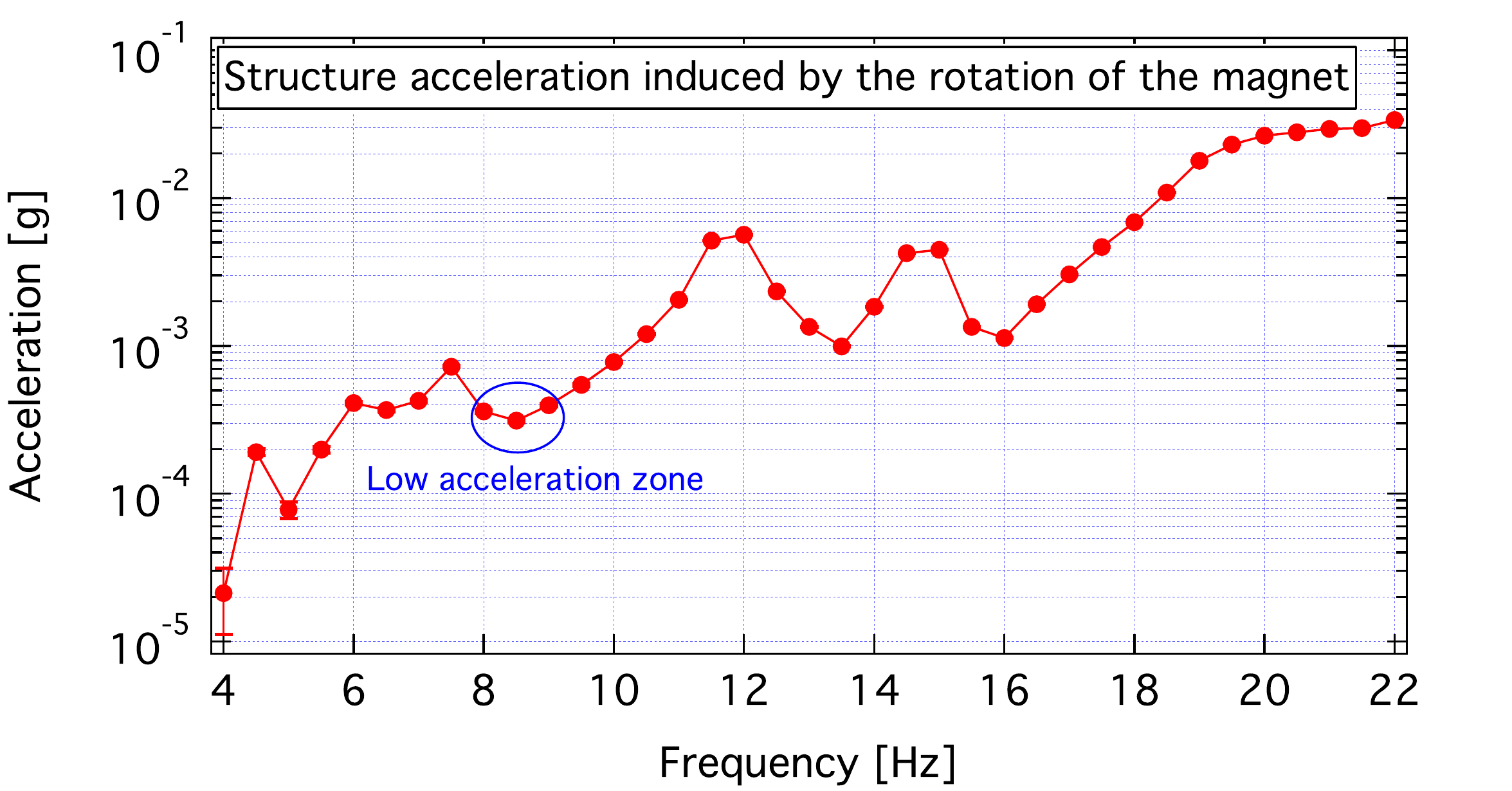}
\caption{First harmonic of the horizontal component of the acceleration of one of the structures supporting the magnets, measured as a function of the rotation frequency of the magnet.} \label{fig:acc-freq}
\end{figure}

We also studied the resonances of the structures supporting the magnets. In Figure~\ref{fig:acc-freq}, we show the horizontal acceleration measured on one of the two structures supporting the magnets, after balancing, as a function of the rotation frequency. It is evident that, as a general trend, the acceleration increases with frequency due to a residual unbalance. As we will see in the following, the measurements of the vacuum birefringence were taken initially with the rotation frequency of the magnets ranging from 3~Hz to 5~Hz, and only in 2016 the frequency $\nu_B$ was increased to 8~Hz to exploit the relative minimum of the noise around 16~Hz.

As a last attempt to reduce the mechanical noise supposedly associated with the rotation of the magnets, we lifted the structures supporting the magnets on anti-vibration feet. In order to comply with the general stability criterion, we connected the two structures with two girders placed down near the floor. We obtained a single 1.6~ton structure that we lifted on four pneumatic FAEBI\textsuperscript{\textregistered} Rubber Air Springs by Bilz. The measurements indicated that the acceleration measured on the bench at high rotation frequencies was significantly reduced at $\nu_B$, but again not at $2\nu_B$. Moreover, due to this modification to the apparatus, the position of the magnets with respect to the table became dependent on the air pressure in the FAEBI\textsuperscript{\textregistered} feet, an issue we will come back to further on.

\subsubsection{Movements of the optical bench}

\begin{figure}[hbt]
\centering
\includegraphics[width=12cm]{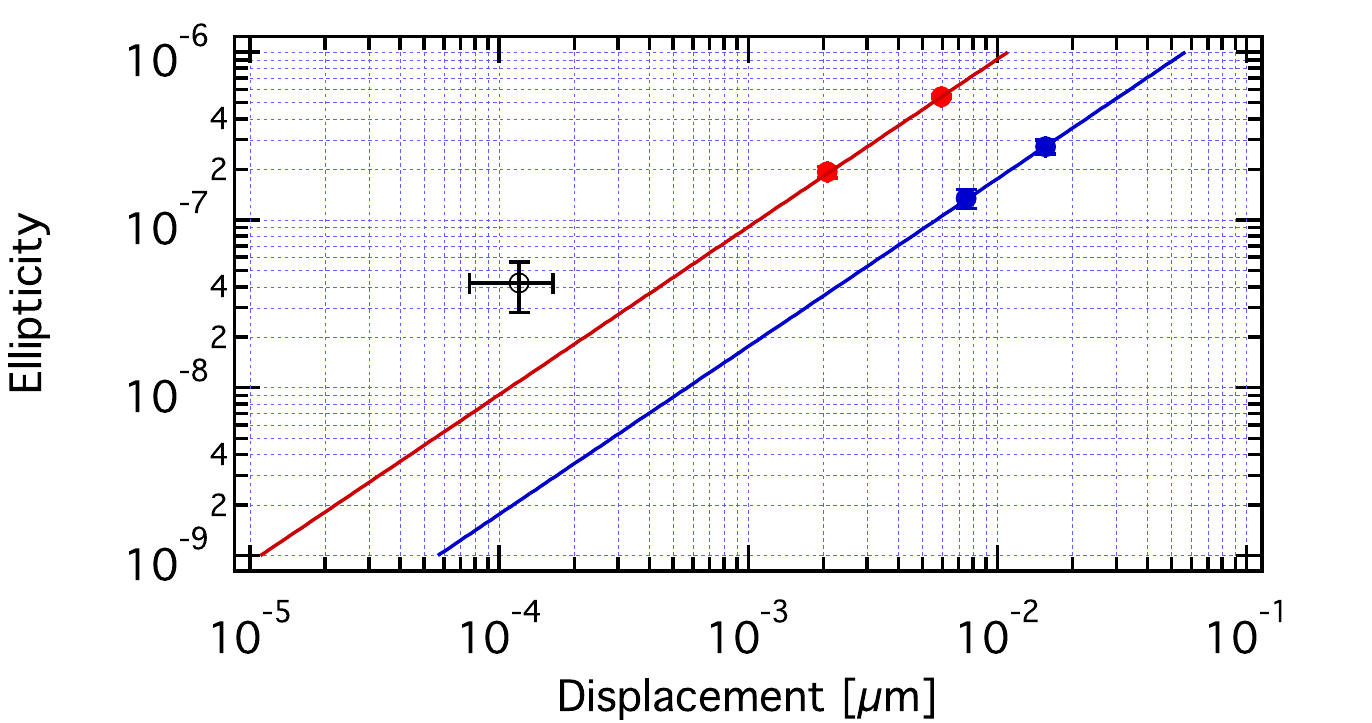}
\caption{Ellipticity as a function of the oscillation amplitude of the optical bench inertially generated by a mass oscillating on the bench. Points on the two linear curves are the 11~Hz components due to the longitudinal (top) and transverse (bottom) oscillation of the mass measured at two amplitudes. The isolated point is the 8~Hz component of longitudinal oscillation related to the rotation of the magnet at 4~Hz.}
\label{fig:MassOscLin}
\end{figure}

In this Section we report on the investigation of the connection between the movements of the optical bench and the ellipticity signals. For this study, we placed a mass of lead of 10~kg on a linear translator mounted horizontally on the optical bench. By substituting the fine thread screw with a piezoelectric ceramics, the mass could be put in oscillation at a chosen frequency in the directions parallel or orthogonal to the light propagating in the FP cavity. While the mass was oscillating, we measured the acceleration of the optical bench in the direction of the oscillation and the spurious ellipticity signals at the frequency of oscillation. The observed ellipticity signals were linear in the displacement, with linearity coefficients $(91\pm2)$~m$^{-1}$ for the direction parallel to the light path and $(17.6\pm1.4)$~m$^{-1}$ for the direction perpendicular, as shown in Figure~\ref{fig:MassOscLin}. From the figure one can see that the 8~Hz spurious ellipticity signal associated with the acceleration due to the rotation of the magnet is too intense to fit in the linear relations, suggesting the existence of a different coupling mechanism between the rotation of the magnets and the ellipticity.

The oscillation of the 10~kg mass generated a modulation also in the correction signal of the feedback system locking the laser to the cavity. This modulation indicates that there was a phase modulation of the electric field reflected by the cavity with respect to the incident beam. A modulation in $\delta$ together with a static ellipticity will be translated into a modulated ellipticity proportional to $\delta$ whenever $\alpha_{\rm EQ}\ne0$ [see equation~(\ref{eq:E_perp})]. 

\begin{figure}[bt]
\centering
\includegraphics[width=8cm]{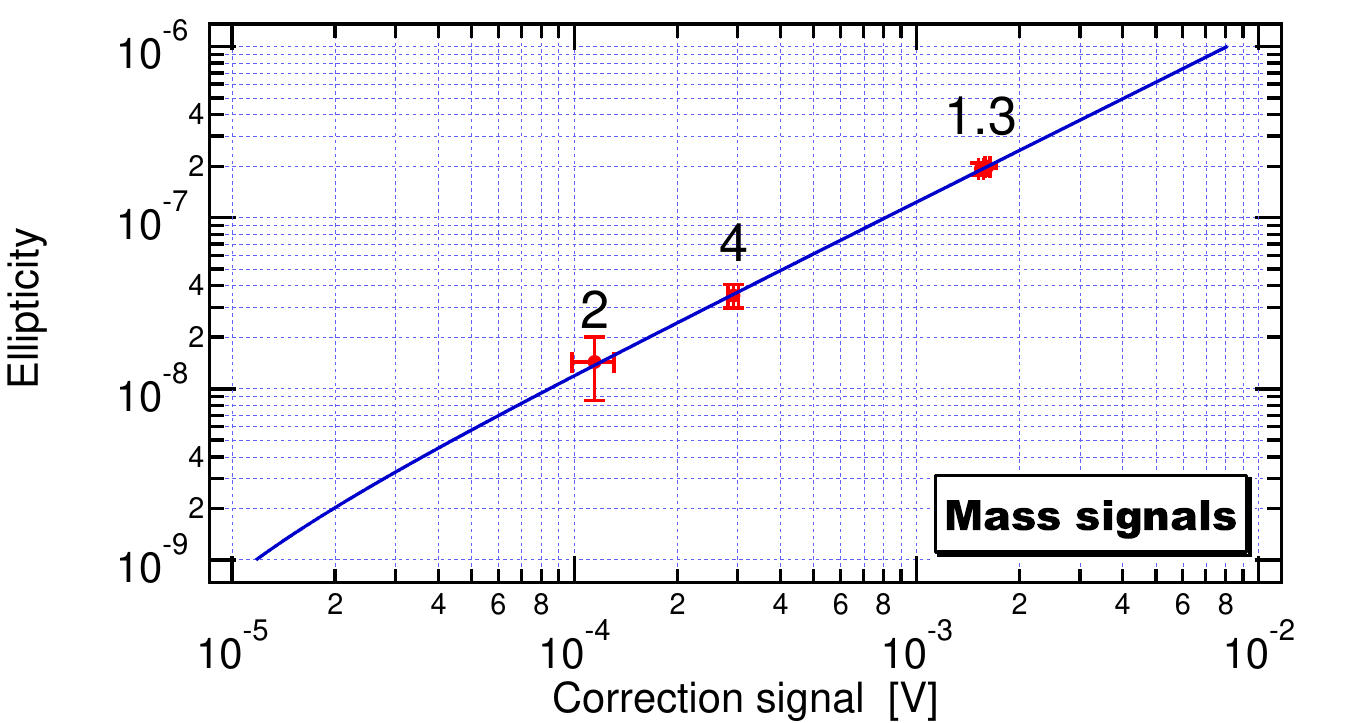}
\includegraphics[width=8cm]{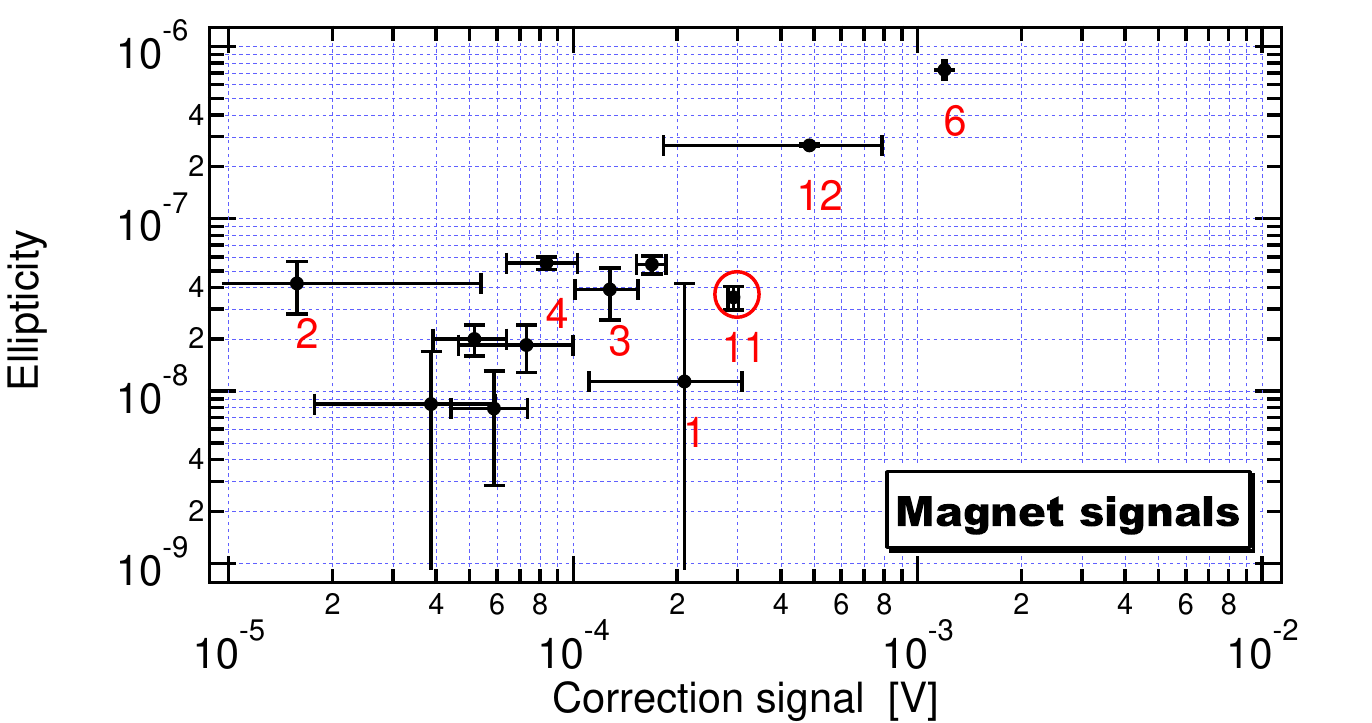}
\caption{Correlations between the ellipticity and the correction signal. Integration time was 1600~s for the ellipticity and 160~s for the correction signal. Left: the experimental points are the amplitudes of the first four harmonics of the mass oscillation frequency (11~Hz). The numbers above each data point indicate the corresponding harmonic. The straight line is the best fit with slope $1.24\times10^{-4}/$V. Right: the experimental points are the amplitudes of the harmonics of the rotation frequency of the magnet (4~Hz). The eleventh harmonic indicated by the red circle is common to the two graphs.}
\label{fig:spurious_correlation}
\end{figure}

In Figure~\ref{fig:spurious_correlation} we show the correlation of the ellipticity and the correction signal of the feedback system locking the laser to the cavity. The measurement was taken while the mass was oscillating at 11~Hz along the cavity direction and both magnets were rotating at 4~Hz. The top panel plots the amplitude of harmonics of the mass oscillation; the bottom panel plots the amplitude of the harmonics of the rotating magnets. While the correlation in the first graph is clear, the correlation shown in the second graph is fuzzy, again indicating that the two noise sources, the oscillating mass and the rotating magnets, had different mechanisms of coupling with the ellipticity.

\subsubsection{Diffused light and `in-phase' spurious peaks}
\label{sec:diffused_light}

\begin{figure}[b!]
\begin{center}
\includegraphics[width=14cm]{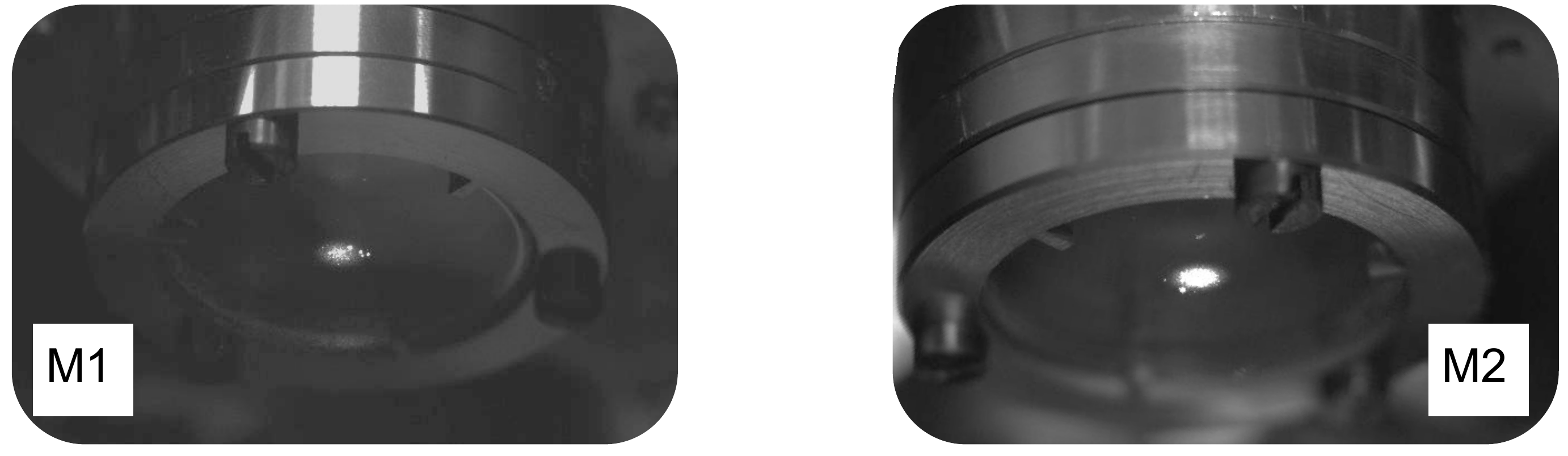}
\caption{Infrared photos of the mirrors with the laser locked to the cavity. The TEM$_{00}$ of the cavity is visible due to diffused light from the mirrors. Small bright spots can also be seen near the edges of the modes.}
\label{fig:spot_infrared}
\end{center}
\end{figure}

A real breakthrough with the spurious signals came only when we started paying attention to diffused light. The sources of diffused light inside the cavity are the intense spots of the light reflecting on the mirrors when the laser is locked. Photographs of these spots can be seen in Figure~\ref{fig:spot_infrared}. Besides the Gaussian beam a few extra bright dots distributed around the main spots can be seen. These dots could be due either to dust or local defects of the mirror surface. As seen from the centre of the mirror, the incidence angle on the inner surface of the glass tube ranged from $\geq88^\circ$ to $\approx89.8^\circ$. This corresponded to an average reflective power of the inner surface of the tube ranging from 0.82 to 1. Since diffused light is a source of noise in a Fabry-Perot cavity \cite{Virgo2012}, its modulation might generate spurious signals. Diffused light is essentially unpolarised light that can traverse both the polariser and the analyser. Its power might have been modulated if the tube vibrated synchronously with the rotation of the magnets. As a matter of fact, by monitoring the infrared radiation coming out laterally from the accessible portion of the glass tube just outside the magnets, we found a power modulation at harmonics of the rotating magnets. On the other hand, the ellipticity and rotation signals in the extinguished beam were extracted through a demodulation process that was insensitive, to first order, to a power modulation. The same was true for the reflected beam and the error signal of the laser frequency-locking system. The point was that diffused light could also be modulated in phase. 

\begin{figure}[hbt]
\centering
\includegraphics[width=12cm]{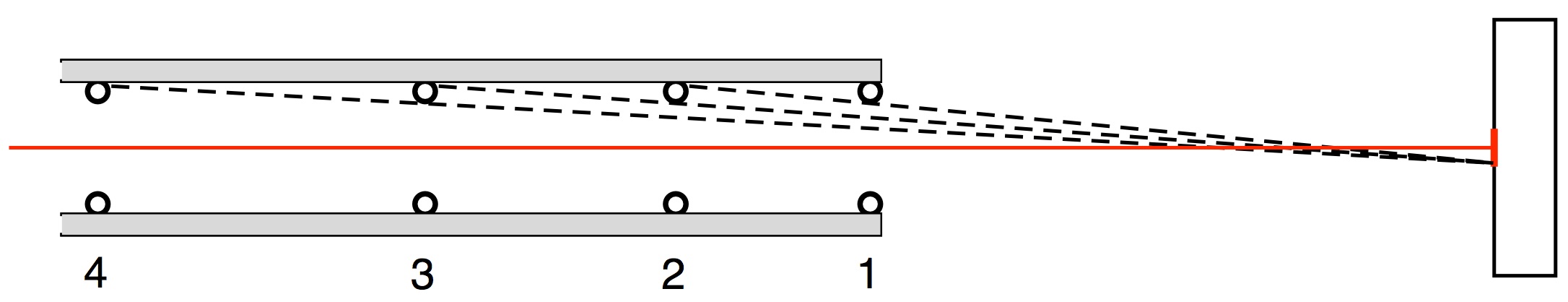}
\caption{Positioning of the baffles inside the tube. The baffles prevent light from a given area on the mirror from reaching the internal surface of the glass tubes.}
\label{fig:baffles}
\end{figure}

\begin{figure}[bht]
\centering
\includegraphics[width=12cm]{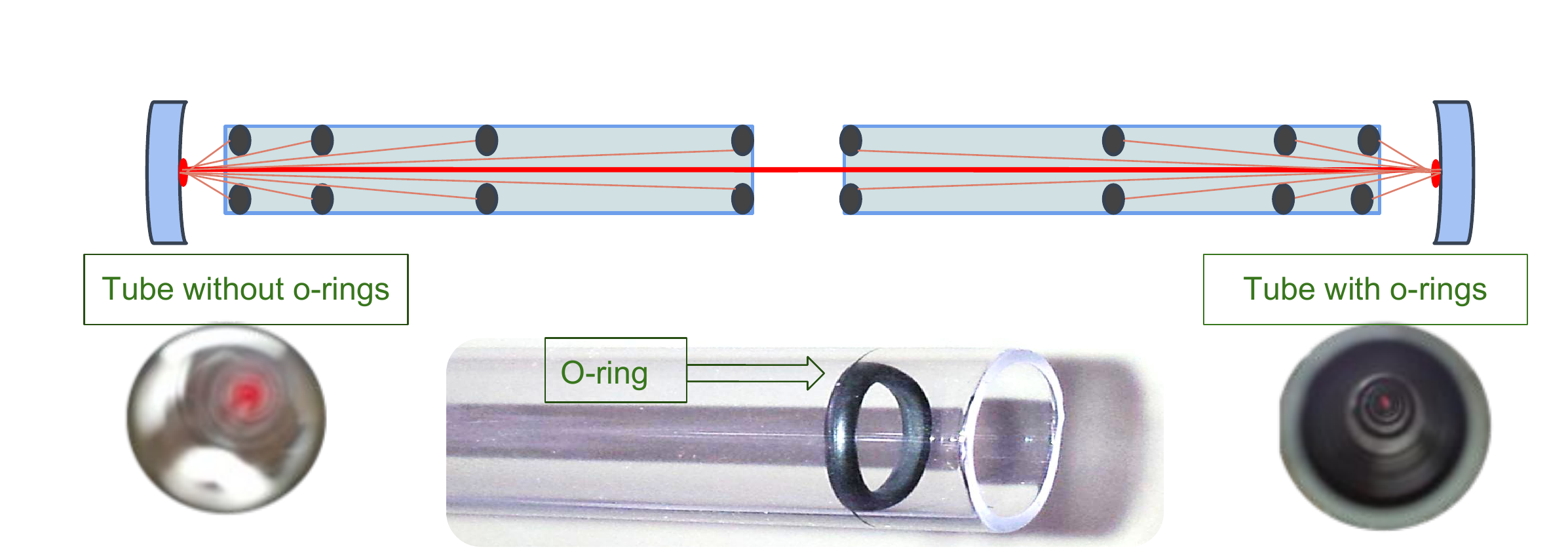}
\caption{Above: schematic view of the baffles inserted inside the tubes. Below: looking through the tube before and after the insertion of the o-rings; a net attenuation of the diffused light is observed.}
\label{fig:oring}
\end{figure}

The first action we took was to place inside the tube (at that time a 12.5~mm internal diameter glass tube) a system of baffles to absorb the diffused light. The irises where Viton o-rings with external diameter equal to the internal diameter of the tube and chord thickness $\sim1$~mm. The sequence of the positions of the o-rings inside the tube was such that the internal surface of the glass tube could not be seen from any position inside a round spot in the centre of the mirror (see Figure~\ref{fig:baffles}). The first o-ring was placed just at the end of the tube near the mirror; the second o-ring intercepted the light that grazed the edge of the first o-ring coming from the periphery of the blind spot; the position of the third o-ring was further away, chosen with the same criterion, and so on. The improvement obtained could be appreciated already by looking through the tube with the naked eye (see Figure~\ref{fig:oring}). The diameter of the blind spot grows with the number of baffles, in principle allowing to screen the whole surface of the mirror; however, as the edges of the o-rings themselves are reflective in grazing incidence, we never used more than 20 o-rings per tube, with blind spot dimensions of the order of twice the waist of the laser light on the mirrors.

\begin{figure}[hbt]
\centering
\includegraphics[width=8cm]{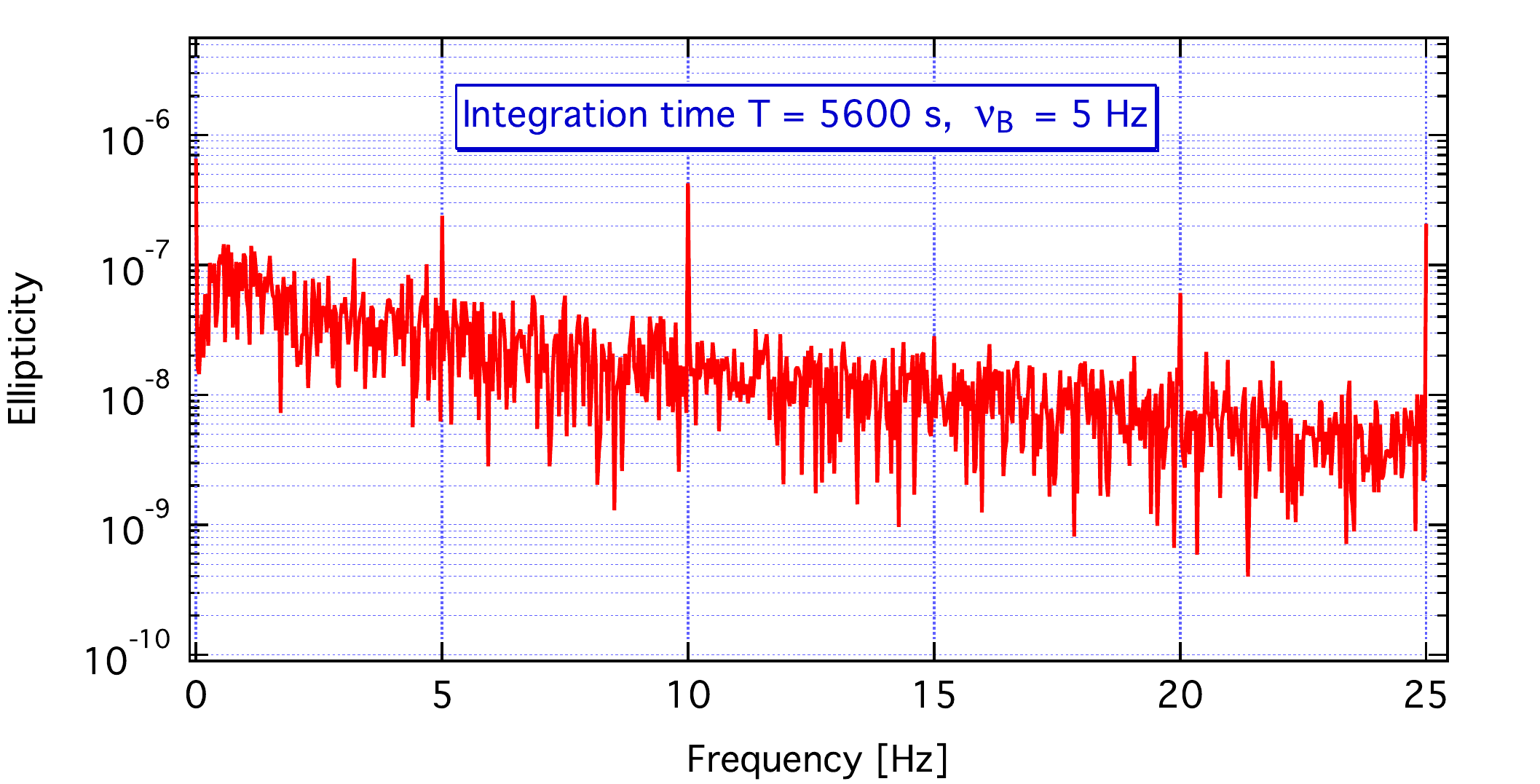}
\includegraphics[width=8cm]{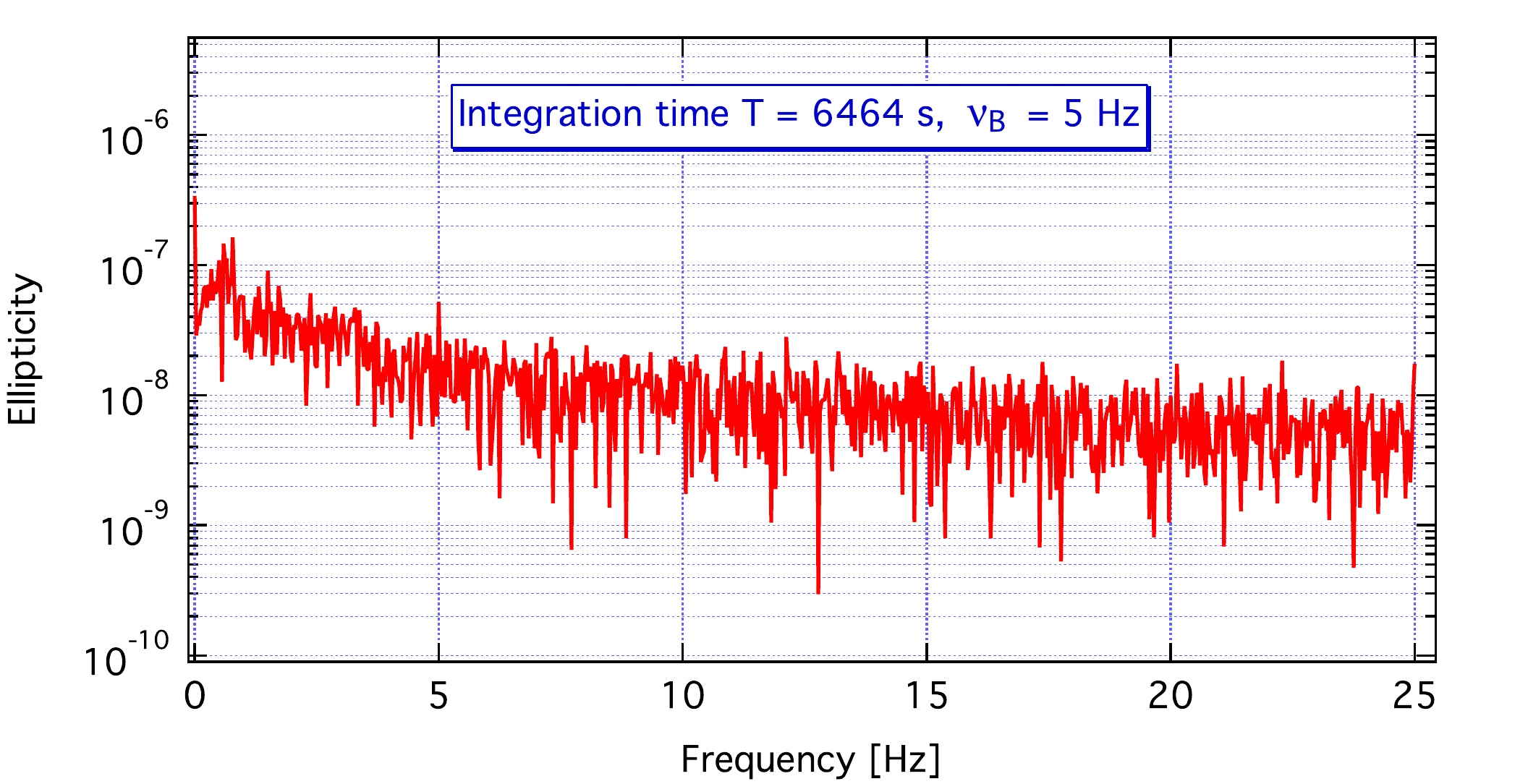}
\caption{Left: ellipticity spectrum before the insertion of the o-rings inside the tubes; signals are observed at harmonics of the magnet rotation frequency $\nu_B=5$~Hz. Right: ellipticity spectrum after installation of the baffles.}
\label{fig:ell-oring}
\end{figure}

The effect of the installation of the o-rings was a sudden reduction of the spurious signals. The spectra reported in Figure~\ref{fig:ell-oring} prove this beyond any doubt: the signal at $2\nu_B$ disappears, being reduced by a factor of at least ten. We note that the peak at $\nu_B$, which is due to a Faraday effect on the mirrors, is reduced but does not disappear, unlike the other harmonics. The remedy we found to the problem of the spurious signals was very effective, indicating that the spurious signals are actually generated by a modulation of the diffused light (amplitude and/or phase). The nature of this modulation was still unclear. In the next section we will show that the movement of the tube induced ellipticity signals, thus suggesting that the movement of the tube modulated the diffused light.

\subsubsection{Magnetic forces on the tube}
\label{sec:tube-acceleration}

\begin{figure}[bht]
\centering
\includegraphics[width=10cm]{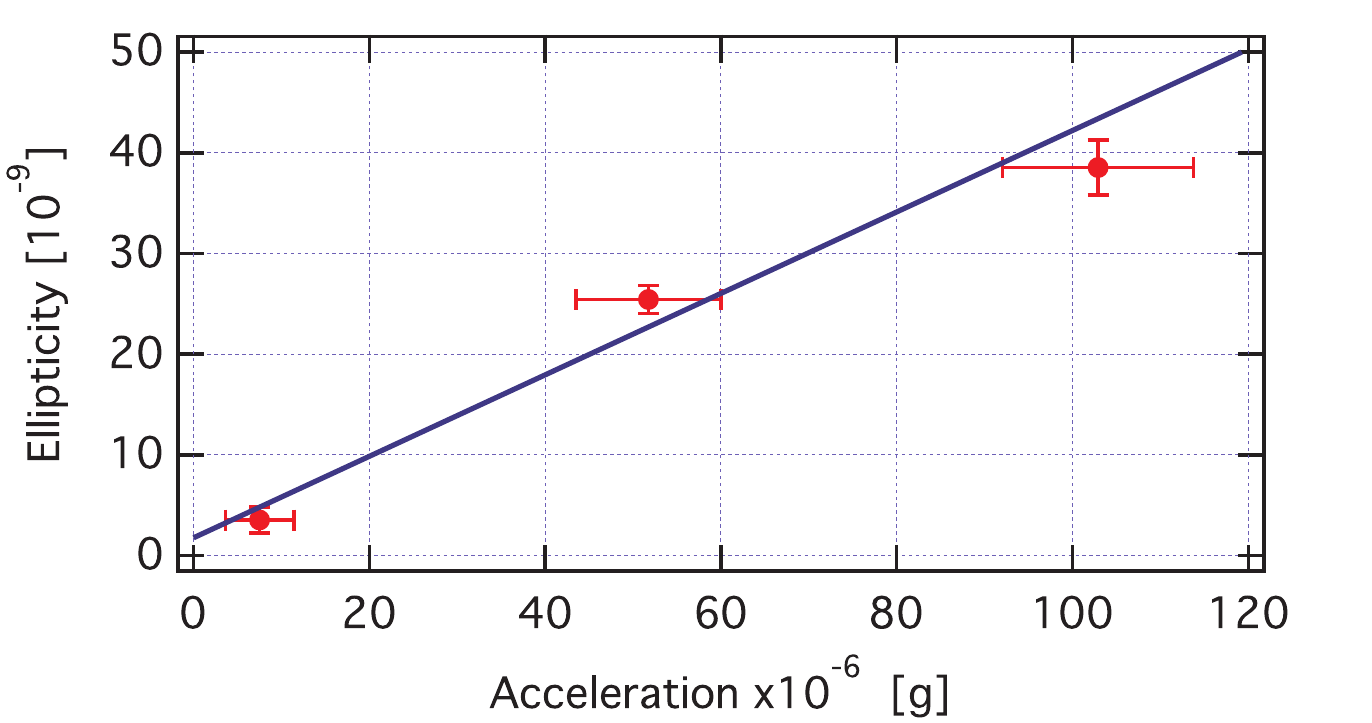}
\caption{Correlation of measured ellipticity and tube acceleration. The spectra are obtained  forcing the movement of the tube with a piezo at $\nu_{\rm piezo}=7.5$~Hz. In the acceleration spectrum three harmonics were observed. The three points shown correspond to the harmonics at 7.5, 15.0 and 22.5~Hz.}
\label{fig:corr_ell_acc}
\end{figure}

A piezoelectric crystal was used to induce an oscillation of the (glass) tube at 7.5~Hz in the transverse direction with a nominal amplitude of $1~\mu$m. The induced acceleration was measured with a three-axes accelerometer fastened at the extremity of the tube on the mirror side. Figure~\ref{fig:corr_ell_acc} shows that the acceleration (or the oscillation amplitude) of the tube along the piezo direction was correlated to the ellipticity. The piezoelectric crystal applied a force between the optical bench and the tube. Given the difference in masses, the results were interpreted in terms of the movement of the tube (not of the bench) with respect to the magnets. We concluded that the movement of the tube generated spurious ellipticity.

With the magnets in rotation, the transverse acceleration of the tube appeared at the second harmonic of the magnet rotation frequency. Both the glass and the ceramic tubes were made of diamagnetic materials inserted in a rotating dipole magnetic field. The magnetic force on a magnetised body is given by
\begin{equation}
\vec{F} = \frac{1}{2\mu_0}\frac{\mu_r-1}{\mu_r^2}\int\limits_{\rm body}\vec{\nabla}B(t)^2\,dV
\label{eq:magforce}
\end{equation}
where the integral extends over the volume of the body and the magnetic susceptibility $\chi=\mu_r-1$ was in the present case small and negative ($\chi\sim-10^{-6}$). 

\begin{figure}[hbt]
\centering
\includegraphics[width=10cm]{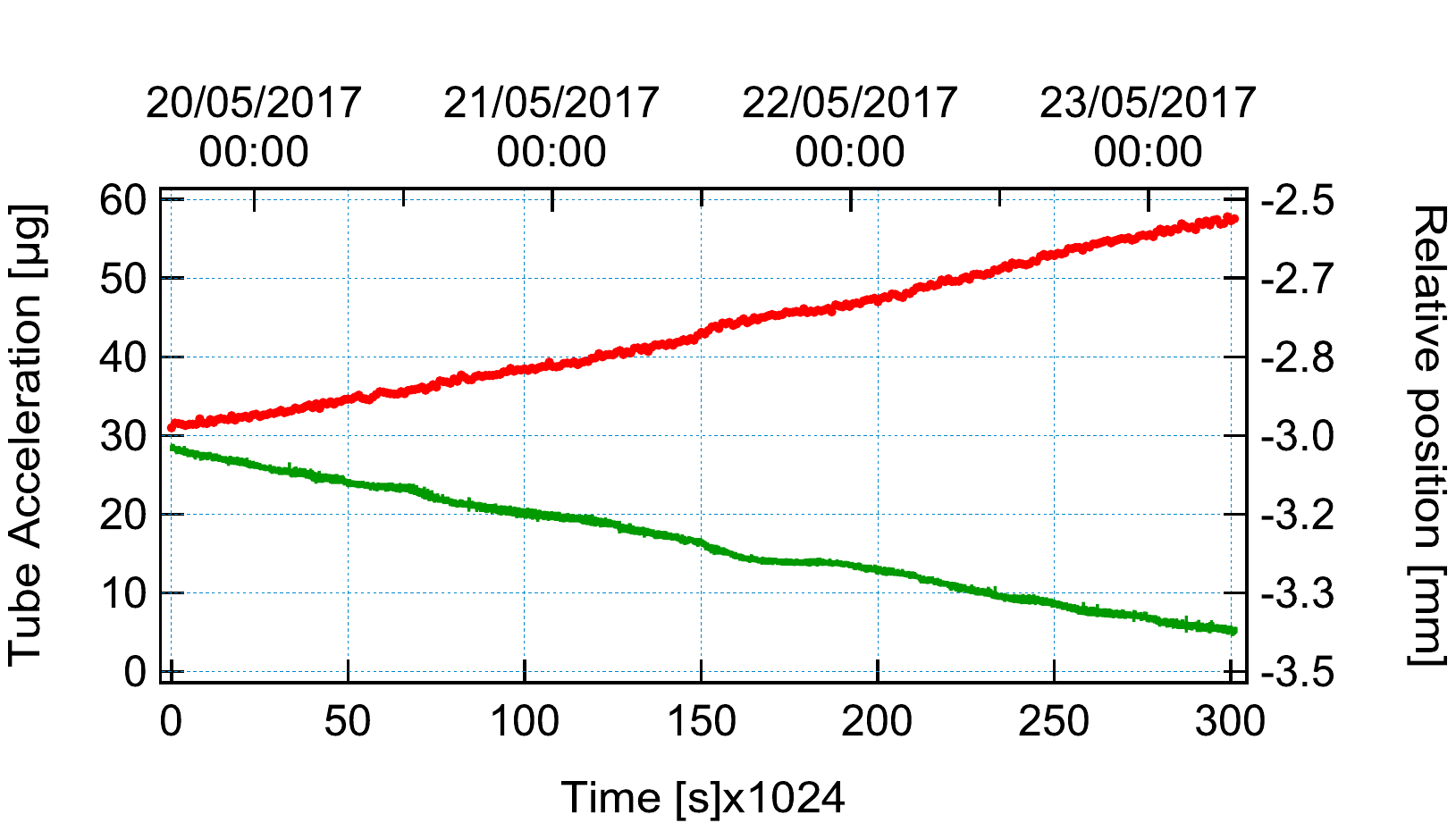}
\caption{
Plot of the $2\nu_B$ component of the transverse acceleration of the tube (red points) and the position (green points) of the tube inside the magnet bore, monitored during a few days. Each point is integrated for a time of 1024~s.}
\label{fig:acc-monitoring}
\end{figure}

Two gradients of the magnetic fields were present: a strong gradient of the field directed parallel to the axis of the magnets at both ends resulting in a longitudinal force, and a smaller radial gradient due to the non ideal dipolar field near the edges of the bore and therefore on the tubes. If the geometry of the magnets and the positioning of the tubes had been ideal, the net force on the tubes would have been zero. The position of the tubes coincided only approximately with the axis of the magnets and therefore the asymmetry of the magnetic field with respect to the tubes' position resulted in a net transverse force on the tubes. This force rotated with the magnet; since the force field described by equation~(\ref{eq:magforce}) has rotational symmetry of order two determined by $B(t)^2$, the main Fourier component of the force was expected at $2\nu_B$, as observed. Furthermore the proportionality of this transverse force with the volume of material inside the magnetic field was verified by progressively inserting a second diamagnetic rod, of outer diameter equal to the inner diameter of the glass tube, inside the glass tube. This last test also excluded the longitudinal gradients as a source of the observed acceleration of the tube. 

The component at $2\nu_B$ of the transverse acceleration of one of the two tubes is shown in Figure~\ref{fig:acc-monitoring} together with the position of the tube inside the magnet bore recorded during a few days in which the newly installed  FAEBI\textsuperscript{\textregistered} rubber supports of the structure of the magnets were still settling. The graph bears a clear evidence of a correlation between the acceleration of the tube and the relative position of the tube with respect to the magnet axis. 

\begin{figure}[bht]
\centering 
\includegraphics[width=8cm]{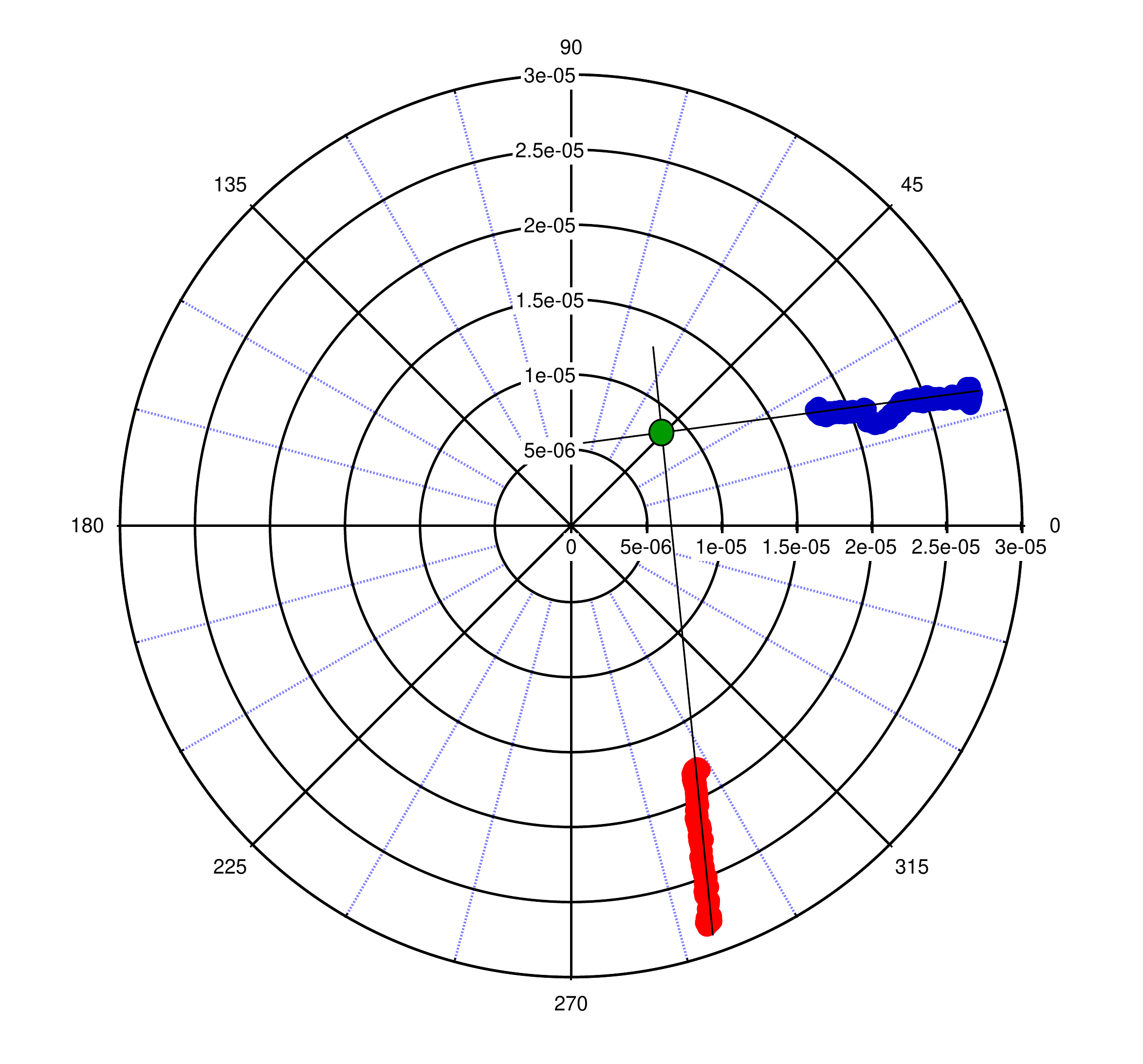}
\caption{Polar plot of the tube acceleration at twice the frequency $\nu_B$. The two transverse acceleration components are shown separately. During this measurement, which lasted a few days, the relative position of magnet and tube was drifting in a straight line about 1~mm long.}
\label{fig:tube_acceleration}
\end{figure}

In the polar plot of Figure~\ref{fig:tube_acceleration} the two components of the acceleration reported in Figure~\ref{fig:acc-monitoring} are shown. The two components both describe straight lines which, though, do not intercept at the origin of the polar plane.

The above tests clearly indicate the existence of transverse magnetic forces on the tube depending on its position inside the magnetic field. 

In a first attempt to solve this problem, the glass tube was partially lined with a paramagnetic sheath to try to compensate the diamagnetism of the glass \cite{LNL2015}. The operation proved to be too difficult and a fine $xy$ positioning system at the two ends of each tube was realised. On the ends of the tubes far from the mirrors the position was defined by two manual 100-threads-per-inch screws. On the mirror ends, NanoPz piezoelectric actuators by Newport with minimum step-size of 10~nm were employed allowing remote operation. 
The accelerometer signals of the two transverse axes were sent to two lock-in amplifiers referenced to the signal of the magnetometer sampling the oscillating stray field of the rotating magnet. The lock-in amplifiers demodulated the two acceleration components at the second harmonic of the reference frequency; a long integration time of about one hundred seconds was employed to extract the average value of the acceleration at $2\nu_B$. Depending on these two lock-in signals, the corresponding NanoPz was actuated to minimise the acceleration. 

\subsubsection{`In-phase' spurious signals conclusion}

To conclude this report on the `in-phase' noise, we want to stress that the fundamental tools for cutting the systematic ellipticity signals at $2\nu_B$ were the reduction of the diffused light inside the vacuum tubes and their precise centering with respect to the magnets to minimise the force give by equation~(\ref{eq:magforce}). Only the implementation of these two techniques allowed us to integrate the ellipticity up to $\approx 5\times10^6$~s without spurious signals appearing. We also explicitly note that these findings were possible only thanks to the long debugging time allowed by the use of permanent magnets.
    
\subsection{\bf Wide band noise}
\label{sec:wide_band_noise}
 
The wide band noise of the PVLAS-FE apparatus, in the absence of systematic signals in phase with the rotating magnets, determined the ultimate sensitivity of the polarimeter. The mechanism that produced wide band noise is still not completely understood but, as it will be shown, its nature is an ellipticity originating from inside the cavity.  As seen in Section~\ref{sec:noise} the estimated ellipticity sensitivity at frequencies in the range 10-20~Hz should have been $S_\Psi^{\rm (expected)}\approx 8\times 10^{-9}/\sqrt{\rm Hz}$ with the PVLAS-FE parameters. Experimentally though, the measured ellipticity noise during data acquisition was $S_{\Psi}^{\rm (PVLAS)}\approx4\times 10^{-7}/\sqrt{\rm Hz}$~@~$\approx 16$~Hz with an approximate frequency dependence proportional to $1/\nu$. The experimental evidence was about a factor 50 worse than the expected sensitivity.
This correspond to optical path difference sensitivities at $\approx16$~Hz of $S^{\rm (expected)}_{\Delta\cal D} \approx 6\times10^{-21}\;$m/$\sqrt{\rm Hz}$ and $S^{\rm (PVLAS)}_{\Delta\cal D} \approx3.5\times10^{-19}\;$m/$\sqrt{\rm Hz}$, respectively.

\begin{figure}[htb]
\centering 
\includegraphics[width=14cm]{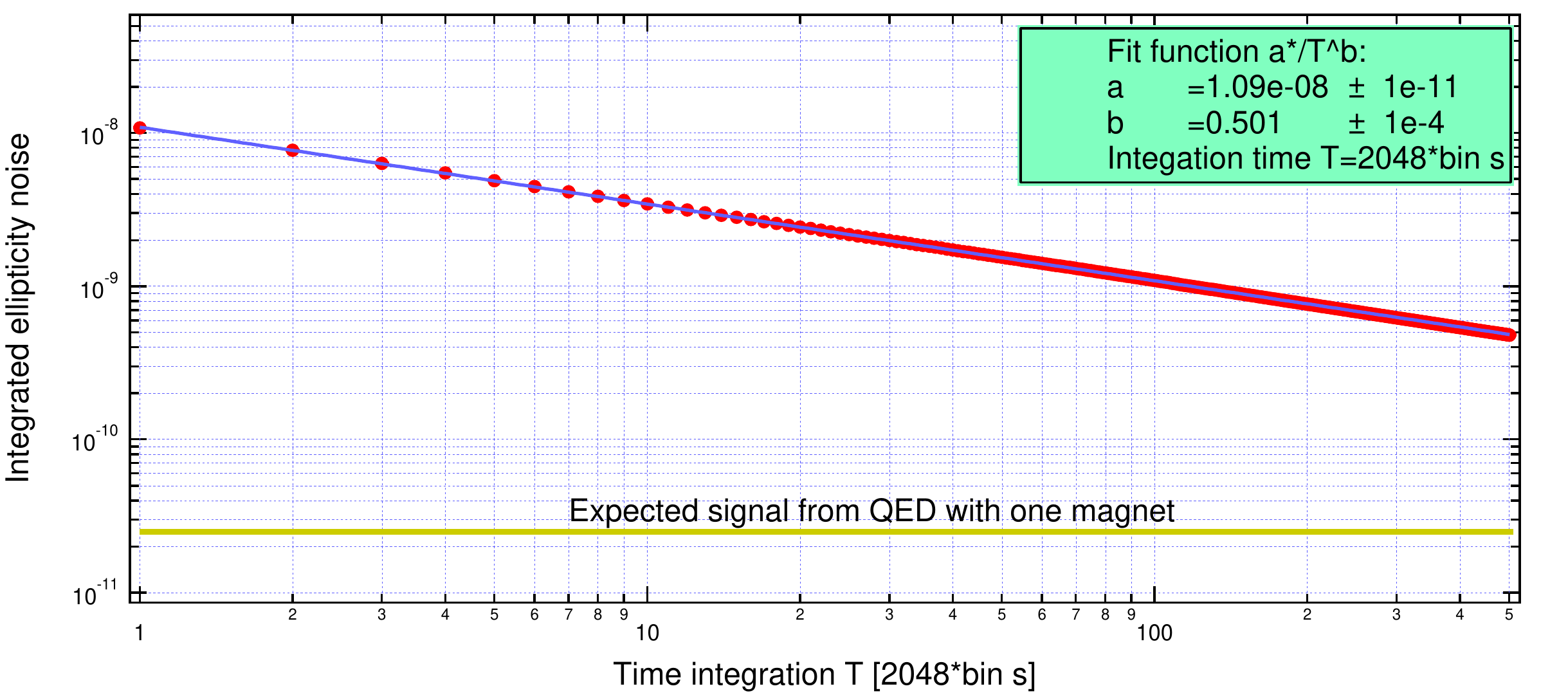}
\caption{Integrated ellipticity noise with one magnet in rotation at 5~Hz as a function of integration time ${\cal T}$ averaged over the frequency ranges $9.6\div9.9986\;$Hz and $10.0016\div10.4\;$Hz excluding a few bins around 10~Hz. The time separation between two consecutive points in the graph is 2048~s. The fit shows a decrease of the integrated noise as 1/$\sqrt{{\cal T}}$ as expected for an uncorrelated noise up to ${\cal T} = 10^6$~s. The ellipticity expected from the vacuum magnetic birefringence with one magnet in rotation is also shown.}
\label{fig:noise_time}
\end{figure}

In Figure~\ref{fig:noise_time} the integrated noise measured over a total time ${\cal T}=10^6$~s is shown with one magnet rotating at 5~Hz. The noise is averaged over a frequency interval $9.6\div9.9986$~Hz and $10.0016\div10.4$~Hz therefore excluding $2\nu_B$. It decreases as a function of ${\cal T}$ as $1/\sqrt{\cal T}$, as expected for uncorrelated noise, and shows no evidence of any significant deviation. The same behaviour was observed for all the other runs. The total run time of the PVLAS-FE apparatus was $\approx5\times10^6$~s with which a $1\,\sigma$ noise floor of $\sigma_\Psi \approx 2\times10^{-10}$ was reached, a factor of about 7 from the expected value $\Psi^{\rm (QED)} = 2.6\times10^{-11}$ (for each magnet). Increasing the integration time by a factor of about 100 to close the gap was unthinkable and work was done to understand the origin of this and improve it. 

\subsubsection{Diffused light and wide band noise}

\begin{figure}[htb]
\begin{center}
\includegraphics[width=14cm]{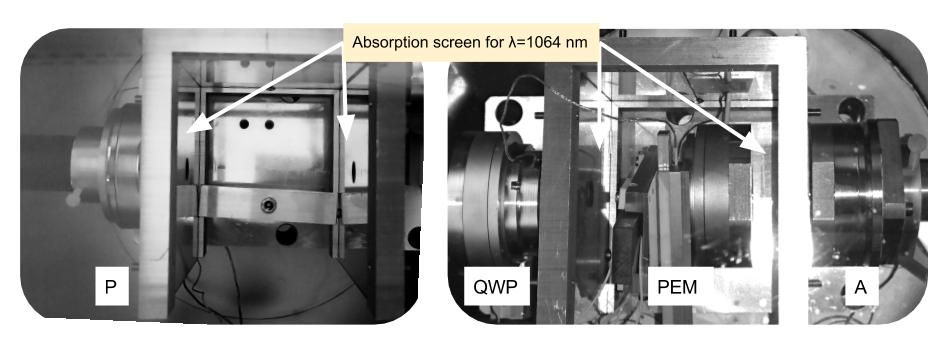}
\caption{Left: input vacuum chamber C1 hosting the polariser (P); Right: output vacuum chamber C2 hosting the analyser (A), the photoelastic modulator (PEM) and the quarter-wave plate (QWP). In the photographs, the frames supporting the absorbing screens can be seen.}
\label{fig:screen absorbent}
\end{center}
\end{figure}

The introduction of the baffles in the glass vacuum tubes, which resulted in a drastic reduction of the `in-phase' signals as discussed in Section~\ref{sec:diffused_light}, did not generate an improvement in the wide-band noise. This was shown in Figure~\ref{fig:ell-oring}. The installation of 1064~nm absorbing glass in the polariser and analyser vacuum chambers (see Figure~\ref{fig:screen absorbent}) did not help either. The conclusion was that the wide-band noise present in the apparatus was not due to diffused light.

\subsubsection{Ambient noise}

Previous experience showed a clear evidence that seismic isolation reduced the ellipticity noise, reason why the whole polarimeter was mounted on a single vibration isolated optical bench \cite{DellaValle:2010oex}. Furthermore, from being structured (presence of wide resonances) when the optical bench was not well seismically isolated, the Fourier ellipticity spectrum acquired a very smooth and time independent aspect. In the previous sections we discussed the `in-phase' ellipticity generated by the rotating magnets compared to the ellipticity induced by an oscillating mass on the bench. The conclusion there was that it was not the movement of the optical bench but the movement of the vacuum tube, generated by the field gradient of the rotating magnet, coupled to the diffused light to induce spurious ellipticity signals.

Residual mechanical wide-band noise present on the optical bench could also have been the source of the observed ellipticity noise. The mechanism of generating ellipticity noise from the optical bench vibrations could have been the induced random movement of the light spots over the cavity mirrors. Indeed the mirrors have a birefringence pattern and each surface point corresponds to a different phase delay \cite{MicossiAPB1993}. 

Using the same linear relation described in Figure~\ref{fig:MassOscLin} applied to the observed mechanical noise measured by an accelerometer mounted on the optical bench would imply a level of ellipticity noise significantly lower than the one observed. In fact, the observed vertical (transverse to the beam) acceleration noise density measured on the optical table is shown in Figure~\ref{fig:acc_ground_bench} on page \pageref{fig:acc_ground_bench} and is about $\approx10^{-8}\;g/\sqrt{\rm Hz}$ at $\approx8$~Hz, corresponding to a bench movement of $\approx 4\times10^{-11}\;{\rm m}/\sqrt{\rm Hz}$. Using the stronger dependence $\frac{d\Psi}{dx}\approx 100$~m$^{-1}$, from Figure~\ref{fig:MassOscLin} on page ~\pageref{fig:MassOscLin}, such a movement corresponds to an ellipticity noise $\approx 4\times10^{-9}/\sqrt{\rm Hz}$ or lower.

We therefore concluded that mechanical vibrations of the bench could not account for the observed sensitivity of the polarimeter.

\begin{figure}[bht]
\begin{center}
\includegraphics[width=8cm]{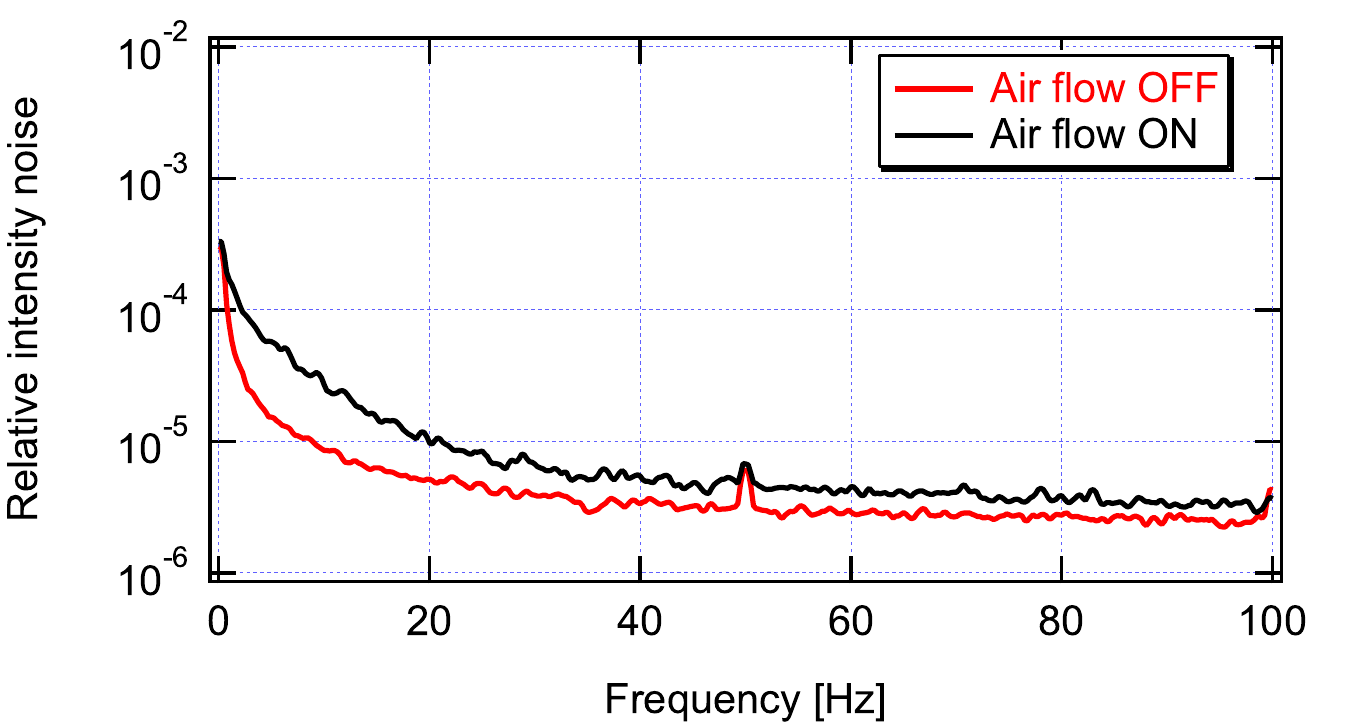}
\includegraphics[width=8cm]{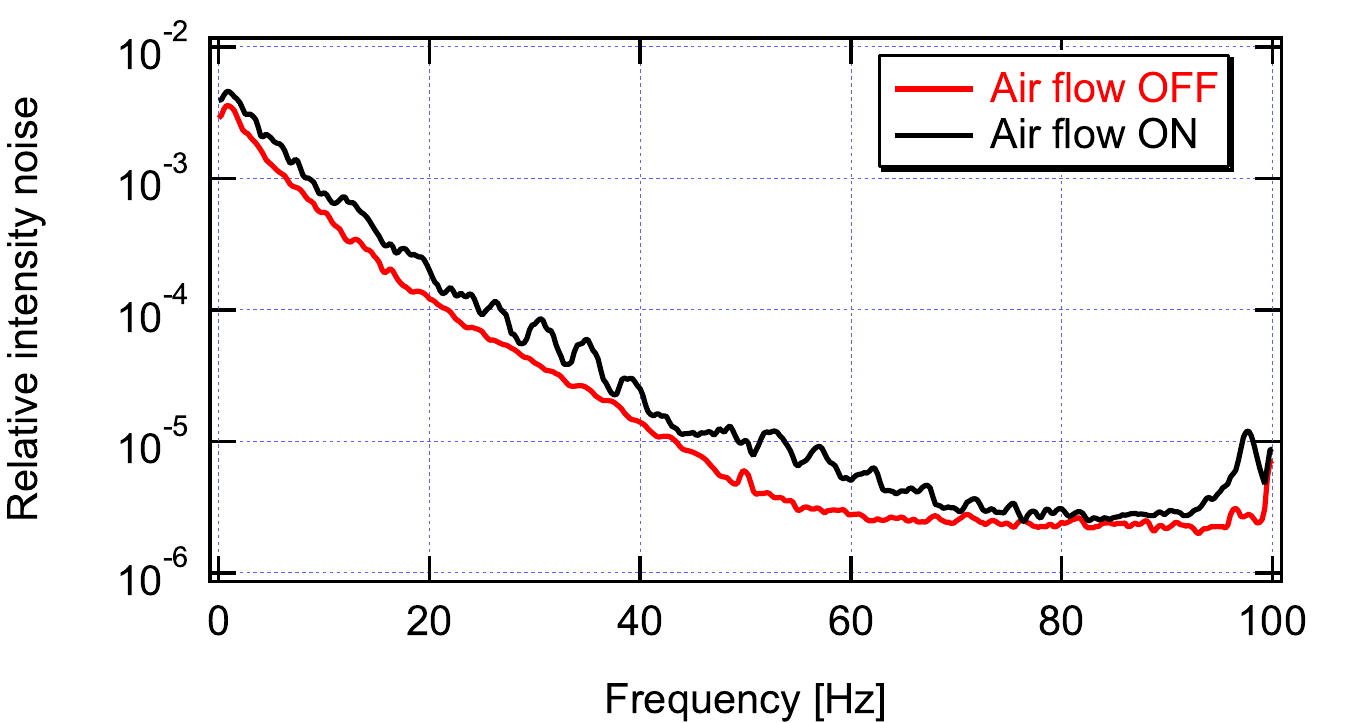}
\caption{Comparison of the relative intensity noise (RIN) of the laser with the air conditioning on and with it off. Left: measurements taken 0.5~m from the laser head. Right: measurements taken at a distance of 2.5~m from the laser head.}
\label{fig:rin_rms_laser}
\end{center}
\end{figure}

\begin{figure}[bht]
\begin{center}
\includegraphics[width=8cm]{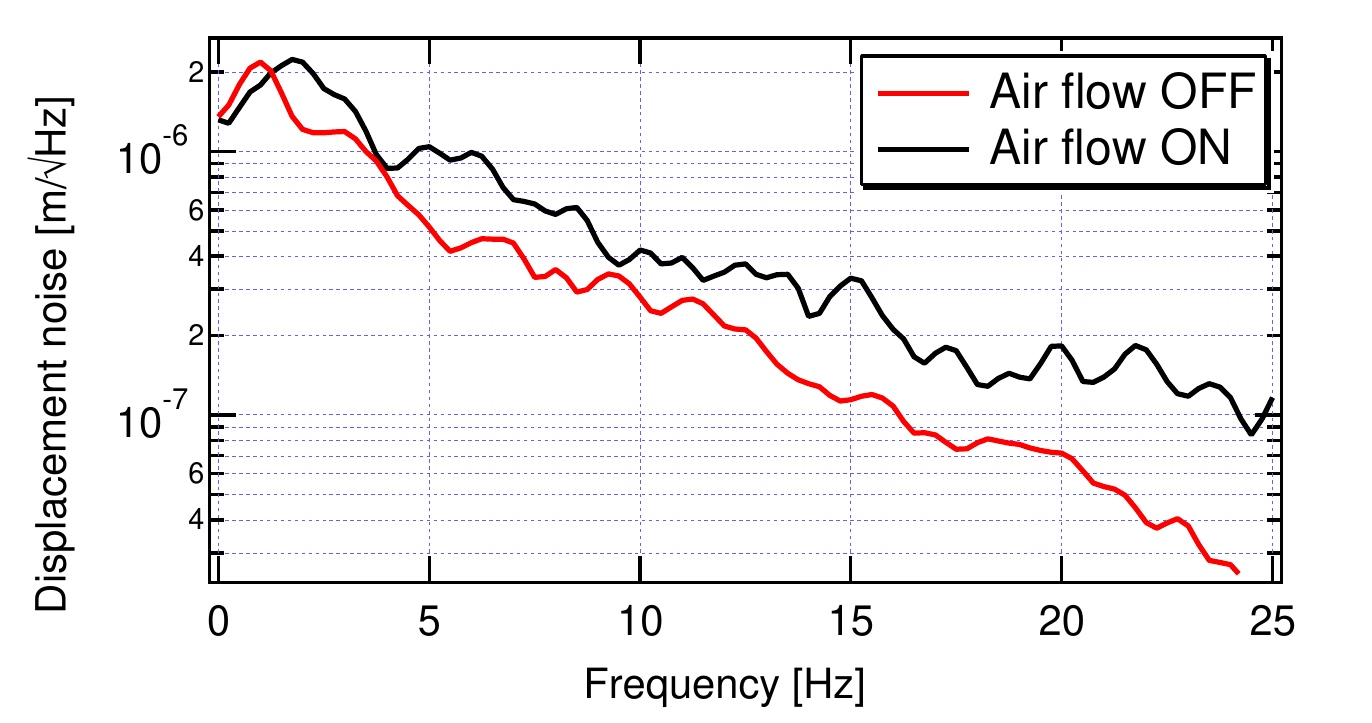}
\includegraphics[width=8cm]{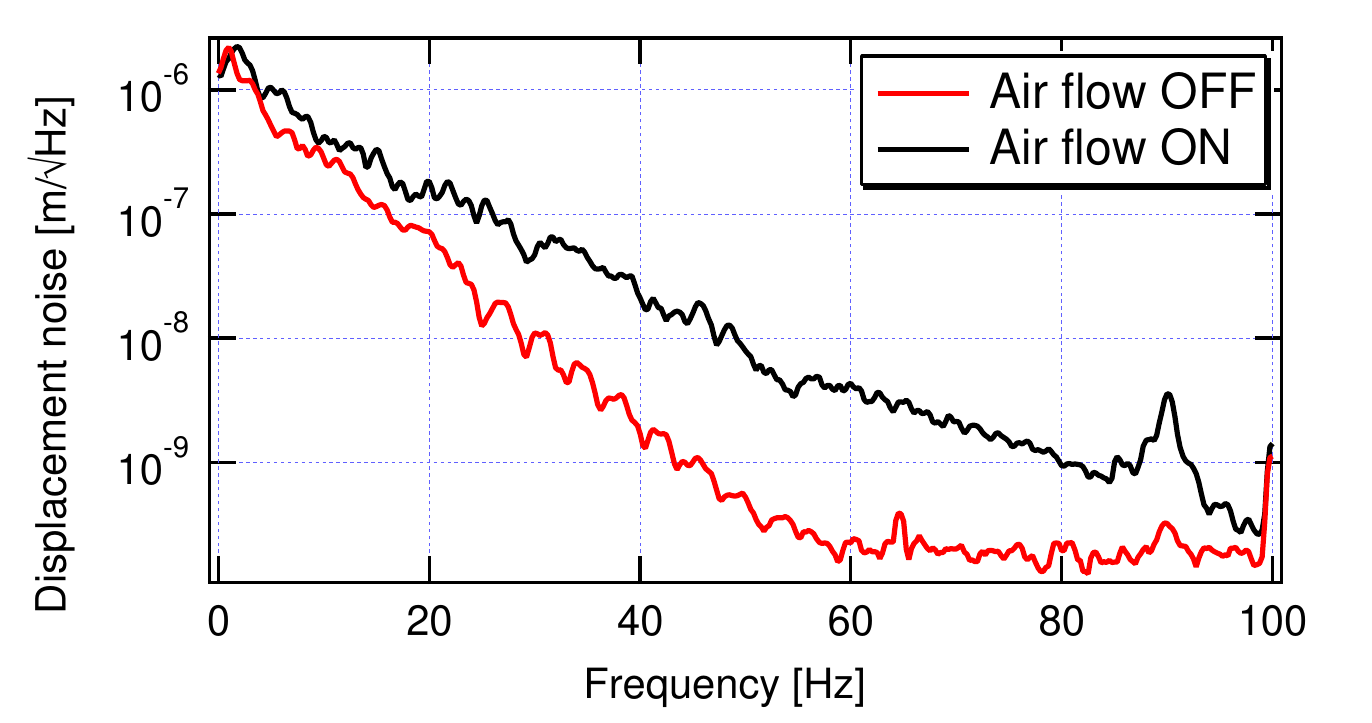}
\caption{Comparison of the laser pointing noise $S_{\rm R} = \sqrt{S_{\rm X}^2+S_{\rm Y}^2}$ measured at about 3~m from the laser with the mode matching lens in position. The two panels show different frequency ranges.}
\label{fig:pointing}
\end{center}
\end{figure}

Acoustic noise and the ventilation in the clean room was also considered as a noise source. The air flow generated turbulences which affected the propagation of the laser entering the polarimeter. A comparison of the intensity noise and pointing noise of the incident laser beam with the air flow ON and OFF was performed. The intensity noise was measured in two different positions: 0.5~m from the laser and after a further distance of about 2~m whereas the pointing noise was measured at about 3~m from the laser with the mode matching lens in place. In Figure~\ref{fig:rin_rms_laser} the intensity noise of the laser beam at these two positions with the air flow ON and OFF are shown. The amplitude noise is considerably worse at the farther position and further worsens with the air flow ON (black curves). 
In Figure~\ref{fig:pointing} the pointing noise $S_{\rm R} = \sqrt{S_{\rm X}^2+S_{\rm Y}^2}$ is shown with the conditioning ON and OFF. A clear difference is visible at all frequencies here too.

\begin{figure}[bht]
\begin{center}
\includegraphics[width=10cm]{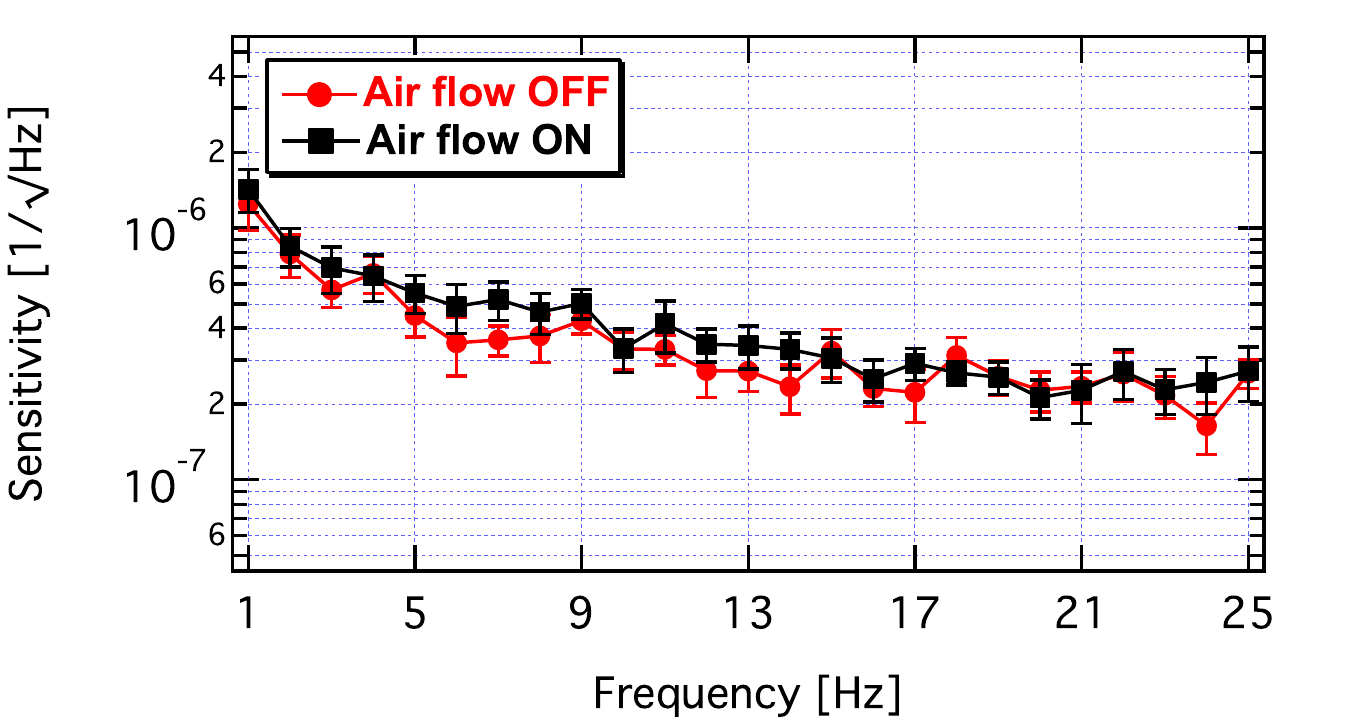}
\caption{Ellipticity sensitivity with air flow on (black points) and off (red points). Each point in the two graphs is the average over the 32 bins in a 1~Hz frequency interval. Integration time for the two graphs is ${\cal T}=4.1\times10^3$~s and ${\cal T}=5.6\times10^3$~s, respectively.}
\label{fig:sensitivity-air-flow}
\end{center}
\end{figure}

In Figure~\ref{fig:sensitivity-air-flow} we show two measurements of the sensitivity in ellipticity in the frequency range of interest as a function of frequency with the air flow ON and OFF. As can be seen, differently from the cases of relative intensity noise and pointing noise, the air flow had little or no influence on the sensitivity in ellipticity. This indicates that neither the input intensity noise nor the input pointing noise were limiting the ellipticity sensitivity.
The condition for the intensity noise to be negligible is given by equation~(\ref{RIN_noise}) and was kept under control. As for the input beam pointing noise it must be noted that the beam stability inside the cavity is defined \emph{only} by the stability of the mirrors thereby excluding the birefringence pattern of the mirrors, coupled to the input pointing noise, as the noise source.  

The dominant ellipticity noise seems to be of a different nature. To complete this series of tests, we also performed measurements with all the other sources of acoustic and vibrational noise, such as the turbo and scroll vacuum pumps, switched off. Again, the sensitivity did not change.



\subsubsection{The role of the finesse}

The design of the PVLAS-FE experiment was fundamentally based on the following considerations:
\begin{enumerate}
\item with a Fabry-Perot cavity, the total acquired ellipticity is $\Psi=N\psi$, where $\psi$ is the ellipticity acquired for a single pass in the birefringent medium. With $\int B^2\,dL\approx10~$T$^2$m and a finesse $\cal{ F}$ $=7\times 10^{5}$ this gives $\Psi=5\times 10^{-11}$. In order to reach a unitary signal to noise ratio in an integration time ${\cal T}=10^{6}$~s, a sensitivity of $5\cdot10^{-8}{/\sqrt{\rm Hz}}$ would be needed;
\item in principle, at a modulation $\eta_0=0.01$, a near shot-noise sensitivity of $\approx 8\cdot10^{-9}/\sqrt{\rm Hz}$ should have been possible. Since shot noise is always very difficult to achieve, the PVLAS-FE apparatus was designed with a contingency factor about ten.
\end{enumerate} 

\begin{figure}[htb]
\centering
\includegraphics[width=10cm]{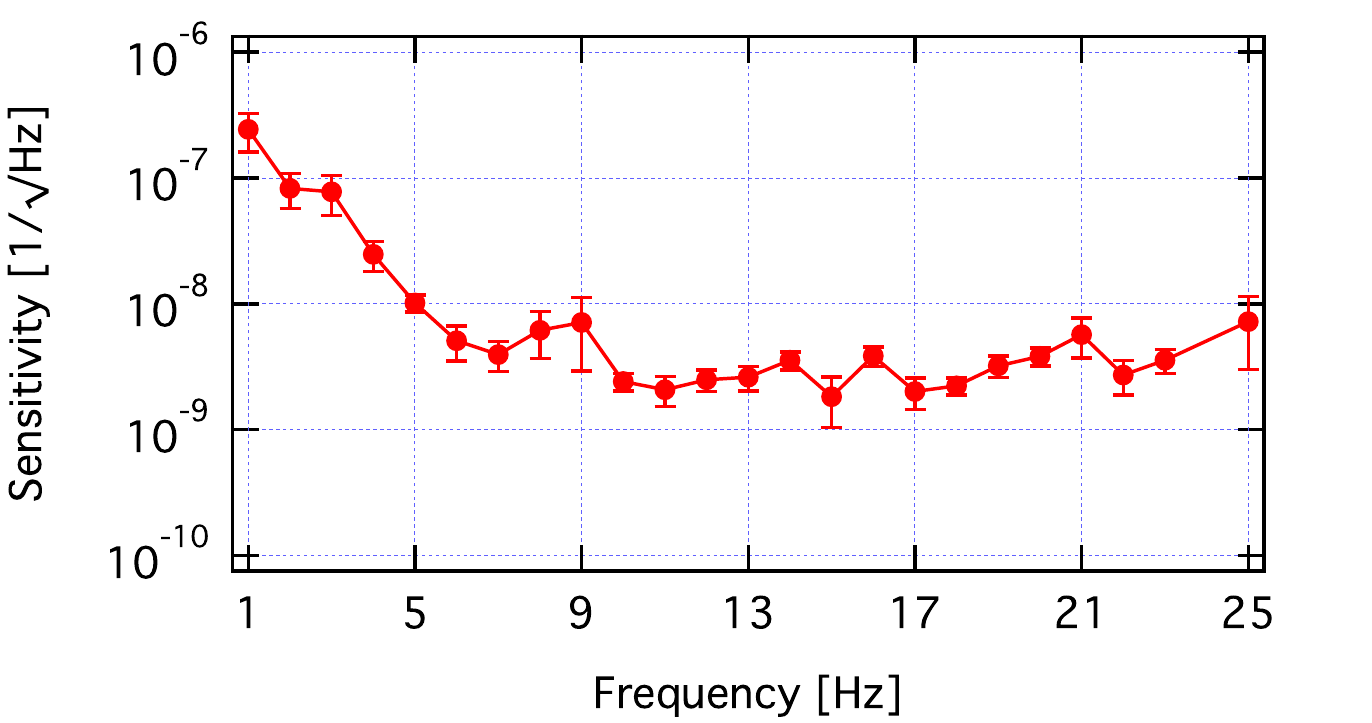} 
\caption{Ellipticity sensitivity measured as a function of the frequency with the optical scheme of the PVLAS-FE experiment but without the FP cavity.} 
\label{fig:noise_no_cavity}
\end{figure} 

As can be seen in Figure~\ref{fig:noise_no_cavity}, without the Fabry-Perot cavity the sensitivity above 7~Hz was limited by the expected noise budget determined in Section~\ref{sec:intrinsic_noise}. Below 7~Hz the noise was limited by the pointing stability of the laser beam (in the presence of the Fabry-Perot the beam stability inside the cavity is determined solely by the mirror stability and not by the input laser beam). In the frequency region from 6~Hz to 25~Hz the sensitivity was flat and reached the expected value from the known noise budget. As already mentioned, the noise sources in Figure~\ref{IntrinsicNoise} are electronic/instrumental noises which can be translated to an ellipticity noise. They do not represent a direct ellipticity noise which will beat with the modulator.

 In principle the Fabry-Perot cavity amplifies an ellipticity generated between the cavity mirrors thereby improving the signal to noise ratio by a factor $N$. This is indeed the case for relatively low finesses but the introduction of the very high finesse Fabry-Perot cavity changes the wide band noise distribution in an unexpected way. In this case the signal to noise ratio reaches a plateau due to a noise $\gamma_{\rm cavity}$ generated inside the cavity which is therefore also multiplied by $N$.

\subsubsection{Ellipticity modulation}

\begin{figure}[bhtp]
\begin{center}
\includegraphics[width=8cm]{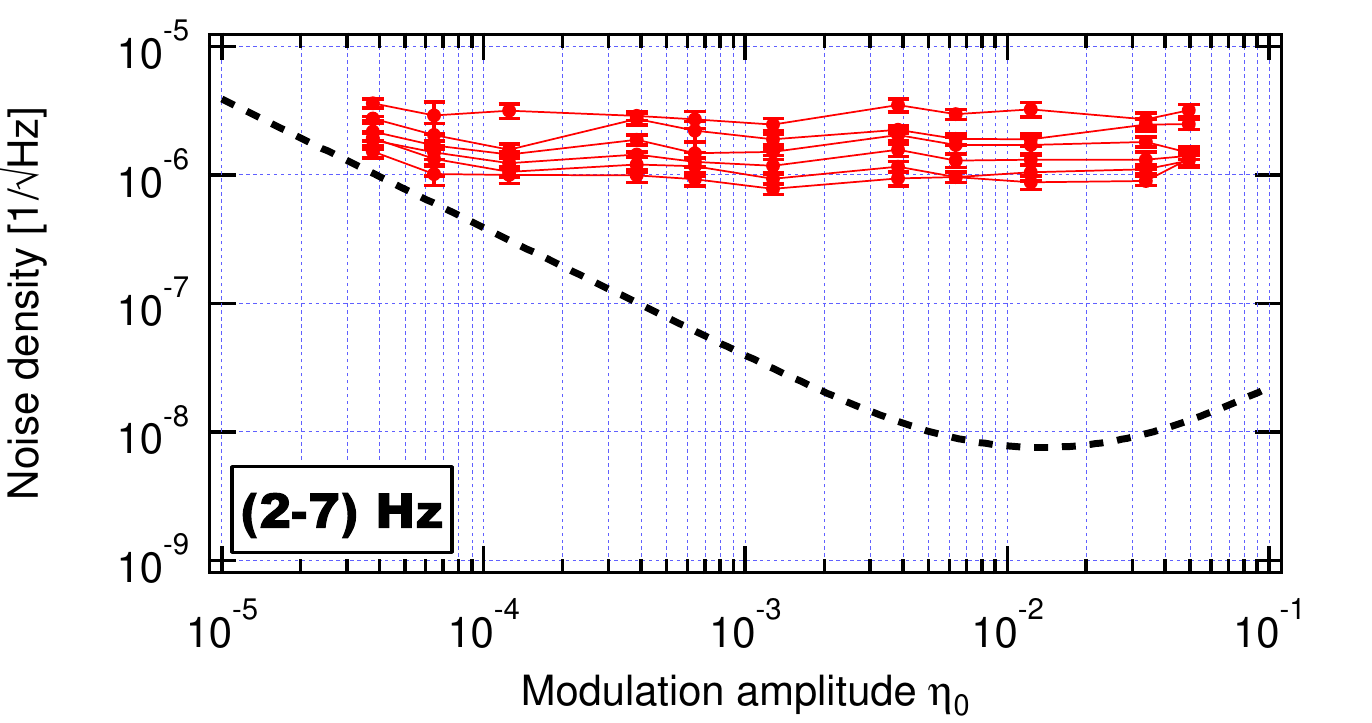}
\includegraphics[width=8cm]{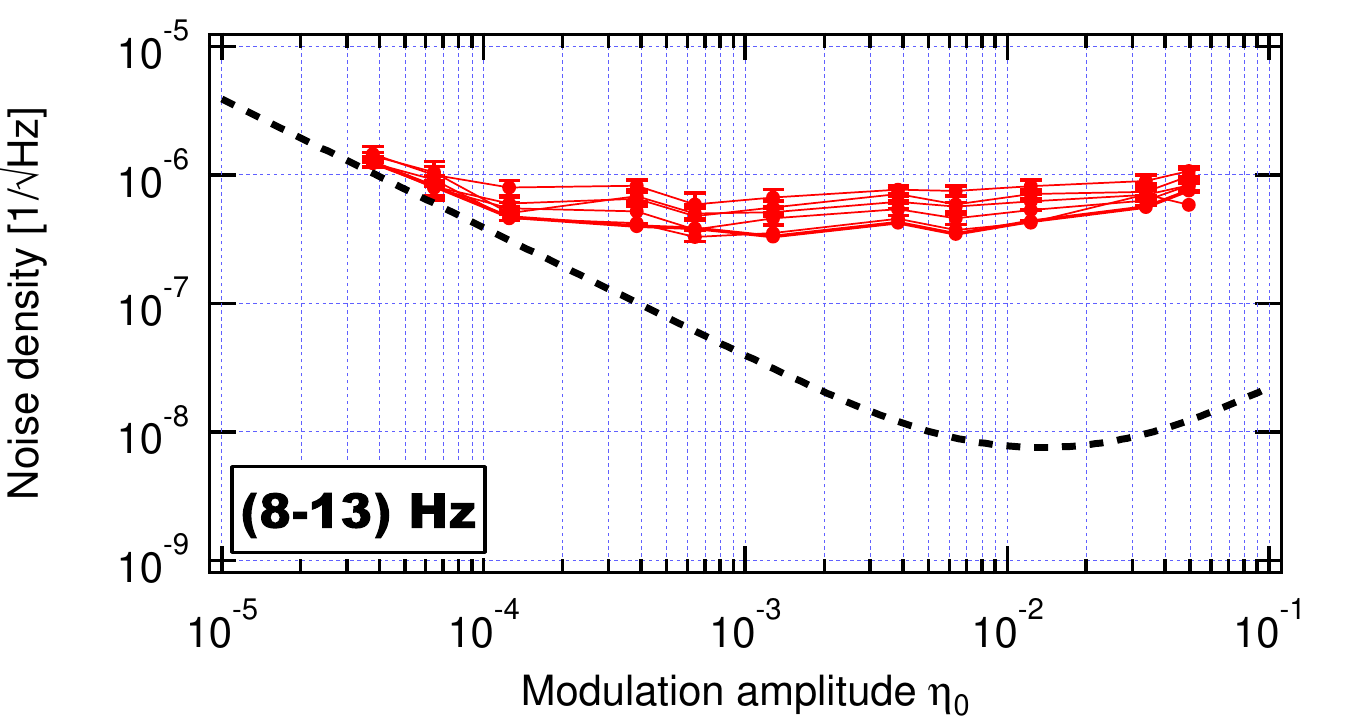}
\includegraphics[width=8cm]{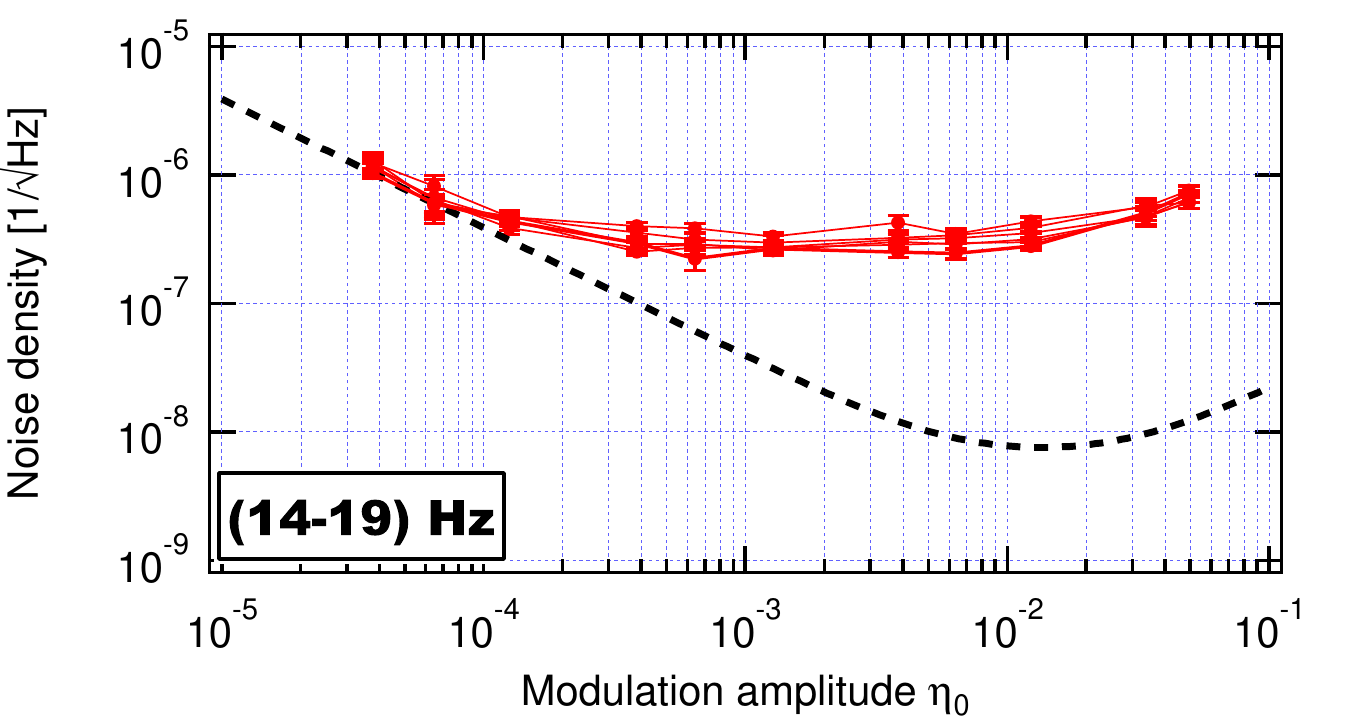}
\includegraphics[width=8cm]{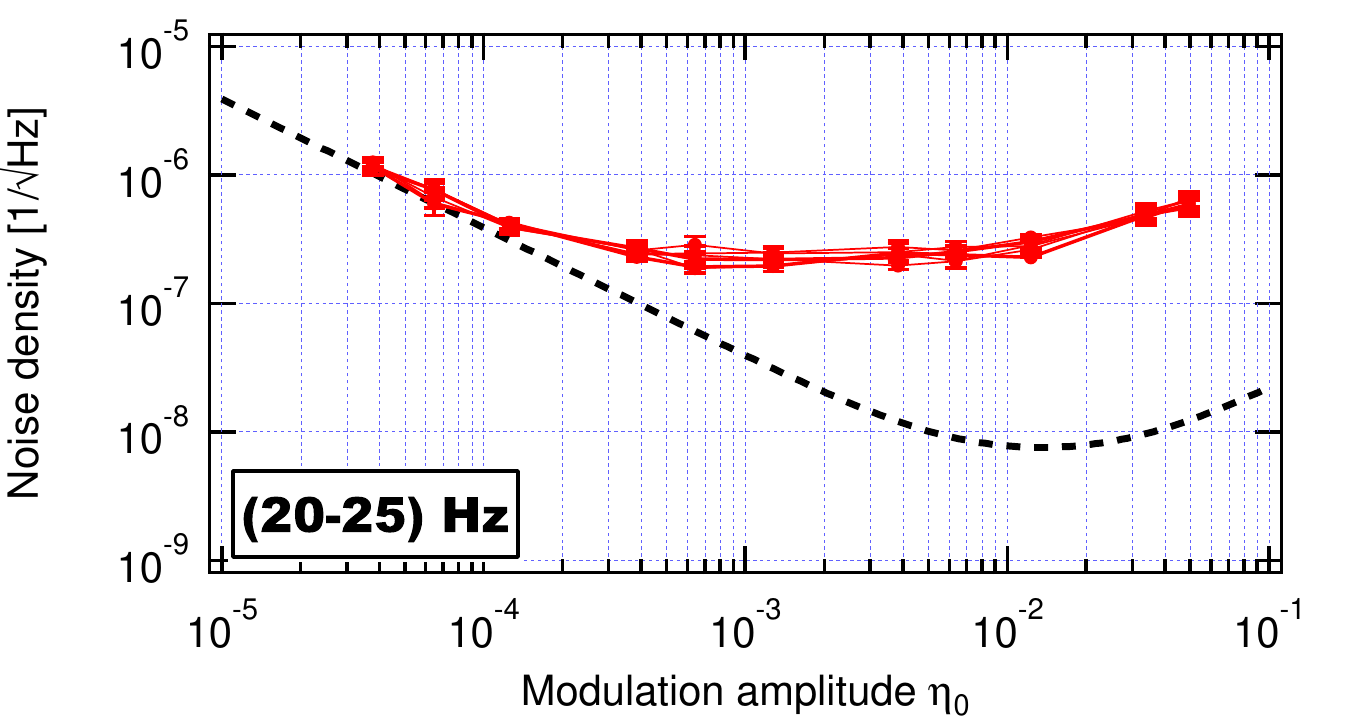}
\includegraphics[width=8cm]{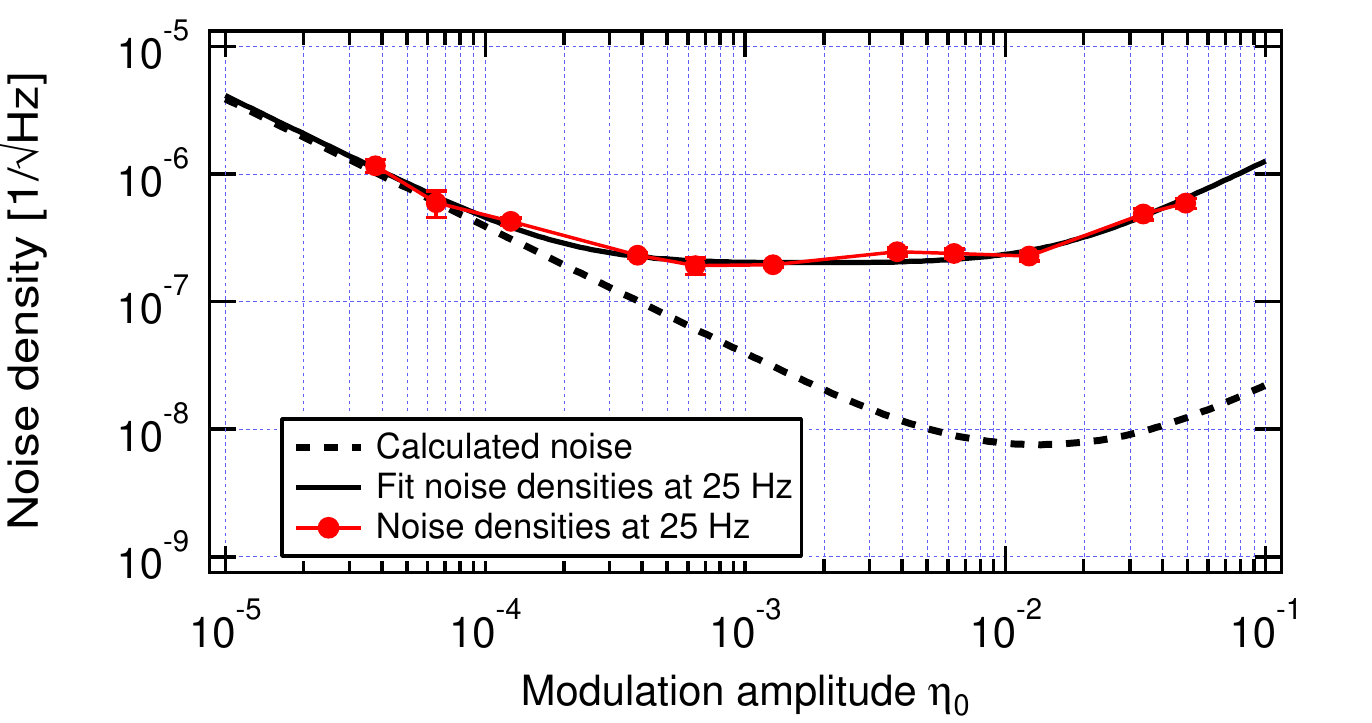}
\includegraphics[width=8cm]{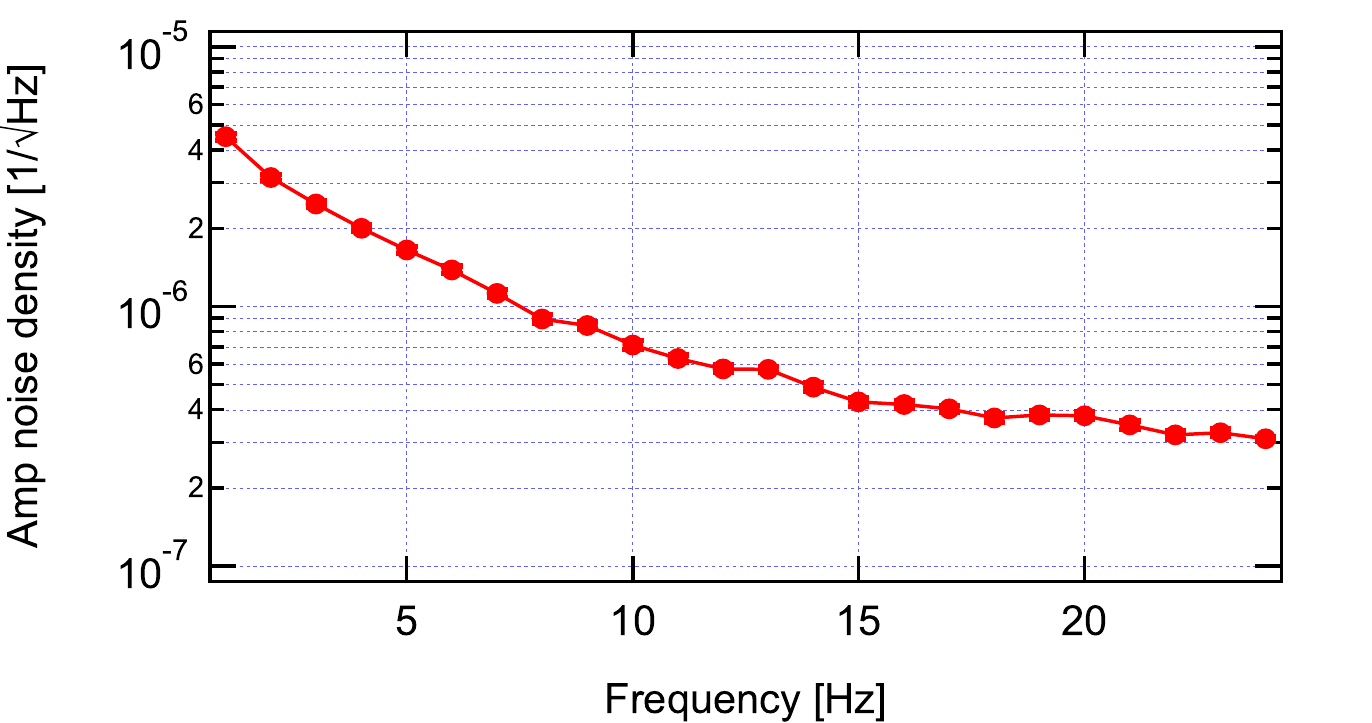}
\caption{First four panels: measured ellipticity noise density (red dots) compared to the theoretical intrinsic noise density $S^{\rm (tot)}_\Psi$ (dashed line, $I_\parallel = 5$~mW) of equation~(\ref{eq:tot_noise}), plotted as a function of the modulation amplitude $\eta_{0}$. Each of the panels presents measurements at six different frequencies, from 2~Hz to 25~Hz, in 1~Hz steps. 
In the fifth panel, a fit of the 25~Hz points is shown. The fitting function is the theoretical noise $S_\Psi^{\rm (tot)}$ of equation~(\ref{eq:tot_noise}) plus a frequency dependent uncorrelated noise $A_\nu$ and with $N^{\rm (RIN)}_{\nu_m}$ left as a free common parameter for the fit of all the 24 curves to describe the measured linear rise in $S_\Psi$ for large values of $\eta_0$.  The sixth and last panel shows the 24 values of $A_\nu$. Assuming the noise originates as an ellipticity noise $S_\Psi$, the data in this last panel have been normalised with $k(\alpha_{\rm EQ}) = 0.65$. The much smaller correction due to the frequency response of the cavity has been neglected.}
\label{fig:sens-mod}
\end{center}
\end{figure}

To understand the ellipticity nature of the wide-band noise and the discrepancy from the expected one, and to determine a possible contribution of the PEM, measurements of the sensitivity in ellipticity were performed as a function of the modulation amplitude $\eta_0$ and are shown in Figure~\ref{fig:sens-mod}. 
The measurements were performed at frequencies ranging from 2 to 25~Hz. Each of the first four panels of the figure presents the data relative to a 6~Hz frequency interval. In the first frequency range from 2 to 7~Hz (top to lower curves), the noise is almost independent of the modulation amplitude and is far from the calculated noise of the polarimeter (dashed curve, $I_\parallel = 5$~mW, see Figure~\ref{IntrinsicNoise} on page~\ref{IntrinsicNoise}). Furthermore there is an improvement as the frequency increases. 
In the interval from 8 to 13~Hz the sensitivity further improves. At small modulation amplitudes, with $\eta_0< 10^{-4}$, the measured noise densities seem to tend to the calculated value. At higher frequencies and higher modulations, the ellipticity noise improves. In the middle two panels the sensitivity as a function of $\eta_0$ deviates from the calculated curve and flattens off for values up to $\eta_0 \approx 10^{-2}$. For $\eta_0 > 10^{-2}$ the sensitivity deteriorates proportionally to $\eta_0$. 
The measured minimum is reached at values of $\eta_{0}\approx10^{-3}$ in contrast to the minimum at 
about $\eta_{0}\approx10^{-2}$ of the calculated curve.

The 24 sensitivity curves were fitted using the function of equation~(\ref{eq:tot_noise}) to which a constant, uncorrelated, frequency-dependent noise $A_\nu$ was added:
\begin{equation}
S_\Psi^{\rm (tot)}{'}(\nu)= \sqrt{S^{\rm (tot)}_\Psi{^2}+A^2_\nu}.
\end{equation}
Free parameters are $N^{\rm (RIN)}_{\nu_m}$ [see equation~(\ref{eq:SRIN})], which is common for all the curves and describes the sensitivity deterioration proportional to the modulation at large values of $\eta_0$, and $A_\nu$. 
A common value $N^{\rm (RIN)}_{\nu_m}=(1.6 \pm 0.3)\times10^{-5}{/\sqrt{\rm Hz}}$ is obtained. An example of these fits is shown for the 25~Hz case in the fifth panel of Figure~\ref{fig:sens-mod}, for which $A_{\rm 25~Hz}=(1.99 \pm 0.05)\times10^{-7}{/\sqrt{\rm Hz}}$. For low values of $\eta_0$ the measured points follow the expected curve which depends only on the readout electronics. Assuming the values of $A_\nu$ represent an ellipticity, they were normalised by $k(\alpha_{\rm EQ}) = 0.65$ and are plotted as a function of frequency in the sixth panel of the same Figure~\ref{fig:sens-mod}. 

The value obtained for the $N^{\rm (RIN)}_{\nu_m}$ from the fit was about a factor 50 greater than the value measured at $\nu_m =50$~kHz corresponding to $N^{\rm RIN}_{\nu_m}\approx 3\times 10^{-7}{/\sqrt{\rm Hz}}$. 
Another contribution to the sensitivity proportional to the modulation $\eta_0$ seemed to be present and dominated for $\eta_0\gg\sigma^2$ just as equation~(\ref{eq:SRIN}) does: $S_{\Psi} = k \eta_0 $ with $k\approx (1.6 \pm 0.3)\times10^{-5}{/\sqrt{\rm Hz}}$. This contribution remains unexplained and may be due to the PEM. 

For intermediate values of $\eta_0$ between $10^{-3}$ and $10^{-2}$ there was a frequency dependent ellipticity noise beating with the modulator just like an ellipticity should. Furthermore, this noise dominated and seemed to be due to the presence of the cavity which generates $\gamma_{\rm cavity}(t)$. Further evidence that the Fabry-Perot was the source of this ellipticity noise will be given in Section~\ref{sec:intrinsic noise}.


\subsubsection{Cavity frequency difference for the two polarisation states}

Two effects related to the frequency difference between the two polarisation states were also considered as possible ellipticity noise sources: a laser-cavity frequency difference noise and a fluctuation of the frequency difference between the two resonances. 

As discussed in Section~\ref{sec:MirrorBirefringence} and as demonstrated by the measurements presented in Section~\ref{sec:meas_birif_cavity}, the presence of a birefringent cavity with $\alpha_{\rm EQ}$ leads to a linear dependence of an ellipticity on the round-trip phase $\delta$ as can be seen from equation (\ref{eq:E_perp_alpha}). 
This can be understood graphically from 
Figure~\ref{BirefringentFP} on page \pageref{BirefringentFP}: when locked near resonance (top of red curve) the ellipticity signal will follow the blue curve which has a non zero derivative as a function of the phase $\delta$. 
A noise in $\delta$ together with a static ellipticity will be translated into an ellipticity noise proportional to $\delta$ whenever $\alpha_{\rm EQ}\ne0$. The noise in $\delta$ will depend on the quality of the locking circuit, namely on the noise of the voltage $V_{\rm E}$ at the error point when the laser is unlocked. With the laser locked, this noise is translated into a frequency noise by the locking circuit and if the gain is high enough (as was the case for the PVLAS-FE circuit) this is the dominant feedback noise. The dependence of the ellipticity on $\delta$ can be observed by modulating the laser frequency around the resonance \cite{DellaValle2016EPJC} with a known $V_{\rm E}(t) = A\cos({2\pi\nu_{\rm off}t})$ and measuring the induced ellipticity $\Psi_{\nu_{\rm off}}$ at $\nu_{\rm off}$. This allows the determination of $\frac{\partial\Psi}{\partial V_{\rm E}}$.
Injecting a modulation signal at the error point of amplitude of $A = 9.9$~mV the resulting ellipticity at $\nu_{\rm off}$ was $\Psi_{\nu_{\rm off}} = 2.0\times 10^{-4}$. At the same error point the measured noise density, with the laser unlocked, was $S_{V_{\rm E}} \approx 3.5\;\mu$V/$\sqrt{\rm Hz}$ with a flat frequency spectrum. The estimated ellipticity noise $S_{\Psi}^{\rm (feedback)}$ induced by $S_{V_{\rm E}}$ is therefore
\begin{equation}
    S_{\Psi}^{\rm (feedback)} = {\partial\Psi \over \partial V_{\rm E}}S_{V_{\rm E}} = \frac{\Psi_{\nu_{\rm off}}}{A}S_{V_{\rm E}} \approx 7\times10^{-8}/\sqrt{\rm Hz}.
    \label{FeedbackNoise}
\end{equation}
This value could not account for the measured noise at least up to 25 Hz, as can be seen in Figure~\ref{fig:sens-mod} bottom right, thereby excluding the locking system as the wide band noise source. Furthermore, if this effect were the noise source it would generate a flat ellipticity spectrum density. 
The ellipticity of equation (\ref{FeedbackNoise}), translated to optical path difference, results in 
\begin{equation}
    S^{\rm (feedback)}_{\Delta{\cal D}} =  \frac{\lambda}{2{\cal F}} S_{\Psi}^{\rm (feedback)} = 5\times10^{-20}~{\rm m}/\sqrt{\rm Hz}.
\end{equation}

In anticipation of a discussion that will be made in Section~\ref{sec:intrinsic noise} aimed at understanding the nature of the wide band noise, we want to show that $S_{\Psi}^{\rm (feedback)}$ is proportional to ${\cal F}$ and therefore that the optical path difference noise induced by the feedback does not depend on the finesse. Three factors contribute to this behaviour of $S_{\Psi}^{\rm (feedback)}$. Firstly the locking error signal slope $D_\nu$ when using the PDH locking scheme is 
\begin{equation}
D_\nu = \frac{\partial{V_{\rm E}}}{\partial{\nu}}\propto I_F\frac{\beta}{2}h_{\rm R}(0)\frac{{\cal F}T}{P\nu_{\rm fsr}} \propto I_F\frac{{\cal F}^2T}{\nu_{\rm fsr}}.
\end{equation}
where $I_F\propto I_{\rm in}$ is a fraction of the incident power reflected by the cavity used for the feedback, $\beta$ is the radio frequency modulation depth for generating the side bands, $T$ is the transmission coefficient of the cavity mirrors which is an intrinsic property of theirs, $P$ are the losses of the cavity and $h_{\rm R}(0) = P/(1-R)$ is the reflection transfer function of the cavity at resonance. Generally the product $T{\cal F} \sim 1$ resulting in $D_\nu\propto {\cal F}$. In the measurements presented in Section~\ref{sec:intrinsic noise} though the finesse ${\cal F}$ of the cavity was varied by increasing the losses $P$ in the cavity without changing $T$. Therefore in what will follow below $D_\nu \propto I_F{\cal F}^2$. Secondly the noise $S_{V_{\rm E}}$ at the error point, with the laser unlocked, is determined by the RIN of the laser at the locking frequency of $\approx 500$~kHz (in our case shot noise is reached above 5~MHz). Therefore $S_{V_{\rm E}}\propto I_F$. Finally the total static ellipticity $\Psi$ due to a single pass ellipticity $\psi$ induced inside the cavity is ($N = 2{\cal F}/\pi$)
\begin{equation}
    \Psi = \psi\,\frac{N}{1+N^2\sin^2{\alpha_{\rm EQ}/2}}.
\end{equation}
with the laser locked to the $\parallel$ polarisation ($\delta = -\alpha_{\rm EQ}/2$ in equation~(\ref{eq:E_perp_alpha})). With the modulation on $\delta = -\alpha_{\rm EQ}/2 - \Delta\delta(t)$ this ellipticity will become
\begin{equation}
    \Psi = \psi\frac{N}{1+N^2\sin^2{\left(\frac{\alpha_{\rm EQ}}{2}+\frac{\Delta\delta(t)}{2}\right)}}.
\end{equation}
and one will observe an ellipticity at the modulation frequency $\nu_{\rm off}$
\begin{equation}
    \Psi_{\nu_{\rm off}} = \left(\frac{\partial\Psi}{\partial\Delta\delta}\right)_{\delta = -\alpha_{\rm EQ}/2}\Delta\delta.
\end{equation}
For $N\alpha_{\rm EQ}/2 < 1$ the result is that $\Psi_{\nu_{\rm off}}\propto N^3$. Putting these considerations together results in
\begin{equation}
    S_\Psi^{\rm (feedback)} = \frac{\partial\Psi}{\partial\nu}\frac{1}{D_\nu}S_{V_{\rm E}}\propto N^3\frac{1}{I_FN^2}I_F\propto N.
    \label{eq:s_feedback}
\end{equation}
thereby scaling with the finesse. This will be important when comparing the optical path difference noises measured at different finesse values.

What was considered above was the \emph{simultaneous} scanning of the two perpendicular resonances due to a laser frequency noise. An ellipticity can also be generated by a \emph{relative} frequency shift of the two resonances due to an optical path difference noise $S_{\Delta\cal D}$.
From equation~(\ref{eq:sfasamento}), the phase difference $\varphi_{\alpha_{\rm EQ}}$ between the two polarisation states due to the cavity birefringence is related to the frequency separation by
\begin{equation}
\tan\varphi_{\alpha_{\rm EQ}} = \frac{(1+R)\sin\frac{\alpha_{\rm EQ}}{2}}{(1-R)\cos\frac{\alpha_{\rm EQ}}{2}}\approx\frac{\alpha_{\rm EQ}}{1-R}=\frac{2\pi}{1-R}\frac{\Delta\nu_{\alpha_{\rm EQ}}}{\nu_{\rm fsr}}=\pi N\frac{\Delta\nu_{\alpha_{\rm EQ}}}{\nu_{\rm fsr}}.
\label{eq:birif}
\end{equation} 
Let us consider a value of $\alpha_{\rm EQ} \approx 2\times10^{-6}$ resulting in $\Delta\nu_{\alpha_{\rm EQ}} \approx 15$~Hz and an ellipticity noise $S_{\Psi} \approx 5\times10^{-7}/\sqrt{\rm Hz}$. Given that the measured ellipticity noise is one half of the phase noise, $S_{\Psi}={S_{\varphi}}/{2}$, one finds a relative frequency noise between the two polarisation states
\begin{equation}
S_{\Delta\nu_{\alpha_{\rm EQ}}} = \frac{2S_{\Psi}\nu_{\rm fsr}}{\pi N} \approx 3\times 10^{-5}~\frac{{\rm Hz}}{\sqrt{\rm Hz}}.
\label{eq:freqnoise}
\end{equation} 
One can estimate the relative frequency noise of equation~(\ref{eq:freqnoise}) as due to a variation of the cavity optical path length ${\cal D}$. With the laser locked to the cavity $\delta_\parallel = 2\pi m$, the round-trip phase $\delta_\perp$ for the perpendicular polarisation is therefore
\begin{equation}
    \delta_\perp = 4\pi\frac{{\cal D}\Delta\nu_{\alpha_{\rm EQ}}}{c}
\end{equation}
A fluctuation $S_{\Delta\nu_{\alpha_{\rm EQ}}}$ of the relative frequency difference between the two resonant frequencies is related to a fluctuation of the cavity length as
\begin{equation}
{\cal D}S_{\Delta\nu_{\alpha_{\rm EQ}}}=S_{\cal D}\Delta\nu_{\alpha_{\rm EQ}}
\end{equation}
From this one obtains
\begin{equation}
S_{\cal D} = {\cal D}\frac{S_{\Delta\nu_{\alpha_{\rm EQ}}}}{\Delta\nu_{\alpha_{\rm EQ}}} \approx 7~\frac{\mu\rm m}{\sqrt{\rm Hz}}.
\label{eq:lengthnoise}
\end{equation} 
The cavity length stability is far better than this value excluding this effect too as a source of wide band noise. Indeed with such a length fluctuation the dynamical range of the locking circuit would not have allowed stable locking.  Again vibrational noise of the mirrors of the cavity could not account for the observed ellipticity noise.

\subsubsection{Power induced noise}
\label{sec:power}

\begin{figure}[bht]
\begin{center}
\includegraphics[width=12cm]{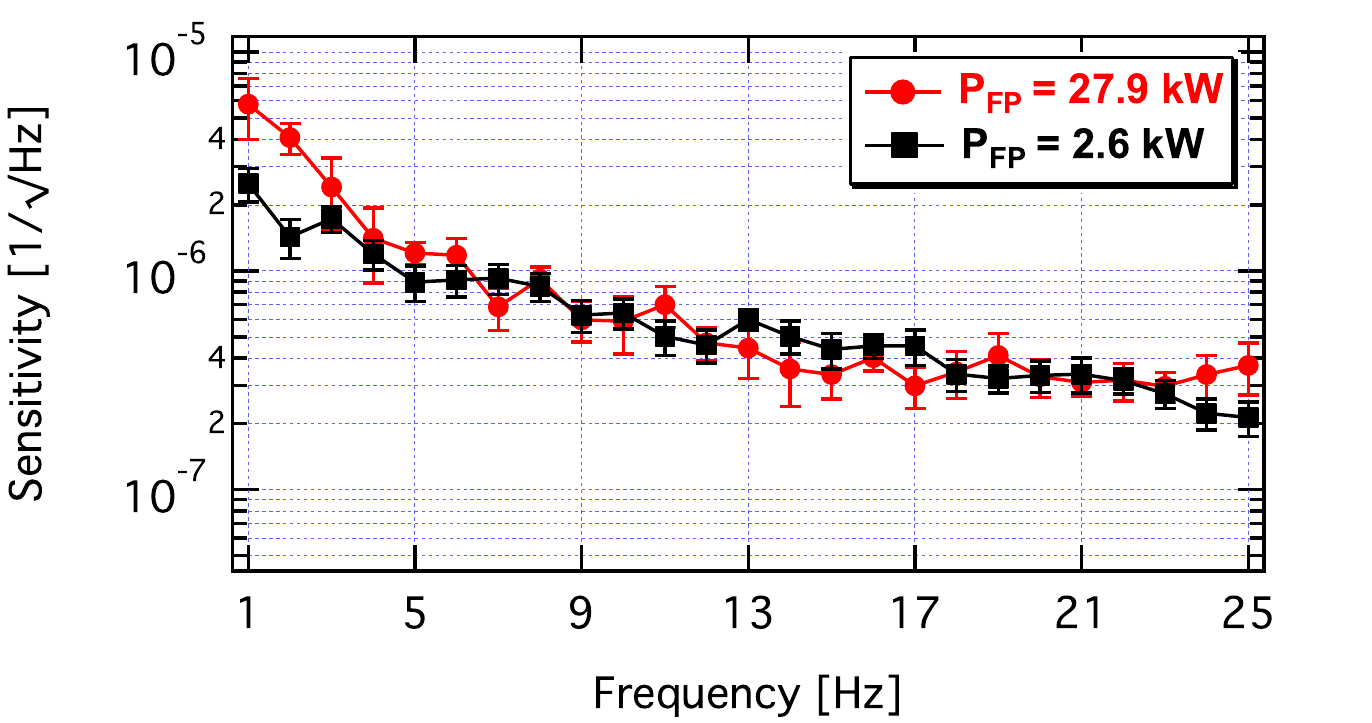}
\caption{Ellipticity sensitivity measured as a function of frequency at two different values of the circulating power in the Fabry-Perot cavity. Integration time was ${\cal T}=7\times10^3$~s for the 2.6~kW spectrum and ${\cal T}=2.6\times10^3$~s for the 27.9~kW spectrum.}
\label{fig:sensitivity-power}
\end{center}
\end{figure}

Thermal noise due to absorbed power on the surface of the mirrors was also investigated. In Figure~\ref{fig:sensitivity-power} the sensitivity with two different circulating powers is shown. Above about 5~Hz, the wide band noise $S_\Psi$ is unaffected by changing the input power by a factor 10. The two sensitivity curves refer to circulating powers of 27.9~kW and 2.6~kW corresponding respectively to  0.6~MW/cm$^2$ and 0.06~MW/cm$^2$ on the mirror surfaces. With the higher power, sub-hertz static ellipticity  instabilities are observed and are thought to be due to stress induced birefringence. During the vacuum birefringence measurements presented in Section~\ref{sec:FE_results} even lower powers were used.

\subsubsection{Intrinsic noise}
\label{sec:intrinsic noise}

\begin{figure}[bht]
\centering
\includegraphics[width=12cm]{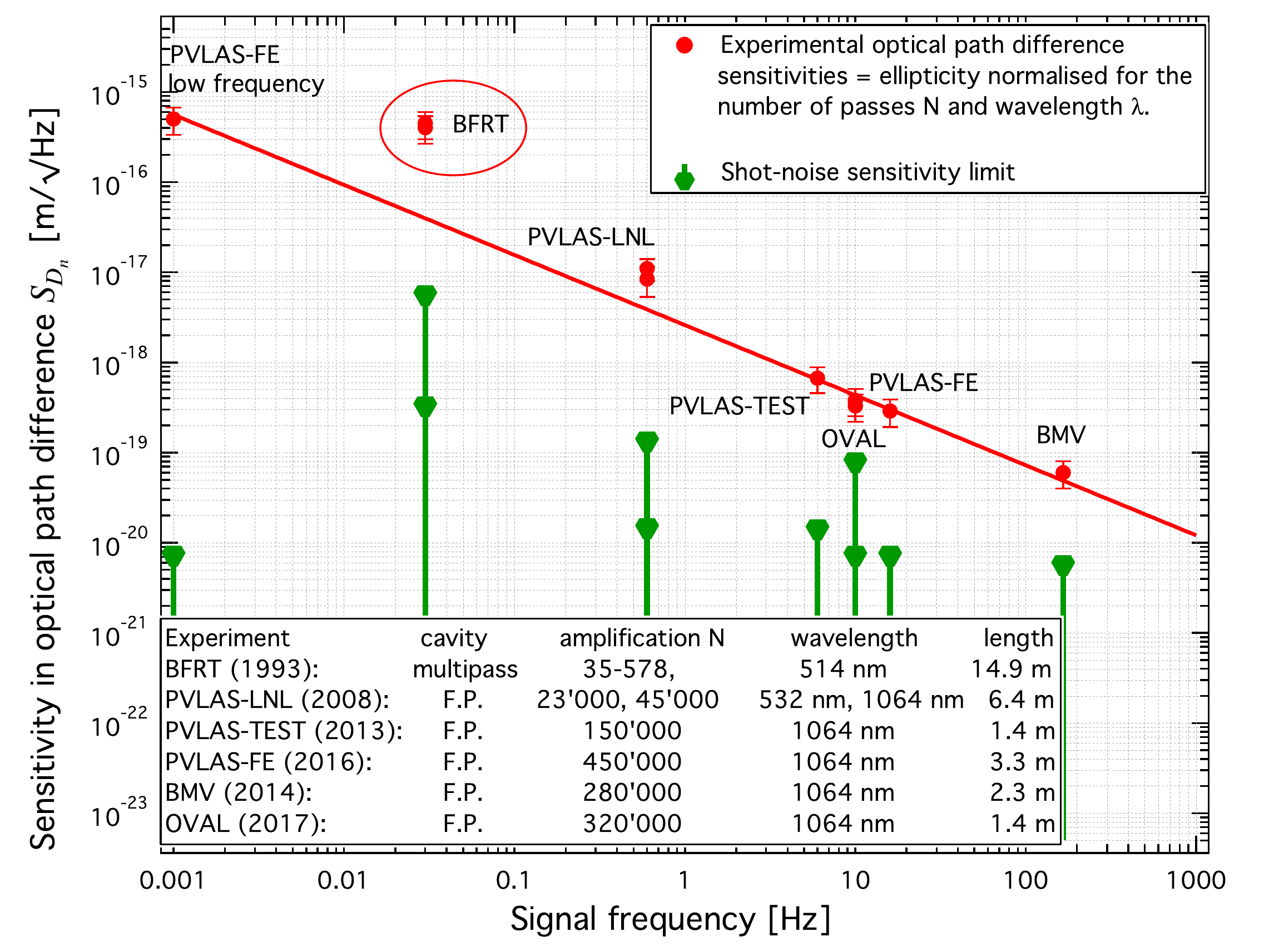}
\caption{Measured optical path difference noise densities in polarimeters set up to measure the vacuum magnetic birefringence plotted as a function of their working frequency. Data were taken from the experiments BFRT \cite{Cameron1993PRD}, PVLAS-LNL \cite{Zavattini2007PRD,Bregant2007PRD}, PVLAS-Test \cite{DellaValle2013NJP}, PVLAS-FE \cite{DellaValle2016EPJC}, BMV \cite{Cadene2014EPJD} and OVAL \cite{Fan2017} and are normalised to the number of passes and to the wavelength. The leftmost point has been measured during the 2015 data taking campaign of the PVLAS-FE experiment. The two almost equivalent points from BFRT were measured with two different cavities, one having 34 passes and the other 578 passes. The error bars are an estimated 50\%.}
\label{fig:noise_global}
\end{figure}

A comparative study of the sensitivities of different experimental efforts was done to have some insight on the source of this wide band noise afflicting them all. Assuming a birefringence noise source coming from inside the cavity the parameter studied was the optical path difference sensitivity $S_{\Delta{\cal D}} = \frac{\lambda}{N\pi}S_\Psi$ having normalised $S_\Psi$ for both the wavelength and, more importantly, for the number of passes $N$ of each experiment.
The sensitivities are shown in Figure~\ref{fig:noise_global} as a function of the working frequency of each experiment. In the case of pulsed fields lasting $t_{\rm pulse}$, as in BMV and OVAL, the frequency was taken as $\nu = 1/(2\pi t_{\rm pulse})$. In the same figure, in green, we report the shot-noise expected sensitivities.

Except for BFRT all experiments seem to lie on a common power law with exponent of approximately $-0.8$ reaching $S_{\Delta{\cal D}}<10^{-18}$~m/$\sqrt{\rm Hz}$ for $\nu\geq 5$~Hz. Note that BFRT is the only experiment which used a multi-pass cavity instead of a Fabry-Perot. We attach no particular meaning to the value of the exponent, but the fit puts in evidence that a common optical path difference noise source due to the cavities is present and is then multiplied by the finesse.

\begin{figure}[htb]
\centering
\includegraphics[width=12cm]{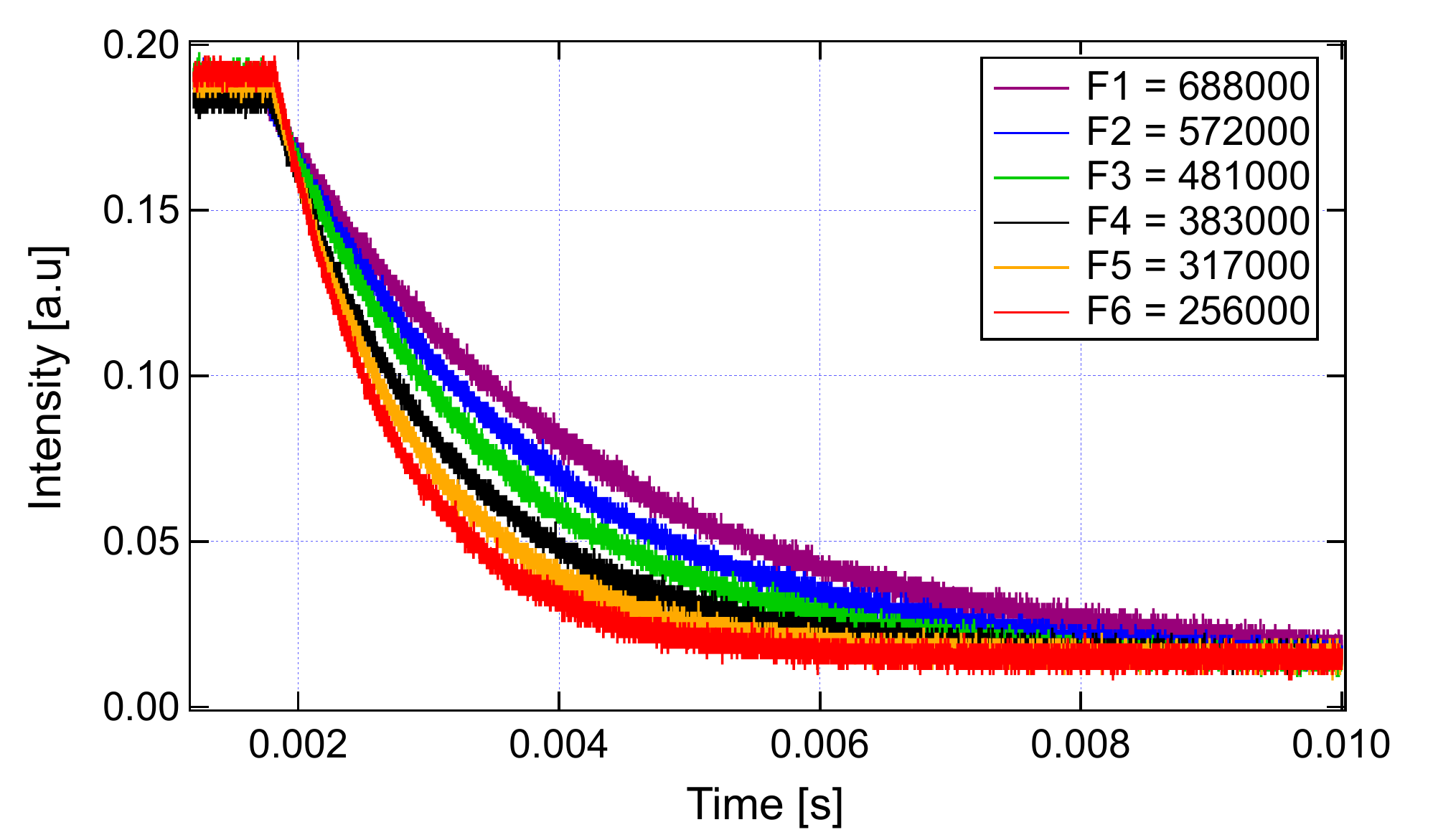}
\caption{Light decay curves for the six different finesses used. F1-F6 indicate the respective values. The finesse was reduced from F1 by clipping the beam introducing losses in the range $0\div10$~ppm. From reference \cite{IntrinsicNoise}, Figure 5.}
\label{fig:finesse}
\end{figure}

To confirm this behaviour, a series of measurements was therefore undertaken with the PVLAS-FE apparatus in which the finesse was changed \cite{IntrinsicNoise}. A low pressure gas was used to generate a reference ellipticity due to the Cotton-Mouton effect and a variable magnetic field applied to the input mirror was used to generate a Faraday rotation on the coatings. Both of these effects are proportional to the number of passes $N$. Rotation and ellipticity measurements were performed, and for each measurement these signals and the noises $S_{\Delta{\cal D},\Delta{\cal A}}$ were determined. The values of the finesse varied from 256000 to 688000 and the relative light power decay curves are shown in Figure~\ref{fig:finesse}. The finesse was reduced by slightly clipping the cavity mode introducing losses ranging from $0\div10$~ppm.

\begin{figure}[htb]
\centering
\includegraphics[width=12cm]{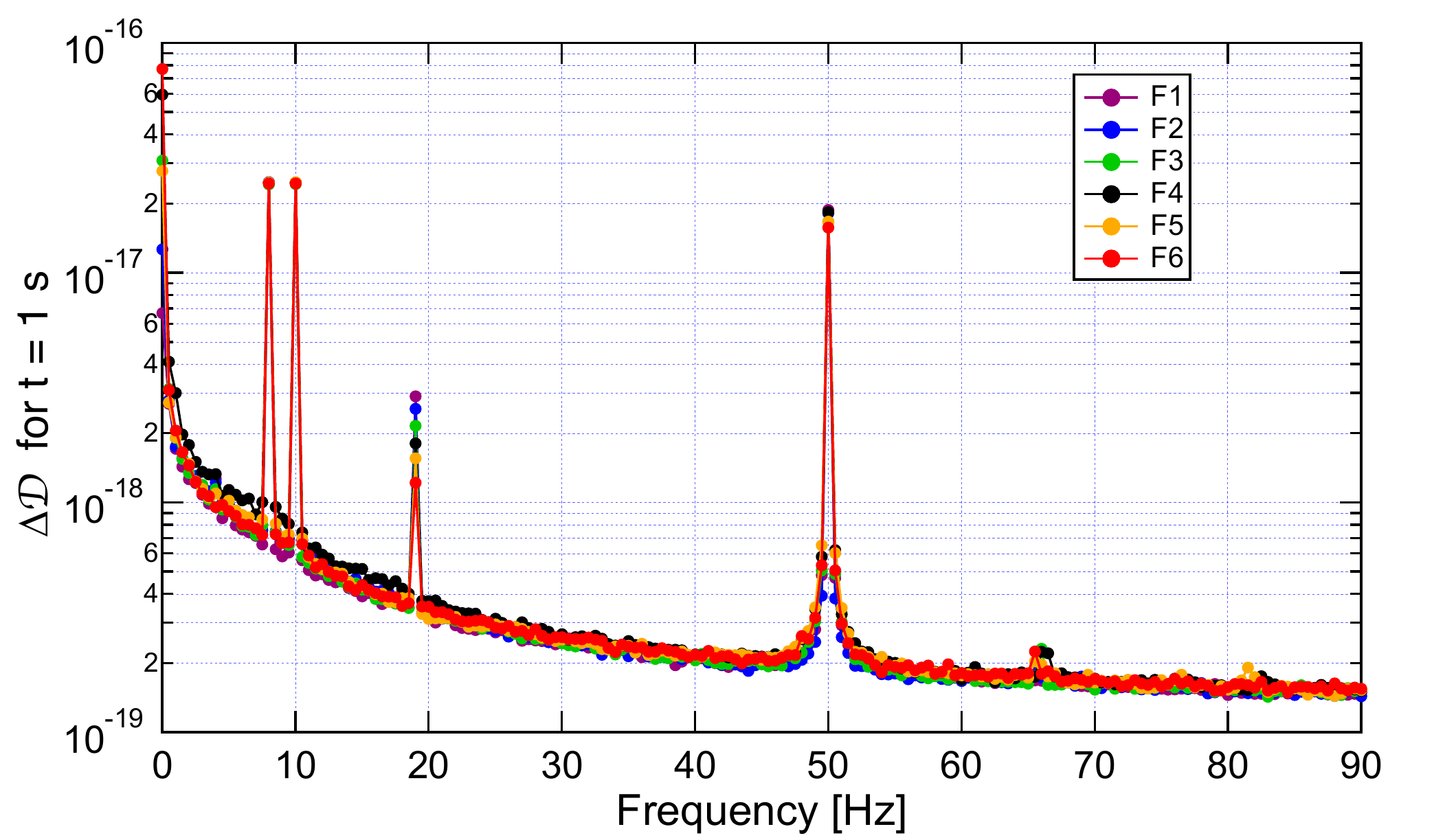}
\caption{Optical path difference spectra for the six finesse values determined by assuming a common ellipticity noise $S_\Psi$ proportional to $N$ and taking into account the frequency response of the cavity. The Cotton-Mouton signals at 8~Hz and 10~Hz are all superimposed as expected from equation~(\ref{eq:OPD_CM}) and the signal at 19~Hz correctly scales with $R_{\Psi',\Phi}$. The noise curves are all superimposed indicating a birefringence noise source originating from inside the cavity. From reference \cite{IntrinsicNoise}, Figure 14.}
\label{fig:intrinsic_noise}
\end{figure}

Assuming that the PVLAS-FE polarimeter was limited by a birefringence noise generated inside the cavity and that a possible noise induced by the feedback is proportional to $N$ [see (equation \ref{eq:s_feedback})], the ellipticity measurements $S_\Psi$, corrected for $k(\alpha_{\rm EQ})$ and for the cavity frequency response at each finesse value, were normalised according to 
\begin{equation}
    S_{\Delta\cal D} = \frac{\lambda}{N\pi}S_\Psi
\end{equation}
to determine the optical path difference sensitivities. In Figure~\ref{fig:intrinsic_noise} the measured optical path difference spectra for all six finesse values are shown normalised to a 1~second duration.
The two magnets were rotating at $\nu_\alpha = 4$~Hz and $\nu_\beta=5$~Hz resulting in two peaks at $2\nu_\alpha = 8$~Hz and $2\nu_\beta = 10$~Hz. The current in the Faraday cell had a frequency $\nu_F = 19$~Hz.

As expected the Cotton-Mouton peaks resulting from
\begin{equation}
    \Delta{\cal D}_{\rm CM} = \Delta n_uP\int{B_{\rm ext}^2\,dL}
\label{eq:OPD_CM}
\end{equation}
were independent of the finesse. Furthermore, the cross talk of the Faraday rotation into an ellipticity, as given by equation~(\ref{i_perp_ell}) due to the cavity birefringence, scaled with the factor $R_{\Psi',\Phi}$ which is proportional to $N$ [see also equation~(\ref{eq:verdet})]:
\begin{equation}
    \Delta{\cal D}^{\rm (spurious)}_{\rm F} = C_{\rm Ver}B_{\rm long}N\frac{\alpha_{\rm EQ}}{2}.
\end{equation}

Interestingly the noise $S_{\Delta{\cal D}}$ behaves just like the Cotton-Mouton signals indicating its nature as a birefringence noise originating from inside the cavity. Furthermore $S_{\Delta{\cal D}}$ determined from these measurements followed very closely the curve in Figure~\ref{fig:noise_global}. Taking for good that that curve 
is a relatively general behaviour of the cavity mirrors, in such experiments a maximum value of the finesse, depending on the limiting shot-noise, can be determined from
\begin{equation}
    {\cal F}_{\rm max} \approx \sqrt{\frac{e}{I_\parallel q}}\frac{\lambda}{2S_{\Delta{\cal D}}}.
    \label{eq:fmax}
\end{equation}
With 10~mW exiting the cavity, $\lambda = 1064$~nm and considering a value $S_{\Delta{\cal D}}\approx 3\times10^{-19}$~m/$\sqrt{\rm Hz}$ one finds ${\cal F}_{\rm max} \approx 8500$. Increasing the finesse of the cavity beyond ${\cal F}_{\rm max}$ will not improve the signal to noise ratio. 

By repeating the same procedure for the rotation measurements one finds that the behaviour of the Cotton-Mouton peaks and the Faraday peak were exchanged as expected. 
It is important to note, though, that the noise $S_{\Delta{\cal A}}$ scaled with the finesse, again indicating that its nature was an ellipticity seen as a rotation noise through the cross talk due to the birefringent cavity. A detailed report of these measurements can be found in Reference~\cite{IntrinsicNoise}. 

\subsubsection{Final discussion on wide band noise: thermal noise issues}
\label{sec:intrinsic_noise}

At present we have no explanation for the source of the observed birefringence noise which limited the sensitivity of the PVLAS-FE experiment, and not only. Note that the value of $S_{\Delta{\cal D}} \sim 10^{-(18\div19)}$~m/$\sqrt{\rm Hz}$ as an order of magnitude recalls thermal noise issues. Indeed 
intrinsic thermal fluctuations of the mirror coatings (not due to the laser power as demonstrated in Section~\ref{sec:power}) could generate stress fluctuations in the plane of the mirror. Through the stress-optical coefficient such stress noise would be translated to birefringence noise \cite{IntrinsicNoise}. In the following we will estimate two typical thermal noise contributions to the optical path difference noise: thermoelastic noise and Brownian noise. The conclusions will strongly support this hypothesis.

Following \cite{BRAGINSKY1999,Braginsky2003}, consider the characteristic diffusive heat transfer length $r_{\rm T}$ 
\begin{equation}
    r_{\rm T} = \sqrt{\frac{\lambda_{\rm T}}{\rho C_{\rm T}2\pi\nu}},
\end{equation}
with $\lambda_{\rm T}$ the heat conductivity, $\rho$ the density and $C_{\rm T}$ the specific heat capacity. The thermodynamic temperature fluctuations of each volume $V\approx r_{\rm T}^3$ are independent of one another and their variance is given by
\begin{equation}
    \sigma^2_{\rm T} = 
    \frac{k_{\rm B}T^2}{\rho C_{\rm T}V}
\end{equation}
where $k_{\rm B}$ is Boltzmann's constant.
The variance of the induced index of refraction fluctuations within such a volume $V$ due to stress variations can be written as
\begin{equation}
    \sigma^2_n = C_{SO}^2Y^2\frac{\sigma^2_{r_{\rm T}}}{r_{\rm T}^2}
\end{equation}
where $C_{SO}$ and $Y$ are respectively the material's stress optical coefficient and Young's modulus and the local relative length variations ${\sigma_{r_{\rm T}}}/{r_{\rm T}}$ will in turn depend on the thermal expansion coefficient $\alpha_{\rm T}$:  ${\sigma^2_{r_{\rm T}}}/{r^2_{\rm T}}= \alpha_{\rm T}^2\sigma^2_T$. Given that the volume $V$ is surrounded by adjacent independently fluctuating volumes, the induced stress along two perpendicular directions, $\parallel$ and $\perp$, will fluctuate independently leading to a variance of birefringence fluctuations. Therefore
\begin{equation}
    \sigma^2_{\Delta n} = C_{SO}^2Y^2\left[\left(\frac{\sigma^2_{r_{\rm T}}}{r_{\rm T}^2}\right)_\parallel+\left(\frac{\sigma^2_{r_{\rm T}}}{r_{\rm T}^2}\right)_\perp\right] = 2\sigma^2_{n} \approx 2C_{SO}^2Y^2\alpha_{\rm T}^2\frac{k_{\rm B}T^2}{\rho C_{\rm T}r_{\rm T}^3}
\end{equation}


This birefringence variance must now be averaged over the volume occupied by the beam's electric field. This is the region which will generate an ellipticity noise in the polarimeter. In the PVLAS cavity the beam had a radius $r_0 = 10^{-3}$~m. The mirrors had a transmission coefficient $T_{\rm PVLAS} = 2.4$~ppm and the number of high index of refraction - low index of refraction film pairs composing the coatings was $N_{\rm film} \approx 20$. Therefore the number of film pairs $\lambda_{\rm film}$ after which the field being reflected has reduced to $1/e$ is given by
\begin{equation}
    \frac{N_{\rm film}}{\lambda_{\rm film}} = -\ln{\sqrt{T_{\rm PVLAS}}}
\end{equation}
resulting in $\lambda_{\rm film} = 3$ corresponding to a geometrical thickness $d_{e}\approx 1~\mu$m.
Averaging over the beam's spot of radius $r_0$ one finds
\begin{equation}
    \sigma^2_{\Delta n} = C_{SO}^2Y^2\alpha_{\rm T}^2\langle\sigma^2_{\rm T}\rangle_{\rm spot} \approx 2C_{SO}^2Y^2\alpha_{\rm T}^2\frac{k_{\rm B}T^2}{\rho C_{\rm T}r_{\rm T}^3}\frac{r_{\rm T}^2}{r_0^2}
\end{equation}
where the ratio $\frac{r_{\rm T}^2}{r_0^2}$ represents the number of independent volumes $V\approx r_{\rm T}^3$ occupying the laser beam surface. This is justified because $d_e\ll r_{\rm T}\ll r_0$ and therefore the beam's electric field only sees a single layer of fluctuating volumes. Indeed for fused silica $\rho^{\rm (FS)} = 2200$~kg/m$^3$, $C_{\rm T}^{\rm (FS)} = 670$~J/(kg K) and $\lambda_{\rm T}^{\rm (FS)} = 1.4$~W/(K m) whereas for tantala (Ta$_2$O$_5$) $\rho^{\rm (Ta)} = 8200$~kg/m$^3$, $C_{\rm T}^{\rm (Ta)} = 300$~J/(kg K) and $\lambda_{\rm T}^{\rm (Ta)} = 0.026 - 15$~W/(K m) (for a film) \cite{Welsch1999}. For the PVLAS cavity $r_0 = 10^{-3}$~m and $d_e\sim1~\mu$m and the narrowest frequency range for which the above condition is satisfied is between $1~{\rm Hz}\div 1.5$~kHz.

The optical path difference fluctuations accumulated by the laser beam upon reflection will therefore be 
\begin{equation}
    \sigma_{\Delta{\cal D}} = 2d_{e}
    \sigma_{\Delta n} = 2d_{e}\sqrt{2}C_{SO}Y\alpha_{\rm T}\sigma_{\rm T} \approx 2d_{e}\sqrt{2}C_{SO}Y\alpha_{\rm T}\sqrt{\frac{k_{\rm B}T^2}{\rho C_{\rm T}r_{\rm T} r_0^2}}. 
\end{equation}
 A more rigorous averaging taking into account the Gaussian profile of the beam and exponential penetration \cite{BRAGINSKY1999,Braginsky2003,Braginsky:2000wc} leads to a temperature spectral density 
\begin{equation}
    S_T = \sqrt{\frac{\sqrt{2}k_{\rm B}T^2}{\pi\rho C_{\rm T}r_{\rm T} r_0^22\pi\nu}} = \sqrt{\frac{\sqrt{2}k_{\rm B}T^2}{\pi r_0^2\sqrt{\rho C_T \lambda_T 2\pi\nu}}}
\end{equation}
and an optical path difference spectral density
\begin{equation}
    S_{\Delta{\cal D}} = 2d_{e}\sqrt{2}C_{SO}Y\alpha_{\rm T} S_T = 
    d_{e}C_{SO}Y\alpha_{\rm T}\sqrt{\frac{8k_{\rm B}T^2}{\pi r_0^2\sqrt{\pi\rho C_T \lambda_T\nu}}}\propto \nu^{-1/4}. 
    \label{eq:th_e}
\end{equation}
Considering the values reported above for the various parameters of fused silica and tantala and using $C_{\rm SO}^{\rm (FS)} = 3\times10^{-12}$~Pa$^{-1}$, $Y^{\rm (FS)} = 70$~GPa, $\alpha_{\rm T}^{\rm (FS)} = 5\times10^{-7}$~K$^{-1}$, $C_{\rm SO}^{\rm (Ta)} \approx 3\times10^{-12}$~Pa$^{-1}$, $Y^{\rm (Ta)} = 150$~GPa and $\alpha_{\rm T}^{\rm (Ta)} = 8\times10^{-6}$~K$^{-1}$ one finds
\begin{equation}
    S_{\Delta{\cal D}}^{\rm (FS)} \sim 4\times10^{-21}~{\rm m}/\sqrt{\rm Hz}\quad @ \quad 1~{\rm Hz}
\end{equation}
whereas for tantala
\begin{equation}
    S_{\Delta{\cal D}}^{\rm (Ta)} \sim (1\div5)\times10^{-19}~{\rm m}/\sqrt{\rm Hz}\quad @\quad 1~{\rm Hz}.
    \label{eq:tantala_noise}
\end{equation}
Not having found a specific value, for tantala we have used the value for fused silica. Generally it is found in literature that $C_{SO}\sim 10^{-12\div-11}$~Pa$^{-1}$ with a particularly large value of $C_{SO} = 95 \times 10^{-12}$~Pa$^{-1}$ for Nb$_2$O$_5$ \cite{Tei-Chen2007}.

\begin{figure}[htb]
\centering
\includegraphics[width=12cm]{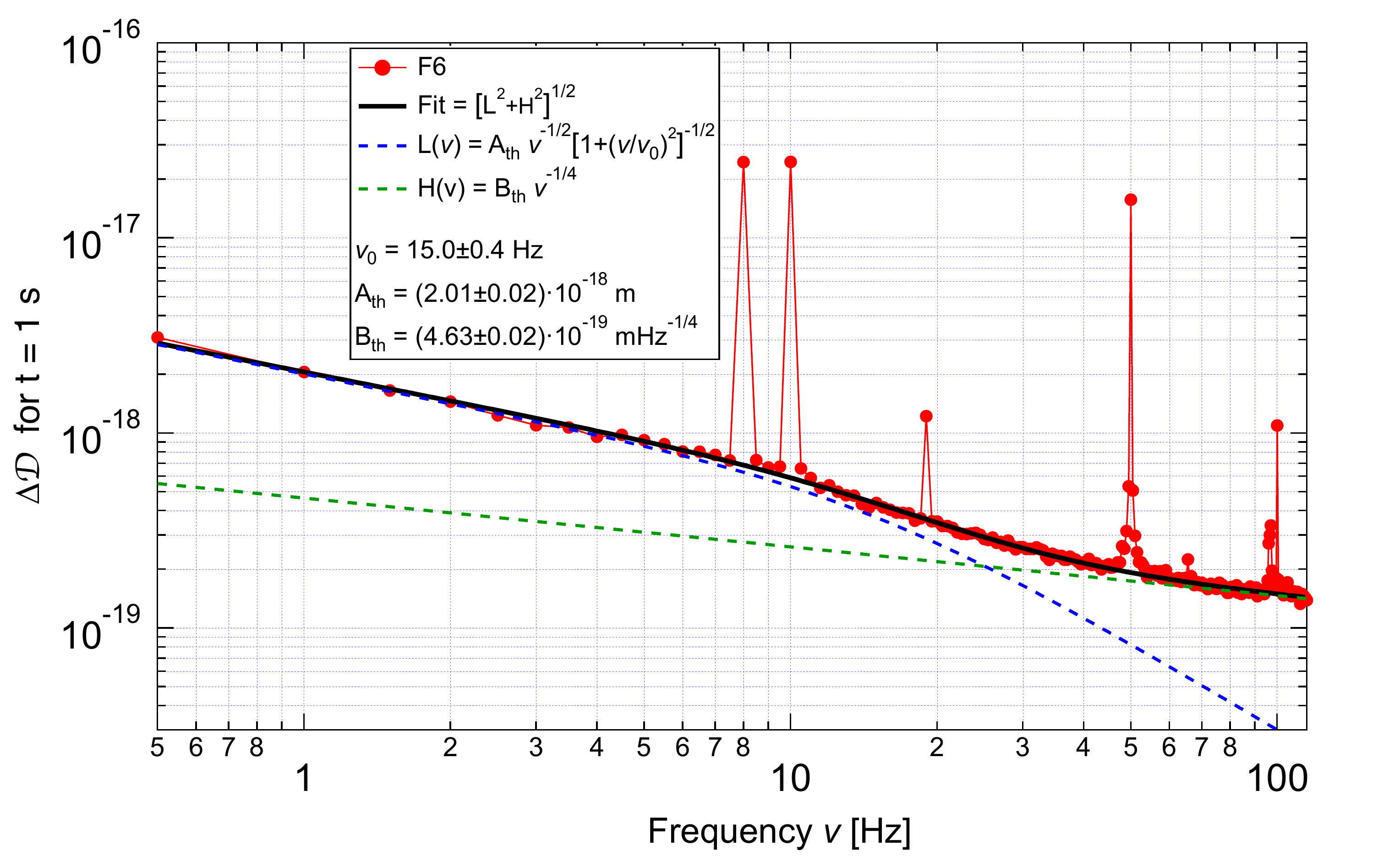}
\caption{Optical path difference spectrum for F6 = 688000 taken from Figure~\ref{fig:intrinsic_noise}. Superimposed is the fit obtained using equation~(\ref{eq:fit_noise}). With a 3.5\% uncertainty on the experimental data the reduced $\chi^2 = 149/160$. 
Superimposed on the graph as dashed lines are the two curves describing the fit. 
}
\label{fig:noise_fit}
\end{figure}

Finally considering equation (\ref{eq:th_e}) and  that Brownian noise typically scales as $\nu^{-1/2}$ the optical difference noise spectrum for F6 reported in Figure~\ref{fig:intrinsic_noise} was fitted with the function
\begin{equation}
    f(\nu) = \sqrt{\left(\frac{A_{\rm th}\nu^{-1/2}}{\sqrt{1+(\nu/\nu_0)^2}}\right)^2+\left(B_{\rm th}\nu^{-1/4}\right)^2}
    \label{eq:fit_noise}
\end{equation}
and is shown in Figure~\ref{fig:noise_fit}. The structure of this function was dictated by the low and high frequency slopes in the log-log curve and by the transition between these two power laws. An (unreported) initial fit  resulted in the powers of the two slopes and in the order of the filter in equation (\ref{eq:fit_noise}). 
With a 3.5\% error for the experimental data the reduced chi-squared was $\chi^2 = 149/160$. The fitting procedure resulted in $A_{\rm th} = (2.01\pm0.02)\times10^{-18}$~m, $\nu_0 = (15.0\pm0.4)$~Hz and $B_{\rm th} = (4.63\pm0.02)\times10^{-19}$~m/${\rm Hz}^{1/4}$. Introducing in the fit function (\ref{eq:fit_noise}) an additive constant does not improve the fit and gives a value compatible with zero. The value of $B_{\rm th}$ is in reasonable agreement with equation~(\ref{eq:tantala_noise}) for tantala. Although PVLAS is sensitive to optical path length differences between two perpendicular polarisations, it is also interesting to note that the parameter $A_{\rm th}$ is in good agreement with the surface displacement spectral density due to Brownian noise for fused silica \cite{Braginsky2003}:
\begin{equation}
    S_{\rm BN}^{\rm (FS)} \approx \sqrt{\frac{4k_{\rm B }T}{2\pi\nu}\frac{\phi_{\rm FS}}{\sqrt{2\pi}Yr_0}}
\end{equation}
provided a loss angle $\phi_{\rm FS}\approx 2.7\times 10^{-7}$ rad which is not an unreasonable value. At present we have no justification for the cut-off.

These considerations have led us to think that the PVLAS polarimeter was indeed limited by intrinsic thermal noise issues. We believe the same is true for any Fabry-Perot based polarimeter.



An attempt was made to cool the mirrors radiatively towards liquid nitrogen temperatures \cite{LNL2016} to verify this hypothesis but the experiment proved to be much more difficult than expected and was abandoned. Furthermore, at best the noise could have decreased inversely with the temperature, insufficient to close the gap to reach the QED effect.

\subsection{Conclusions on the commissioning: lessons learned}

In this section we have discussed `in-phase' and wide band noise issues and how the careful debugging for the `in-phase' systematics allowed integration for $5\times10^6$~s without observing systematic peaks at the signal frequency $2\nu_{\rm B}$. Let us describe the procedure we followed in the presence of an unexpected signal at twice the rotation frequency of the magnets to distinguish between a systematic and a physical birefringence signal. First of all, to qualify as a magnetic birefringence effect generated by the rotating magnets, the signal had to occupy a single bin in a Fourier spectrum of the ellipticity at a sub-microhertz frequency resolution just as the stray magnetic field shown in Figure \ref{fig:fft_mag} on page \pageref{fig:fft_mag}. Moreover, the amplitude of the peak had to be the same for the two magnets rotating at two different frequencies, whereas the phase had to coincide with the phase measured during the calibration (possibly apart from the sign). These requirements ruled out the spurious effects which we encountered. 

The systematic peaks of the PVLAS-FE experiment were proven to be under control down to the level of the integrated noise floor of the experiment (see next Section~\ref{sec:FE_results}). A longer integration might have let more systematics emerge from noise, requiring more debugging. (For this reason any integration time longer than $\sim10^6$~s is likely to make debugging impossible.) For the sake of discussion, let us treat the hypothetical case of the presence of systematic effects at the level of the QED signal. In section \ref{sec:stray field} we showed that systematic peaks at the rotation frequency $\nu_B$ of the magnets were always present in the ellipticity spectra. These peaks were due to the Faraday effects generated in the reflecting coating of the mirrors by the axial component of the magnetic stray fields of the magnets at the positions of the beam spots on the mirrors. The resulting polarisation rotation is then transformed into ellipticity through the cavity birefringence. We never observed such rotations at the second harmonics of the rotation frequencies of the magnets, but they could emerge lowering the noise floor due to small imperfections in the geometry of the magnets. In this case, the ellipticity signal of each magnet would be composed of the QED signal and of a spurious Faraday vector, and one would have only two equations with three unknowns. For this eventuality two complementary approaches could be envisaged. First of all, the magnetic fields at the positions of the mirrors should be accurately mapped. The axial field component at the position of the beam spot could then be canceled by means of two additional coils either in feedback or using the field map. A second possibility to disentagle the three contributions is to add a rotation modulator to the optical setup. In fact, the cavity birefringence transforms ellipticities into rotations (and {\it vice versa}) in a known way. This would then provide two more equations making the algebraic system solvable.

We have also discussed that the wide band noise seems to be compatible with thermal noise issues in the mirrors of the Fabry-Perot which generate an optical path difference noise $S_{\Delta{\cal D}}$ well described by the function $f(\nu)$ in equation~(\ref{eq:fit_noise}). Given that vacuum magnetic birefringence generates a difference in optical path length, finesse values above ${\cal F}_{\rm max}$ in equation~(\ref{eq:fmax}) will not improve the signal to noise ratio of the polarimeter. Given equation~(\ref{eq:fit_noise}) the necessary integration time to reach a SNR = 1 at a signal frequency $\nu$ is therefore
\begin{equation}
    {\cal T}(\nu) = \left[\frac{f(\nu)}{3A_eB_{\rm ext}^2L_{B}}\right]^2.
\end{equation}
independently of the finesse. Even assuming to double the highest employed rotation frequency of the magnets in the current setup (a non trivial mechanical project), at $\nu\approx30$~Hz one has ${\cal T}\approx40$~Ms, more than one year of continuous acquisition.

\section{Measurements of the vacuum magnetic birefringence and dichroism}
\label{sec:FE_results}

In this section we present the polarimetric measurements performed on vacuum during the years 2014-2016 in the attempt to test its magneto-optical properties. The results of this activity represent the best limits on the magnetic birefringence and dichroism of vacuum at low energy stemming from laboratory measurements. The more recent measurements are presented here for the first time in some detail, whereas the previous ones, that have already been published elsewhere \cite{DellaValle2014PRD,DellaValle2016EPJC}, will be only summarised.

Note that, besides their direct measurements, from ellipticity (rotation) measurements rotation (ellipticity) information can also be retrieved through equation~(\ref{i_perp_ell}) [or (\ref{i_perp_rot})]. According to these equations, one can interpret the measured ellipticity as due to a dichroism and {\em vice versa}:
\begin{equation}
\Delta\kappa'=\frac{2}{N\alpha_{\rm EQ}}\Delta n=\frac{\Delta n}{R_{\Phi',\Psi}} \qquad\qquad
\Delta n'=-\frac{2}{N\alpha_{\rm EQ}}\Delta\kappa=\frac{\Delta\kappa}{R_{\Psi',\Phi}}.
\label{eq:DichroismFromEllipticity}
\end{equation}
Moreover, both the birefringence and dichroism limits presented here will be also interpreted as limits on the existence of ALPS and millicharged particles.

\begin{figure}[hbt]
\begin{center}
\includegraphics[width=8cm]{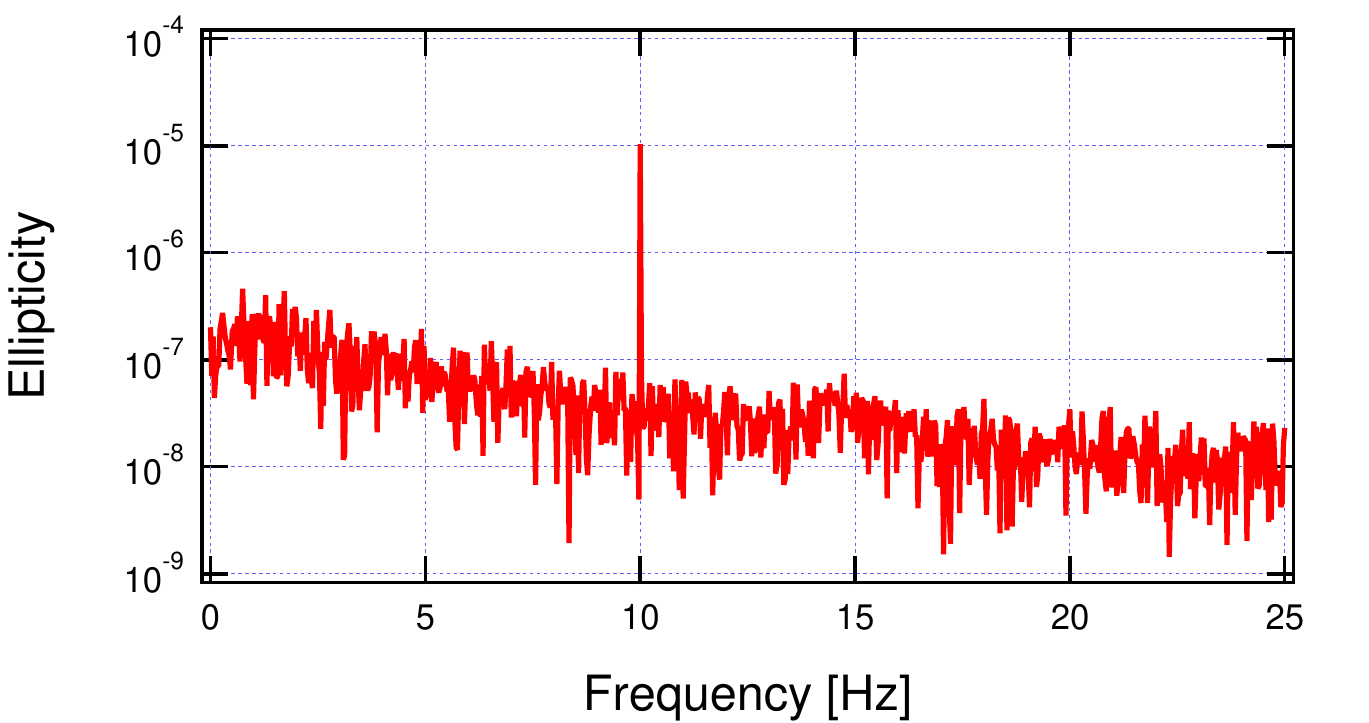}
\includegraphics[width=8cm]{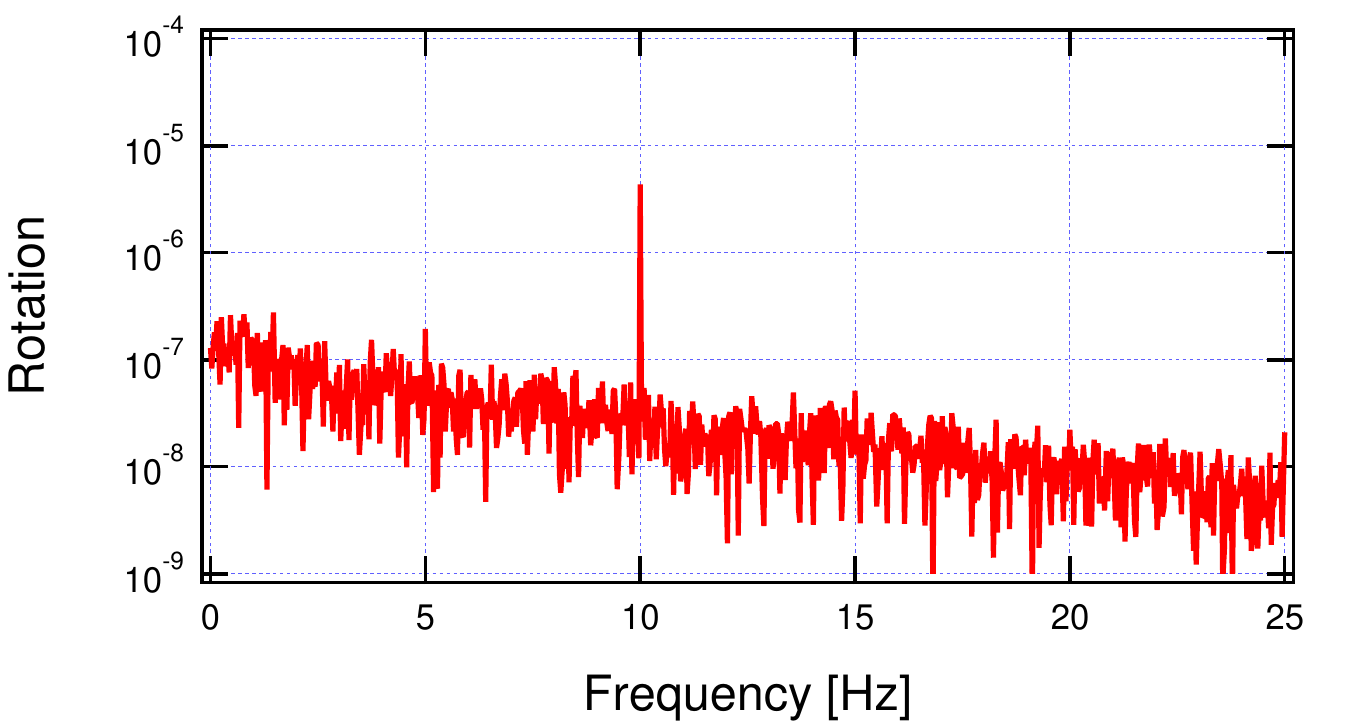}
\end{center}
\caption{Cotton-Mouton effect measurements for 228~$\mu$bar of Ar gas: Fourier spectra of the extinguished power demodulated at the modulator frequency $\nu_m$. A single magnet was rotating at $\nu_B=5$~Hz. Left: ellipticity spectrum. Right: rotation spectrum. Integration time was ${\cal T}=128$~s for both spectra.}
\label{fig:CM_spectra2016}
\end{figure}

Data taking started in 2014 in the absence of systematics as soon as Viton o-rings were in place inside the glass tubes. In the course of its three-years activity, the apparatus was calibrated more than once per run and the optical setup was maintained optimised. All the runs performed between two such optimisations were considered to be in the same experimental conditions and with the same characteristics; the runs can be grouped and labelled with the solar year. As an example of a Cotton-Mouton calibration of the polarimeter described in Section~\ref{sec:calibration-CM}, Figure~\ref{fig:CM_spectra2016} shows one of the 2016 calibrations of the polarimeter performed using 228~$\mu$bar of Ar gas. The ratio of the amplitudes of the rotation to the ellipticity [see equation~(\ref{R_Phi_Psi})] gives $\alpha_{\rm EQ}=1.9~\mu$rad, with an attenuation factor $k(\alpha_{\rm EQ})=0.85$. The corresponding frequency distance between the two Airy curves of the two orthogonal polarisation states was 14~Hz. From these data a value for the unitary magnetic birefringence of Ar gas at room temperature was extracted resulting in  
\begin{equation}
\Delta n_u({\rm Ar})=(7.6\pm0.5)\times10^{-15}~{\rm T}^{-2}~{\rm atm}^{-1}
\end{equation}
in agreement with the literature (see Table~\ref{tab:ListCM} on page~\pageref{tab:ListCM}).

\begin{sidewaystable}
\begin{center}
\begin{tabular}{l|r@{,~}l|c|r|r|r|c|c}
\hline\hline
Run & \multicolumn{2}{c|}{$\nu_{\alpha},\nu_{\beta}$~(Hz)} & Quantity & \multicolumn{1}{c|}{${\cal T}$~(ks)} & \multicolumn{1}{c|}{In-phase} & \multicolumn{1}{c|}{Quadrature} & Noise floor $\sigma$ & $S_{2\nu_B}~\left(1/\sqrt{\rm{Hz}}\right)$ \\
\hline\hline
\multicolumn{9}{c}{Year 2014, $I_{\rm out}\approx70$~mW, $I_{\rm FP}\approx30$~kW, ${\cal F}=670\,000$, $\alpha_{\rm EQ}=4.5~\mu$rad, $\Delta\nu_{\alpha_{\rm EQ}}=33$~Hz, $k(\alpha_{\rm EQ})=0.5$ \cite{DellaValle2014PRD}} \\
\hline
1          & 3.0 & 3.0  & $\Delta n$      & 370  & $+1.8\times10^{-21}$ & $-0.05\times10^{-21}$ & $1.8\times10^{-21}$ & $1.1\times10^{-18}$ \\
2          & 2.5 & 2.5  & $\Delta n$      & 239  & $-0.6\times10^{-21}$ & $-2.0\times10^{-21}$  & $1.9\times10^{-21}$ & $1.0\times10^{-18}$ \\
$3\alpha$  & 3.0 & --   & $\Delta n$      & 143  & $-4.6\times10^{-21}$ & $+6.2\times10^{-21}$  & $4.6\times10^{-21}$ & $1.7\times10^{-18}$ \\
$3\beta$   & --~ & 2.4  & $\Delta n$      & 143  & $+0.1\times10^{-21}$ & $+11\times10^{-21}$   & $4.9\times10^{-21}$ & $1.9\times10^{-18}$ \\
\hline
1'         & 3.0 & 3.0  & $\Delta\kappa'$ &      & $+1.8\times10^{-21}$ &                       & $1.8\times10^{-21}$ & $1.1\times10^{-18}$ \\
2'         & 2.5 & 2.5  & $\Delta\kappa'$ &      & $-0.6\times10^{-21}$ &                       & $1.9\times10^{-21}$ & $1.0\times10^{-18}$ \\
$3\alpha'$ & 3.0 & --   & $\Delta\kappa'$ &      & $-4.6\times10^{-21}$ &                       & $4.6\times10^{-21}$ & $1.7\times10^{-18}$ \\
$3\beta'$  & --~ & 2.4  & $\Delta\kappa'$ &      & $+0.1\times10^{-21}$ &                       & $4.9\times10^{-21}$ & $1.9\times10^{-18}$ \\
\hline\hline
\multicolumn{9}{c}{Year 2015, $I_{\rm out}\approx9$~mW, $I_{\rm FP}\approx3.7$~kW, ${\cal F}=700\,000$, $\alpha_{\rm EQ}=3.3~\mu$rad, $\Delta\nu_{\alpha_{\rm EQ}}=24$~Hz, $k(\alpha_{\rm EQ})=0.65$ \cite{DellaValle2016EPJC}} \\
\hline
$4\alpha$  & 4.0 & --   & $\Delta n$      & 1000 & systematic peak      &                       &                     &                     \\
$4\beta$   & --~ & 5.0  & $\Delta n$      & 1000 & $-0.6\times10^{-22}$ & $+2.4\times10^{-22}$  & $4.5\times10^{-22}$ & $4.5\times10^{-19}$ \\
$5\alpha$  & 5.0 & --   & $\Delta n$      & 890  & $-3.8\times10^{-22}$ & $+9.3\times10^{-22}$  & $5.0\times10^{-22}$ & $4.7\times10^{-19}$ \\
$5\beta$   & --~ & 6.25 & $\Delta n$      & 890  &    systematic peak   &                       &                     &                     \\
6          & 5.0 & 5.0  & $|\Delta\kappa|$  & 700  & $-0.3\times10^{-22}$ & $-8.8\times10^{-22}$  & $6.0\times10^{-22}$ & $2.2\times10^{-19}$ \\
\hline
$4\beta'$  & --~ & 5.0  & $\Delta\kappa'$ &      & $-0.9\times10^{-22}$ &                       & $6.2\times10^{-22}$ & $6.2\times10^{-19}$ \\
$5\alpha'$ & 5.0 & --   & $\Delta\kappa'$ &      & $-5.2\times10^{-22}$ &                       & $6.8\times10^{-22}$ & $6.4\times10^{-19}$ \\
6'         & 5.0 & 5.0  & $\Delta n'$     &      & $+0.4\times10^{-22}$ &                       & $8.2\times10^{-22}$ & $3.0\times10^{-19}$ \\
\hline\hline
\multicolumn{9}{c}{Year 2016, $I_{\rm out}\approx2.5$~mW, $I_{\rm FP}\approx1$~kW, ${\cal F}=700\,000$, $\alpha_{\rm EQ}=1.9~\mu$rad, $\Delta\nu_{\alpha_{\rm EQ}}=14$~Hz, $k(\alpha_{\rm EQ})=0.85$} \\
\hline
$7\alpha$  & 8.0 & --   & $\Delta n$      & 1600 & $+6.0\times10^{-22}$ & $+2.1\times10^{-22}$  & $3.0\times10^{-22}$ & $3.8\times10^{-19}$ \\
$7\beta$   & --~ & 8.5  & $\Delta n$      & 1600 & $-0.7\times10^{-22}$ & $-5.4\times10^{-22}$  & $2.8\times10^{-22}$ & $3.7\times10^{-19}$ \\
\hline
$7\alpha'$ & 8.0 & --   & $\Delta\kappa'$ &      & $+14\times10^{-22}$  &                       & $7.1\times10^{-22}$ & $8.8\times10^{-19}$ \\
$7\beta'$  & --~ & 8.5  & $\Delta\kappa'$ &      & $-1.7\times10^{-22}$ &                       & $6.7\times10^{-22}$ & $8.6\times10^{-19}$ \\
\hline\hline
\end{tabular}
\caption{List of all the PVLAS-FE runs in vacuum. The values of the magnetic birefringence and dichroism are already corrected for the factor $k(\alpha_{\rm EQ})$ and the frequency response of the cavity. The primed values are obtained through the use of equations~(\ref{eq:DichroismFromEllipticity}). Only the absolute value of the dichroism in run 6 is given because its sign relative to the values $\Delta\kappa$' is unknown. Note that the data integration time ${\cal T}$ differs from the run time in that the data from each magnet when $\nu_\alpha\ne\nu_\beta$ are independent. The noise floor $\sigma$ represents the standard deviation of the integrated noise, obtained by fitting the Rayleigh distributions for each run (see Section~\ref{DataAnalysis}). The $S_{2\nu_B}$ column reports the average sensitivity for each run. The header of each section of the table lists the parameters of the run. The power circulating in the cavity is calculated as $I_{\rm FP}=I_{\rm out}/T$.}
\label{tab:GlobalRun}
\end{center}
\end{sidewaystable}

Table~\ref{tab:GlobalRun} lists all the runs performed. In the table, the `parasite' runs 1', 2' and 3' are presented here for the first time. The 2015 runs $4\alpha$ and $5\beta$ were since they presented structures at $2\nu_B$ occupying several bins due to an incorrectly centered vacuum tube. They are shown in Figure~11 of Reference~\cite{DellaValle2016EPJC}

The automatic locking circuit was realised at the beginning of 2015, greatly improving the duty cycle of the measurements. In turn, this resulted in an improved overall stability of the polarimeter limiting thermal drifts. In the course of this year, the acquisition rate passed from 32/40~samples/turn to 16/20~samples/turn.

\begin{figure}[bht]
\begin{center}
\includegraphics[width=7.5cm]{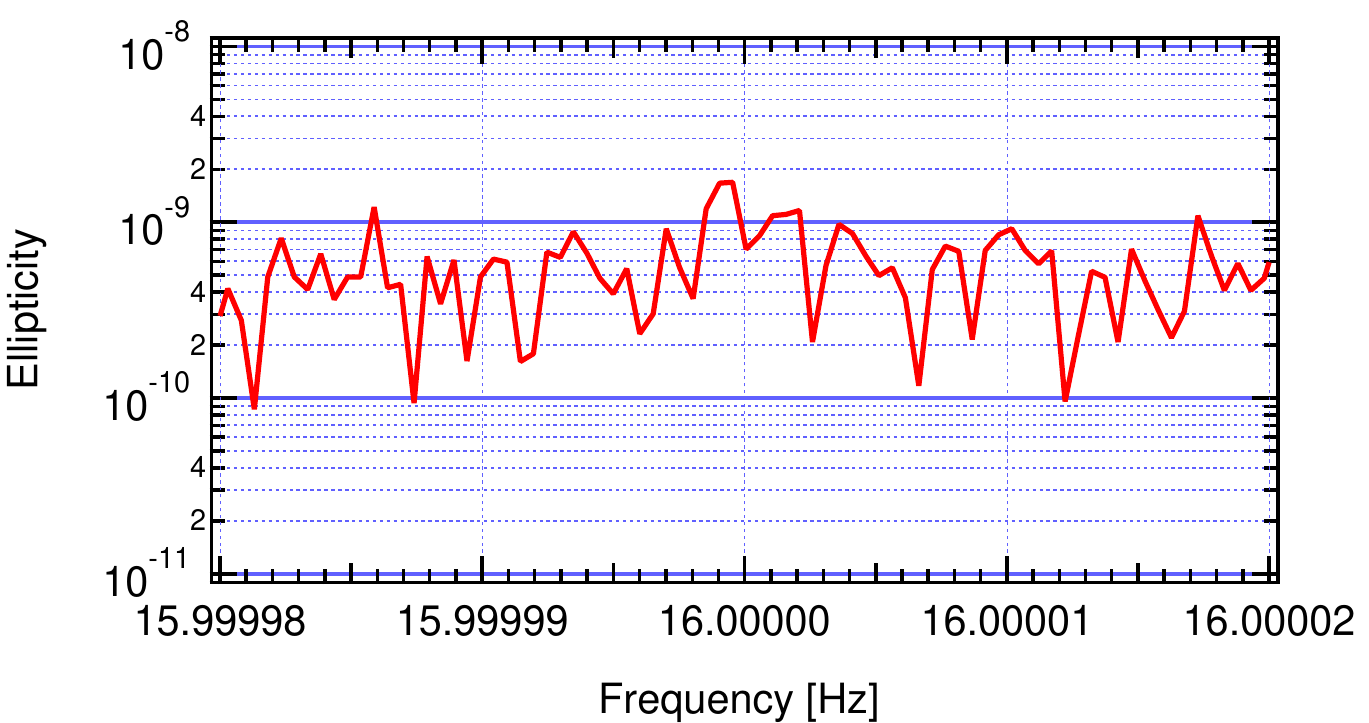}
\includegraphics[width=7.5cm]{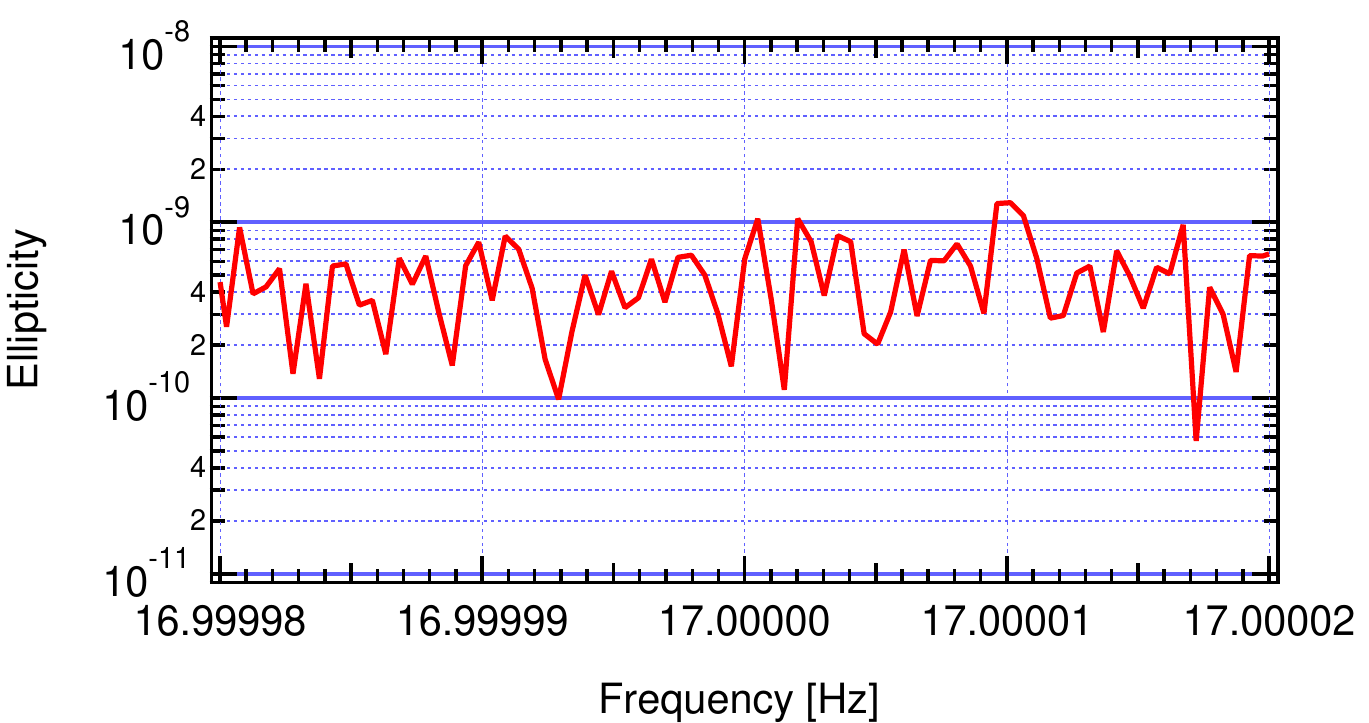}
\includegraphics[width=7.5cm]{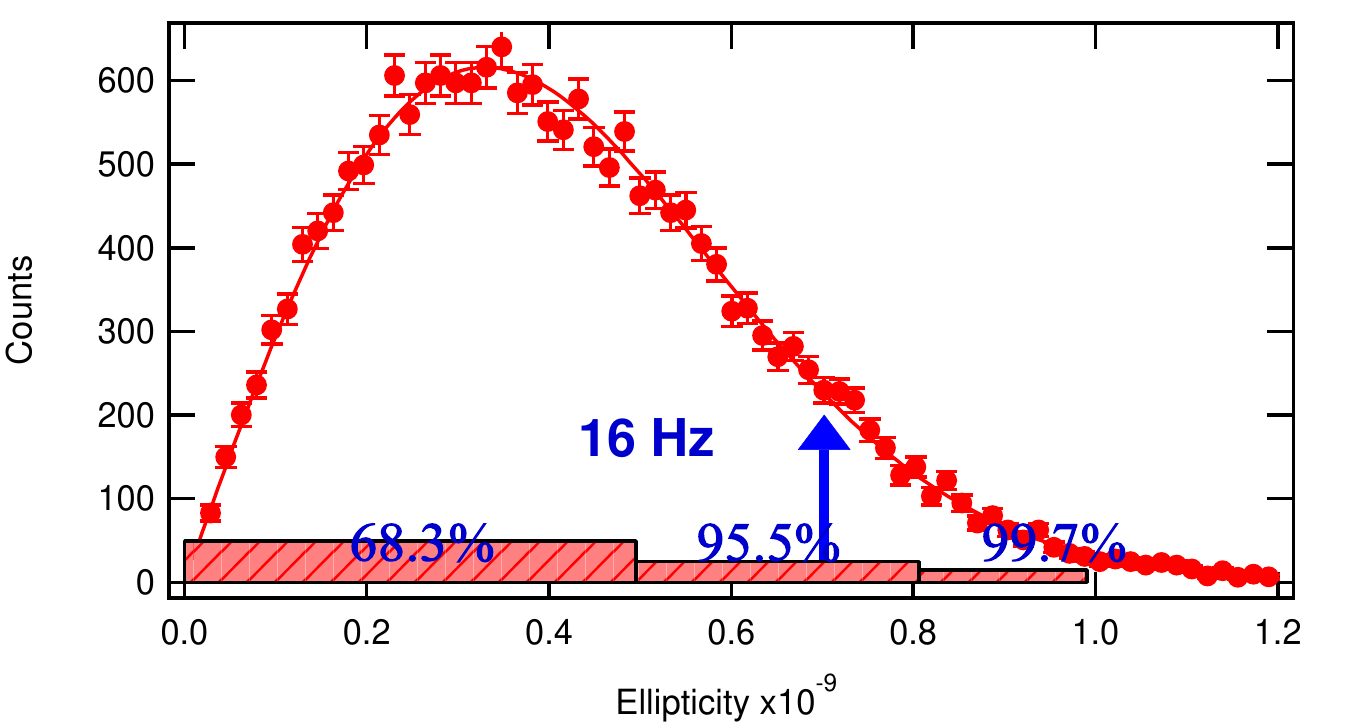}
\includegraphics[width=7.5cm]{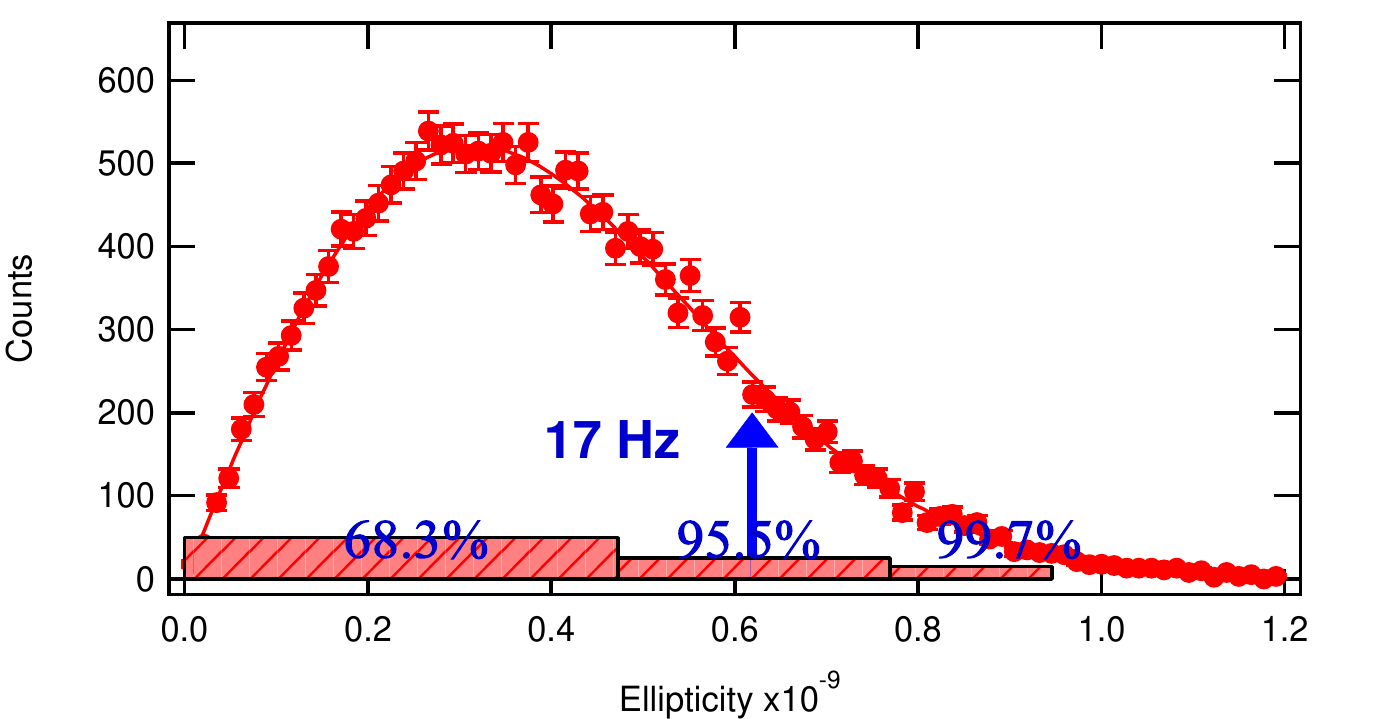}
\includegraphics[width=7.5cm]{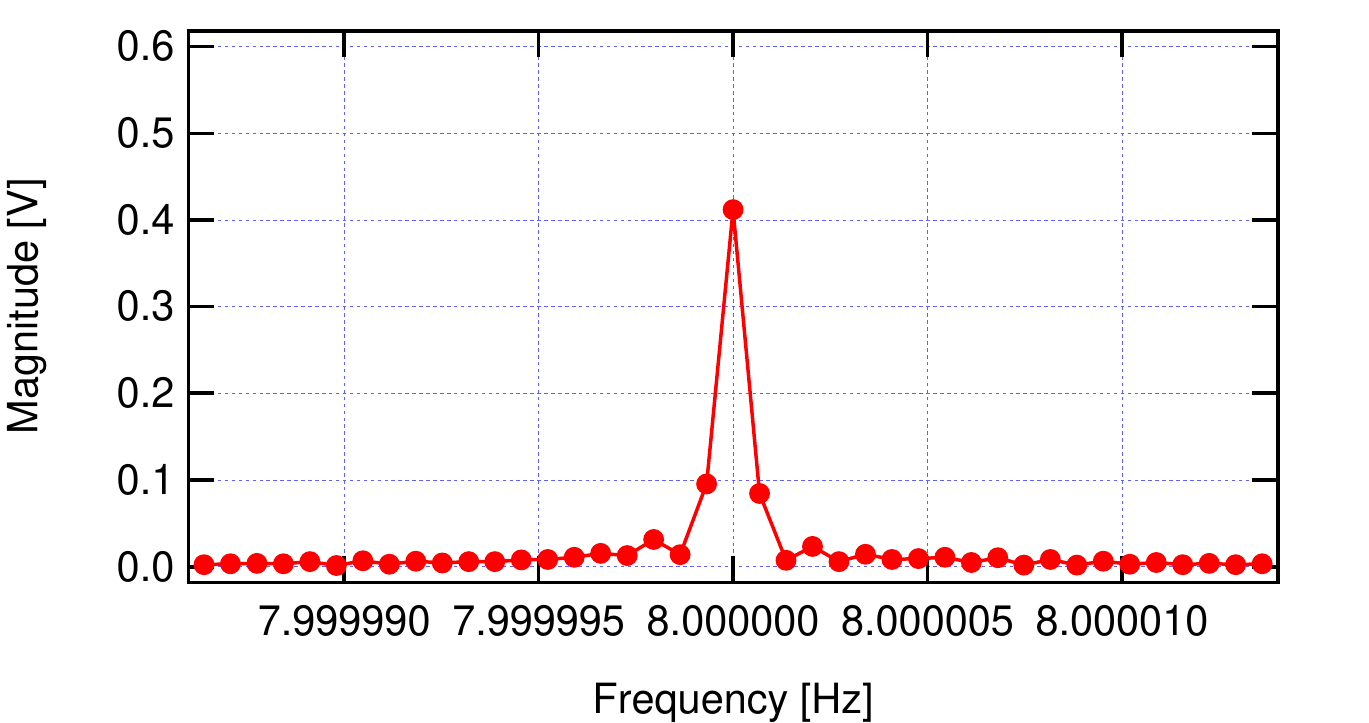}
\includegraphics[width=7.5cm]{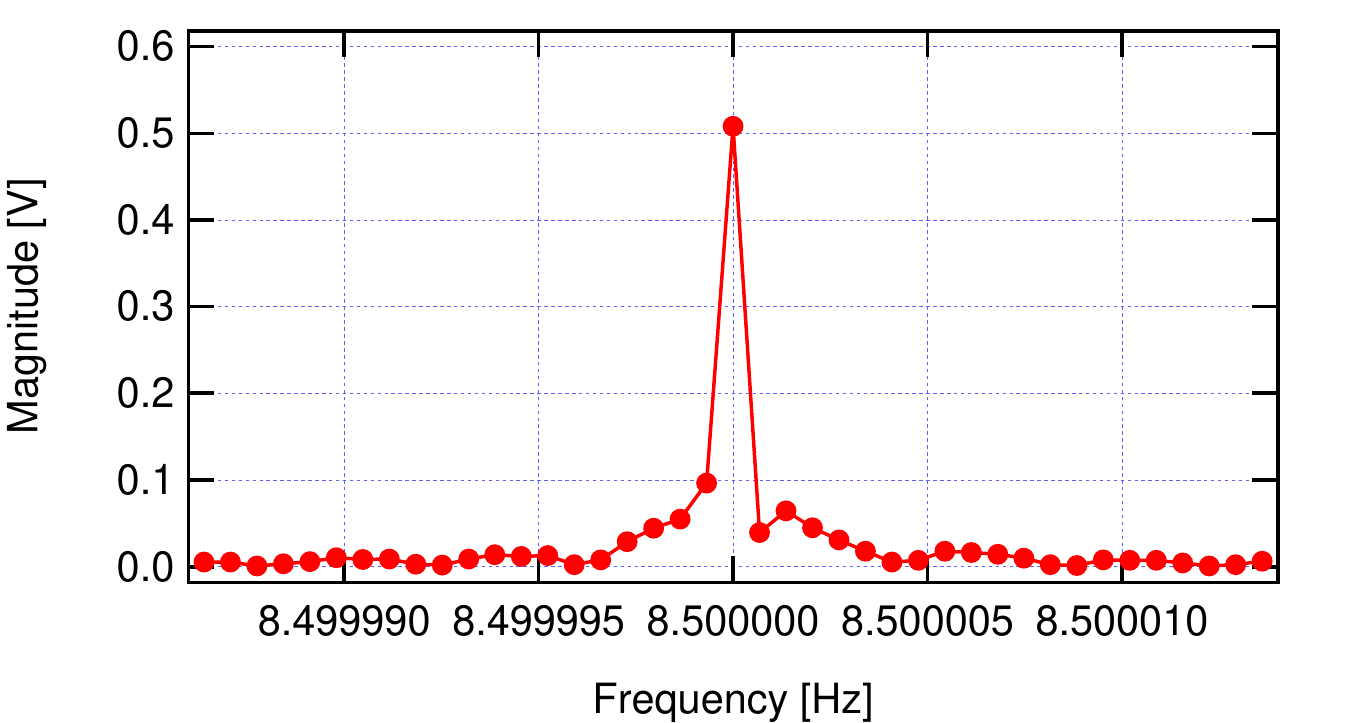}
\end{center}
\caption{2016 ellipticity data runs: Top row: unprojected Fourier transform of the ellipticity signal in a narrow interval around $2\nu_B$. Middle row: histograms of the data fitted with a Rayleigh distribution. The vertical arrows indicate the values of the bin at $2\nu_B$; the strips at the bottom of the plots mark integrated probabilities. Bottom row: Fourier spectra of the magnetic field of the two rotating magnets.}
\label{fig:Run2016}
\end{figure}

In 2016 a few upgrades of the apparatus were made. The glass vacuum tubes were replaced with more rigid ceramic (silicon nitride) ones. In these tubes no baffle could be inserted to block the diffused light, but the intrinsic roughness of their inner wall made this point less crucial. In an attempt to improve the signal to noise ratio, the rotation frequency of the magnets was increased up to 23~Hz. However, the higher the rotation frequency, the worse were the vibrations of the structures supporting the magnets (see Figure~\ref{fig:acc-freq} on page \pageref{fig:acc-freq}). As a compromise, the rotation frequencies were chosen as $\nu_{\alpha}=8$~Hz and $\nu_{\beta}=8.5$~Hz. 
Also during this year, a number of data blocks featuring an excess of wideband noise was discarded, reducing the useful integration time from 2.0~Ms to 1.6~Ms. 

In Figure~\ref{fig:Run2016}, top and middle rows, the FFT of the ellipticity data of the 2016 runs and the relative noise histograms are shown, respectively. The bottom row shows the Fourier transforms of the stray magnetic field of the two rotating magnets. These refer to the whole data set lasting $2\times 10^6$~s. As can be seen the signal occupies a single bin. An ellipticity signal, if present, must appear in a single bin of the Fourier spectrum, just as the magnetic field. The slight pedestal is due to the finite resolution of the two signal generators, resulting in a very slow relative phase drift requiring small phase adjustments during the run. During long acquisition runs the relative angular phase of the magnets was adjusted every few days.

\section{Vacuum measurement results and time evolution}
\label{FinalResults}

\subsection{\bf Limits on vacuum magnetic birefringence and dichroism}

The results listed in Table~\ref{tab:GlobalRun} can be averaged to give the final limits on vacuum magnetic birefringence and dichroism of the PVLAS-FE experiment, for a total run time of $\approx5\times10^6$~s:
\begin{eqnarray}
\label{BirefringencePVLAS}
\Delta n^{\rm{ (PVLAS-FE)}}&=&(12\pm17)\times10^{-23}\qquad @~B=2.5{\rm{~T}}\\
|\Delta\kappa|^{\rm{ (PVLAS-FE)}}&=&(10\pm28)\times10^{-23}\qquad @~B=2.5{\rm{~T}}.
\label{DichroismPVLAS}
\end{eqnarray}
These values represent the current best limits on these quantities obtained by optical means. The value for the dichroism is reported as an absolute value because its sign, depending on the sign of $\alpha_{\rm EQ}$ in equation~(\ref{R_Phi_Psi}), was never determined but was common to all measurements. Therefore the relative signs of all the  $\Delta\kappa$' values are consistent. The value of $\Delta\kappa$ in run 6 of 2015 with the QWP inserted is reported in Table~\ref{tab:GlobalRun} with an absolute value because its sign relative to all the $\Delta\kappa$' values was unknown. In calculating $|\Delta\kappa|^{\rm (PVLAS-FE)}$ the sign of $\Delta\kappa$ resulting in the larger central value was used.

We note that the vacuum magnetic birefringence predicted by the Euler and Kockel Lagrangian is $\Delta n^{\rm{(EK)}}=2.5\times10^{-23}$ at $B_{\rm ext}=2.5{\rm{~T}}$, i.e. the integrated noise level of the PVLAS-FE measurement is a factor seven larger than the predicted effect. The practical impossibility to beat the noise in the actual scheme was a show stopper and spurred for new ideas \cite{Zavattini:2016sqz}.

\begin{figure}[hbt]
\begin{center}
\includegraphics[width=14cm]{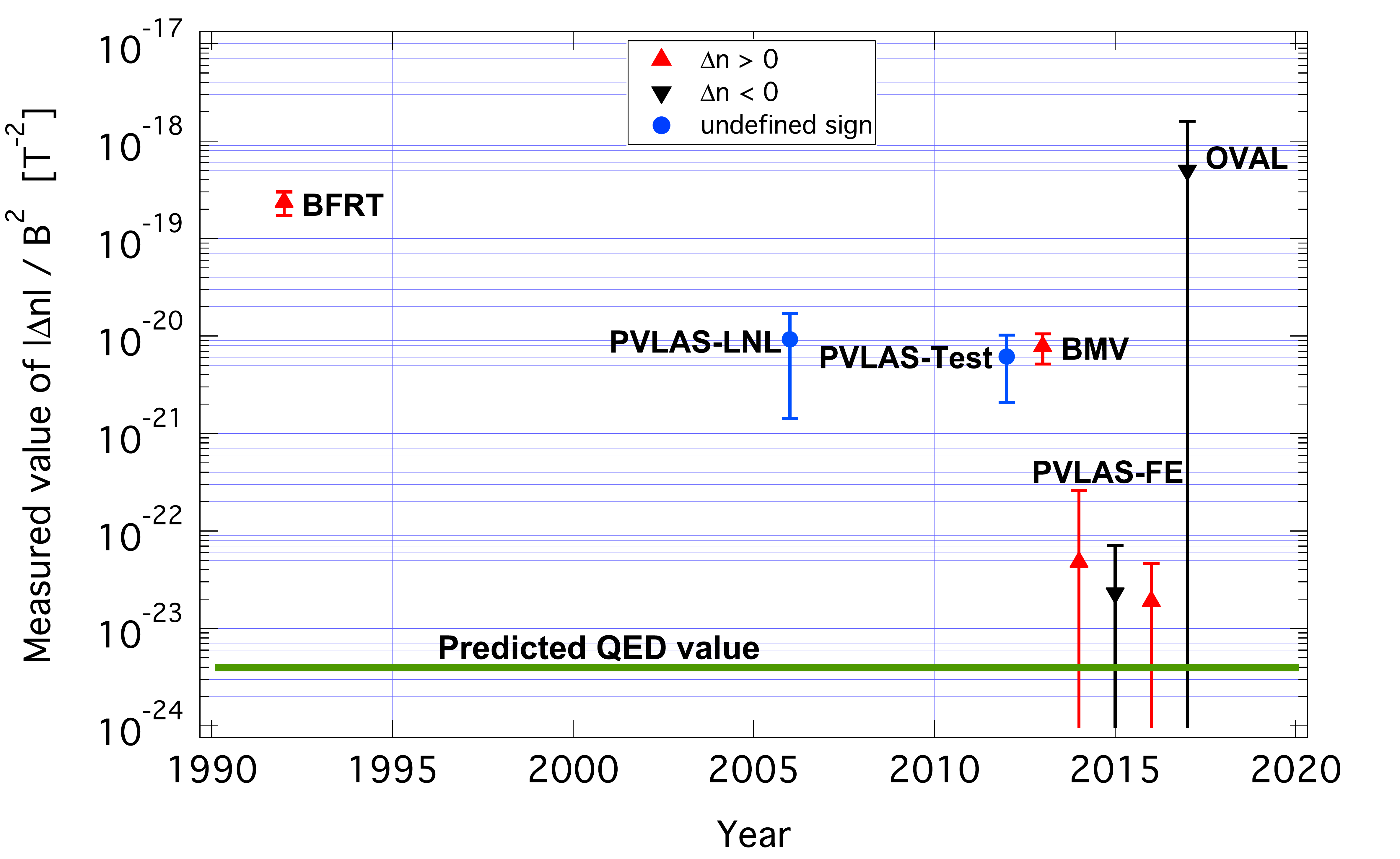}
\end{center}
\caption{Historical time evolution of the measurement of vacuum magnetic birefringence normalised to $B_{\rm{ ext}}^2$. Error bars correspond to one $\sigma$. The values derive from the following references: BFRT \cite{Cameron1993PRD}; PVLAS-LNL \cite{Zavattini2007PRD,Bregant2007PRD}, PVLAS-Test \cite{DellaValle2013NJP}, BMV \cite{Cadene2014EPJD}, PVLAS-FE \cite{DellaValle2014PRD,DellaValle2016EPJC}, OVAL \cite{Fan2017}.}
\label{fig:TimeEvolution}
\end{figure}

The historical time evolution of the measurement of vacuum magnetic birefringence normalised to $B^2_{\rm ext}$ from different experiments is shown in Figure~\ref{fig:TimeEvolution}. This normalisation allows the comparison of the limits of the different experiments trying to measure VMB. In the figure, the PVLAS-FE experiment appears with three points, representing the integrated progression of this measurement by the experiment. The first two points correspond to already published papers \cite{DellaValle2014PRD,DellaValle2016EPJC} whereas the third, including the 2016 data, represents the final result of the experiment concerning VMB. The global 2016 value reported in Figure~\ref{fig:TimeEvolution} is\begin{equation}
    \frac{\Delta n^{\rm (PVLAS-FE)}}{B^2_{\rm ext}} = (+19\pm27)\times10^{-24}\;{\rm T}^{-2}
\end{equation}

\subsection{\bf Limits on hypothetical particles}
\subsubsection{Axion Like Particles}

\begin{figure}[bht]
\begin{center}
\includegraphics[width=10cm]{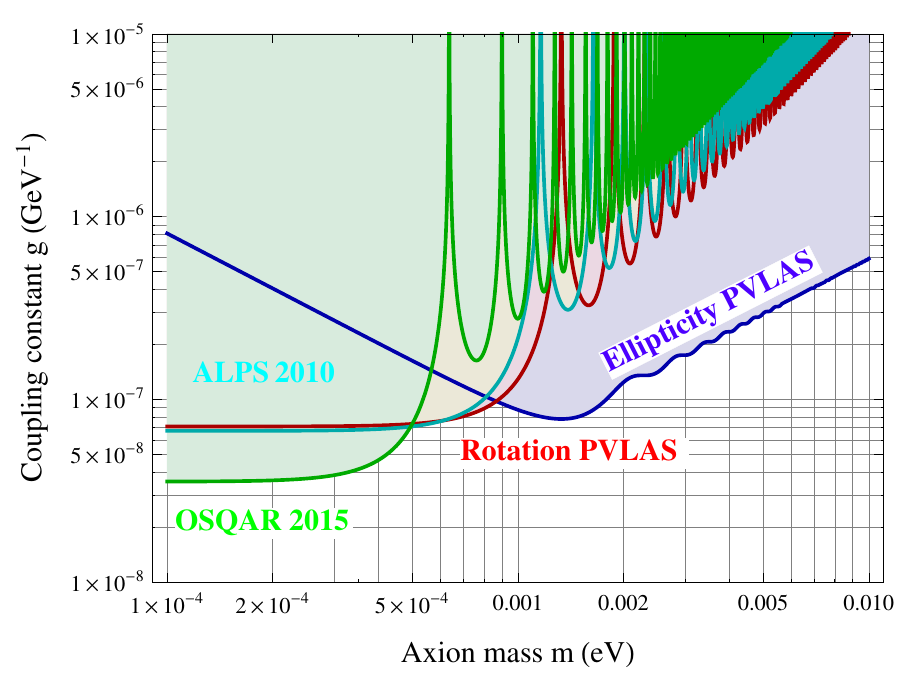}
\end{center}
\caption{Laboratory limits on the existence of ALPs particles at 95\% c.l. The shaded regions of the graph are excluded. The figure also shows the measurements by the OSQAR \cite{OSQAR2015} and the ALPS \cite{Ehret:2010mh} collaborations.}
\label{ALPsLimits}
\end{figure}

The results of the polarimetric  measurements of the PVLAS-FE experiment can be used to draw exclusion plots in the plane $(m_a,g_a)$ for Axion Like Particles. The birefringence value of equation~(\ref{BirefringencePVLAS}), through equation~(\ref{pseudo}), translates into the curve labelled as `Ellipticity PVLAS' in Figure~\ref{ALPsLimits}. The region excluded by this curve dominates the exclusion plot at axion masses $m_a\ge1$~meV. For the dichroism, one must note that the line labelled `Rotation PVLAS' in the same figure does not derive from the dichroism value of equation~(\ref{DichroismPVLAS}). In fact, as seen in equation~(\ref{dichroism}), the dichroism generated by ALPs has a non trivial dependence on the length of the magnetic field region. We distinguish two magnet configurations, according to whether the magnets rotated synchronously or not. It turns out that the best ALPs rotation limits are set almost solely by the single run 6 in Table~\ref{tab:GlobalRun}, with $L_B=1.64$~m. The small mass limit of the PVLAS rotation curve is $7.2\times10^{-8}$~GeV$^{-1}$. Below $0.5$~meV the limit given by the OSQAR experiment \cite{OSQAR2015} is more stringent by about a factor two. One must remind the reader that the whole region down to the level $g_a\sim10^{-10}$~GeV$^{-1}$ has already been excluded by the CAST solar helioscope \cite{CAST}. However, the CAST results depend on the model assumed for axion production and emission by the sun, whereas the limits of Figure~\ref{ALPsLimits} come from model independent laboratory experiments,

In the small mass limit with $m_a\ll\sqrt{\frac{4\omega}{L_B}} = 10^{-3}$~eV, where the coupling constant does not depend on the mass of the ALP, the value of $g_a$ can be determined from (1 T~$=\sqrt{\frac{\hbar^3 c^3}{e^4 \mu_0}}= 195$~eV$^2$ and 1 m$= \frac{e}{\hbar c}= 5.06\times 10^6$ eV$^{-1}$)
\begin{equation}
g_a = \sqrt{\frac{\omega}{2} \frac{\Delta\kappa}{L_B}}\frac{4}{B_{\rm ext}}.
\label{g_low_m}
\end{equation}
One can therefore do slightly better by taking the weighted average of $\frac{\Delta\kappa}{L_B}$ for the single and double magnet configurations.
By averaging Runs $3\alpha'$, $3\beta'$,  $4\beta'$ and $5\alpha'$ of Table~\ref{tab:GlobalRun}, all divided by $L_B$, and Run 6 divided by $2L_B$ and inserting it in the expression (\ref{g_low_m}) one finds
\begin{equation}
\left\langle\frac{\Delta\kappa}{L_B}\right\rangle = (1.0\pm2.6)\times 10^{-22}~{\rm m}^{-1}
\end{equation}
having used $L_B=0.82$~m corresponding to the length of one magnet. The resulting 95\%~c.l. limit on $g_a$ is therefore
\begin{equation}
g_a^{\rm (95\%)} 
< 6.4\times 10^{-8}~{\rm GeV}^{-1}.
\label{g_limit}
\end{equation}

\subsubsection{Millicharged Particles}

\begin{figure}[bht]
\begin{center}
\includegraphics[width=8cm]{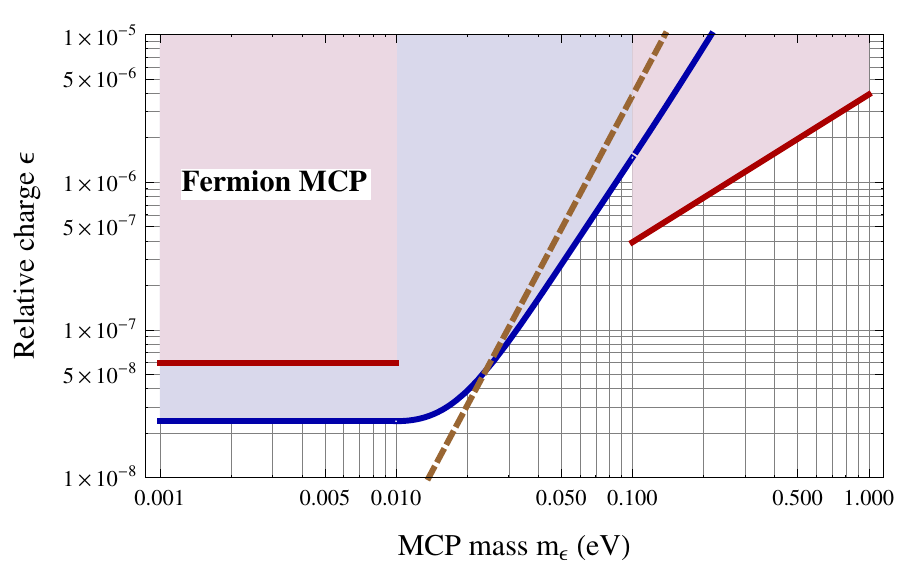}
\includegraphics[width=8cm]{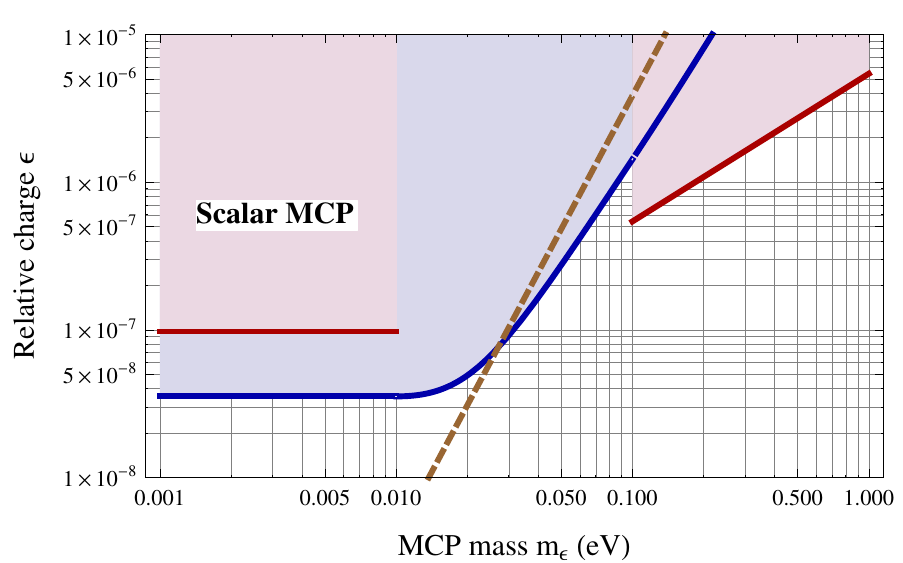}
\end{center}
\caption{Exclusion plots for millicharged particles at 95\%~c.l. deriving from equations~(\ref{BirefringencePVLAS}) and (\ref{DichroismPVLAS}). Left panel: fermion MCP. Right panel: scalar MCP. The excluded region is above the curves. The limit derived from rotation dominates at small masses, whereas the birefringence limit is effective at large masses. The two branches of the birefringence curve are not connected in the mass range around $\chi=1$ (dashed line), where $\Delta n$ changes sign. A cubic spline joins the two branches of the dichroism curve.}
\label{MCPsLimits}
\end{figure}

Figure~\ref{MCPsLimits} shows the PVLAS-FE exclusion plots on the existence of millicharged particles. Two independent limits are derived from the birefringence and the dichroism values of equations~(\ref{BirefringencePVLAS}) and (\ref{DichroismPVLAS}), the latter being more stringent in the low-mass range ($m_\epsilon\leq0.1$~eV), whereas the former is dominating the high-mass range. We explicitly note that the fermion exclusion plot also applies to all types of neutrinos, limiting their charge to be less than $\approx3\times10^{-8}e$ for masses smaller than 10~meV.

\section{Conclusions}

The PVLAS-FE experiment officially ended on December 31$^{\rm st}$ 2017 after 25 years. The present paper represents the final results of the experiment with the magnetically induced birefringence and dichroism limits summarised in equations~(\ref{BirefringencePVLAS}) and (\ref{DichroismPVLAS}). Unfortunately implementing further improvements to the PVLAS-FE setup of at least a factor ten to reach the predicted VMB value was not possible given the wide band noise intrinsic to the presence of the Fabry-Perot cavity which we believe to be of thermal origin (see Section~\ref{sec:intrinsic_noise}).  Nor was it conceivable to integrate a factor fifty times longer to only reach a SNR$ = 1$.

A total run time of $5\times10^6$~s was made possible thanks to the use of permanent magnets which, for the first time, allowed a detailed debugging of the setup. As was discussed in Section~\ref{sec:diffused_light} one of the key issues was the coupling between the diffused light inside the vacuum tube passing through the magnets and the induced movement of the tubes due to the small transverse gradient of the rotating magnetic field. The reduction of the diffused light with the use of baffles and a careful monitoring of the acceleration of the tubes allowed their centering thereby reducing systematic signals to below the achieved noise floor. 

The phase coherence of the rotating magnets was another key factor allowing to distinguish between a physical signal and a mechanically induced disturbance which appeared at times. Indeed one of the requirements of a physical signal was that it occupy a single bin in the demodulated Fourier spectrum even after a very long integration time ($\Delta\nu_{\rm bin}\lesssim 1\;\mu$Hz). 

From the experience presented in this paper and assuming to have all systematics under control, a new experiment to measure VMB using a Fabry-Perot based polarimeter will need to take into account the intrinsic optical path difference noise given in equation~(\ref{eq:fit_noise}). The key factor, we believe, is to improve the optical path difference source $\int{B^2_{\rm ext}\,dL}$. A few possible ideas could be:
\begin{itemize}
    \item
    use a relatively low magnetic field $\approx 1\div2$~T but increase the magnetic field length. This would allow the use of permanent or normal conducting magnets. Such a solution could then be applied to a gravitational wave antenna detector with long arms \cite{Calloni,Grote} where the optics is in continuous development as briefly introduced in Section~\ref{sec:ligo}. In this situations $\int B_{\rm ext}^2\,dL\sim 200$~T$^2$m would be necessary considering an integration time ${\cal T} = 10^6$~s. One extremely interesting feature of this solution is the independent measurement of $n_\parallel$ and $n_\perp$ allowing the direct determination of the parameters $\eta_1$ and $\eta_2$ in the general Lagrangian (\ref{eq:pM}) \cite{Calloni};
    \item
    continue to develop pulsed magnets as in BMV and OVAL. Given a pulse width $t_{\rm pulse}\approx 10$~ms and assuming therefore that the optical path difference noise is given by equation~(\ref{eq:fit_noise}), the integrated peak noise per pulse will be $\Delta{\cal D}_{\rm pulse}\sim 2 B_{\rm th} t_{\rm pulse}^{-1/4}\approx3\times10^{-19}$~m/pulse with $B_{\rm th} = (4.63\pm0.02)\times10^{-19}$~mHz$^{-1/4}$. Assuming a pulse rate $R_{\rm pulse}\approx0.1$~Hz \cite{Fan2017} and a running time $t_{\rm run}\sim$~ 1 month (a longer running time would not allow debugging in the presence unexpected peaks) results in
    \begin{equation}
        \int{B_{\rm ext}^2\,dL} = \frac{\Delta{\cal D}_{\rm pulse}}{3A_e\sqrt{R_{\rm pulse}t_{\rm run}}}\sim 150~{\rm T}^2{\rm m};
    \end{equation}
    \item use a constant field superconducting magnet and modulate the induced ellipticity using the polarisation. As discussed in \cite{Zavattini:2016sqz} one possibility to achieve this is to insert two co-rotating half-wave plates inside the Fabry-Perot, one at each end of the magnetic field. Due to the losses introduced by these wave plates, clearly this would reduce the maximum finesse to about $1000\div5000$. The intrinsic optical path difference noise give by equation~(\ref{eq:fit_noise}) of $10^{-18}$~m/$\sqrt{\rm Hz}$ @ 4~Hz could still be reached provided the intensity exiting the cavity is $I_{\rm out}\approx 50$~mW. Assuming the usual integration time of $10^6$~s results in a requirement for the magnetic field $\int{B^2_{\rm ext}\,dL}\approx 250$~T$^2$m.
\end{itemize}
This last configuration seems to be the most attractive in that such superconducting magnets already exist at CERN. In particular the use of a spare LHC dipole magnet with a maximum field of $B_{\rm ext} = 9$~T and a field length of 14.3~m, resulting in $\int{B^2_{\rm ext}\,dL}\approx 1158$~T$^2$m, would in principle allow a SNR = 1 in less then a day. Following this line a new collaboration is coalescing and a Letter of Intent has been submitted to CERN  \cite{LOI2018} and initial testing to demonstrate the feasibility is underway. 

\section*{Acknowledgements}
\addcontentsline{toc}{section}{\protect\numberline{}Acknowledgements}%
This paper is dedicated to the memory of Emilio (Mimmo) Zavattini, Erseo Polacco and Guido Petrucci. They were colleagues, teachers, friends, and more.

The PVLAS experiment was financed by the Istituto Nazionale di Fisica Nucleare (INFN), Italy and by the Italian Ministry of Research (MIUR).

\bibliography{PVLAS}

\begin{thebibliography}{100}
\expandafter\ifx\csname url\endcsname\relax
  \def\url#1{\texttt{#1}}\fi
\expandafter\ifx\csname urlprefix\endcsname\relax\def\urlprefix{URL }\fi
\expandafter\ifx\csname href\endcsname\relax
  \def\href#1#2{#2} \def\path#1{#1}\fi

\bibitem{Hanneke:2008tm}
D.~Hanneke, S.~Fogwell, G.~Gabrielse, {New Measurement of the Electron Magnetic
  Moment and the Fine Structure Constant}, Phys. Rev. Lett. 100 (2008) 120801.
\newblock \href {https://doi.org/10.1103/PhysRevLett.100.120801}
  {\path{doi:10.1103/PhysRevLett.100.120801}}.

\bibitem{Bennett:2008dy}
G.~W. Bennett, et~al., {An Improved Limit on the Muon Electric Dipole Moment},
  Phys. Rev. D 80 (2009) 052008.
\newblock \href {https://doi.org/10.1103/PhysRevD.80.052008}
  {\path{doi:10.1103/PhysRevD.80.052008}}.

\bibitem{PhysRev.72.241}
W.~E. Lamb, R.~C. Retherford, Fine structure of the hydrogen atom by a
  microwave method, Phys. Rev. 72 (1947) 241--243.
\newblock \href {https://doi.org/10.1103/PhysRev.72.241}
  {\path{doi:10.1103/PhysRev.72.241}}.

\bibitem{Schumacher1999}
M.~Schumacher, Delbr\"uck scattering, Rad. Phys. Chem. 56 (1999) 101--111.
\newblock \href {https://doi.org/10.1016/S0969-806X(99)00289-3}
  {\path{doi:10.1016/S0969-806X(99)00289-3}}.

\bibitem{Heisenberg1936}
W.~Heisenberg, H.~Euler, {Folgerungen aus der Diracschen Theorie des Positrons,
  English translation by W. Korolevski and H. Kleinert: {\em Consequences of
  Dirac's theory of positrons}}, Z. Phys. 98 (1936) 714--732.
\newblock \href {http://arxiv.org/abs/physics/0605038}
  {\path{arXiv:physics/0605038}}, \href {https://doi.org/10.1007/BF01343663}
  {\path{doi:10.1007/BF01343663}}.

\bibitem{Euler:1935}
H.~Euler,
  \href{http://www.neo-classical-physics.info/uploads/3/4/3/6/34363841/euler_-_scattering_of_light_by_light.pdf}{{\"Uber
  die Streuung von Licht an Licht nach der Diracschen Theorie, English
  translation by D. H. Delphenich: {\em On the scattering of light by light in
  Dirac's theory}}}, Ann. Phys. 26 (1936) 398--448.
\newblock \href {https://doi.org/10.1002/andp.19364180503}
  {\path{doi:10.1002/andp.19364180503}}.
\newline\urlprefix\url{http://www.neo-classical-physics.info/uploads/3/4/3/6/34363841/euler_-_scattering_of_light_by_light.pdf}

\bibitem{EulerKockel1935}
H.~Euler, B.~Kockel,
  \href{http://www.neo-classical-physics.info/uploads/3/4/3/6/34363841/euler-koeckel_-_scattering_of_light_by_light.pdf}{{\"U}ber
  die streuung von licht an licht nach der diracschen theorie, english
  translation by d. h. delphenich: {\em The scattering of light by light in
  Dirac's theory}}, Naturwiss. 23 (1935) 246--247.
\newblock \href {https://doi.org/10.1007/BF01493898}
  {\path{doi:10.1007/BF01493898}}.
\newline\urlprefix\url{http://www.neo-classical-physics.info/uploads/3/4/3/6/34363841/euler-koeckel_-_scattering_of_light_by_light.pdf}

\bibitem{Toll1952}
J.~S. Toll, The dispersion relation for light and its applications to problems
  involving electron pairs, in: PhD Thesis, Princeton University, Princeton,
  NJ, 1952, unpublished.

\bibitem{Erber1961}
T.~Erber, Velocity of light in a magnetic field, Nature 190 (1961) 25--27.
\newblock \href {https://doi.org/10.1038/190025a0}
  {\path{doi:10.1038/190025a0}}.

\bibitem{Klein1964_BR}
J.~J. Klein, B.~P. Nigam, {Birefringence of the Vacuum}, Phys. Rev. 135 (1964)
  B1279--B1280.
\newblock \href {https://doi.org/10.1103/PhysRev.135.B1279}
  {\path{doi:10.1103/PhysRev.135.B1279}}.

\bibitem{Baier:1967zza}
R.~Baier, P.~Breitenlohner, {Photon Propagation in External Fields}, Acta Phys.
  Austriaca 25 (1967) 212--223.

\bibitem{Baier:1967zzc}
R.~Baier, P.~Breitenlohner, {The Vacuum refraction Index in the presence of
  External Fields}, Nuovo Cim. B 47 (1967) 117--120.
\newblock \href {https://doi.org/10.1007/BF02712312}
  {\path{doi:10.1007/BF02712312}}.

\bibitem{atlas2017}
M.~Aaboud, G.~Aad, B.~Abbott, et~al., {Evidence for light-by-light scattering
  in heavy-ion collisions with the ATLAS detector at the LHC}, Nature Phys. 13
  (2017) 852--858.
\newblock \href {https://doi.org/10.1038/nphys4208}
  {\path{doi:10.1038/nphys4208}}.

\bibitem{atlas2019}
G.~Aad, et~al., Observation of light-by-light scattering in ultraperipheral
  $\mathrm{Pb}+\mathrm{Pb}$ collisions with the atlas detector, Phys. Rev.
  Lett. 123 (2019) 052001.
\newblock \href {https://doi.org/10.1103/PhysRevLett.123.052001}
  {\path{doi:10.1103/PhysRevLett.123.052001}}.

\bibitem{Mignani2016}
R.~P. Mignani, V.~Testa, D.~Gonz\'alez~Caniulef, R.~Taverna, R.~Turolla,
  S.~Zane, K.~Wu, {Evidence for vacuum birefringence from the first
  optical-polarimetry measurement of the isolated neutron star RXJ1856.5-3754},
  Mon. Not. R. Astron. Soc. 465 (2016) 492--500.
\newblock \href {https://doi.org/10.1093/mnras/stw2798}
  {\path{doi:10.1093/mnras/stw2798}}.

\bibitem{Iacopini:1979ci}
E.~Iacopini, E.~Zavattini, {Experimental Method to Detect the Vacuum
  Birefringence Induced by a Magnetic Field}, Phys. Lett. 85B (1979) 151.
\newblock \href {https://doi.org/10.1016/0370-2693(79)90797-4}
  {\path{doi:10.1016/0370-2693(79)90797-4}}.

\bibitem{Nezrick1998}
F.~Nezrick, Experiment to measure the effect of a magnetic field on the speed
  of light in vacuum - fermilab p-877, Nuclear Physics B - Proceedings
  Supplements 72 (1999) 198 -- 200, proceedings of the 5th IFT Workshop on
  Axions.
\newblock \href {https://doi.org/10.1016/S0920-5632(98)00525-8}
  {\path{doi:10.1016/S0920-5632(98)00525-8}}.

\bibitem{Chen2007}
S.-J. Chen, H.-H. Mei, W.-T. Ni, Q \& a experiment to search for vacuum
  dichroism, pseudoscalar-photon interaction and millicharged fermions, Mod.
  Phys. Lett. A 22 (2007) 2815--2831.
\newblock \href {https://doi.org/10.1142/S0217732307025844}
  {\path{doi:10.1142/S0217732307025844}}.

\bibitem{Cadene2014EPJD}
A.~Cad{\`e}ne, P.~Berceau, M.~Fouch{\'e}, R.~Battesti, C.~Rizzo, Vacuum
  magnetic linear birefringence using pulsed fields: status of the bmv
  experiment, Eur. Phys. J. D 68 (2014) 16.
\newblock \href {https://doi.org/10.1140/epjd/e2013-40725-9}
  {\path{doi:10.1140/epjd/e2013-40725-9}}.

\bibitem{Fan2017}
X.~Fan, S.~Kamioka, T.~Inada, T.~Yamazaki, T.~Namba, S.~Asai, J.~Omachi,
  K.~Yoshioka, M.~Kuwata-Gonokami, A.~Matsuo, K.~Kawaguchi, K.~Kindo,
  H.~Nojiri, The oval experiment: a new experiment to measure vacuum magnetic
  birefringence using high repetition pulsed magnets, Eur. Phys. J. D 71 (2017)
  308.
\newblock \href {https://doi.org/10.1140/epjd/e2017-80290-7}
  {\path{doi:10.1140/epjd/e2017-80290-7}}.

\bibitem{Sarazin2016EPJD}
X.~Sarazin, F.~Couchot, A.~Djannati-Atai, O.~Guilbaud, S.~Kazamias, M.~Pittman,
  M.~Urban, Refraction of light by light in vacuum, Eur. Phys. J. D 70 (2016).
\newblock \href {https://doi.org/10.1140/epjd/e2015-60428-5}
  {\path{doi:10.1140/epjd/e2015-60428-5}}.

\bibitem{Moulin1996}
F.~Moulin, D.~Bernard, F.~Amiranoff, Photon-photon elastic scattering in the
  visible domain, Z. Phys. C 72 (1996) 607--611.
\newblock \href {https://doi.org/10.1007/s002880050282}
  {\path{doi:10.1007/s002880050282}}.

\bibitem{Bernard2000}
D.~Bernard, F.~Moulin, F.~Amiranoff, A.~Braun, J.~P. Chambaret, G.~Darpentigny,
  G.~Grillon, S.~Ranc, F.~Perrone, {Search for Stimulated Photon-Photon
  Scattering in Vacuum}, Eur. Phys. J. D 10 (2000) 141--145.
\newblock \href {https://doi.org/10.1007/s100530050535}
  {\path{doi:10.1007/s100530050535}}.

\bibitem{Watson1929}
W.~H. Watson, The effect of a transverse magnetic field on the propagation of
  light in vacuo, Proc. R. Soc. Lond. A 125 (1929) 345--351, communicated by E.
  Rutherford.
\newblock \href {https://doi.org/10.1098/rspa.1929.0172}
  {\path{doi:10.1098/rspa.1929.0172}}.

\bibitem{Hughes1930}
A.~L. Hughes, G.~E.~M. Jauncey, An attempt to detect collisions of photons,
  Phys. Rev. 36 (1930) 773--777.
\newblock \href {https://doi.org/10.1103/physrev.36.773}
  {\path{doi:10.1103/physrev.36.773}}.

\bibitem{Vavilov1930}
S.~Vavilov, {On the attempt to detect collisions of photons}, Phys. Rev. 36
  (1930) 1590.
\newblock \href {https://doi.org/10.1103/PhysRev.36.1590}
  {\path{doi:10.1103/PhysRev.36.1590}}.

\bibitem{Farr1932}
C.~C. Farr, C.~J. Banwell, Velocity of propagation of light {\em in vacuo} in a
  transverse magnetic field, Proc. R. Soc. Lond. A 137 (1932) 275--282.
\newblock \href {https://doi.org/10.1098/rspa.1932.0135}
  {\path{doi:10.1098/rspa.1932.0135}}.

\bibitem{Banwell1940}
C.~J. Banwell, C.~C. Farr, Further investigation of the velocity of propagation
  of light {\em in vacuo} in a transverse magnetic field, Proc. R. Soc. Lond. A
  175 (1940) 1--25.
\newblock \href {https://doi.org/10.1098/rspa.1940.0040}
  {\path{doi:10.1098/rspa.1940.0040}}.

\bibitem{Jones1960}
R.~V. Jones, Velocity of light in a magnetic field, Nature 186 (1960) 706.
\newblock \href {https://doi.org/10.1038/186706a0}
  {\path{doi:10.1038/186706a0}}.

\bibitem{Jones1961}
R.~V. Jones, The velocity of light in a transverse magnetic field, Proc. R.
  Soc. Lond. A 260 (1961) 47--60, communicated by T. R. Merton.
\newblock \href {https://doi.org/10.1098/rspa.1961.0012}
  {\path{doi:10.1098/rspa.1961.0012}}.

\bibitem{Battesti2013}
R.~Battesti, C.~Rizzo, {Magnetic and electric properties of quantum vacuum},
  Rep. Prog. Phys. 76 (2013) 016401.
\newblock \href {https://doi.org/10.1088/0034-4885/76/1/016401}
  {\path{doi:10.1088/0034-4885/76/1/016401}}.

\bibitem{Ejlli2017}
A.~Ejlli, Progress towards a first measurement of the magnetic birefringence of
  vacuum with a polarimeter based on a fabry-perot cavity, in: PhD Thesis,
  University of Ferrara, Ferrara, Italy, 2017, unpublished.

\bibitem{Halpern:1933dya}
O.~Halpern, {Scattering processes produced by electrons in negative energy
  states}, Phys. Rev. 44 (1933) 855--856.
\newblock \href {https://doi.org/10.1103/PhysRev.44.855.2}
  {\path{doi:10.1103/PhysRev.44.855.2}}.

\bibitem{Puri}
S.~P. Puri, Classical electrodynamics, Narosa, New Delhi, India, 2011.

\bibitem{Sauter}
F.~Sauter,
  \href{http://www.neo-classical-physics.info/uploads/3/4/3/6/34363841/sauter_-_electron_in_homo_electric_field.pdf}{{\"Uber
  das Verhalten eines Elektrons im homogenen elektrischen Feld nach der
  relativistischen Theorie Diracs, English translation by D. H. Delphenich:
  {\em On the behavior of an electron in a homogeneous electric field in
  Dirac's relativistic theory }}}, Z. Phys. 69 (1931) 742--764.
\newblock \href {https://doi.org/10.1007/BF01339461}
  {\path{doi:10.1007/BF01339461}}.
\newline\urlprefix\url{http://www.neo-classical-physics.info/uploads/3/4/3/6/34363841/sauter_-_electron_in_homo_electric_field.pdf}

\bibitem{Weisskopf1936}
V.~Weisskopf,
  \href{https://neo-classical-physics.info/uploads/3/4/3/6/34363841/weisskopf_-_electrodynamics.pdf}{\"uber
  die elektrodynamik des vakuums auf grund der quantentheorie des elektrons,
  english translation by d. h. delphenich: {\em On the Electrodynamics of
  Vacuum on the Basis of the Quantum Theory of the Electron}}, K. Dan. Vidensk.
  Selsk., Mat. Fys. Medd. 14~(6) (1936).
\newline\urlprefix\url{https://neo-classical-physics.info/uploads/3/4/3/6/34363841/weisskopf_-_electrodynamics.pdf}

\bibitem{Karplus:1950zza}
R.~Karplus, M.~Neuman, {Non-Linear Interactions between Electromagnetic
  Fields}, Phys. Rev. 80 (1950) 380--385.
\newblock \href {https://doi.org/10.1103/PhysRev.80.380}
  {\path{doi:10.1103/PhysRev.80.380}}.

\bibitem{Schwinger:1951nm}
J.~S. Schwinger, {On gauge invariance and vacuum polarization}, Phys. Rev. 82
  (1951) 664--679.
\newblock \href {https://doi.org/10.1103/PhysRev.82.664}
  {\path{doi:10.1103/PhysRev.82.664}}.

\bibitem{Robertson2019}
S.~Robertson, Optical kerr effect in vacuum, Phys. Rev. A 100 (2019) 063831.
\newblock \href {https://doi.org/10.1103/PhysRevA.100.063831}
  {\path{doi:10.1103/PhysRevA.100.063831}}.

\bibitem{Breit1934}
G.~Breit, J.~A. Wheeler, Collision of two light quanta, Phys. Rev. 46 (1934)
  1087--1091.
\newblock \href {https://doi.org/10.1103/PhysRev.46.1087}
  {\path{doi:10.1103/PhysRev.46.1087}}.

\bibitem{Karplus:1950zz}
R.~Karplus, M.~Neuman, {The scattering of light by light}, Phys. Rev. 83 (1951)
  776--784.
\newblock \href {https://doi.org/10.1103/PhysRev.83.776}
  {\path{doi:10.1103/PhysRev.83.776}}.

\bibitem{DeTollis:1965vna}
B.~De~Tollis, {The scattering of photons by photons}, Nuovo Cim. 35 (1965)
  1182--1193.
\newblock \href {https://doi.org/10.1007/BF02735534}
  {\path{doi:10.1007/BF02735534}}.

\bibitem{Klein:1964zza}
J.~J. Klein, B.~P. Nigam, {Dichroism of the Vacuum}, Phys. Rev. 136 (1964)
  B1540--B1542.
\newblock \href {https://doi.org/10.1103/PhysRev.136.B1540}
  {\path{doi:10.1103/PhysRev.136.B1540}}.

\bibitem{Adler1970}
S.~L. Adler, J.~N. Bahcall, C.~G. Callan, M.~N. Rosenbluth, {Photon splitting
  in a strong magnetic field}, Phys. Rev. Lett. 25 (1970) 1061--1065.
\newblock \href {https://doi.org/10.1103/PhysRevLett.25.1061}
  {\path{doi:10.1103/PhysRevLett.25.1061}}.

\bibitem{BialynickaBirula1970}
Z.~Bialynicka-Birula, I.~Bialynicki-Birula, {Nonlinear effects in Quantum
  Electrodynamics. Photon propagation and photon splitting in an external
  field}, Phys. Rev. D 2 (1970) 2341--2345.
\newblock \href {https://doi.org/10.1103/PhysRevD.2.2341}
  {\path{doi:10.1103/PhysRevD.2.2341}}.

\bibitem{Adler1971}
S.~L. Adler, {Photon splitting and photon dispersion in a strong magnetic
  field}, Ann. Phys. NY 67 (1971) 599--647.
\newblock \href {https://doi.org/10.1016/0003-4916(71)90154-0}
  {\path{doi:10.1016/0003-4916(71)90154-0}}.

\bibitem{Stoneham1979}
R.~J. Stoneham, {Photon splitting in the magnetised vacuum}, J. Phys. A: Math.
  Gen. 12 (1979) 2187--2203.
\newblock \href {https://doi.org/10.1088/0305-4470/12/11/028}
  {\path{doi:10.1088/0305-4470/12/11/028}}.

\bibitem{Baier1996}
V.~N. Baier, A.~I. Milstein, R.~Z. Shaisultanov, Photon splitting in a very
  strong magnetic field, Phys. Rev. Lett. 77 (1996) 1691--1694.
\newblock \href {https://doi.org/10.1103/PhysRevLett.77.1691}
  {\path{doi:10.1103/PhysRevLett.77.1691}}.

\bibitem{Adler1996PRL}
S.~L. Adler, C.~Schubert, Photon splitting in a strong magnetic field:
  Recalculation and comparison with previous calculations, Phys. Rev. Lett. 77
  (1996) 1695--1698.
\newblock \href {https://doi.org/10.1103/PhysRevLett.77.1695}
  {\path{doi:10.1103/PhysRevLett.77.1695}}.

\bibitem{Akhmadaliev1998}
S.~Z. Akhmadaliev, et~al., Experimental investigation of high-energy photon
  splitting in atomic fields, Phys. Rev. Lett. 89 (1998) 061802.
\newblock \href {https://doi.org/10.1103/PhysRevLett.89.061802}
  {\path{doi:10.1103/PhysRevLett.89.061802}}.

\bibitem{Ritus:1975cf}
V.~I. Ritus, {The Lagrange Function of an Intensive Electromagnetic Field and
  Quantum Electrodynamics at Small Distances}, Sov. Phys. JETP 42 (1975)
  774--782, [{\em Pisma Zh. Eksp. Teor. Fiz.} 69 (1975) p.~1517].

\bibitem{Born1933Nature}
M.~Born, Modified field equations with a finite radius of the electron, Nature
  132 (1933) 282.
\newblock \href {https://doi.org/10.1038/132282a0}
  {\path{doi:10.1038/132282a0}}.

\bibitem{BornInfeld1933}
M.~Born, L.~Infeld, Foundations of the new field theory, Nature 132 (1933)
  1004.
\newblock \href {https://doi.org/10.1038/1321004b0}
  {\path{doi:10.1038/1321004b0}}.

\bibitem{BornInfeld1934}
M.~Born, L.~Infeld, {Foundations of the new field theory}, Proc. Roy. Soc.
  Lond. A 144 (1934) 425--451.
\newblock \href {https://doi.org/10.1098/rspa.1934.0059}
  {\path{doi:10.1098/rspa.1934.0059}}.

\bibitem{Iacopini:1982cm}
E.~Iacopini, E.~Zavattini, {Vacuum polarization effects in the
  ($\mu^{-4}$-He)$^+$ atom and the Born-Infeld electromagnetic theory}, Nuovo
  Cim. B 78 (1983) 38--52.
\newblock \href {https://doi.org/10.1007/BF02721380}
  {\path{doi:10.1007/BF02721380}}.

\bibitem{Fouche2016}
M.~Fouch\'e, R.~Battesti, C.~Rizzo, {Limits on nonlinear electrodynamics},
  Phys. Rev. D 93 (2016) 093020.
\newblock \href {https://doi.org/10.1103/PhysRevD.93.093020}
  {\path{doi:10.1103/PhysRevD.93.093020}}.

\bibitem{Davila2014}
J.~M. D\'avila, C.~Schubert, M.~A. Trejo, {Photonic processes in Born Infeld
  theory}, Int. J. Mod. Phys. A 29 (2014) 1450174.
\newblock \href {https://doi.org/10.1142/S0217751X14501747}
  {\path{doi:10.1142/S0217751X14501747}}.

\bibitem{Primakoff}
H.~Primakoff, {Photo-Production of Neutral Mesons in Nuclear Electric Fields
  and the Mean Life of the Neutral Meson}, Phys. Rev. 81 (1951) 899.
\newblock \href {https://doi.org/10.1103/PhysRev.81.899}
  {\path{doi:10.1103/PhysRev.81.899}}.

\bibitem{Wilczek:1982rv}
F.~Wilczek, {Axions and Family Symmetry Breaking}, Phys. Rev. Lett. 49 (1982)
  1549--1552.
\newblock \href {https://doi.org/10.1103/PhysRevLett.49.1549}
  {\path{doi:10.1103/PhysRevLett.49.1549}}.

\bibitem{Chikashige:1980ui}
Y.~Chikashige, R.~N. Mohapatra, R.~D. Peccei, {Are There Real Goldstone Bosons
  Associated with Broken Lepton Number?}, Phys. Lett. B 98 (1981) 265--268.
\newblock \href {https://doi.org/10.1016/0370-2693(81)90011-3}
  {\path{doi:10.1016/0370-2693(81)90011-3}}.

\bibitem{Peccei:1977ur}
R.~D. Peccei, H.~R. Quinn, {Constraints Imposed by CP Conservation in the
  Presence of Instantons}, Phys. Rev. D 16 (1977) 1791--1797.
\newblock \href {https://doi.org/10.1103/PhysRevD.16.1791}
  {\path{doi:10.1103/PhysRevD.16.1791}}.

\bibitem{Peccei:1977hh}
R.~D. Peccei, H.~R. Quinn, {CP Conservation in the Presence of Instantons},
  Phys. Rev. Lett. 38 (1977) 1440--1443.
\newblock \href {https://doi.org/10.1103/PhysRevLett.38.1440}
  {\path{doi:10.1103/PhysRevLett.38.1440}}.

\bibitem{Weinberg1978}
S.~Weinberg, {A New Light Boson?}, Phys. Rev. Lett. 40 (1978) 223--226.
\newblock \href {https://doi.org/10.1103/PhysRevLett.40.223}
  {\path{doi:10.1103/PhysRevLett.40.223}}.

\bibitem{Wilczek1978}
F.~Wilczek, {Problem of Strong $P$ and $T$ Invariance in the Presence of
  Instantons}, Phys. Rev. Lett. 40 (1978) 279--282.
\newblock \href {https://doi.org/10.1103/PhysRevLett.40.279}
  {\path{doi:10.1103/PhysRevLett.40.279}}.

\bibitem{PDG2019}
A.~Ringwald, L.~J. Rosenberg, G.~Rybka,
  \href{http://pdg.lbl.gov/2019/reviews/rpp2018-rev-axions.pdf}{{Axions and
  Other Similar Particles}}, 2019 update to M. Tanabashi and others, Particle
  Data Group. Review of Particle Physics. In: {\em Phys. Rev. D} 98 (2018), p.
  030001. DOI: 10.1103/PhysRevD.98.030001.
\newline\urlprefix\url{http://pdg.lbl.gov/2019/reviews/rpp2018-rev-axions.pdf}

\bibitem{axion_searches}
P.~W. Graham, I.~G. Irastorza, S.~K. Lamoreaux, A.~Lindner, K.~A. van Bibber,
  Experimental searches for the axion and axion-like particles, Ann. Rev. Nucl.
  Part. Sci. 65 (2015) 485--514.
\newblock \href {https://doi.org/10.1146/annurev-nucl-102014-022120}
  {\path{doi:10.1146/annurev-nucl-102014-022120}}.

\bibitem{Dine:1981rt}
M.~Dine, W.~Fischler, M.~Srednicki, {A Simple Solution to the Strong CP Problem
  with a Harmless Axion}, Phys. Lett. B 104 (1981) 199--202.
\newblock \href {https://doi.org/10.1016/0370-2693(81)90590-6}
  {\path{doi:10.1016/0370-2693(81)90590-6}}.

\bibitem{Zhitnitsky:1980tq}
A.~R. Zhitnitsky, {On Possible Suppression of the Axion Hadron Interactions},
  Sov. J. Nucl. Phys. 31 (1980) 260, [Yad. Fiz. 31, 497-504 (1980)].

\bibitem{Kuster:1105861}
{Kuster, M. and Raffelt, G. and Beltr\'an, B.}, {Axions: theory, cosmology, and
  experimental searches}, Springer, Berlin, 2008.
\newblock \href {https://doi.org/10.1007/978-3-540-73518-2}
  {\path{doi:10.1007/978-3-540-73518-2}}.

\bibitem{Irastorza:2018dyq}
I.~G. Irastorza, J.~Redondo, {New experimental approaches in the search for
  axion-like particles}, Prog. Part. Nucl. Phys. 102 (2018) 89--159.
\newblock \href {https://doi.org/10.1016/j.ppnp.2018.05.003}
  {\path{doi:10.1016/j.ppnp.2018.05.003}}.

\bibitem{PhysRevLett.120.151301}
N.~Du, N.~Force, R.~Khatiwada, E.~Lentz, R.~Ottens, L.~J. Rosenberg, G.~Rybka,
  G.~Carosi, N.~Woollett, D.~Bowring, A.~S. Chou, A.~Sonnenschein, W.~Wester,
  C.~Boutan, N.~S. Oblath, R.~Bradley, E.~J. Daw, A.~V. Dixit, J.~Clarke, S.~R.
  O'Kelley, N.~Crisosto, J.~R. Gleason, S.~Jois, P.~Sikivie, I.~Stern, N.~S.
  Sullivan, D.~B. Tanner, G.~C. Hilton, Search for invisible axion dark matter
  with the axion dark matter experiment, Phys. Rev. Lett. 120 (2018) 151301.
\newblock \href {https://doi.org/10.1103/PhysRevLett.120.151301}
  {\path{doi:10.1103/PhysRevLett.120.151301}}.

\bibitem{Semertzidis:1990qc}
Y.~Semertzidis, R.~Cameron, G.~Cantatore, A.~C. Melissinos, J.~Rogers,
  H.~Halama, A.~Prodell, F.~Nezrick, C.~Rizzo, E.~Zavattini, {Limits on the
  Production of Light Scalar and Pseudoscalar Particles}, Phys. Rev. Lett. 64
  (1990) 2988--2991.
\newblock \href {https://doi.org/10.1103/PhysRevLett.64.2988}
  {\path{doi:10.1103/PhysRevLett.64.2988}}.

\bibitem{Bakalov1998HI}
D.~Bakalov, et~al., {The measurement of vacuum polarization: The PVLAS
  experiment}, in: R.~Calabrese, V.~Guidi, H.-J. Kluge, L.~Moi (Eds.),
  {proceedings of the Third Euroconference on “Atomic Physics with Stored
  Highly Charged Ions”, Ferrara, Italy, 22-26 September 1997}, Vol. 114,
  1998, pp. 103--113, hyperfine Interact. 114 (1998).
\newblock \href {https://doi.org/10.1023/A:1012610102642}
  {\path{doi:10.1023/A:1012610102642}}.

\bibitem{Zavattini2007PRD}
E.~Zavattini, G.~Zavattini, G.~Ruoso, G.~Raiteri, E.~Polacco, E.~Milotti,
  V.~Lozza, M.~Karuza, U.~Gastaldi, G.~Di~Domenico, F.~Della~Valle, R.~Cimino,
  S.~Carusotto, G.~Cantatore, M.~Bregant, New pvlas results and limits on
  magnetically induced optical rotation and ellipticity in vacuum, Phys. Rev. D
  77 (2008) 032006.
\newblock \href {https://doi.org/10.1103/PhysRevD.77.032006}
  {\path{doi:10.1103/PhysRevD.77.032006}}.

\bibitem{Bregant2007PRD}
M.~Bregant, G.~Cantatore, S.~Carusotto, R.~Cimino, F.~Della~Valle,
  G.~Di~Domenico, U.~Gastaldi, M.~Karuza, V.~Lozza, E.~Milotti, E.~Polacco,
  G.~Raiteri, G.~Ruoso, E.~Zavattini, G.~Zavattini, Limits on low energy
  photon-photon scattering from an experiment on magnetic vacuum birefringence,
  Phys. Rev. D 78 (2008) 032006.
\newblock \href {https://doi.org/10.1103/PhysRevD.78.032006}
  {\path{doi:10.1103/PhysRevD.78.032006}}.

\bibitem{VanBibber:1987rq}
K.~Van~Bibber, N.~R. Dagdeviren, S.~E. Koonin, A.~Kerman, H.~N. Nelson,
  {Proposed experiment to produce and detect light pseudoscalars}, Phys. Rev.
  Lett. 59 (1987) 759--762.
\newblock \href {https://doi.org/10.1103/PhysRevLett.59.759}
  {\path{doi:10.1103/PhysRevLett.59.759}}.

\bibitem{Hoogeveen:1990vq}
F.~Hoogeveen, T.~Ziegenhagen, {Production and Detection of Light Bosons Using
  Optical Resonators}, Nucl. Phys. B358 (1991) 3--26.
\newblock \href {https://doi.org/10.1016/0550-3213(91)90528-6}
  {\path{doi:10.1016/0550-3213(91)90528-6}}.

\bibitem{Sikivie:1012940}
P.~Sikivie, D.~B. Tanner, K.~Van~Bibber, {Resonantly Enhanced Axion-Photon
  Regeneration}, Phys. Rev. Lett. 98 (2007) 172002.
\newblock \href {https://doi.org/10.1103/PhysRevLett.98.172002}
  {\path{doi:10.1103/PhysRevLett.98.172002}}.

\bibitem{Ruoso:1992nx}
G.~Ruoso, et~al., {Limits on light scalar and pseudoscalar particles from a
  photon regeneration experiment}, Z. Phys. C 56 (1992) 505--508.
\newblock \href {https://doi.org/10.1007/BF01474722}
  {\path{doi:10.1007/BF01474722}}.

\bibitem{Ehret:2010mh}
K.~Ehret, et~al., {New ALPS Results on Hidden-Sector Lightweights}, Phys. Lett.
  B 689 (2010) 149--155.
\newblock \href {https://doi.org/10.1016/j.physletb.2010.04.066}
  {\path{doi:10.1016/j.physletb.2010.04.066}}.

\bibitem{OSQAR2015}
R.~Ballou, G.~Deferne, M.~Finger, M.~Finger, L.~Flekova, J.~Hosek, v.~S. Kunc,
  K.~Macuchova, K.~A. Meissner, P.~Pugnat, M.~Schott, A.~Siemko, M.~Slunecka,
  M.~\v~Sulc, C.~Weinsheimer, J.~Zicha, New exclusion limits on scalar and
  pseudoscalar axionlike particles from light shining through a wall, Phys.
  Rev. D 92 (2015) 092002.
\newblock \href {https://doi.org/10.1103/PhysRevD.92.092002}
  {\path{doi:10.1103/PhysRevD.92.092002}}.

\bibitem{Maiani:1986md}
L.~Maiani, R.~Petronzio, E.~Zavattini, {Effects of Nearly Massless, Spin Zero
  Particles on Light Propagation in a Magnetic Field}, Phys. Lett. B 175 (1986)
  359--363.
\newblock \href {https://doi.org/10.1016/0370-2693(86)90869-5}
  {\path{doi:10.1016/0370-2693(86)90869-5}}.

\bibitem{Cameron1993PRD}
R.~Cameron, et~al., {Search for nearly massless, weakly coupled particles by
  optical techniques}, Phys. Rev. D 47 (1993) 3707--3725.
\newblock \href {https://doi.org/10.1103/PhysRevD.47.3707}
  {\path{doi:10.1103/PhysRevD.47.3707}}.

\bibitem{Sikivie:1983ip}
P.~Sikivie, {Experimental Tests of the Invisible Axion}, Phys. Rev. Lett. 51
  (1983) 1415--1417.
\newblock \href
  {https://doi.org/10.1103/PhysRevLett.51.1415,10.1103/PhysRevLett.52.695.2}
  {\path{doi:10.1103/PhysRevLett.51.1415,10.1103/PhysRevLett.52.695.2}}.

\bibitem{Raffelt:1987im}
G.~Raffelt, L.~Stodolsky, {Mixing of the Photon with Low Mass Particles}, Phys.
  Rev. D 37 (1988) 1237--1249.
\newblock \href {https://doi.org/10.1103/PhysRevD.37.1237}
  {\path{doi:10.1103/PhysRevD.37.1237}}.

\bibitem{Ahlers:2006iz}
M.~Ahlers, H.~Gies, J.~Jaeckel, A.~Ringwald, {On the Particle Interpretation of
  the PVLAS Data: Neutral versus Charged Particles}, Phys. Rev. D 75 (2007)
  035011.
\newblock \href {https://doi.org/10.1103/PhysRevD.75.035011}
  {\path{doi:10.1103/PhysRevD.75.035011}}.

\bibitem{Gies:2006ca}
H.~Gies, J.~Jaeckel, A.~Ringwald, {Polarized Light Propagating in a Magnetic
  Field as a Probe of Millicharged Fermions}, Phys. Rev. Lett. 97 (2006)
  140402.
\newblock \href {https://doi.org/10.1103/PhysRevLett.97.140402}
  {\path{doi:10.1103/PhysRevLett.97.140402}}.

\bibitem{Tsai:1975iz}
W.-y. Tsai, T.~Erber, {The Propagation of Photons in Homogeneous Magnetic
  Fields: Index of Refraction}, Phys. Rev. D 12 (1975) 1132--1137.
\newblock \href {https://doi.org/10.1103/PhysRevD.12.1132}
  {\path{doi:10.1103/PhysRevD.12.1132}}.

\bibitem{Tsai:1974fa}
W.-y. Tsai, T.~Erber, {Photon Pair Creation in Intense Magnetic Fields}, Phys.
  Rev. D 10 (1974) 492--499.
\newblock \href {https://doi.org/10.1103/PhysRevD.10.492}
  {\path{doi:10.1103/PhysRevD.10.492}}.

\bibitem{Riess_1998}
A.~G. Riess, A.~V. Filippenko, P.~Challis, A.~Clocchiatti, A.~Diercks, P.~M.
  Garnavich, R.~L. Gilliland, C.~J. Hogan, S.~Jha, R.~P. Kirshner,
  B.~Leibundgut, M.~M. Phillips, D.~Reiss, B.~P. Schmidt, R.~A. Schommer, R.~C.
  Smith, J.~Spyromilio, C.~Stubbs, N.~B. Suntzeff, J.~Tonry, Observational
  evidence from supernovae for an accelerating universe and a cosmological
  constant, Astron. J. 116 (1998) 1009--1038.
\newblock \href {https://doi.org/10.1086/300499} {\path{doi:10.1086/300499}}.

\bibitem{Perlmutter_1999}
S.~Perlmutter, G.~Aldering, G.~Goldhaber, R.~A. Knop, P.~Nugent, P.~G. Castro,
  S.~Deustua, S.~Fabbro, A.~Goobar, D.~E. Groom, I.~M. Hook, A.~G. Kim, M.~Y.
  Kim, J.~C. Lee, N.~J. Nunes, R.~Pain, C.~R. Pennypacker, R.~Quimby,
  C.~Lidman, R.~S. Ellis, M.~Irwin, R.~G. McMahon, P.~Ruiz-Lapuente, N.~Walton,
  B.~Schaefer, B.~J. Boyle, A.~V. Filippenko, T.~Matheson, A.~S. Fruchter,
  N.~Panagia, H.~J.~M. Newberg, W.~J. Couch, T.~S.~C. Project, Measurements of
  $\omega$ and $\lambda$ from 42 high-redshift supernovae, Astrophys. Jour. 517
  (1999) 565--586.
\newblock \href {https://doi.org/10.1086/307221} {\path{doi:10.1086/307221}}.

\bibitem{Caldwell2009}
R.~R. Caldwell, M.~Kamionkowski, The physics of cosmic acceleration, Ann. Rev.
  Nucl. Part. Sci. 59 (2009) 397--429.
\newblock \href {https://doi.org/10.1146/annurev-nucl-010709-151330}
  {\path{doi:10.1146/annurev-nucl-010709-151330}}.

\bibitem{JAIN20101479}
B.~Jain, J.~Khoury, Cosmological tests of gravity, Ann. Phys. NY 325 (2010)
  1479 -- 1516.
\newblock \href {https://doi.org/10.1016/j.aop.2010.04.002}
  {\path{doi:10.1016/j.aop.2010.04.002}}.

\bibitem{PhysRevLett.93.171104}
J.~Khoury, A.~Weltman, Chameleon fields: Awaiting surprises for tests of
  gravity in space, Phys. Rev. Lett. 93 (2004) 171104.
\newblock \href {https://doi.org/10.1103/PhysRevLett.93.171104}
  {\path{doi:10.1103/PhysRevLett.93.171104}}.

\bibitem{PhysRevD.69.044026}
J.~Khoury, A.~Weltman, Chameleon cosmology, Phys. Rev. D 69 (2004) 044026.
\newblock \href {https://doi.org/10.1103/PhysRevD.69.044026}
  {\path{doi:10.1103/PhysRevD.69.044026}}.

\bibitem{PhysRevD.76.085010}
P.~Brax, C.~van~de Bruck, A.-C. Davis, D.~F. Mota, D.~Shaw, Testing chameleon
  theories with light propagating through a magnetic field, Phys. Rev. D 76
  (2007) 085010.
\newblock \href {https://doi.org/10.1103/PhysRevD.76.085010}
  {\path{doi:10.1103/PhysRevD.76.085010}}.

\bibitem{Burrage2018}
C.~Burrage, J.~Sakstein, Tests of chameleon gravity, Living Rev. Relativ. 21
  (2018) 1.
\newblock \href {https://doi.org/10.1007/s41114-018-0011-x}
  {\path{doi:10.1007/s41114-018-0011-x}}.

\bibitem{Jones1941}
R.~C. Jones, A new calculus for the treatment of optical systems. i.
  description and discussion of the calculus, J. Opt. Soc. Am. 31 (1941)
  488--493.
\newblock \href {https://doi.org/10.1364/JOSA.31.000488}
  {\path{doi:10.1364/JOSA.31.000488}}.

\bibitem{DellaValle2014OE}
F.~D. Valle, E.~Milotti, A.~Ejlli, U.~Gastaldi, G.~Messineo, L.~Piemontese,
  G.~Zavattini, R.~Pengo, G.~Ruoso, Extremely long decay time optical cavity,
  Opt. Express 22 (2014) 11570--11577.
\newblock \href {https://doi.org/10.1364/OE.22.011570}
  {\path{doi:10.1364/OE.22.011570}}.

\bibitem{Rosenberg1964}
R.~Rosenberg, C.~B. Rubinstein, D.~R. Herriott, Resonant optical faraday
  rotator, Appl. Opt. 3 (1964) 1079--1083.
\newblock \href {https://doi.org/10.1364/AO.3.001079}
  {\path{doi:10.1364/AO.3.001079}}.

\bibitem{Pace1994}
P.~Pace, Studio e realizzazione di un ellissometro basato su una cavità
  fabry-perot in aria nell’ambito dell’esperimento pvlas, in: Laurea
  Thesis, University of Trieste, Trieste, Italy, 1994, unpublished.

\bibitem{Jacob1995APL}
D.~Jacob, M.~Vallet, F.~Bretenaker, A.~Le~Floch, R.~Le~Naour, Small faraday
  rotation measurement with a fabry–p\'erot cavity, Appl. Phys. Lett. 66
  (1995) 3546--3548.
\newblock \href {https://doi.org/10.1063/1.113811}
  {\path{doi:10.1063/1.113811}}.

\bibitem{Zavattini2006APB}
G.~Zavattini, G.~Cantatore, R.~Cimino, G.~Di~Domenico, F.~Della~Valle,
  M.~Karuza, E.~Milotti, G.~Ruoso, On measuring birefringences and dichroisms
  using fabry–p\'erot cavities, Appl. Phys. B 83 (2006) 571--577.
\newblock \href {https://doi.org/10.1007/s00340-006-2189-y}
  {\path{doi:10.1007/s00340-006-2189-y}}.

\bibitem{DellaValle2016EPJC}
F.~Della~Valle, A.~Ejlli, U.~Gastaldi, G.~Messineo, E.~Milotti, R.~Pengo,
  G.~Ruoso, G.~Zavattini, The pvlas experiment: measuring vacuum magnetic
  birefringence and dichroism with a birefringent fabry-perot cavity, Eur.
  Phys. J. C 76 (2016) 24.

\bibitem{Bouchiat:1982}
M.~A. Bouchiat, L.~Pottier, Light-polarization modifications in a multipass
  cavity, Appl. Phys. B 29 (1982) 43--54.
\newblock \href {https://doi.org/10.1007/BF00694368}
  {\path{doi:10.1007/BF00694368}}.

\bibitem{Carusotto:1989}
S.~Carusotto, E.~Polacco, E.~Iacopini, G.~Stefanini, E.~Zavattini, F.~Scuri,
  The ellipticity introduced by interferential mirrors on a linearly polarized
  light beam orthogonally reflected, Appl. Phys. B 48 (1989) 231--234.
\newblock \href {https://doi.org/10.1007/BF00694350}
  {\path{doi:10.1007/BF00694350}}.

\bibitem{MicossiAPB1993}
P.~Micossi, F.~Della~Valle, E.~Milotti, E.~Zavattini, C.~Rizzo, G.~Ruoso,
  {Measurement of the birefringence properties of the reflecting surface of an
  interferential mirror}, Appl. Phys. B 57 (1993) 95--98.
\newblock \href {https://doi.org/10.1007/BF00425990}
  {\path{doi:10.1007/BF00425990}}.

\bibitem{Brandi1997APB}
F.~Brandi, F.~Della~Valle, A.~De~Riva, P.~Micossi, F.~Perrone, C.~Rizzo,
  G.~Ruoso, G.~Zavattini, Measurement of the phase anisotropy of very high
  reflectivity interferential mirrors, Appl. Phys. B 65 (1997) 351--355.
\newblock \href {https://doi.org/10.1007/s003400050283}
  {\path{doi:10.1007/s003400050283}}.

\bibitem{Pound:1946}
R.~V. Pound, Electronic frequency stabilization of microwave oscillators, Rev.
  Sci. Instrum. 17 (1946) 490--505.
\newblock \href {https://doi.org/10.1063/1.1770414}
  {\path{doi:10.1063/1.1770414}}.

\bibitem{Drever:1983qsr}
R.~W.~P. Drever, J.~L. Hall, F.~V. Kowalski, J.~Hough, G.~M. Ford, A.~J.
  Munley, H.~Ward, {Laser phase and frequency stabilization using an optical
  resonator}, Appl. Phys. B 31 (1983) 97--105.
\newblock \href {https://doi.org/10.1007/BF00702605}
  {\path{doi:10.1007/BF00702605}}.

\bibitem{Ejlli2018}
A.~Ejlli, F.~Della~Valle, G.~Zavattini, Polarisation dynamics of a birefringent
  fabry-perot cavity, Appl. Phys. B 124 (2018) 22.
\newblock \href {https://doi.org/10.1007/s00340-018-6891-3}
  {\path{doi:10.1007/s00340-018-6891-3}}.

\bibitem{Uehara1995}
N.~Uehara, K.~Ueda, Accurate measurement of ultralow loss in a high-finesse
  fabry-perot interferometer using the frequency response functions, Appl.
  Phys. B 61 (1995) 9--15.
\newblock \href {https://doi.org/10.1007/BF01090966}
  {\path{doi:10.1007/BF01090966}}.

\bibitem{Rizzo:1997cm}
C.~Rizzo, A.~Rizzo, D.~M. Bishop, The cotton-mouton effect in gases: Experiment
  and theory, Int. Rev. Phys. Chem. 16 (1997) 81--111.
\newblock \href {https://doi.org/10.1080/014423597230316}
  {\path{doi:10.1080/014423597230316}}.

\bibitem{Bregant2009CM_He}
M.~Bregant, G.~Cantatore, S.~Carusotto, R.~Cimino, F.~Della~Valle,
  G.~Di~Domenico, U.~Gastaldi, M.~Karuza, V.~Lozza, E.~Milotti, E.~Polacco,
  G.~Raiteri, G.~Ruoso, E.~Zavattini, G.~Zavattini, New precise measurement of
  the cotton-mouton effect in helium, Chem. Phys. Lett. 471 (2009) 322--325.
\newblock \href {https://doi.org/10.1016/j.cplett.2009.02.035}
  {\path{doi:10.1016/j.cplett.2009.02.035}}.

\bibitem{PhysRevA.88.043815}
A.~Cad\`ene, D.~Sordes, P.~Berceau, M.~Fouch\'e, R.~Battesti, C.~Rizzo, Faraday
  and cotton-mouton effects of helium at $\lambda=1064$~nm, Phys. Rev. A 88
  (2013) 043815.
\newblock \href {https://doi.org/10.1103/PhysRevA.88.043815}
  {\path{doi:10.1103/PhysRevA.88.043815}}.

\bibitem{DellaValle2014PRD}
F.~Della~Valle, E.~Milotti, A.~Ejlli, G.~Messineo, L.~Piemontese, G.~Zavattini,
  U.~Gastaldi, R.~Pengo, G.~Ruoso, {First results from the new PVLAS apparatus:
  A new limit on vacuum magnetic birefringence}, Phys. Rev. D 90 (2014) 092003.
\newblock \href {https://doi.org/10.1103/PhysRevD.90.092003}
  {\path{doi:10.1103/PhysRevD.90.092003}}.

\bibitem{DellaValle2014CPL}
F.~Della~Valle, A.~Ejlli, U.~Gastaldi, G.~Messineo, E.~Milotti, R.~Pengo,
  L.~Piemontese, G.~Ruoso, G.~Zavattini, {Measurement of the Cotton Mouton
  effect of water vapour}, Chem. Phys. Lett. 592 (2014) 288--291.
\newblock \href {https://doi.org/10.1016/j.cplett.2013.12.049}
  {\path{doi:10.1016/j.cplett.2013.12.049}}.

\bibitem{Bregant2005CM_Ne}
M.~Bregant, G.~Cantatore, S.~Carusotto, R.~Cimino, F.~Della~Valle,
  G.~Di~Domenico, U.~Gastaldi, M.~Karuza, E.~Milotti, E.~Polacco, G.~Ruoso,
  E.~Zavattini, G.~Zavattini, A precise measurement of the cotton-mouton effect
  in neon, Chem. Phys. Lett. 410 (2005) 288--292.
\newblock \href {https://doi.org/10.1016/j.cplett.2005.05.087}
  {\path{doi:10.1016/j.cplett.2005.05.087}}.

\bibitem{Bregant2009Erratum}
M.~Bregant, G.~Cantatore, S.~Carusotto, R.~Cimino, F.~D. Valle, G.~D. Domenico,
  U.~Gastaldi, M.~Karuza, V.~Lozza, E.~Milotti, E.~Polacco, G.~Raiteri,
  G.~Ruoso, E.~Zavattini, G.~Zavattini, Erratum to `measurement of the
  cotton-mouton effect in krypton and xenon at 1064nm with the pvlas apparatus'
  [chem. phys. lett. 392 (2004) 276] and `a precise measurement of the
  cotton-mouton effect in neon' [chem. phys. lett. 410 (2005) 288], Chem. Phys.
  Lett. 477 (2009) 415.
\newblock \href {https://doi.org/10.1016/j.cplett.2009.06.094}
  {\path{doi:10.1016/j.cplett.2009.06.094}}.

\bibitem{MEI2009216}
H.-H. Mei, W.-T. Ni, S.-J. Chen, S.~shi Pan, Measurement of the cotton-mouton
  effect in nitrogen, oxygen, carbon dioxide, argon, and krypton with the q \&
  a apparatus, Chem. Phys. Lett. 471 (2009) 216 -- 221.
\newblock \href {https://doi.org/10.1016/j.cplett.2009.02.048}
  {\path{doi:10.1016/j.cplett.2009.02.048}}.

\bibitem{Brandi1998JOSAB}
F.~Brandi, F.~D. Valle, P.~Micossi, A.~M.~D. Riva, G.~Zavattini, F.~Perrone,
  C.~Rizzo, G.~Ruoso, Cotton-mouton effect of molecular oxygen: a novel
  measurement, J. Opt. Soc. Am. B 15 (1998) 1278--1281.
\newblock \href {https://doi.org/10.1364/JOSAB.15.001278}
  {\path{doi:10.1364/JOSAB.15.001278}}.

\bibitem{Bregant2004CM_KrXe}
M.~Bregant, G.~Cantatore, S.~Carusotto, R.~Cimino, F.~Della~Valle,
  G.~Di~Domenico, U.~Gastaldi, M.~Karuza, E.~Milotti, E.~Polacco, G.~Ruoso,
  E.~Zavattini, G.~Zavattini, Measurement of the cotton-mouton effect in
  krypton and xenon at 1064~nm with the pvlas apparatus, Chem. Phys. Lett. 392
  (2004) 276--280.
\newblock \href {https://doi.org/10.1016/j.cplett.2004.05.064}
  {\path{doi:10.1016/j.cplett.2004.05.064}}.

\bibitem{Cadene2015JCP}
A.~Cad\`ene, M.~Fouch\'e, A.~Riv\`ere, R.~Battesti, S.~Coriani, A.~Rizzo,
  C.~Rizzo, Circular and linear magnetic birefringences in xenon at $\lambda =
  1064$~nm, J. Chem. Phys. 142 (2015) 124313.
\newblock \href {https://doi.org/10.1063/1.4916049}
  {\path{doi:10.1063/1.4916049}}.

\bibitem{KAGRA}
T.~Akutsu, et~al., Kagra: 2.5 generation interferometric gravitational wave
  detector, Nature Astron. 3 (2019) 35--40.
\newblock \href {https://doi.org/10.1038/s41550-018-0658-y}
  {\path{doi:10.1038/s41550-018-0658-y}}.

\bibitem{LIGO}
B.~Abbott, et~al., Prospects for observing and localizing gravitational-wave
  transients with advanced ligo, advanced virgo and kagra, Living Rev. Relativ.
  21 (2018) 3.
\newblock \href {https://doi.org/10.1007/s41114-018-0012-9}
  {\path{doi:10.1007/s41114-018-0012-9}}.

\bibitem{GWOpenscience}
G.~W. O.~S. Center,
  \href{https://www.gw-openscience.org/detector_status/day/20200226/}{Geo-ligo-virgo
  gravitational-wave strain of 26 february 2020}.
\newline\urlprefix\url{https://www.gw-openscience.org/detector_status/day/20200226/}

\bibitem{GrassiStrini:1979kp}
A.~M. Grassi~Strini, G.~Strini, G.~Tagliaferri, {Testability of nonlinear
  electrodynamics}, Phys. Rev. D 19 (1979) 2330--2335.
\newblock \href {https://doi.org/10.1103/PhysRevD.19.2330}
  {\path{doi:10.1103/PhysRevD.19.2330}}.

\bibitem{Calloni}
G.~Zavattini, E.~Calloni, Probing for new physics and detecting non-linear
  vacuum qed effects using gravitational wave interferometer antennas, Eur.
  Phys. J. C 62 (2009) 459--466.
\newblock \href {https://doi.org/10.1140/epjc/s10052-009-1079-y}
  {\path{doi:10.1140/epjc/s10052-009-1079-y}}.

\bibitem{Grote}
H.~Grote, {On the possibility of vacuum QED measurements with gravitational
  wave detectors}, Phys. Rev. D 91 (2015) 022002.
\newblock \href {https://doi.org/10.1103/PhysRevD.91.022002}
  {\path{doi:10.1103/PhysRevD.91.022002}}.

\bibitem{Iacopini:2301707}
E.~Iacopini, P.~Lazeyras, M.~Morpuirgo, E.~Picasso, B.~Smith, E.~Zavattini,
  E.~Polacco, \href{https://cds.cern.ch/record/2301707}{{Experimental
  determination of vacuum polarization effects on a laser light-beam
  propagating in a strong magnetic field}}, Tech. Rep. Proposal D2, CERN,
  Geneva (1980).
\newline\urlprefix\url{https://cds.cern.ch/record/2301707}

\bibitem{Ni:1991iu}
W.-T. Ni, S.-C. Chen, S.-K. King, S.-S. Pan, K.~Tsubono, N.~Mio, K.~Narihara,
  {Test of quantum electrodynamics using ultrahigh sensitive interferometers},
  Mod. Phys. Lett. A 6 (1991) 3671--3678.
\newblock \href {https://doi.org/10.1142/S0217732391004243}
  {\path{doi:10.1142/S0217732391004243}}.

\bibitem{BAKALOV1994}
D.~Bakalov, G.~Cantatore, G.~Carugno, S.~Carusotto, P.~Favaron, F.~D. Valle,
  I.~Gabrielli, U.~Gastaldi, E.~Iacopini, P.~Micossi, E.~Milotti, R.~Onofrio,
  R.~Pengo, F.~Perrone, G.~Petrucci, E.~Polacco, C.~Rizzo, G.~Ruoso,
  E.~Zavattini, G.~Zavattini, Pvlas: Vacuum birefringence and production and
  detection of nearly massless, weakly coupled particles by optical
  techniques., in: C.~Arpesella, E.~Bellotti, A.~Bottino (Eds.), TAUP 93.
  Proceedings of the 3rd International Workshop on Theoretical and
  Phenomenological Aspects of Underground Physics, Assergi, Gran Sasso, Italy,
  19-23 September 1993, Vol.~35, 1994, pp. 180 -- 182, nucl. Phys. B (Proc.
  Suppl.) 35 (1994),.
\newblock \href {https://doi.org/10.1016/0920-5632(94)90243-7}
  {\path{doi:10.1016/0920-5632(94)90243-7}}.

\bibitem{Lee:1995wp}
S.~A. Lee, W.~M. Fairbank, Jr., W.~H. Toki, J.~L. Hall, T.~S. Jaffery,
  P.~Colestock, V.~Cupps, H.~Kautzky, M.~Kuchnir, F.~Nezrick, {Measurement of
  the magnetically induced QED birefringence of the vacuum and an improved
  laboratory search for axions: Proposal}, Tech. Rep. FERMILAB-PROPOSAL-0877
  (1995).

\bibitem{Ni:1996dg}
W.-T. Ni, {Cryogenic test and interferometric test of QED and search for axions
  and anomalous spin-dependent interactions}, in: {Proceedings of the 1995
  Taiwan International Conference on Intermediate and High Energy Physics,
  Dubna, Russia, 26-28 June 1995}, Vol.~34, pp. 962--967, chin. J. Phys. 34
  (1996).

\bibitem{Pengo1998QED}
R.~Pengo, et~al., Magnetic birefringence of vacuum: the pvlas experiment, in:
  E.~Zavattini, D.~Bakalov, C.~Rizzo (Eds.), Frontier Tests of QED and Physics
  of the Vacuum, 9-15 June 1998, Sandanski, Bulgaria, Heron Press, Sofia, 1998.

\bibitem{Battesti2008}
R.~Battesti, B.~Pinto Da~Souza, S.~Batut, C.~Robilliard, G.~Bailly, C.~Michel,
  M.~Nardone, L.~Pinard, O.~Portugall, G.~Tr{\'e}nec, J.-M. Mackowski, G.~L.
  Rikken, J.~Vigu{\'e}, C.~Rizzo, The bmv experiment: a novel apparatus to
  study the propagation of light in a transverse magnetic field, Eur. Phys. J.
  D 46 (2008) 323--333.
\newblock \href {https://doi.org/10.1140/epjd/e2007-00306-3}
  {\path{doi:10.1140/epjd/e2007-00306-3}}.

\bibitem{Iacopini:1980ws}
E.~Iacopini, B.~Smith, G.~Stefanini, E.~Zavattini, {On a Sensitive Ellipsometer
  to Detect the Vacuum Polarization Induced by a Magnetic Field}, Nuovo Cim. 61
  B (1981) 21--37.
\newblock \href {https://doi.org/10.1007/BF02721700}
  {\path{doi:10.1007/BF02721700}}.

\bibitem{Carusotto:2301709}
S.~Carusotto, F.~Scuri, F.~Stefanini, E.~Iacopini, P.~Lazeyras, M.~Morpuirgo,
  E.~Picasso, B.~Smith, E.~Zavattini, E.~Polacco,
  \href{https://cds.cern.ch/record/2301709}{{Addendum to the Proposal D2:
  Experimental determination of vacuum polarization effects on a laser
  light-beam propagating in a strong magnetic field}}, Tech. Rep. Proposal
  D2-Add, CERN, Geneva (1983).
\newline\urlprefix\url{https://cds.cern.ch/record/2301709}

\bibitem{Bakalov1998QSO}
D.~Bakalov, et~al., {Experimental method to detect the magnetic birefringence
  of vacuum}, Quantum Semiclass. Opt.: J. Eur. Opt. Soc. B 10 (1998) 239--250.
\newblock \href {https://doi.org/10.1088/1355-5111/10/1/027}
  {\path{doi:10.1088/1355-5111/10/1/027}}.

\bibitem{Cantatore1995RSI}
G.~Cantatore, F.~Della~Valle, E.~Milotti, P.~Pace, E.~Zavattini, E.~Polacco,
  F.~Perrone, C.~Rizzo, G.~Zavattini, G.~Ruoso, Frequency locking of a nd:yag
  laser using the laser itself as the optical phase modulator, Rev. Sci.
  Instrum. 66 (1995) 2785--2787.
\newblock \href {https://doi.org/10.1063/1.1145555}
  {\path{doi:10.1063/1.1145555}}.

\bibitem{DeRiva1996RSI}
A.~M. De~Riva, G.~Zavattini, S.~Marigo, C.~Rizzo, G.~Ruoso, G.~Carugno,
  R.~Onofrio, S.~Carusotto, M.~Papa, F.~Perrone, E.~Polacco, G.~Cantatore,
  F.~Della~Valle, P.~Micossi, E.~Milotti, P.~Pace, E.~Zavattini, Very high $q$
  frequency-locked fabry-perot cavity, Rev. Sci. Instrum. 67 (1996) 2680--2684.
\newblock \href {https://doi.org/10.1063/1.1147094}
  {\path{doi:10.1063/1.1147094}}.

\bibitem{Zavattini2012IJMPA}
G.~Zavattini, U.~Gastaldi, R.~Pengo, G.~Ruoso, F.~Della~Valle, E.~Milotti,
  Measuring the magnetic birefringence of vacuum: the pvlas experiment, in:
  M.~Asorey, M.~Bordag, E.~Elizalde (Eds.), Selected Papers from the 10th
  Conference on Quantum Field Theory Under the Influence of External Conditions
  (QFEXT11) Spain, 18-24 September, 2011, Vol.~27, 2012, p. 1260017, int. J.
  Mod. Phys. A 27 (2012),.
\newblock \href {https://doi.org/10.1142/S0217751X12600172}
  {\path{doi:10.1142/S0217751X12600172}}.

\bibitem{Zavattini2013JPCS}
G.~Zavattini, F.~D. Valle, U.~Gastaldi, G.~Messineo, E.~Milotti, R.~Pengo,
  L.~Piemontese, G.~Ruoso, The pvlas experiment: detecting vacuum magnetic
  birefringence, in: L.~Di\'osi, H.-T. Elze, L.~Fronzoni, J.~Halliwell,
  E.~Prati, G.~Vitiello, J.~Yearsley (Eds.), 6th International Workshop
  DICE2012 Spacetime – Matter – Quantum Mechanics: From the Planck Scale to
  Emergent Phenomena 17–21 September 2012, Castiglioncello (Tuscany), Italy,
  Vol. 442, 2013, p. 012057, j. Phys.: Conf. Series 442 (2013),.
\newblock \href {https://doi.org/10.1088/1742-6596/442/1/012057}
  {\path{doi:10.1088/1742-6596/442/1/012057}}.

\bibitem{DellaValle:2013dwa}
F.~Della~Valle, G.~Di~Domenico, U.~Gastaldi, E.~Milotti, G.~Messineo, R.~Pengo,
  L.~Piemontese, G.~Ruoso, G.~Zavattini, {The new PVLAS apparatus for detection
  of magnetic birefringence of vacuum}, in: F.~Cervelli, G.~Chiarelli,
  F.~Forti, M.~Grassi, A.~Scribano (Eds.), {Proceedings of the 12th Pisa
  Meeting on Advanced Detectors, La Biodola, Isola d'Elba, Italy, 20-26 May
  2012}, Vol. 718, 2013, pp. 495--496, nucl. Instrum. Meth. Phys. Res. A 718
  (2013),.
\newblock \href {https://doi.org/10.1016/j.nima.2012.11.084}
  {\path{doi:10.1016/j.nima.2012.11.084}}.

\bibitem{DellaValle2013NJP}
F.~Della~Valle, U.~Gastaldi, G.~Messineo, E.~Milotti, R.~Pengo, L.~Piemontese,
  G.~Ruoso, G.~Zavattini, {Measurements of vacuum magnetic birefringence using
  permanent dipole magnets: the PVLAS experiment}, New J. Phys. 15 (2013)
  053026.
\newblock \href {https://doi.org/10.1088/1367-2630/15/5/053026}
  {\path{doi:10.1088/1367-2630/15/5/053026}}.

\bibitem{Carusotto:136650}
S.~Carusotto, E.~Polacco, E.~Iacopini, G.~Stefanini, E.~Zavattini, {Measurement
  of the magnetic birefringence in oxygen and nitrogen gases}, Opt. Commun. 42
  (1982) 104--108.

\bibitem{Carusotto:1983tt}
S.~Carusotto, E.~Iacopini, E.~Polacco, F.~Scuri, G.~Stefanini, E.~Zavattini,
  {Measurement of the magnetic birefringence of noble gases}, J. Opt. Soc. Am.
  1 (1984) 635--640.
\newblock \href {https://doi.org/10.1364/JOSAB.1.000635}
  {\path{doi:10.1364/JOSAB.1.000635}}.

\bibitem{Morpurgo:1720288}
\href{https://cds.cern.ch/record/1720288}{Superconducting magnet record}, {CERN
  Bulletin} 28 (1982) 1.
\newline\urlprefix\url{https://cds.cern.ch/record/1720288}

\bibitem{Pengo:1999ud}
R.~Pengo, G.~Petrucci, S.~Marigo, R.~G. Scurlock, J.~H.~P. Watson, {An original
  rotating cryostat for the experiment PVLAS}, in: D.~Dew-Hughes, R.~G.
  Scurlock, J.~H.~P. Watson (Eds.), {Proceedings of the 17th International
  Cryogenic Engineering Conference (ICEC 17): Bournemouth, UK, 14-17 Jul 1998},
  IOP, Bristol, UK, 1998, pp. 851--854.

\bibitem{Claudet1974}
G.~Claudet, A.~Lacaze, P.~Roubeau, J.~Verdier, {The design and operation of a
  refrigerator system using superfluid helium}, in: K.~Mendelssohn (Ed.),
  {Proceedings of the 5th International Cryogenic Engineering Conference (ICEC
  5): Kyoto, Japan, 7-10 May 1974}, IPC Science and Technology Press,
  Guildford, Surrey, UK, 1974, pp. 265--267.

\bibitem{Multi1999}
\href{https://ec.staubli.com/AcroFiles/Catalogues/News/Archiv/SZ_News-1999_(en)_hi.pdf}{Multilams
  first choice in scientific experiments}, {MC$^{\circledR}$/HCK$^{\circledR}$
  News} March (1999) 13.
\newline\urlprefix\url{https://ec.staubli.com/AcroFiles/Catalogues/News/Archiv/SZ_News-1999_(en)_hi.pdf}

\bibitem{Bregant2002RSI}
M.~Bregant, G.~Cantatore, F.~Della~Valle, G.~Ruoso, G.~Zavattini, Frequency
  locking to a high-finesse fabry-perot cavity of a frequency doubled nd:yag
  laser used as the optical phase modulator, Rev. of Sci. Instrum. 73 (2002)
  4142--4144.
\newblock \href {https://doi.org/10.1063/1.1519933}
  {\path{doi:10.1063/1.1519933}}.

\bibitem{IDM5}
E.~Zavattini, et~al., {Axion-like Dark Matter candidate particles}, in:
  N.~J.~C. Spooner, V.~Kundryavtsev (Eds.), {Proceedings of the Fifth
  International Workshop, Edinburgh, UK, 6–10 September 2004}, World
  Scientific, Singapore, 2005, pp. 420--425.
\newblock \href {https://doi.org/10.1142/9789812701848_0063}
  {\path{doi:10.1142/9789812701848_0063}}.

\bibitem{Nove2005}
E.~Zavattini, et~al., {Experimental observation of magnetically induced linear
  dichroism of vacuum}, in: M.~Baldo-Ceolin (Ed.), {XI International Workshop
  on Neutrino Telescopes, 22-25 February 2005, Venice, Italy}, Papergraf,
  Padova, Italy, 2005, pp. 433--449.

\bibitem{ZAVATTINI2007NPBPS}
E.~Zavattini, G.~Zavattini, G.~Ruoso, E.~Polacco, E.~Milotti, M.~Karuza,
  U.~Gastaldi, G.~D. Domenico, F.~D. Valle, R.~Cimino, S.~Carusotto,
  G.~Cantatore, M.~Bregant, Pvlas: probing vacuum with polarized light, in:
  S.~Narison (Ed.), Proceedings of the 12th High-Energy Physics International
  Conference on Quantum ChromoDynamics (QCD 05), 4-8 July 2005, Montpellier,
  France, Vol. 164, 2007, pp. 264 -- 269, nucl. Phys. B (Proc. Suppl.) 164
  (2007),.
\newblock \href {https://doi.org/10.1016/j.nuclphysbps.2006.11.096}
  {\path{doi:10.1016/j.nuclphysbps.2006.11.096}}.

\bibitem{Zavattini:2006zz}
E.~Zavattini, et~al., {Observation of vacuum birefringence induced by a
  transverse magnetic field}, in: A.~Sissakian, G.~Kozlov, E.~Kolganova (Eds.),
  {High Energy Physics ICHEP'06. Proceedings of the 33rd International
  Conference, Moscow, Russia, 26 July - 2 August 2006}, World Scientific,
  Singapore, 2007, pp. 218--221.
\newblock \href {https://doi.org/10.1142/9789812790873_0024}
  {\path{doi:10.1142/9789812790873_0024}}.

\bibitem{Zavattini2006PRL}
E.~Zavattini, G.~Zavattini, G.~Ruoso, E.~Polacco, E.~Milotti, M.~Karuza,
  U.~Gastaldi, G.~Di~Domenico, F.~Della~Valle, R.~Cimino, S.~Carusotto,
  G.~Cantatore, M.~Bregant, Experimental observation of optical rotation
  generated in vacuum by a magnetic field, Phys. Rev. Lett. 96 (2006) 110406.
\newblock \href {https://doi.org/10.1103/PhysRevLett.96.110406}
  {\path{doi:10.1103/PhysRevLett.96.110406}}.

\bibitem{Zavattini9}
G.~Zavattini, et~al., {PVLAS}, in: M.~Baldo-Ceolin (Ed.), {XII International
  Workshop on Neutrino Telescopes, 6-9 March 2007, Venice, Italy}, Papergraf,
  Padova, Italy, 2007, pp. 373--395.

\bibitem{PRL_Erratum}
E.~Zavattini, G.~Zavattini, G.~Ruoso, E.~Polacco, E.~Milotti, M.~Karuza,
  U.~Gastaldi, G.~Di~Domenico, F.~Della~Valle, R.~Cimino, S.~Carusotto,
  G.~Cantatore, M.~Bregant, Editorial note: Experimental observation of optical
  rotation generated in vacuum by a magnetic field [phys. rev. lett. 96, 110406
  (2006)], Phys. Rev. Lett. 99 (2007) 129901.
\newblock \href {https://doi.org/10.1103/PhysRevLett.99.129901}
  {\path{doi:10.1103/PhysRevLett.99.129901}}.

\bibitem{HALBACH19801}
K.~Halbach, Design of permanent multipole magnets with oriented rare earth
  cobalt material, Nucl. Instrum. Meth. 169 (1980) 1 -- 10.
\newblock \href {https://doi.org/10.1016/0029-554X(80)90094-4}
  {\path{doi:10.1016/0029-554X(80)90094-4}}.

\bibitem{Kogelnik:66}
H.~Kogelnik, T.~Li, Laser beams and resonators, Appl. Opt. 5 (1966) 1550--1567.
\newblock \href {https://doi.org/10.1364/AO.5.001550}
  {\path{doi:10.1364/AO.5.001550}}.

\bibitem{Iacopini:1983rp}
E.~Iacopini, G.~Stefanini, E.~Zavattini, {Effects of a Magnetic Field on the
  Optical Properties of Dielectric Mirrors}, Appl. Phys. A 32 (1983) 63--67.
\newblock \href {https://doi.org/10.1007/BF00617830}
  {\path{doi:10.1007/BF00617830}}.

\bibitem{Virgo2012}
V.~Collaboration, {Advanced Virgo Technical Design Report} VIR–0128A–12
  (2012).

\bibitem{LNL2015}
G.~P. Buso, F.~Della~Valle, A.~Ejlli, U.~Gastaldi, S.~Marigo, E.~Milotti,
  R.~Pengo, G.~Ruoso, F.~Spizzo, G.~Zavattini, {Magnetic Strain in the PVLAS
  Experiment and Spurious Signals}, in: {LNL Annual Report 2015}, 2016, pp.
  176--177.

\bibitem{DellaValle:2010oex}
F.~Della~Valle, G.~Di~Domenico, U.~Gastaldi, E.~Milotti, R.~Pengo, G.~Ruoso,
  G.~Zavattini, {Towards a direct measurement of vacuum magnetic birefringence:
  PVLAS achievements}, Opt. Commun. 283 (2010) 4194--4198.
\newblock \href {https://doi.org/10.1016/j.optcom.2010.06.065}
  {\path{doi:10.1016/j.optcom.2010.06.065}}.

\bibitem{IntrinsicNoise}
G.~Zavattini, F.~Della~Valle, A.~Ejlli, W.-T. Ni, U.~Gastaldi, E.~Milotti,
  R.~Pengo, G.~Ruoso, Intrinsic mirror noise in fabry-perot based polarimeters:
  the case for the measurement of vacuum magnetic birefringence, Eur. Phys. J.
  C 78 (2018).
\newblock \href {https://doi.org/10.1140/epjc/s10052-018-6063-y}
  {\path{doi:10.1140/epjc/s10052-018-6063-y}}.

\bibitem{BRAGINSKY1999}
V.~B. Braginsky, M.~L. Gorodetsky, S.~P. Vyatchanin, Thermodynamical
  fluctuations and photo-thermal shot noise in gravitational wave antennae,
  Physics Letters A 264 (1999) 1 -- 10.
\newblock \href {https://doi.org/10.1016/S0375-9601(99)00785-9}
  {\path{doi:10.1016/S0375-9601(99)00785-9}}.

\bibitem{Braginsky2003}
V.~B. Braginsky, S.~P. Vyatchanin, Thermodynamical fluctuations in optical
  mirror coatings, Physics Letters A 312 (2003) 244--255.
\newblock \href {https://doi.org/10.1016/S0375-9601(03)00473-0}
  {\path{doi:10.1016/S0375-9601(03)00473-0}}.

\bibitem{Welsch1999}
E.~Welsch, K.~Ettrich, D.~Ristau, U.~Willamowski, Absolute measurement of
  thermophysical and optical thin-film properties by photothermal methods for
  the investigation of laser damage, Int. J. Thermophys. 20 (1999) 965--976.
\newblock \href {https://doi.org/10.1023/A:1022603823628}
  {\path{doi:10.1023/A:1022603823628}}.

\bibitem{Braginsky:2000wc}
V.~B. Braginsky, M.~L. Gorodetsky, S.~P. Vyatchanin, {Thermorefractive noise in
  gravitational wave antennae}, Phys. Lett. A 271 (2000) 303--307.
\newblock \href {https://doi.org/10.1016/S0375-9601(00)00389-3}
  {\path{doi:10.1016/S0375-9601(00)00389-3}}.

\bibitem{Tei-Chen2007}
T.-C. Chen, et~al., Determination of stress-optical and thermal-optical
  coefficients of nb$_2$o$_5$ thin film material, J. Appl. Phys. 101 (2007)
  043513.
\newblock \href {https://doi.org/10.1063/1.2435796}
  {\path{doi:10.1063/1.2435796}}.

\bibitem{LNL2016}
F.~D. Valle, A.~Ejlli, F.~Evangelisti, A.~Friso, G.Galeazzi, U.~Gastaldi,
  S.~Marigo, E.~Milotti, R.~Pengo, G.~Ruoso, G.~Zavattini, {Tests for the
  radiative cooling of mirrors for the PVLAS experiment}, in: {LNL Annual
  Report 2016}, 2017, pp. 154--155.

\bibitem{Zavattini:2016sqz}
G.~Zavattini, F.~Della~Valle, A.~Ejlli, G.~Ruoso, {A polarisation modulation
  scheme for measuring vacuum magnetic birefringence with static fields}, Eur.
  Phys. J. C 76 (2016) 294, [Erratum: Eur. Phys. J. C 77, (2017), p. 873].
\newblock \href {https://doi.org/10.1140/epjc/s10052-016-4139-0,
  10.1140/epjc/s10052-017-5448-7} {\path{doi:10.1140/epjc/s10052-016-4139-0,
  10.1140/epjc/s10052-017-5448-7}}.

\bibitem{CAST}
V.~Anastassopoulos, et~al., New cast limit on the axion–photon interaction,
  Nature Phys. 13 (2017) 584--590.
\newblock \href {https://doi.org/10.1038/nphys4109}
  {\path{doi:10.1038/nphys4109}}.

\bibitem{LOI2018}
R.~Ballou, F.~Della~Valle, A.~Ejlli, U.~Gastaldi, H.~Grote, v.~S. Kunc,
  K.~Meissner, E.~Milotti, W.-T. Ni, S.-s. Pan, R.~Pengo, P.~Pugnat, G.~Ruoso,
  A.~Siemko, M.~\v~Sulc, G.~Zavattini,
  \href{https://cds.cern.ch/record/2649744}{{Letter of Intent to measure Vacuum
  Magnetic Birefringence: the VMB@CERN experiment}}, Tech. Rep.
  CERN-SPSC-2018-036. SPSC-I-249, CERN, Geneva (2018).
\newline\urlprefix\url{https://cds.cern.ch/record/2649744}

\end{thebibliography}

\end{document}